\pdfoutput=1
\documentclass[useAMS]{mn2e}
\usepackage[tight]{subfigure}
\usepackage{longtable}
\usepackage{lscape}
\usepackage{graphicx}
\usepackage{amssymb}
\usepackage{url}

\title[Stellar population gradients in BCGs]{Stellar population gradients in brightest cluster galaxies}
\author[Loubser $\&$ S\'{a}nchez-Bl\'{a}zquez]{S. I. Loubser$^{1}$\thanks{E-mail:
Ilani.Loubser@nwu.ac.za (SIL)}, P. S\'{a}nchez-Bl\'{a}zquez$^{2}$\\
$^{1}$Centre for Space Research, North-West University, Potchefstroom 2520, South Africa\\
$^{2}$Departamento de F\'{\i}sica Te\'orica, M\'odulo C15, Universidad Aut\'onoma de Madrid, E28049, Cantoblanco, Spain}
\begin{document}


\pagerange{\pageref{firstpage}--\pageref{lastpage}} \pubyear{2011}

\maketitle

\label{firstpage}

\begin{abstract}
We present the stellar population and velocity dispersion gradients for a sample of 24 brightest cluster galaxies (BCGs) in the nearby Universe for which we have obtained 
high quality long-slit spectra at the Gemini telescopes. With the aim of studying the possible connection between the formation of the BCGs and their host clusters, we explore the relations between the stellar population gradients and properties of the host clusters as well as the possible connections between the stellar population gradients and other properties of the galaxies. We find mean stellar population gradients (negative $\Delta$[Z/H]/$log\ r$ gradient of --0.285$\pm$0.064; small positive $\Delta$log (age)/$log\ r$ gradient of 0.069$\pm$0.049; and null $\Delta$[E/Fe]/$log\ r$ gradient of --0.008$\pm$0.032) that are consistent with those of normal massive elliptical galaxies. However, we find a trend between metallicity gradients and velocity dispersion (with a negative slope of --1.616$\pm$0.539) that is not found for the most massive ellipticals. Furthermore, we find trends between the metallicity gradients and $K$-band luminosities (with a slope of 0.173$\pm$0.081) as well as the distance from the BCG to the X-ray peak of the host cluster (with a slope of --7.546$\pm$2.752). The latter indicates a possible relation between the formation of the cluster and that of the central galaxy. 
\end{abstract}

\begin{keywords}
galaxies: evolution -- galaxies: elliptical and lenticular, cD -- galaxies: stellar content
\end{keywords}

\section{Introduction}

The formation of the supermassive galaxies at the centres of galaxy clusters, called brightest cluster galaxies (BCGs), is still one of the most challenging problems in galaxy formation studies. The  properties of these galaxies can be a result of their special location at the bottom of the cluster potential well, or because BCGs occupy the massive end of the galaxy luminosity function (i.e.\ the properties can be driven by environment or mass). The comparison of the properties of BCGs with other ellipticals and with the properties of the cluster they reside in can help us elucidate which has the bigger influence on the evolution of these systems.

While BCGs morphologically resemble elliptical galaxies, their central surface brightnesses tend to be lower, and they follow a steeper Faber-Jackson relation (Faber $\&$ Jackson 1976) than ordinary ellipticals (see, e.g. Von der Linden et al.\ 2007). They also have a luminosity function that differs from the usual Schechter (1976) function that holds for normal cluster members (e.g. Hansen et al.\ 2005). The stars in BCGs have similar ages and metallicities than non-BCGs of the same mass, although some studies hint towards higher [$\alpha/Fe$] measurements in BCGs (von der Linden et al. 2007; Loubser et al. 2009).

Hierarchical models of galaxy formation predict that BCGs form through the (mainly late) assembling of small galaxies, and that the formation history of the BCG is closely linked to
that of the host cluster (e.g., De Lucia \& Blaizot 2007). Some observational results support this view: a significant alignment between the elongations of BCGs and their host clusters is observed in both the optical (Carter $\&$ Metcalfe 1980; Struble 1990; Plionis et al.\ 2003) and X-ray bands (Hashimoto, Henry $\&$ Boehringer 2008), and correlations between the BCG luminosity and cluster properties (for example X-ray temperature; Edge $\&$ Stewart 1991) have been reported. There is also a (weak) correlation between BCG mass and the mass of their host clusters which does not change significantly with redshift out to $z\simeq0.8$ (Edge 1991; Collins $\&$ Mann 1998; Burke, Collins $\&$ Mann 2000; Brough et 
al.\ 2007; Stott et al.\ 2008; Whiley et al.\ 2008). Furthermore, an increase of BCG mass and size from z $\sim$ 0.7 to $z \sim 0.04$ by a factor of 2 and 4, respectively, has been found (e.g. Bernardi 2009; Valentinuzzi et al.\ 2010), although compare with Whiley et al.\ (2008) and Collins et al.\ (2009) who did not find an increase in mass since z $\sim$ 1. However, it should be noted that the measurement of the scale size of BCGs is notoriously difficult (Lauer et al.\ 2007; Stott et al.\ 2011 and references therein).

Because the models predict that the majority of mergers happen at recent times, the gas content of the accreted galaxies is believed to be 
low (e.g.\ Dubinski 1998; Conroy, Wechsler $\&$ Kravtsov 2007; De Lucia $\&$ Blaizot 2007) and these mergers would therefore not change the central ages and metallicities of the BCGs. This would explain the lack of large differences between the stellar populations of BCGs and normal galaxies. However, these merger or accretion events are expected to change stellar population gradients, as dry minor mergers would deposit metal poor stars outwards (Kawata et al. 2006). 

Stellar population gradients in ordinary early-type galaxies have been studied using line-strength 
indices by various authors (Gorgas, Efstathiou $\&$ Arag\'{o}n-Salamanca 1990; Gonz\'{a}lez 1993; 
Fisher, Illingworth $\&$ Franx 1995a; Fisher, Franx $\&$ Illingworth 1995b; Cardiel, Gorgas $\&$ Arag\'{o}n-Salamanca 1998a; 
Mehlert et al.\ 2003; S\'{a}nchez-Bl\'{a}zquez, Gorgas $\&$ Cardiel 2006b; S\'{a}nchez-Bl\'{a}zquez 
et al.\ 2007; and many more). A few studies have investigated the stellar population gradients in small 
samples of BCGs (Gorgas et al. 1990; Davidge $\&$ Grinder 1995 ; Fisher et al.\ 1995b; Carter, 
Bridges $\&$ Hau 1999; Mehlert et al.\ 2003; S\'anchez-Bl\'azquez et al.\ 2006b; Brough et al.\ 2007, 
Spolaor et al.\ 2009). Most of them either found no difference between the stellar population gradients 
of BCGs and those of the ordinary ellipticals (Mehlert et al.\ 2003), or slightly flatter metallicity 
gradients for the latter (e.g.\ Gorgas et al.\ 1990).

However, a systematic study of the relations between the stellar population gradients and other 
properties of the cluster where they reside -- e.g.\ the richness (parametrised with the X-ray 
luminosity) or the cluster mass -- has not yet been carried out. This study presents a sample that 
is greater, in number, to all previous studies dealing with stellar population gradients in BCGs. 
Furthermore, our BCG sample has the advantage of having detailed X-ray values of the host cluster 
properties available from the literature, and significantly improved signal-to-noise (S/N) compared to previous samples (Crawford et al.\ 1999, and references therein).

This paper is part of a series of papers investigating an overall sample of 49 BCGs in the nearby 
Universe for which we have obtained high S/N ratio, long-slit spectra on the Gemini and WHT telescopes. 
The spatially resolved kinematics, central Single Stellar Population (SSP)-parameters, and the UV-upturn 
of the BCGs were presented in Loubser et al.\ 2008, Loubser et al.\ 2009, and Loubser $\&$ 
S\'{a}nchez-Bl\'{a}zquez 2011a (hereafter Paper 1, 2 and 3). The Mg$_{2}$ gradients of a sub-sample 
of BCGs with sufficient S/N ratios were presented in Loubser $\&$ S\'{a}nchez-Bl\'{a}zquez 2011b 
(Paper 4). 

Here, we explore the possible correlation between the stellar population gradients and other 
properties of the clusters with the aim of clarifying if the galaxy gradients are shaped by the cluster 
assembly process. We fit SSP models, and present the stellar population gradients and their correlations 
with other properties, for a sub-sample of the BCGs with sufficient S/N ratios. We review the sample 
and data reduction, and discuss the data analysis in Section \ref{data}. The index and SSP gradients are 
derived and discussed in Sections \ref{indices} and \ref{SSP}, respectively. We then correlate the 
findings with the host cluster properties in Section \ref{cluster}, and conclude our findings in 
Section \ref{conclusions}.

\section{Sample and data reduction}
\label{data}

The sample selection, spectroscopic data, and the reduction procedures were presented in Papers 1 and 2, and will not be repeated here. The procedure to derive the gradients from 
spatially binned spectra was presented in Paper 4, but will be briefly reviewed here for completeness. The galaxy spectra of the entire sample of 49 galaxies were binned in the spatial direction to a minimum S/N of 40 per \AA{} in the H$\beta$ region of the spectrum, ensuring a maximum error of approximately 12 per cent on the measurement of this index. Twenty-six galaxies have four or more spatial bins when the central 0.5 arcsec, to each side, are excluded (thus in total the central 1.0 arcsec was excluded, comparable to the seeing). Those galaxies that did not meet this criterion were eliminated from the final sample. Thus, the gradients were investigated for 24 galaxies (2 were excluded because of emission contamination -- see below).

This subsample comprises of the dominant galaxies closest to the X-ray peaks in the centres of clusters and, for consistency, we call these galaxies BCGs to comply with recent literature (for example De Lucia $\&$ Blaizot 2007; Von der Linden et al.\ 2007; Brough et al.\ 2007)\footnote{According to the above definition, for a small fraction of clusters the BCG might not strictly be the brightest galaxy in the cluster.}. 

\begin{figure}
   \centering
\includegraphics[scale=0.35,angle=-90]{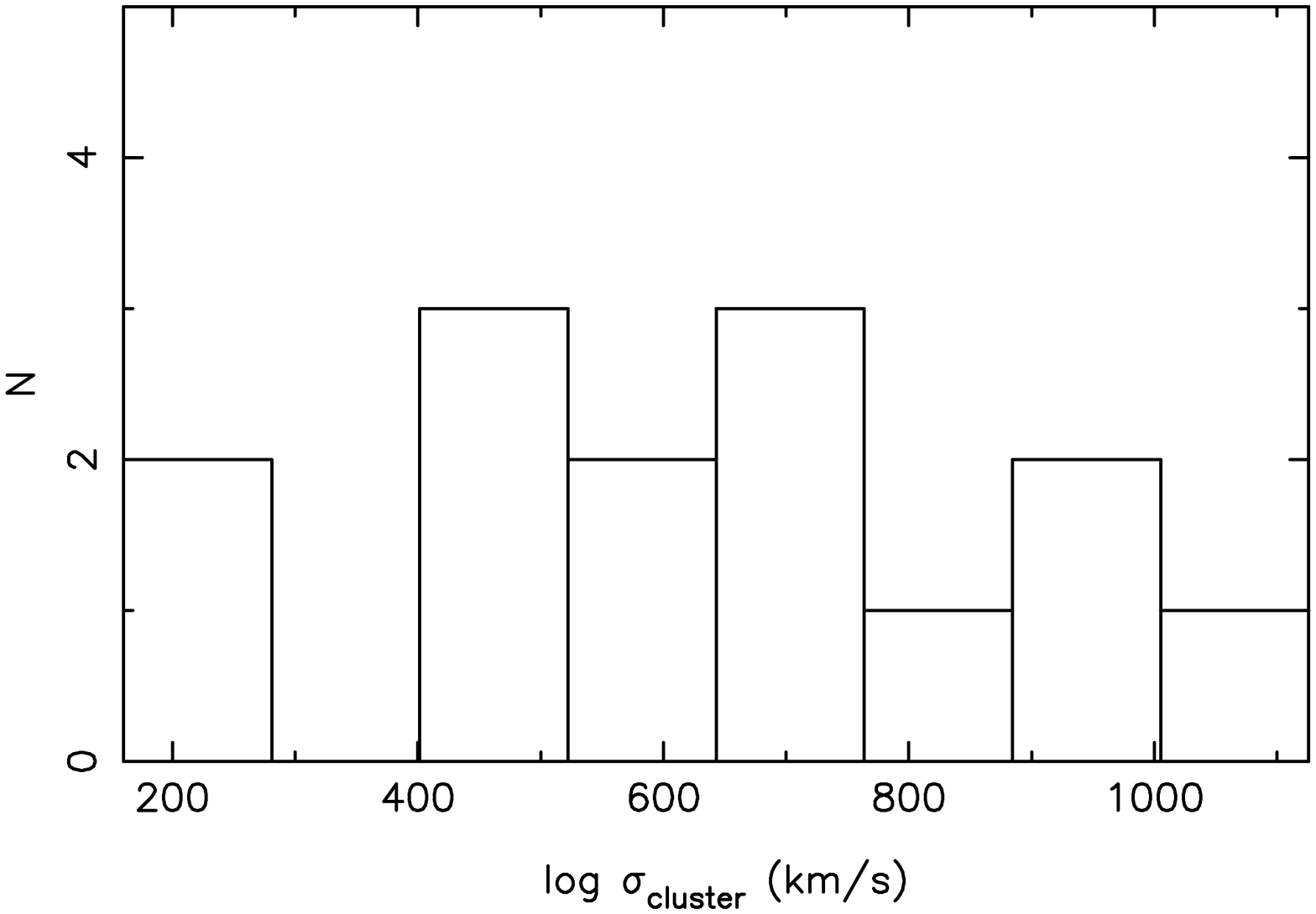}
\includegraphics[scale=0.35,angle=-90]{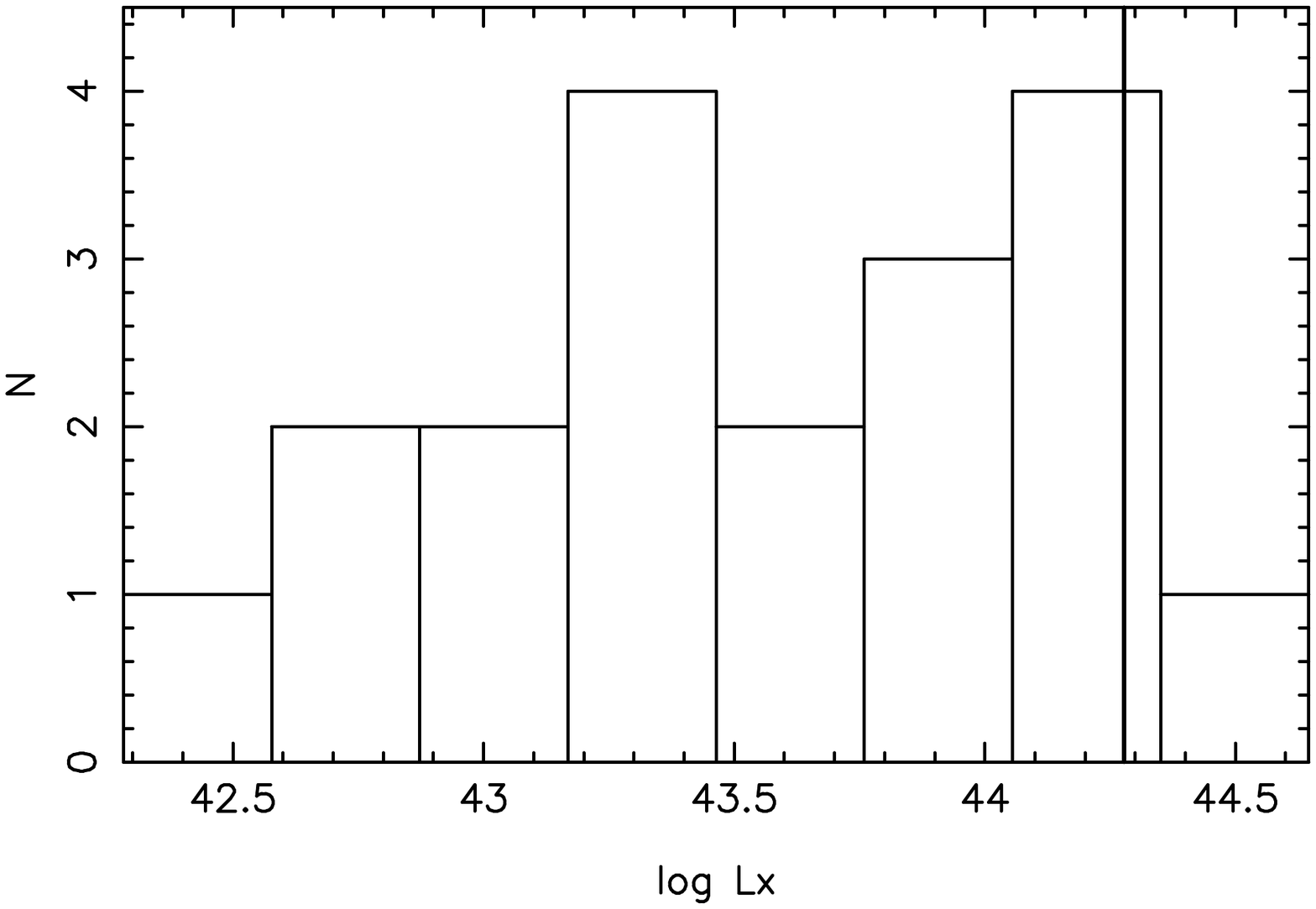}
\caption{The distribution of the host cluster velocity dispersions and X-ray luminosities (all measured in the 0.1 -- 2.4 keV band; where known from the literature). The vertical line in the second histogram represent the X-ray luminosity used in Brough et al.\ (2002) to separate X-ray luminous and less luminous clusters. For the references and other host cluster properties, see Table \ref{Xray}.}
\label{fig:ClusterHist}
\end{figure}

Because some studies have found a possible difference between the evolution of the BCGs in high and low X-ray luminosity clusters (see Brough et al.\ 2002, 2005), it is useful to sample a range of cluster luminosities. Figure~\ref{fig:ClusterHist} shows the distribution of cluster velocity dispersion and the X-ray luminosities for a subsample (14 and 19 respectively) of the clusters studied here for which we have found data in the literature. As can be seen, even in this reduced sample, the coverage in cluster velocity dispersion is quite large and range from $\sigma_{cluster}$ = 240 km s$^{-1}$ (corresponding to galaxy groups) to $\sigma_{cluster}$ = 1038 km s$^{-1}$. We have 5 clusters with X-ray luminosity around or above the limit imposed by Brough et al.\ (2002) to separate X-ray luminous and less-luminous clusters. These authors found that BCGs in X-ray luminous clusters have uniform absolute magnitudes (after correction for passive evolution) over redshifts 0.02 $< z<$ 0.8, suggesting a lack of stellar mass evolution beyond that expected by passive evolution, while those BCGs in less X-ray luminous clusters show significant scatter suggesting an increase in the mass up to a factor of $\sim$ 4 since z $\sim$ 1. According to this, most of our galaxies are in the range where a significant mass evolution is expected.

The observations of this sub-sample were all obtained with Gemini using the multi-object spectrograph (GMOS) in the long-slit mode. Although we tried to place the slit along the major axis, in many cases (11 out of 24), there were no suitable bright guide stars in the area in front of the mask plane, which forced the slit to be rotated to an intermediate axis (see Table \ref{table:objects} for details). 

To use model predictions based on the Lick/IDS absorption index system (Burnstein et al.\ 1984; Faber et al.\ 1985; Gorgas et al.\ 1993; Worthey et al.\ 1994), spectra were degraded to the wavelength dependent resolution of the Lick/IDS spectrograph where necessary (Worthey $\&$ Ottaviani 1997). Indices were also corrected for the broadening caused by the velocity dispersion of the galaxies (see the procedures described in Paper 2). We measured line-strength indices from the flux-calibrated spectra and calculated the index errors according to the error equations presented in Cardiel et al.\ (1998b). We have also thoroughly investigated the possible influence of scattered light on the images by interpolating the counts measured across three unexposed regions in the GMOS 2D images. The scattered light signature is not necessarily spatially flat and was subtracted, with the normal background subtraction, from the images. We then compared the index measurements from the corrected and uncorrected images, and found that this contribution to the overall incident light is negligible within the spatial radii that we use here. 

We identified and corrected the spatially-resolved galaxy spectra contaminated by emission by using a combination of the \textsc{ppxf} (Cappellari $\&$ Emsellem 2004) and \textsc{gandalf} (Sarzi et al.\ 2006) routines\footnote{We make use of the corresponding \textsc{ppxf} and \textsc{gandalf idl} (Interactive Data Language) codes which can be retrieved at \url{http:/www.leidenuniv.nl/sauron/}.}, with the MILES stellar library (S\'{a}nchez-Bl\'{a}zquez et al.\ 2006c). We follow the same procedure as fully described in Paper 2. We found, and successfully removed, weak emission lines in NGC0541; NGC3311; NGC1713; NGC6173; and NGC7012. The emission in ESO349-010 and MCG-02-12-039 could not be completely removed without introducing erroneous features in some of the indices, and were excluded from further analysis (a reliable Mg$_{2}$ index gradient could be measured for MCG-02-12-039, but no H$\beta$ gradient, and thus no stellar population parameters).

The half-light radii ($r_{\rm e}$) were calculated from the 2MASS catalogue, except ESO303-005 for which it was calculated by fitting a de Vaucouleurs profile to a cut in the spatial direction of a 2D spectrum, using the \textsc{vaucoul} task in the \textsc{reduceme} package\footnote{Reduceme is an astronomical data reduction package, specialising in the analysis of long-slit spectroscopy data. It was developed by N.\ Cardiel and J.\ Gorgas (Cardiel 1999) (\url{http://www.ucm.es/info/Astrof/software/reduceme/reduceme.html}).}. The half-light radii were computed from the 2MASS K-band 20th magnitude arcsec$^{−2}$ isophotal radius using the formula by Jarrett et al.\ (2003): $log (r_{e}) \sim log (r_{K20}) - 0.4$. For old stellar populations, these half-light radii do not differ much from those derived using the optical bands (Jarrett et al. 2003), and the typical errors on $r_{e}$ propagated from $r_{K20}$ is less than 0.08 and therefore negligible\footnote{However, see Stott et al.\ (2011) for an extensive discussion on the uncertainties of measuring $r_{e}$ for BCGs. For our purpose, we only use the $r_{e}$ measurements to obtain an estimate of how far our measured gradients reach compared to the extend of the galaxy, and it therefore does not change the results of our gradient correlations.}. We projected this radius on to the major axis (MA), taking into account the ellipticity, as:
\begin{equation}
a_{\rm e} = \frac{r_{\rm e}(1-\epsilon)}{1-\epsilon \ \mid cos(\mid PA - MA \mid)\mid},
\end{equation}
with $\epsilon$ the ellipticity (data from NED), $r_{\rm e}$ the radius containing half the light of the galaxy (computed from the 2MASS $K$-band 20th magnitude arcsec$^{-2}$ isophotal radius as described in Paper 1), PA the slit position axis, and MA the major axis. The fractions of $a_{\rm e}$ which the radial profiles measured in this work spans, are listed in Table \ref{table:objects}. The central radial velocity and velocity dispersion values of the galaxies were derived for regions with the size of $a_{\rm e}/8$ (see Paper 1), and with the galaxy centres defined as the luminosity peaks. The index values at each radius were measured at the local velocity and velocity dispersion for that radius (averaged over the bin) as presented in Paper 1. The luminosity-weighted centres of the spatial bins were used to plot the gradients in this paper.

Note that, despite our derived gradients reaching on average a fraction of 0.4 of the $a_{e}$ of the galaxy, Brough et al.\ (2007) showed that SSP gradients in BCGs do not deviate from a power law up to three times $a_{e}$. Therefore, our results are valid even if very large radii are not reached. 

\begin{table*}
\begin{tabular}{l r r r c r c} 
\hline Object & \multicolumn{1}{c}{Slit PA} & \multicolumn{1}{c}{MA} & \multicolumn{1}{c}{$r_{\rm e}$} & $\epsilon$ & \multicolumn{1}{c}{$a_{\rm e}$} & Fraction of $a_{\rm e}$ \\
       &  \multicolumn{1}{c}{}  & \multicolumn{1}{c}{} & (arcsec) & & (arcsec) & \\
\hline					
ESO146-028$\star$ & 154 & 154$\pm$1 & 12.4 & 0.35$\pm$0.05 & 12.4$\pm$1.8 & 0.4\\
ESO303-005 & 55 & -- & 9.1 & 0.25$\pm$0.04 & 9.1$\pm$0.1 &  0.4 \\
ESO349-010 & 14 & 155$\pm$14 & 15.3 & 0.52$\pm$0.10 & 12.3$\pm$2.8 & 0.6 \\								
ESO488-027 & 88 & 68$\pm$13 & 10.3 & 0.15$\pm$0.02 & 10.1$\pm$2.7 & 0.4 \\								
ESO552-020$\star$ & 148 & 148$\pm$5 & 18.3 & 0.43$\pm$0.05 & 18.3$\pm$3.1 & 0.5 \\
GSC555700266 & 204 & -- & 10.6 & 0.30$\pm$0.00 & 10.6$\pm$0.0 & 0.2 \\
IC1633$\star$ & 97 & 97$\pm$10 & 23.9 & 0.17$\pm$0.04 & 23.9$\pm$6.1 & 0.7 \\
IC4765 & 287 & 123$\pm$6 & 28.0 & 0.46$\pm$0.06 & 27.1$\pm$5.2 &  0.1 \\
IC5358 & 40 & 114$\pm$7 & 17.4 & 0.60$\pm$0.06 & 8.3$\pm$1.3 &  0.6 \\
LEDA094683$\star$ & 226 & 46$\pm$4 & 7.3 & 0.33$\pm$0.03 & 7.3$\pm$1.1 & 0.3 \\
MCG-02-12-039$\star$ & 166 & 180$\pm$13 & 15.2 & 0.19$\pm$0.03 & 15.1$\pm$3.5 & 0.4 \\
NGC0533 & 350 & 50$\pm$1 & 23.7 & 0.39$\pm$0.05 & 18.0$\pm$3.3  & 0.2 \\
NGC0541$\star$ & 64 & 69$\pm$7 & 15.1 & 0.06$\pm$0.05 & 15.1$\pm$17.9 & 0.2 \\
NGC1399 & 222 & -- & 42.2 & 0.06$\pm$0.02 & 42.2$\pm$19.9 & 0.6 \\ 			
NGC1713 & 330 & 39$\pm$5 & 15.5 & 0.14$\pm$0.03 & 14.0$\pm$4.6 &  0.4 \\
NGC2832 & 226 & 172$\pm$10 & 21.2 & 0.17$\pm$0.10 & 19.6$\pm$16.3 &  0.5 \\ 
NGC3311 & 63 & -- & 26.6 & 0.17$\pm$0.04 & 26.6$\pm$8.9 & 0.2 \\									 
NGC4839$\star$ & 63 & 64$\pm$3 & 17.2 & 0.52$\pm$0.07 & 17.2$\pm$3.4 &0.3 \\
NGC6173$\star$ & 139 & 138$\pm$3 & 15.0 & 0.26$\pm$0.05 & 15.0$\pm$4.1 & 0.1\\
NGC6269 & 306 & 80$\pm$4 & 14.1 & 0.20$\pm$0.01 & 13.1$\pm$1.1 & 0.3 \\ 
NGC7012$\star$ & 289 & 100$\pm$5 & 15.8 & 0.44$\pm$0.05 & 15.6$\pm$2.6 &0.5 \\
PGC030223 & 145 & 1$\pm$10 & 8.4 & 0.00$\pm$0.07 & 8.4$\pm$88.2 & 0.4 \\
PGC004072 & 204 & 83$\pm$0 & 10.2 & 0.33$\pm$0.00 & 8.2$\pm$0.0 & 0.2 \\
PGC072804$\star$ & 76 & 76$\pm$13 & 7.6 & 0.16$\pm$0.06 & 7.6$\pm$4.2 & 0.5 \\ 
UGC02232 & 60 &  -- & 9.7 & 0.00$\pm$0.08 & 9.7$\pm$3.9 &  0.4 \\
UGC05515 & 293 & 83$\pm$7 & 12.0 & 0.13$\pm$0.06 & 11.8$\pm$7.8 &0.2 \\
\hline
\end{tabular}
\caption{Galaxies observed with the Gemini North and South telescopes (see Paper 1 for more detailed properties of the galaxies). The position angle (PA) is given as deg E of N. The galaxies marked with a $\star$ is where the slit PA was placed within 15 degrees from the major axis (MA; where the MA was known). Typical errors on $r_{e}$ is less than one per cent. The last column lists the fraction of the effective half-light radii spanned by the radial profiles measured in this work. ESO349-010 and MCG-02-12-039 are included in the table even though the emission lines could not be fully removed.}
\label{table:objects}
\end{table*}

\section{Index gradients}
\label{indices}

Lick indices were measured according to the definitions of Trager et al.\ (1998). We plot the H$\beta$, Mg$_{b}$, Fe5270 and Fe5335 index measurements in magnitudes, obtained by using \begin{equation}
 I'=2.5\log\left[\frac{I}{(\Delta\lambda_c)}+1 \right], 
\end{equation} where ($\Delta\lambda_{c}$) represents the width of the central bandpass, and $I$ and $I'$ are the index measurement in \AA{} and magnitudes respectively.

As a consistency test, we compare the index measurements of the central bins with the central measurements (in $a_{\rm e}/8$ apertures as presented in Paper 2) in Appendix \ref{previous}. The index gradients of one of the galaxies (NGC4839) could also be compared to previous measurements in the literature from Fisher et al.\ (1995b) and Mehlert et al.\ (2000), as shown in Figure \ref{fig:Previous_comp}. The profiles compare well within the errors and small differences can be as a result of spectral resolution and slit width.

We represent the index gradients as $I^{'}=a+b\log(\frac{r}{a_{e}})$, where $I^{'}$ is the index (in magnitudes) and $a_{e}$ is the effective radius as presented in Table \ref{table:objects}. Gradients are taken to be significantly different from zero if the slope is greater than three times the 1$\sigma$ error on that gradient. Gradients (and the error on the gradient) were obtained by a linear least-squares fit using a Marquardt-Levenberg algorithm (Marquardt 1963) to the radial profiles. To take the errors on the individual values into account, statistical t-tests were also performed on all the slopes to assess if a real slope is present or if it is zero (as a null hypothesis). For our degrees of freedom, a $t$-value larger than 1.96 already gives a probability of lower than five percent that the correlation between the two variables is by chance. $P$ is the probability of being wrong in concluding that there is a true correlation (i.e.\ the probability of falsely rejecting the null hypothesis). 

We find only two galaxies with significant H$\beta$ gradients (shown in Figures \ref{Fig:Hbeta1} and \ref{Fig:Hbeta2} in Appendix \ref{GradsFigures} and Table \ref{table:HbetaGrads}). When we separate the gradients of the galaxies where the slit was placed within 15 degrees of the MA (where the MA was known), we find a mean H$\beta$ gradient and standard error of 0.001 $\pm$ 0.010. Similar results are found for normal elliptical galaxies (Mehlert et al.\ 2003; S\'{a}nchez-Bl\'{a}zquez et al.\ 2006b; Kuntschner et al.\ 2006, 2010). When we calculate the mean including all the derived H$\beta$ gradients, regardless of slit orientation, we find  0.003 $\pm$ 0.012. The H$\beta$ gradient of ESO488-027 showed an asymmetrical profile (see Section \ref{SSP}).

We find six galaxies with significant Mg$_{b}$ gradients, nine with significant Fe5270 gradients, and five with significant Fe5335 gradients (shown in Figures \ref{Fig:Indices} to \ref{Fig:Indices4} in Appendix \ref{GradsFigures} and Table \ref{table:HbetaGrads}). When we calculate the mean including all the derived Mg$_{b}$, Fe5270 and Fe5335 gradients, regardless of slit orientation, we find --0.017 $\pm$ 0.022, --0.014 $\pm$ 0.015 and --0.010 $\pm$ 0.021 respectively. We again note asymmetrical profiles for some of the galaxies.



\begin{table*}
\centering
\begin{tiny}
\begin{tabular}{l r@{$\pm$}l r r r@{$\pm$}l r r r@{$\pm$}l r r r@{$\pm$}l r r} 
\hline Object & \multicolumn{2}{c}{H$\beta$ gradient} & \multicolumn{1}{c}{$t$} & \multicolumn{1}{c}{$P$} & \multicolumn{2}{c}{Mg$_{b}$ gradient} & \multicolumn{1}{c}{$t$} & \multicolumn{1}{c}{$P$} & \multicolumn{2}{c}{Fe5270 gradient} & \multicolumn{1}{c}{$t$} & \multicolumn{1}{c}{$P$} & \multicolumn{2}{c}{Fe5335 gradient} & \multicolumn{1}{c}{$t$} & \multicolumn{1}{c}{$P$} \\
\hline					
ESO146-028 & 0.006 & 0.012 & 0.45 & 0.663 & 0.002 & 0.015 & 0.11 & 0.917 & --0.001 & 0.006 & --0.21 & 0.841 & --0.047 & 0.019 & --2.48 & 0.035\\
ESO303-005 & 0.022 & 0.010 & 2.15 & 0.098 & --0.035 & 0.026 & --1.33 & 0.254 & --0.024 & 0.013 & --1.90 & 0.130 & --0.032 & 0.039 & --0.82 & 0.457\\
ESO488-027 & --0.001 & 0.007 & --0.08 & 0.942 & --0.002 & 0.015 & --0.12 & 0.905 & --0.021 & 0.007$\star$ & --3.07 & 0.008 & 0.019 & 0.008 & 2.47 & 0.026\\
ESO552-020 & 0.014 & 0.007 & 1.86 & 0.088 & --0.014 & 0.010 & --1.42 & 0.180 & 0.001 & 0.009 & 0.04 & 0.967 & --0.010 & 0.008 & --1.30 & 0.217\\
GSC555700266 & 0.015 & 0.005$\star$ & 3.29 & 0.030 & 0.004 & 0.017 & 0.22 & 0.839 & --0.009 & 0.013 & --0.69 & 0.527 & --0.007 & 0.019 & --0.34 & 0.751\\
IC1633 & --0.003 & 0.003 & --1.07 & 0.290 & --0.026 & 0.004$\star$ & --7.28 & $<$0.0001 & --0.009 & 0.003$\star$ & --2.88 & 0.006 & --0.009 & 0.004 & --2.16 & 0.035\\
IC4765 & 0.004 & 0.002 & 1.06 & 0.251 & --0.003 & 0.013 & --0.26 & 0.819 & --0.007 & 0.016 & --0.40 & 0.731 & 0.016 & 0.015 & 1.07 & 0.397\\
IC5358 & 0.010 & 0.006 & 1.77 & 0.100 & --0.020 & 0.008 & --2.73 & 0.017 & --0.027 & 0.014 & --2.01 & 0.065 & --0.008 & 0.012 & --0.63 & 0.541\\
Leda094683 & --0.004 & 0.009 & --0.43 & 0.689 & 0.040 & 0.044 & 0.90 & 0.419 & --0.016 & 0.009 & --1.77 & 0.152 & 0.024 & 0.012 & 2.06 & 0.109\\
NGC0533 &  0.035 & 0.014 & 2.56 & 0.043 & --0.014 & 0.013 & --1.05 & 0.334 & --0.001 & 0.004 & --0.12 & 0.907 & 0.007 & 0.014 & 0.53 & 0.614\\
NGC0541 & --0.001 & 0.005 & --0.05 & 0.960 & --0.004 & 0.007 & --0.56 & 0.599 & --0.001 & 0.003 & --0.20 & 0.847 & 0.008 & 0.005 & 1.63 & 0.164\\
NGC1399 & --0.001 & 0.002 & --0.13 & 0.901 & --0.045 & 0.002$\star$ & --20.61 & $<$0.0001 & --0.013 & 0.002$\star$ & --8.50 & $<$0.0001 & --0.030 & 0.003$\star$ & --11.26 & $<$0.0001\\
NGC1713 & 0.001 & 0.004 & 0.28 & 0.784 & --0.035 & 0.007$\star$ & --5.15 & 0.002 & --0.018 & 0.005$\star$ & --3.67 & 0.003 & --0.019 & 0.004$\star$ & --4.40 & 0.001\\
NGC2832 &  0.001 & 0.006 & 0.02 & 0.983 & --0.065 & 0.017$\star$ & --3.76 & 0.002 & --0.032 & 0.008$\star$ & --4.24 & 0.001 & --0.057 & 0.012$\star$ & --4.75 & 0.001\\
NGC3311 &  --0.015 & 0.004$\star$ & --4.23 & 0.008 & 0.026 & 0.019 & 1.41 & 0.217 & 0.004 & 0.006 & 0.63 & 0.556 & --0.006 & 0.005 & --1.07 & 0.331\\
NGC4839 & 0.008 & 0.003 & 3.09 & 0.018 & --0.015 & 0.010 & --1.47 & 0.238 & --0.012 & 0.003$\star$ & --3.74 & 0.007 & --0.007 & 0.004 & --1.87 & 0.104\\
NGC6173 & --0.006 & 0.020 & --0.32 & 0.780 & --0.014 & 0.027 & --0.53 & 0.651 & --0.014 & 0.006 & --2.56 & 0.124 & 0.008 & 0.012 & 0.64 & 0.567\\
NGC6269 & 0.010 & 0.011 & 0.98 & 0.364 & --0.016 & 0.007 & --2.22 & 0.068 & --0.013 & 0.005 & --2.59 & 0.041 & --0.016 & 0.011 & --1.44 & 0.201\\
NGC7012 & 0.007 & 0.007 & 1.05 & 0.309 & --0.025 & 0.007$\star$ & --3.59 & 0.002 & --0.024 & 0.007$\star$ & --3.39 & 0.004 & --0.011 & 0.006 & --1.90 & 0.076\\
PGC004072 & 0.011 & 0.009 & 1.31 & 0.237 & --0.022 & 0.005$\star$ & --4.14 & 0.054 & --0.013 & 0.017 & --0.76 & 0.528 & 0.003 & 0.027 & 0.11 & 0.923\\
PGC030223  & 0.002 & 0.018 & 0.08 & 0.943 & --0.028 & 0.014 & --2.07 & 0.085 & --0.062 & 0.019$\star$ & --3.31 & 0.016 & --0.041 & 0.007$\star$ & --5.53 & 0.002\\
PGC072804 & --0.014 & 0.009 & --1.55 & 0.181 & --0.048 & 0.019 & --2.55 & 0.051 & 0.001 & 0.002 & 0.59 & 0.581 & 0.006 & 0.007 & 0.83 & 0.442\\
UGC02232 &  --0.013 & 0.016 & --0.82 & 0.456 & --0.013 & 0.016 & --0.84 & 0.449 & 0.008 & 0.020 & 0.38 & 0.723 & --0.024 & 0.029 & --0.85 & 0.444\\
UGC05515 & --0.009 & 0.040 & --0.22 & 0.845 & --0.024 & 0.020 & --1.22 & 0.346 & --0.030 & 0.009$\star$ & --3.29 & 0.081 & --0.013 & 0.004$\star$ & --3.19 & 0.086\\
\hline
Mean & \multicolumn{2}{c}{0.003}  & & & \multicolumn{2}{c}{--0.017}  & & & \multicolumn{2}{c}{--0.014}  & & & \multicolumn{2}{c}{--0.010}  & &\\
Std. dev & \multicolumn{2}{c}{0.012}  & & & \multicolumn{2}{c}{0.022}  & & & \multicolumn{2}{c}{0.015}  & & & \multicolumn{2}{c}{0.021}  & &\\
Std. err & \multicolumn{2}{c}{0.002}  & & & \multicolumn{2}{c}{0.005}  & & & \multicolumn{2}{c}{0.003}  & & & \multicolumn{2}{c}{0.004}  & & \\
\hline
\end{tabular}
\end{tiny}
\caption[]{The BCG H$\beta$, Mg$_{b}$, Fe5270 and Fe5335 gradients. A $\star$ indicates the cases where the gradient is bigger than three times the error on that gradient.}
\label{table:HbetaGrads}
\end{table*}

\section{SSP-equivalent parameter gradients}
\label{SSP}

To calculate the ages, metallicities ([Z/H]), and $\alpha$-enhancement ratios ([E/Fe]), we compare our derived line-strength indices with the predictions of Thomas, Maraston $\&$ Bender (2003) and Thomas, Maraston $\&$ Korn (2004) and using the Korn, Maraston $\&$ Thomas (2005) model atmospheres, with a slightly modified method from the one presented by Trager et al.\ (2000). Variations of the indices with chemical partitions departing from solar were included, where the ``E'' group contains O, Ne, Mg, Si, S, Ar, Ca, Ti, Na and N. Ages, metallicities and $\alpha$-enhancement ratios were derived for the BCG sample using the indices $<$Fe$>$\footnote{$<$Fe$>$ =(Fe5270 + Fe5335)/2, (Gonz\'alez 1993).}, H$\beta$ and Mg$_{\rm b}$. Then, a $\chi^{2}$-minimisation was applied to find the combination of SSPs that best reproduced the four indices simultaneously. Errors on the parameters were calculated by performing 50 Monte--Carlo simulations in which, each time, the indices were displaced by an amount given by a Gaussian probability distribution with a width equal to the errors on these indices.


We take the stellar population gradients as the slope of a linear fit, inversely weighted by the errors, to the relation $I^{'}=a+b\log (r)$ where $I^{'}$ is the SSP-equivalent parameter. This enables a direct comparison with previous studies, for example Brough et al.\ (2007). We fit the stellar population gradients using the same procedure as used for the index gradients, and once again exclude the central 0.5 arcsec (to each side) from the fit and take the gradient to be significant if it is larger than three times the 1$\sigma$ error on that gradient. The SSP gradients are tabulated in Table \ref{table:SSPGrads}, and shown in Figures \ref{fig:Profiles} to \ref{fig:Profiles6} in Appendix \ref{GradsFigures}. We only indicate the gradient fit in the plots (with a line) if it is bigger that 3 times the 1$\sigma$ error on the gradient. The age gradients of ESO146-028 and ESO303-005 showed asymmetrical profiles. We have investigated these galaxies (with respect to their emission, slit placement, radial kinematic profiles (see paper 2), central stellar populations and host cluster X-ray properties) without finding any reason for the asymmetric gradients. Since these galaxies have few data points, the two galaxy sides were not fitted separately.

Some galaxies (ESO552-020, IC1633, IC5358, NGC0533, NGC1399, NGC2832 -- none of which show nebular emission) and NGC7012 (where the emission was removed) reached the upper age limit of the stellar population models used. Note that the oldest ages in certain models, such as the ones used here, are older than the current age of the Universe (see discussion in S\'{a}nchez-Bl\'{a}zquez et al.\ 2009). However, all the interpretations of this study are based on relative differences in ages which are much more reliable than absolute values. 

\begin{table*}
\centering
\begin{scriptsize}
\begin{tabular}{l r@{$\pm$}l r r r@{$\pm$}l r r r@{$\pm$}l r r} 
\hline Object & \multicolumn{2}{c}{$\log$ (age) gradient} & \multicolumn{1}{c}{$t$} & \multicolumn{1}{c}{$P$} & \multicolumn{2}{c}{[E/Fe] gradient} & \multicolumn{1}{c}{$t$} & \multicolumn{1}{c}{$P$} & \multicolumn{2}{c}{[Z/H] gradient} & \multicolumn{1}{c}{$t$} & \multicolumn{1}{c}{$P$}\\
\hline					
ESO146-028 & 0.012 & 0.301 & 0.04 & 0.969 & 0.261 & 0.155 & 1.69 & 0.126 & --0.107 & 0.166 & --0.65 & 0.534\\
ESO303-005 & 0.587  & 0.598  & 0.98 & 0.381 & 0.136 & 0.235 & 0.58 & 0.592 & --0.573 & 0.360 & --1.59 & 0.187\\
ESO488-027 &  0.122 & 0.147 & 0.83 & 0.421 & --0.049 & 0.085 & --0.57 & 0.577 & --0.226 & 0.081 & --2.77 & 0.015\\
ESO552-020 &  --0.054 & 0.040 & --1.35 & 0.202 & --0.003 & 0.049 & --0.07 & 0.946 & --0.120 & 0.070 & --1.73 & 0.110\\
GSC555700266 &  0.369 & 0.526 & 0.70 & 0.522 & 0.044 & 0.146 & 0.30 & 0.781 & --0.267 & 0.418 & --0.64 & 0.558\\
IC1633 & 0.094 & 0.043 & 2.21 & 0.031 & --0.061 & 0.031 & --1.98 & 0.053 & --0.301 & 0.048$\star$ & --6.23 & $<$0.0001\\
IC4765 & --0.164 & 0.039$\star$ & --4.21 & 0.052 & --0.075 & 0.059 & --1.28 & 0.329 & 0.136 & 0.182 & 0.75 & 0.532\\
IC5358 &  --0.084 & 0.112 & --0.75 & 0.467 & 0.185 & 0.128 & 1.45 & 0.172 & --0.429 & 0.136$\star$ & --3.17 & 0.007\\
Leda094683 & 0.261 & 0.380 & 0.69 & 0.530 & --0.058 & 0.063 & --0.93 & 0.405 & --0.761 & 0.493 & --1.54 & 0.198\\
NGC0533 &  --0.471 & 0.196 & --2.41 & 0.043 & --0.033 & 0.063 & --0.52 & 0.615 & --0.141 & 0.071 & --1.98 & 0.083\\
NGC0541 & --0.014 & 0.128 & --0.11 & 0.920 & --0.083 & 0.041 & --2.02 & 0.100 & 0.046 & 0.146 & 0.32 & 0.763\\
NGC1399 &  0.104 & 0.025$\star$ & 4.22 & $<$0.0001 & 0.035 & 0.022 & 1.62 & 0.108 & --0.501 & 0.030$\star$ & --16.51 & $<$0.0001\\
NGC1713 &  0.059 & 0.064 & 0.93 & 0.372 & 0.003 & 0.053 & 0.05 & 0.959 & --0.475 & 0.080$\star$ & --5.93 & $<$0.0001\\
NGC2832 &  0.044 & 0.070 & 0.62 & 0.540 & 0.109 & 0.053 & 2.08 & 0.051 & --0.576 & 0.149$\star$ & --3.87 & 0.010\\
NGC3311 &  0.255 & 0.128 & 2.00 & 0.102 & 0.070 & 0.131 & 0.54 & 0.616 & --0.053 & 0.167 & --0.32 & 0.764\\
NGC4839 & --0.081 & 0.055 & --1.48 & 0.181 & 0.035 & 0.039 & 0.89 & 0.403 & --0.122 & 0.074 & --1.65 & 0.144\\
NGC6173 & 0.094 & 0.138 & 0.68 & 0.532 & --0.196 & 0.067 & --2.91 & 0.044 & --0.010 & 0.072 & --0.14 & 0.898\\
NGC6269 & --0.009 & 0.393 & --0.02 & 0.983 & 0.127 & 0.087 & 1.45 & 0.197 & --0.284 & 0.289 & --0.99 & 0.363\\
NGC7012 & 0.020 & 0.112 & 0.18 & 0.858 & 0.081 & 0.061 & 1.33 & 0.202 & --0.404 & 0.129$\star$ & --3.13 & 0.006\\
PGC004072 & 0.141& 0.317 & 0.44 & 0.701 & --0.045 & 0.195 & --0.23 & 0.839 & --0.306 & 0.165 & --1.85 & 0.205\\
PGC030223  & --0.166 & 0.273 & --0.61 & 0.566 & --0.009 & 0.135 & --0.06 & 0.951 & --0.725 & 0.185$\star$ & --3.93 & 0.008\\
PGC072804 & 0.638 & 0.208$\star$ & 3.06 & 0.028 & --0.358 & 0.128 & --2.80 & 0.038 & --0.685 & 0.215$\star$ & --3.18 & 0.025\\
UGC02232 &  --0.220 & 0.240 & --0.92 & 0.428 & --0.442 & 0.036$\star$ & --12.15 & 0.001 & 0.614 & 0.440 & 1.40 & 0.257\\
UGC05515 &  0.131 & 0.358 & 0.36 & 0.751 & 0.123 & 0.069 & 1.78 & 0.216 & --0.578 & 0.001$\star$ & --628.13 & $<$0.0001\\
\hline
Mean & \multicolumn{2}{c}{0.069}  & & & \multicolumn{2}{c}{--0.008} & & & \multicolumn{2}{c}{--0.285} & &  \\
Std. dev & \multicolumn{2}{c}{0.241}  & & & \multicolumn{2}{c}{0.157} & & & \multicolumn{2}{c}{0.315} & &  \\
Std. err & \multicolumn{2}{c}{0.049}  & & & \multicolumn{2}{c}{0.032} & & & \multicolumn{2}{c}{0.064} & &  \\
\hline
\end{tabular}
\end{scriptsize}
\caption[]{Age, metallicity and $\alpha$-enhancement gradients for our sample of BCGs. A $\star$ indicates where the gradient is bigger than three times the error on that gradient.}
\label{table:SSPGrads}
\end{table*}

\section{Results}
In the following sections, we will investigate if the stellar population gradients of BCGs are 
shaped more by the galaxy properties or by the environment. To do this, we will explore
the correlations between the stellar population gradients and the properties of both, the galaxies 
themselves, and their host clusters.

\subsection{Metallicity gradients and correlations with other galaxy properties}

The distribution of metallicity gradients for our sample of BCG galaxies is represented in Figure \ref{fig:Z_Hist}, while the numerical values together with the errors are listed in Table~\ref{table:SSPGrads}. It is immediately seen that our sample of galaxies show a large variety in metallicity gradients. This has already been shown in other studies of especially resolved stellar populations in BCGs (e.g.\ Gorgas et al.\ 1990; Brough et al.\ 2007), but we are showing that this relation also applies to a much larger sample here. The variety of the metallicity gradients for the BCG galaxies contrasts with their small dispersion in their aperture luminosities (Sandage 1972; Gunn $\&$ Oke 1975; Hoessel $\&$ Schneider 1985; Postman $\&$ Lauer 1995).

From Table \ref{table:SSPGrads} it can be seen that we find only nine significant metallicity gradients. When we include all the derived metallicity gradients, we find a mean gradient of --0.285$\pm$0.064 (with a standard deviation of 0.315), which compares very well to --0.31 $\pm$ 0.05  found by Brough et al.\ (2007). For normal ellipticals, S\'{a}nchez-Bl\'{a}zquez et al.\ (2007), Gorgas et al.\ (1990) and Fisher et al.\ (1995b) found mean [Z/H] gradients of $\Delta$[Z/H]/$\log r = -0.306 \pm 0.133, -0.23 \pm 0.09$ and $ -0.25 \pm 0.10$, respectively, , that is, the mean values of the metallicity gradients are very similar in BCGs and ordinary ellipticals. 

\begin{figure}
   \centering
   \includegraphics[scale=0.5]{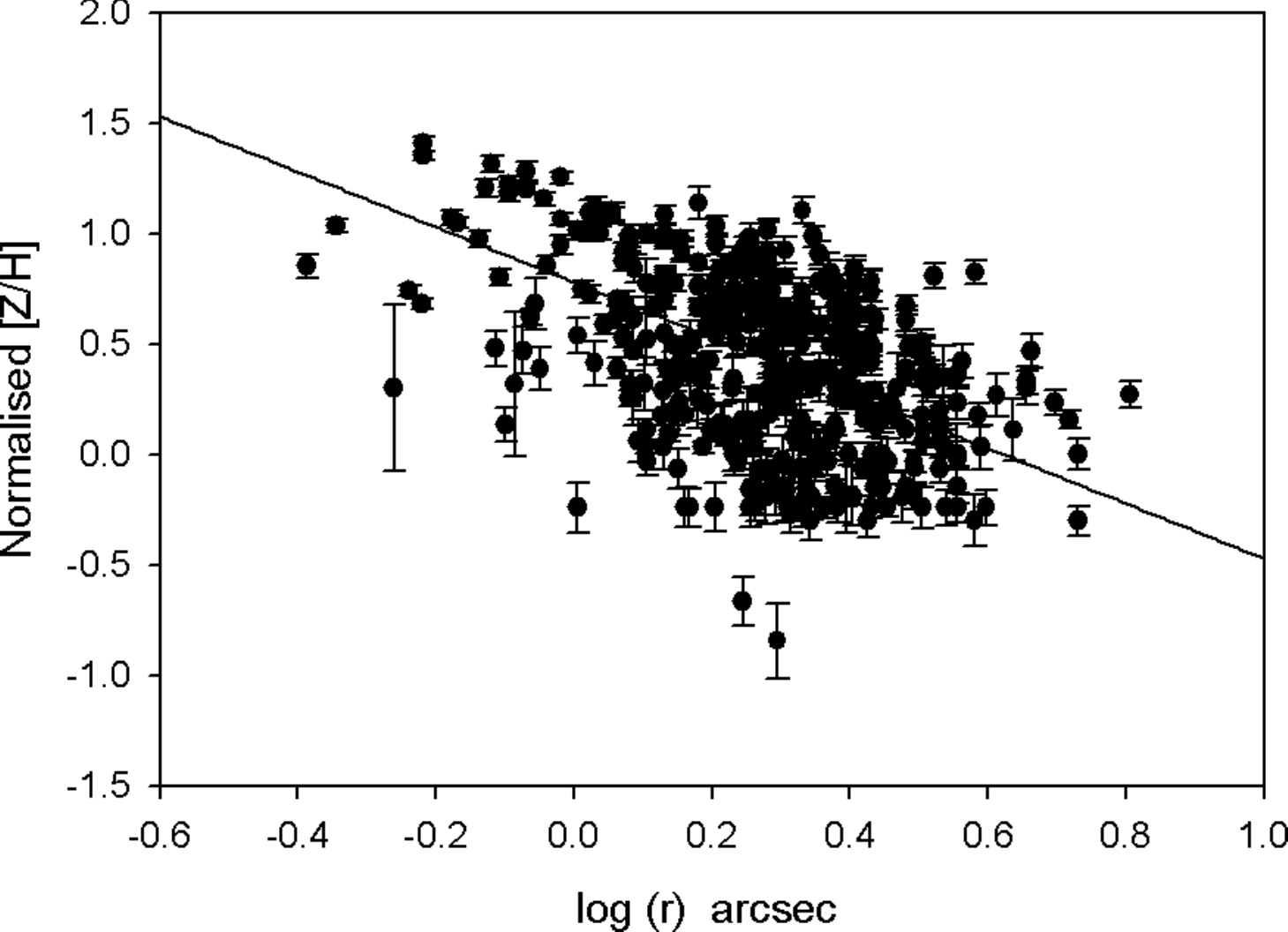}
   \caption{Metallicity measurements normalised to an average metallicity.}
   \label{fig:Normalised_Z}
\end{figure}

In addition of doing 24 individual fits and averaging all these gradients, we also normalised the data to the same average metallicity value on the vertical scale (0.3) and then made a single fit to all the points simultaneously (see Figure \ref{fig:Normalised_Z}). We then find a mean gradient of --1.251$\pm$0.076 (with $P$ = $<$ 0.0001).

\begin{figure*}
   \centering
   \includegraphics[scale=0.7]{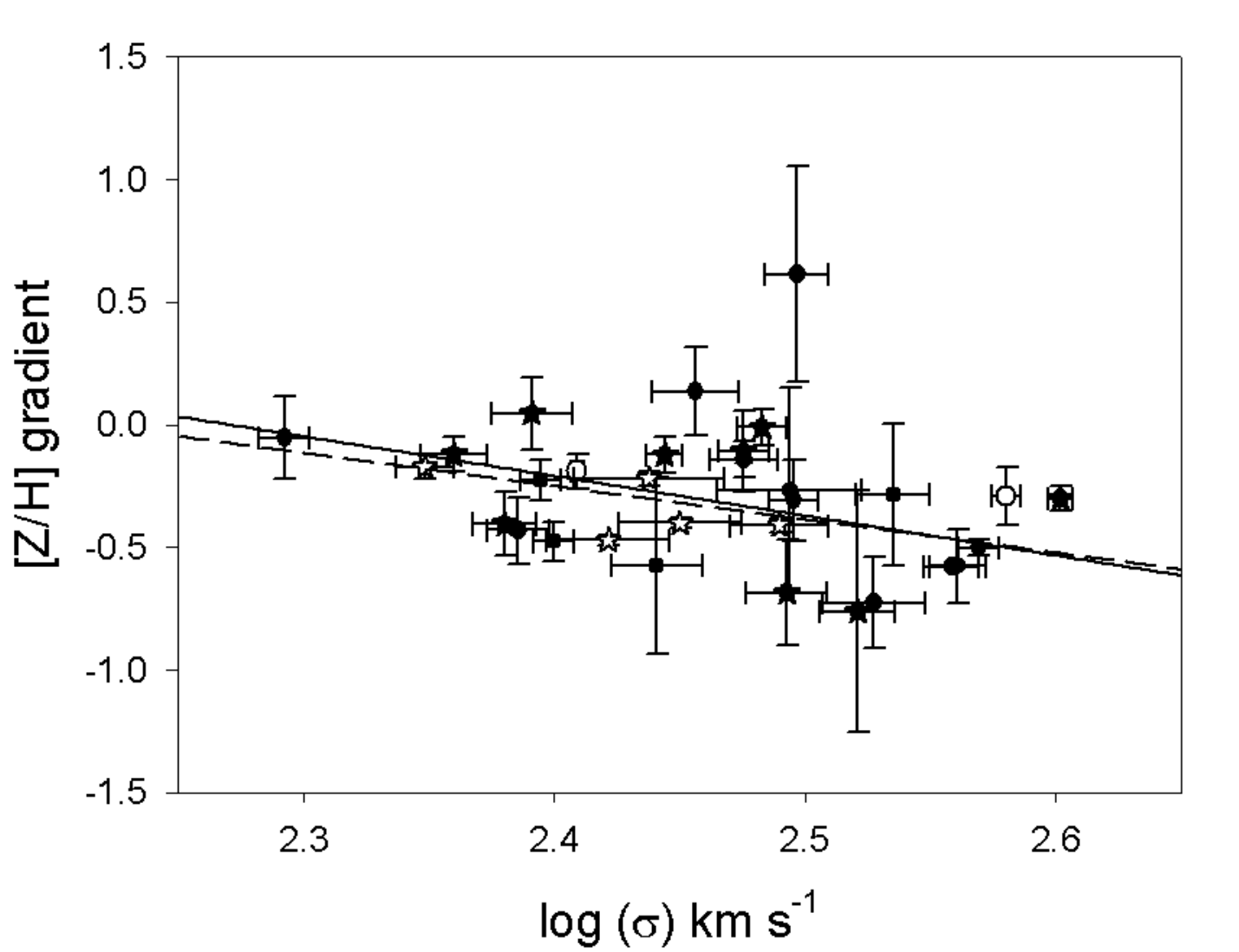}
   \caption{Metallicity gradients plotted against central velocity dispersion. The BCGs in our sample are shown with black symbols (those observed within 15 degrees of the MA (where known) are indicated with stars, and the others with circles. The BCGs from other studies are shown with empty symbols (those from Mehlert et al.\ (2003) with circles and those from Brough et al.\ (2007) with stars). The fitted metallicity -- mass correlation for our sample is shown with a solid black line, and the correlation fitted to all the data with a dashed line.}
   \label{fig:Mass_Z}
\end{figure*}

\subsubsection*{Metallicity gradient -- velocity dispersion relation}

A correlation between the metallicity gradient and the mass of galaxies is predicted by the classical dissipative collapse model (Larson 1974). Because dissipation efficiency increases
with mass, we would expect to find stronger gradients in more massive galaxies. However, a relation between the metallicity gradient and the mass of the galaxy -- usually parametrised by the central velocity dispersion -- has not been found in studies of normal elliptical galaxies (for example Kuntschner et al.\ 2006; S\'anchez-Bl\'azquez et al.\ 2006b, 2007, although 
this relation may exists in cluster galaxies, see, for example, Forbes, S\'anchez-Bl\'azquez $\&$ Proctor 2005). A stronger metallicity gradient is found in intermediate luminosity galaxies while the relation between 
the metallicity gradient and the velocity dispersion gets flatter for more massive galaxies. The change in slope occurs at the transition between those galaxies supported by rotation 
and those galaxies dominated by random motions (see S\'anchez-Bl\'azquez et al.\ 2007).

In this section we explore if BCGs behave differently from ordinary ellipticals. To do this, we plot the 24 metallicity gradients against central velocity dispersion (measured in the central $a_{\rm e}/8$ apertures) in Figure \ref{fig:Mass_Z}.
We find a correlation between the two parameters with a 95 per cent confidence level
(a t-test gives a probability of seven per cent that the two parameters are not correlated).
A linear fit weighted by the errors in both, the velocity dispersion and the metallicity gradients
gives a slope of --1.616$\pm$0.539.  
It can be seen that, contrary to the previous findings for elliptical galaxies -- in the same mass range -- not in the centres of clusters, the gradients seem to become steeper as the velocity dispersion increases. As an additional test, we also calculate the non-parametric Spearman rank coefficient (which is independent of the errors) as $R_{S}=-0.41$ (for $N=$24).


We also plot, in Figure \ref{fig:Mass_Z}, two BCGs from Mehlert et al.\ (2003; observed along the MA) and five from Brough et al.\ (2007; their sixth galaxy was excluded because it falls out of our mass range with a velocity dispersion of $\log \sigma=2.18$ km s$^{-1}$, and these galaxies were also observed along the major or intermediate axes) on the same graph. When we include these seven BCGs in the linear fit, regardless of slit orientation, we find a correlation consistent with a slope of --1.354 $\pm$ 0.427. Similarly, we find a probability of four per cent ($P$=0.0036) that these two parameters are not related according to a statistical t-test at 95 per cent confidence level, i.e.\ the correlation gets even stronger.

Brought et al.\ (2007) also found, for their smaller sample, a relation between the metallicity gradient and the velocity dispersion similar to the one observed here, but claimed it to be an artifact of the small size of the sample. However, we show here that the correlation persists even with a much larger sample of galaxies.

As we have mentioned before, for normal elliptical galaxies, Davidge (1992) and S\'anchez-Bl\'azquez et al.\ (2007) found
correlations between metallicity gradient and the anisotropy\footnote{The amount of flattening due to rotation in a galaxy depends
on the balance between ordered and random motions, and this can be quantified using the anisotropy parameter, which is defined as
$(V_{\rm max}/\sigma_{0})^{\ast} =$ ($V_{\rm max}/\sigma_{0}$)/$\sqrt{\epsilon/1-\epsilon}$          
(Kormendy 1982). The rotational velocity ($V_{\rm max}$) is half the difference between the minimum and maximum peaks of the 
rotation curve, and $\sigma_{0}$ is the measured central velocity dispersion (as derived in Paper~1).}.
We looked for this correlation in our sample of BCGs galaxies, using the $V_{\rm max}$ and $\sigma_{0}$ 
derived in Paper 1. Contrary to the works cited above, we do not find a correlation between these 
two parameters. Similarly to our study, Gorgas et al.\ (1990) also did not find such a correlation in 
a sample that combined BCGs and S0s. Whether these differences are due to the presence or absence of 
BCGs in samples remain to be seen with larger samples.

\subsubsection*{Metallicity gradients vs luminosity}
While the velocity dispersion is a proxy for the dynamical mass of the galaxy, the luminosity in the $K$-band is
sensitive to the predominantly red population in massive elliptical galaxies and hence, is very well correlated with the stellar mass (Brinchmann $\&$ Ellis 2000). 
To check if the correlation found in the previous section is robust, we also check the relation between the metallicity gradients and the luminosity in the $K$-band.

\begin{figure}
 \centering
\includegraphics[scale=0.55]{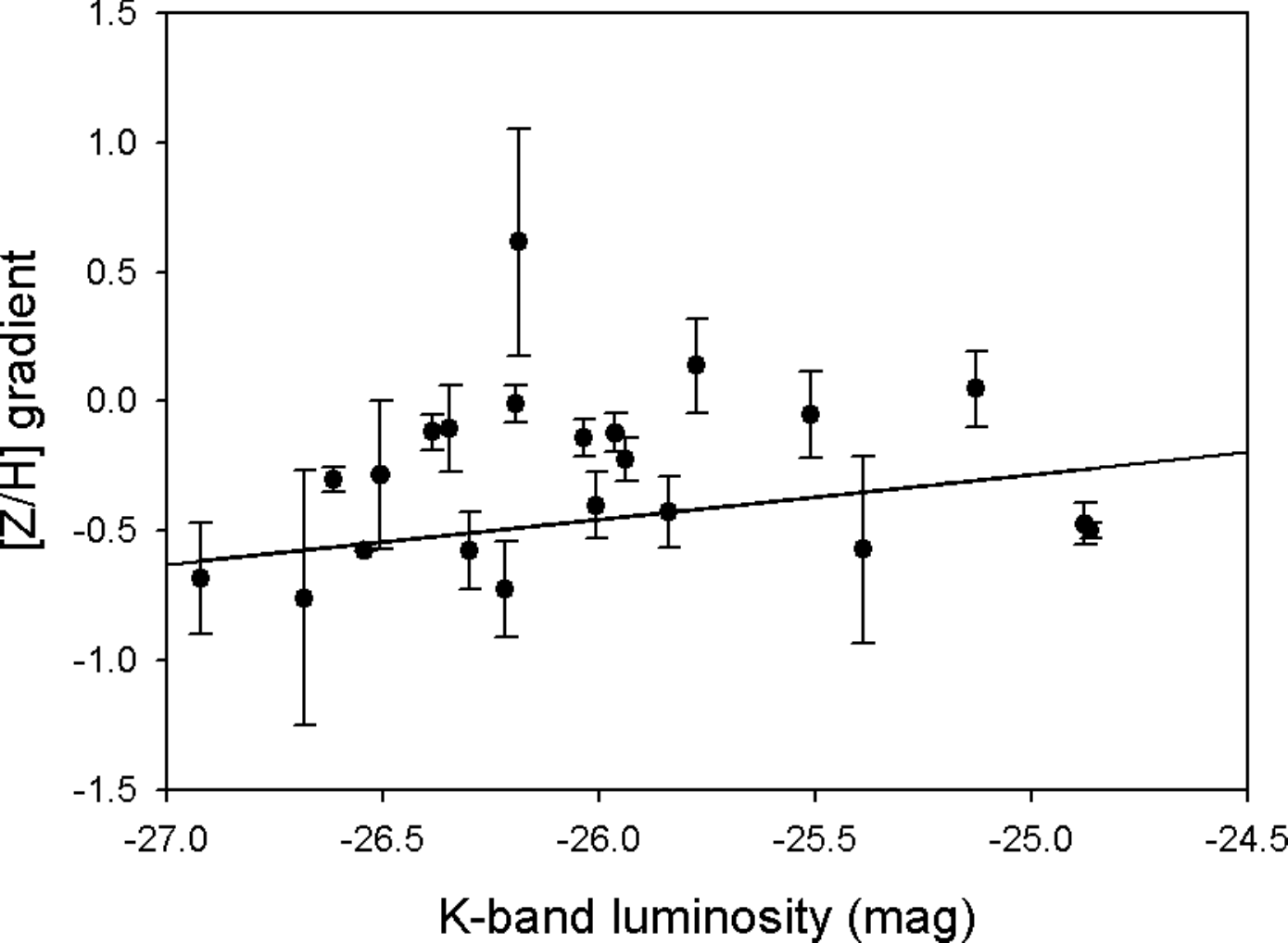}
\caption{BCG metallicity gradients plotted against $K$-band luminosity.}
\label{fig:Mg2Grads}
\end{figure}

We plot the 2MASS absolute $K$-band luminosity against the metallicity gradients in Figure \ref{fig:Mg2Grads}. We do a linear fit inversely weighted by the errors on the gradients for all the BCGs, regardless of slit orientation, and find a correlation with a slope of 0.173 $\pm$ 0.081. We find a probability of five per cent ($P$=0.05) that these two parameters are not related according to a statistical t-test at 95 per cent confidence level. The non-parametric Spearman rank coefficient (which is independent of the errors) is $R_{S}=0.31$ (for $N=$22). We do not correct the $K$-magnitudes for passive evolution. Using the Bruzual $\&$ Charlot (2003) stellar population synthesis code with the assumption that the galaxies are 10 Gyr old and formed in an instantaneous burst, this correction is only --0.2 mag for a galaxy at $z\sim0.054$ (in the $K$-band). Thus, it will not make a significant difference to whether or not a correlation is found. At the low redshift of this sample (mean $z=0.037\pm0.003$), the $K$-corrections to the magnitudes are negligible.

\subsection{Age gradients}

From Table \ref{table:SSPGrads} it can be seen that we find only three significant age gradients, and a mean gradient of 0.069 $\pm$ 0.049 (and a standard deviation of 0.241), which is consistent with a zero age gradient as also found by Brough et al.\ (2007) for BCGs (0.01 $\pm$ 0.04). Mehlert et al.\ (2003) and S\'{a}nchez-Bl\'{a}zquez et al.\ (2007) also find null or very shallow age gradients for their samples of elliptical galaxies, and Spolaor et al.\ (2010) also finds a mean age gradient of zero for their literature compilation of high mass early-type galaxies. Fisher et al.\ (1995b) and Carter et al.\ (1998) found that their smaller samples of BCGs were younger in the centres than in their outer parts, but Cardiel et al. (1998a) found this only in cooling flow clusters.


\subsection{$\alpha$-enhancement gradients}

\begin{figure*}
   \centering
   \includegraphics[scale=0.6]{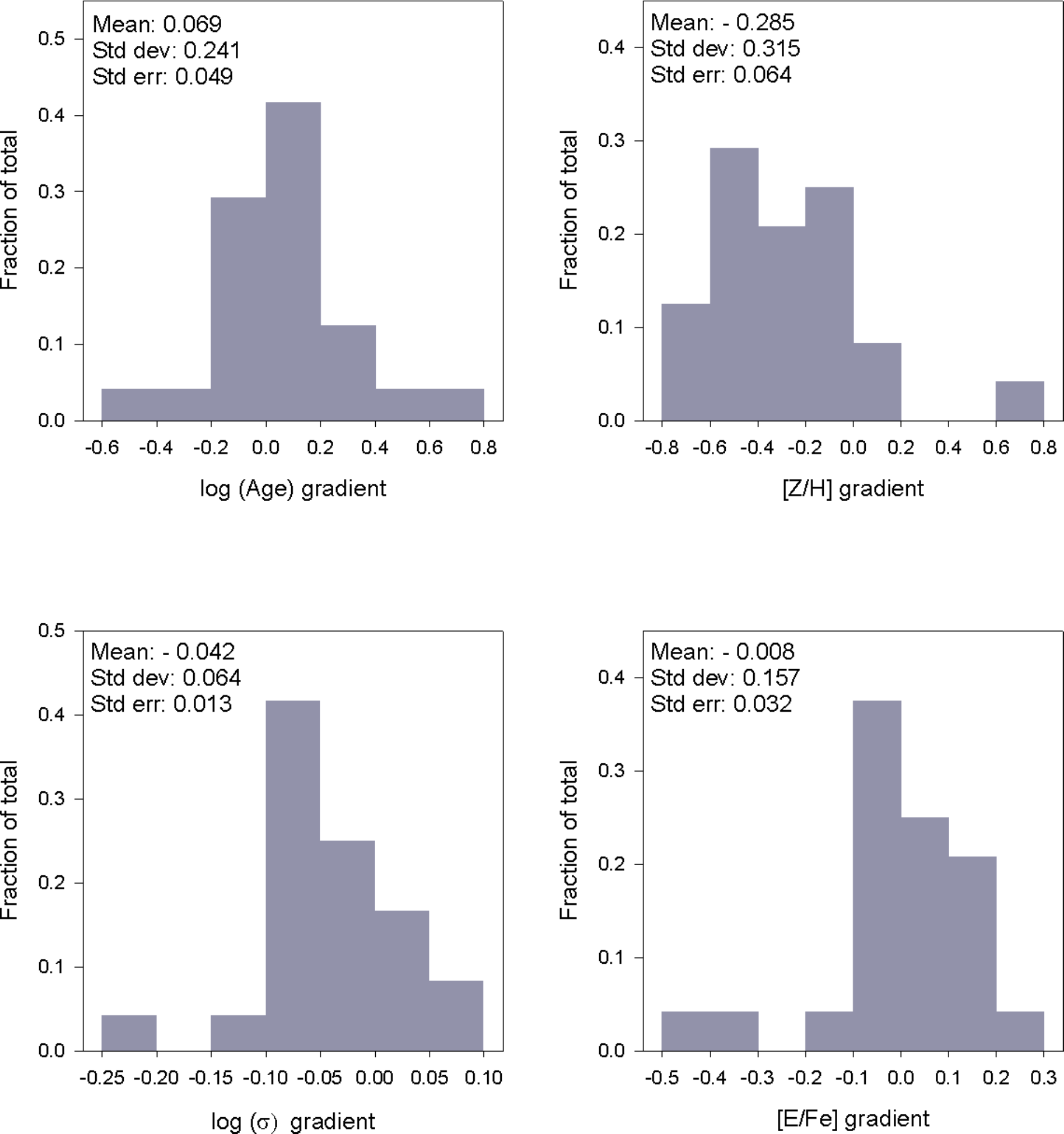}
   \caption{Histograms indicating the mean $\log$ (age), $\alpha$-enhancement, metallicity and velocity dispersion gradients of our BCG sample.}
   \label{fig:Z_Hist}
\end{figure*}

In agreement with previous studies of both BCGs and normal elliptical galaxies, we find null $\alpha$-enhancement gradients in our sample of galaxies (with a mean gradient of --0.008 $\pm$ 0.032 and with a standard deviation of 0.157), as shown in Figure ~\ref{fig:Z_Hist}\footnote{The distribution of the $\alpha$-enhancement gradients in Figure ~\ref{fig:Z_Hist} can be non-symmetrical, and if the non-symmetrical 1$\sigma$ Gaussian error is read from the histogram then the error on the mean changes negligibly, and the mean is still zero within the errors (whether measured symmetrically or non-symmetrically).}. For individual galaxies, we only find one with an statistically significant gradient.




\medskip We have used the well-calibrated indices H$\beta$, Mg$_{\rm b}$ and $<$Fe$>$ to determine SSP-equivalent parameters, as it is known that stellar population parameters, in particular metallicity, depend strongly on the indices used to determine it (S\'{a}nchez-Bl\'{a}zquez et al.\ 2006a, Pipino et al.\ 2010). However, we have tested whether using other indices affected the derived gradients. For example, we also used the indices H$\gamma_{\rm A}$, H$\gamma_{\rm F}$, C$_{2}$4668, Mg$_{\rm b}$ with the same models to derive log age, [Z/H], and [E/Fe]. We then find zero significant age gradients, nine significant metallicity gradients, and one significant $\alpha$-enhancement gradient (mostly for the same galaxies for which we found significant gradients using the H$\beta$, Mg$_{\rm b}$ and $<$Fe$>$ indices). Using H$\gamma_{\rm A}$, H$\gamma_{\rm F}$, C$_{2}$4668 and Mg$_{\rm b}$, we find a mean [Z/H] gradient of --0.289 $\pm$ 0.089, which is in agreement with --0.31 $\pm$ 0.05  found by Brough et al.\ (2007). It does not influence any of the results drawn from the SSP gradients in this study, as the gradients agree within the errors. We also did a boot-strap Monte Carlo simulation (with replacement) to assess the effect of non Gaussian errors on the significance quoted for the correlations. We found new correlations for [Z/H] gradients vs.\ velocity dispersion --2.184 $\pm$ 0.202 (instead of --1.616 $\pm$ 0.539) and [Z/H] gradients vs.\ K-magnitude 0.109 $\pm$ 0.015 (instead of 0.173 $\pm$ 0.081). 

\subsection{Velocity dispersion gradients}

We also characterise the velocity dispersion gradients as $\Delta \log (\sigma)/\Delta \log (r)$ to enable a direct comparison with previous studies, for example Brough et al.\ (2007). We fit the gradients using the same procedure as used for the indices and SSP parameters, and once again exclude the central 0.5 arcsec (to each side) from the fit and take the gradient to be significant if it is larger than three times the 1$\sigma$ error on that gradient. The velocity dispersion gradients are tabulated in Table \ref{table:SigmaGrads}.

We find negative velocity dispersion gradients for 17 BCGs in this sample (of which six are bigger than three times the 1$\sigma$ error), and seven positive velocity dispersion gradients (of which two are bigger than three times the 1$\sigma$ error). In a much smaller sample of BCGs, Brough et al.\ (2007) found five out of their six velocity dispersion gradients were consistent with being zero, and one negative gradient. As discussed in Paper 1, positive velocity dispersion gradients may imply rising mass-to-light (M/L) ratios with distance from the centre of the galaxies.  



\begin{table}
\centering
\begin{tabular}{l r@{$\pm$}l r r} 
\hline Object & \multicolumn{2}{c}{$\log$ ($\sigma$) gradient} & \multicolumn{1}{c}{$t$} & \multicolumn{1}{c}{$P$} \\
\hline					
ESO146-028 & --0.043 & 0.031 & --1.39 & 0.204 \\
ESO303-005 & --0.052 & 0.026 & --1.98 & 0.119\\
ESO488-027 & --0.033 & 0.016 & --2.06 & 0.059 \\
ESO552-020 & 0.082 & 0.012$\star$ & 6.74 & $<$0.0001\\
GSC555700266 & 0.003 & 0.040 & 0.07 & 0.948\\
IC1633 & --0.070 & 0.007$\star$ & --10.59 & $<$0.0001 \\
IC4765 & 0.042 & 0.014 & 2.94 & 0.099 \\
IC5358 & --0.026 & 0.032 & --0.83 & 0.421 \\
Leda094683 & --0.067 & 0.027 & --2.51 & 0.066\\
NGC0533 &  --0.042 & 0.013$\star$ & --3.15 & 0.014\\
NGC0541 & --0.074 & 0.026 & --2.89 & 0.034\\
NGC1399 & --0.137 & 0.006$\star$ & --24.22 & $<$0.0001 \\
NGC1713 & --0.097 & 0.017$\star$ & --5.79 & $<$0.0001 \\
NGC2832 &  --0.045 & 0.018 & --2.52 & 0.021 \\
NGC3311 &  0.049 & 0.014$\star$ & 3.39 & 0.020 \\
NGC4839 & --0.043 & 0.005$\star$ & --8.65 & $<$0.0001\\
NGC6173 & --0.038 & 0.016 & --2.31 & 0.147 \\
NGC6269 & --0.023 & 0.015 & --1.49 & 0.188 \\
NGC7012 & 0.021 & 0.025 & 0.84 & 0.413 \\
PGC004072 & --0.084 & 0.047 & --1.81 & 0.213\\
PGC030223  & --0.338 & 0.039$\star$ & --8.78 & $<$0.0001\\
PGC072804 & 0.032 & 0.042 & 0.77 & 0.478\\
UGC02232 &  0.046 & 0.094 & 0.49 & 0.657 \\
UGC05515 & --0.063 & 0.082 & --0.76 & 0.525\\
\hline
Mean & \multicolumn{2}{c}{--0.042}  & & \\
Std. dev & \multicolumn{2}{c}{0.082}  & & \\
Std. err & \multicolumn{2}{c}{0.017}  & & \\
\hline
\end{tabular}
\caption[]{BCG stellar velocity dispersion gradients. A $\star$ indicates where the gradient is bigger than three times the error on that gradient.}
\label{table:SigmaGrads}
\end{table}

\section{Cluster properties}
\label{cluster}

The correlation between the metallicity gradient and the central velocity dispersion, which is absent in normal elliptical galaxies, suggests that the position of the galaxy in the centre of the cluster potential well may have a stronger influence in the spatial distribution of line-strength indices than the mass of the galaxy. We explore the dependence of the stellar population gradients on the properties of the cluster, by investigating the following host cluster properties: X-ray luminosity $L_{\rm X}$; X-ray temperature $T_{\rm X}$; Cluster velocity dispersion $\sigma_{cluster}$; the BCG offset from the X-ray peak $R_{\rm off}$; and whether or not the host cluster is a cooling flow cluster or not. The literature values used here are also given in Table 7 in Paper 2, but are repeated here for completeness (Table \ref{Xray}). All the values are from spectra observed in the 0.1 -- 2.4 keV band with $ROSAT$, and using the same cosmology, namely the Einstein de--Sitter model of $H_{0}$ = 50 km s$^{-1}$ Mpc$^{-1}$, $\Omega_{\rm m}$ = 1 and $\Omega_{\Lambda}$ = 0. As a test, the X-ray luminosity values were converted to the concordance cosmological model ($\Omega_{\rm m}$ = 0.3 and $\Omega_{\Lambda}$ = 0.7) by calculating the appropriate cosmological luminosity distances of the clusters. We found that the assumed cosmological model does not influence the relative correlations between the parameters, and the previously most often used Einstein de--Sitter X-ray luminosity values are used to plot the correlations (as originally published from the \textit{ROSAT} data). The other X-ray properties used here, such as X-ray temperature, do not depend on the cosmological model assumed (White, Jones $\&$ Forman 1997; Bohringer et al.\ 2004). The X-ray offset $R_{\rm off}$ was calculated using $H_{0}$ = 75 km s$^{-1}$ Mpc$^{-1}$ in Mpc. For the majority of the clusters this offset was not already published, and it was then calculated from the BCG and published X-ray peak coordinates. However, this was not possible for those clusters, e.g.\ Coma, where a BCG is not in the centre and where the coordinates of a corresponding local X-ray maximum were not available.

\begin{figure}
 \centering
\mbox{\subfigure{\includegraphics[scale=0.55]{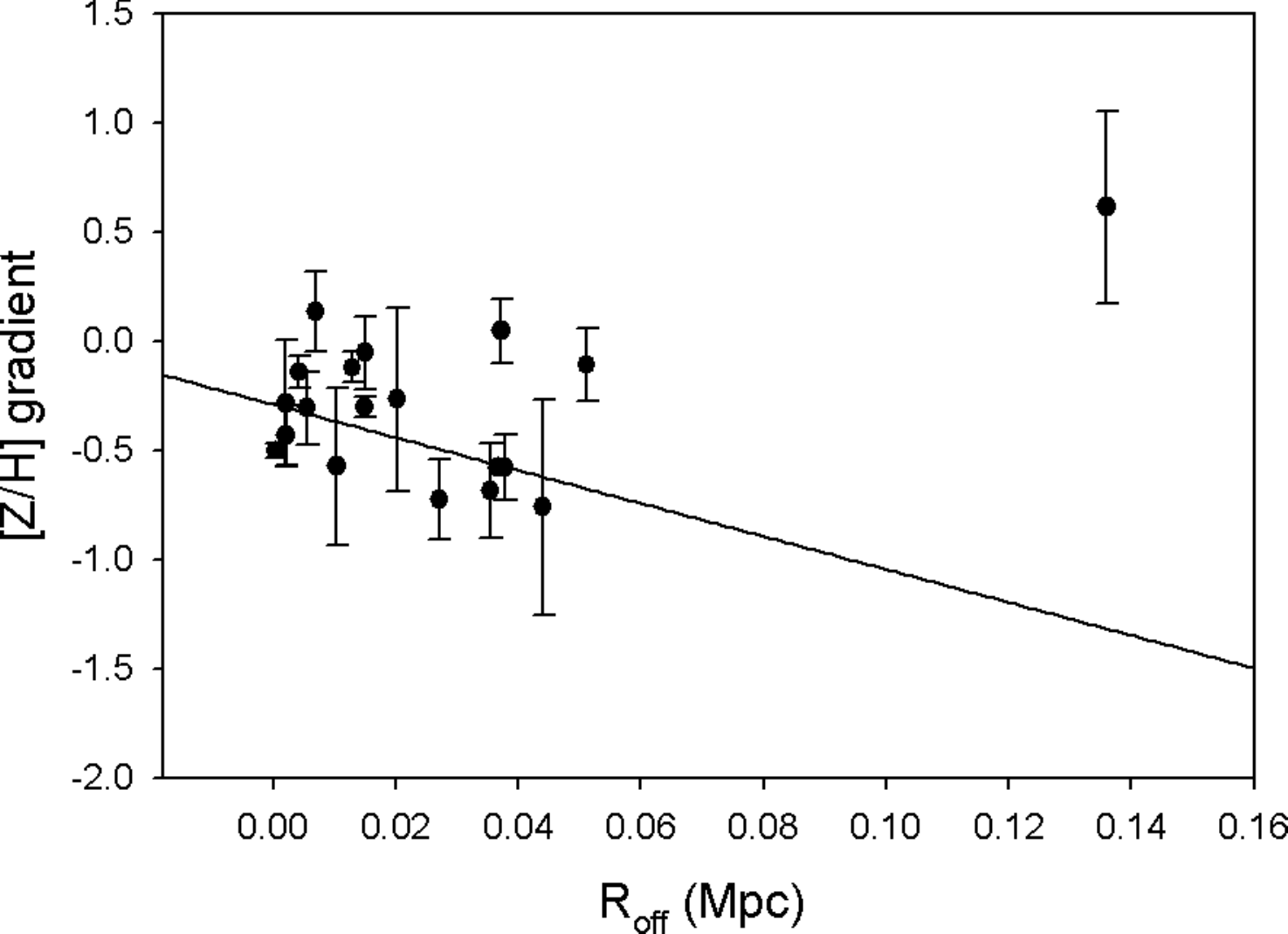}}}
         \mbox{\subfigure{\includegraphics[scale=0.55]{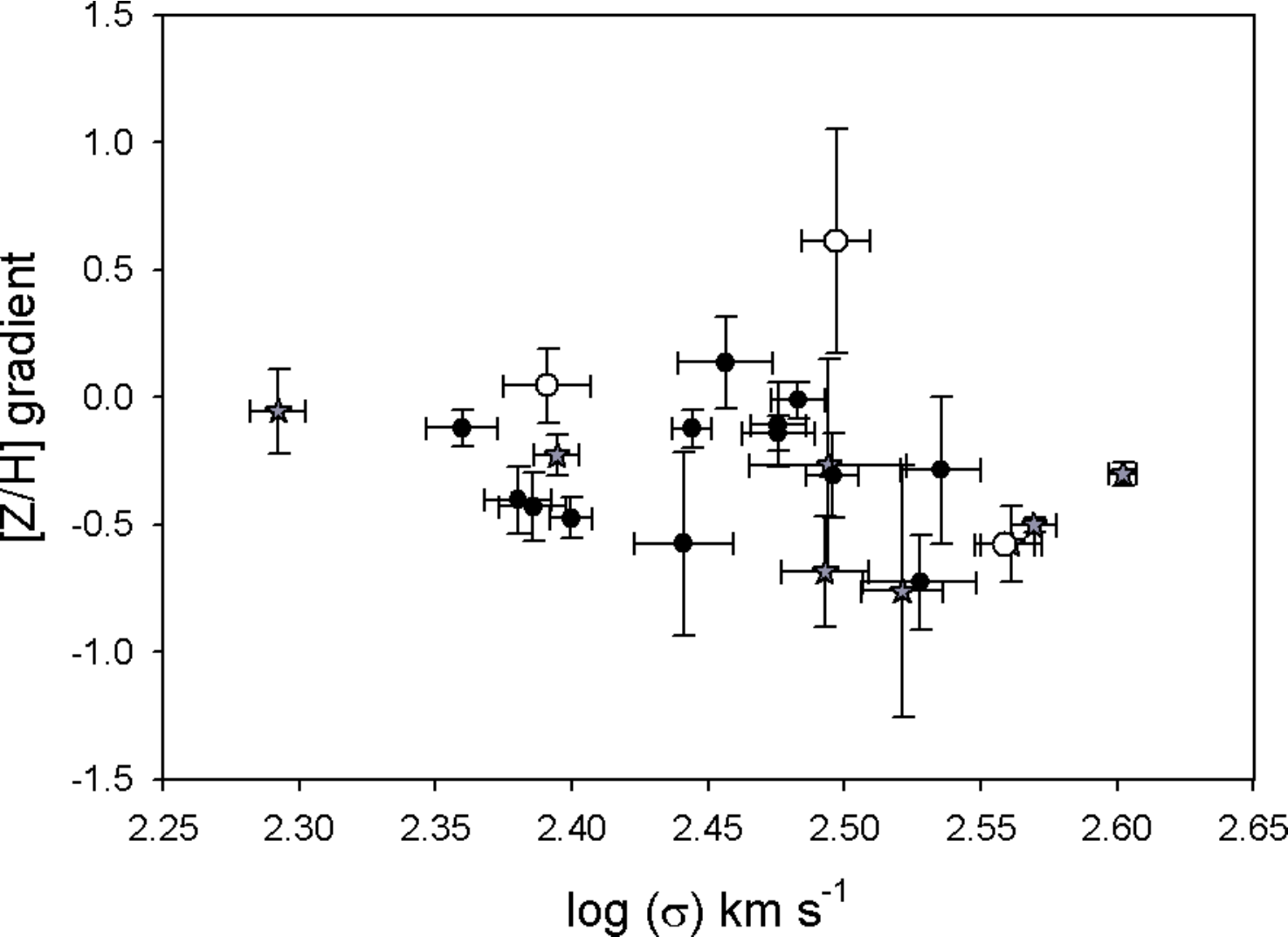}}}
   \caption{The upper plot shows the metallicity gradients against projected distance between BCG and X-ray peak of the host cluster. The lower plot shows the metallicity gradients against the stellar velocity dispersion of the galaxies where different symbols indicate whether or not the BCG is hosted by a cooling flow cluster (the grey stars indicate the cooling flow clusters, the empty circles indicate the non-cooling flow clusters, and the black filled circles indicate the clusters where the cooling flow status is not known).}
   \label{fig:ClusterGrads}
\end{figure}

Numerical simulations predict that the offset of the BCG from the peak of the cluster X-ray emission ($R_{\rm off}$) is an indication of how close the cluster is to the dynamical equilibrium state, and that this decreases as the cluster evolves (Katayama et al.\ 2003). We plot the metallicity gradients against $R_{\rm off}$ in Figure \ref{fig:ClusterGrads}, and do a linear fit inversely weighted by the errors for all the BCGs, regardless of slit orientation, to find a correlation with a slope of --7.546 $\pm$ 2.752. We find a probability of less than one per cent ($P$=0.009) that these two parameters are not related according to a statistical t-test at 95 per cent confidence level. This agrees with our findings for the Mg$_{2}$ gradients in Paper 4, and we find no other correlations with host cluster properties. The non-parametric Spearman rank coefficient (which is independent of the errors) is rather weak at $R_{S}=-0.016$ (for $N=$19). We again did a boot-strap Monte Carlo simulation (with replacement) to assess the effect of non Gaussian errors on the significance quoted for the correlations, and found the correlation: [Z/H] gradients vs.\ $R_{\rm off}$ --5.124 $\pm$ 0.696 (instead of --7.546 $\pm$ 2.752).

\begin{table*}
\centering
\begin{footnotesize}
\begin{tabular}{l l c r r c r c r c r c}
\hline Galaxy  & Cluster & \multicolumn{2}{c}{L$_{\rm X} \times 10^{44}$} & \multicolumn{2}{c}{T$_{\rm X}$} & \multicolumn{2}{c}{Cooling Flow} & \multicolumn{2}{c}{$\sigma_{\rm cluster}$} & \multicolumn{2}{c}{R$_{\rm off}$} \\ 
               &         & (erg s$^{-1}$) & ref & (keV) & ref &  & ref & (km s$^{-1}$) & ref& (Mpc) & ref \\
\hline ESO146-028 & RXCJ2228.8-6053 & 0.17 & b & -- & -- & -- & -- & -- & -- & 0.051 & cb\\
 ESO303-005 & RBS521 & 0.79 & b & --& -- & -- & -- & -- & -- & 0.010 & cb \\
 ESO488-027 & A0548 & 0.21 & b & 3.10$\pm$0.10 & d & $\checkmark$ & w & 853$_{-51}^{+62}$ & w & $\star$ & cb\\
 ESO552-020 & CID 28 & 0.16 & b & --& -- & -- & -- & -- & -- & 0.013 & cb \\
 GSC555700266 & A1837 & 1.28 & b & 4.20$\pm$0.24 & f & $\checkmark$ & w & 596 & w & 0.020 & cb\\
 IC1633 & A2877 & 0.20 & b & 3.27$\pm$0.14 & g & $\checkmark$ & w & 738 & w & 0.015 & cb\\
 IC4765 & A S0805 & 0.03 & b & 1.40$\pm$0.30 & h & -- & -- & -- & -- & 0.007 & cb\\
 IC5358 & A4038 & 1.92$\pm$0.04 & a & 2.61$\pm$0.05 & i & $\checkmark$ & c & 891 & m & 0.002 & cb\\
 Leda094683 & A1809 & -- & -- & 3.70 & w &  $\checkmark$ & w & 249 & w & 0.044 & p\\
 NGC0533 & A0189B & 0.04 & b & 1.08$\pm$0.05 & j & -- & -- & -- & -- & 0.004 & cb \\
 NGC0541 & A0194 & 0.14 & b & 2.87$_{-0.29}^{+0.33}$ & k & X & w & 480$_{-38}^{+48}$ & w & 0.037 & cb \\
 NGC1399 & RBS454 & 0.08$\pm$0.01 & a & -- & -- & $\checkmark$ & w & 240 & w & $<$0.001 & cb\\
 NGC1713 & CID 27 & -- & -- & -- & -- & -- & -- & -- & -- & -- \\
 NGC2832 & A0779 & 0.07 & b & 2.97$\pm$0.39 & l & $\checkmark$ & w & 503$_{-63}^{+100}$ & w & 0.038 & cl\\
 NGC3311 & A1060 & 0.56$\pm$0.03 & a & 3.15$\pm$0.03 & n & $\checkmark$ & w & 608$_{-38}^{+47}$ & w & 0.015 & pe\\
 NGC4839 & A1656 & -- & -- & -- & -- & -- & -- & -- & $\star$ & --\\
 NGC6173 & A2197 & -- & -- & -- & -- & -- & -- & -- & $\star$ & --\\
 NGC6269 & AWM5 & 0.36 & c & 2.16$_{-0.08}^{+0.10}$ & o & -- & -- & -- & -- & 0.002 & cc \\
 NGC7012 & A S0921 & -- & -- & -- & -- & -- & -- & -- & -- & -- \\
 PGC004072 & A0151 & 0.99 & b & -- & -- & -- & -- & 715 & s & 0.006 & cb\\
 PGC030223 & A0978 & 0.50 & b & -- & -- & -- & -- & 498 & st & 0.027 & cb \\
 PGC072804 & A2670 & 2.70 & b & 3.95$_{-0.12}^{+0.14}$ & q & $\checkmark$ & w & 1038$_{-52}^{+60}$ & w & 0.035 & cb\\ 
 UGC02232 & A0376 & 1.36 & c & 3.69$\pm$0.16 & r & X & e & 903 & w & 0.136 & cc\\
 UGC05515 & A0957 & 0.81 & b & 2.9 & w & X & w & 669 & w & 0.037 & cb \\
\hline
\end{tabular}
\end{footnotesize}
\caption[]{\begin{footnotesize} X-ray properties and velocity dispersions of the host clusters. The cluster velocity dispersion ($\sigma_{\rm cluster}$) values are in km s$^{-1}$ and the projected distance between the galaxy and the cluster X-ray peak ($R_{\rm off}$) is in Mpc. The presence of a cooling flow is indicated with a $\checkmark$, the absence of a cooling flow with a X, and -- indicates that the cooling flow status is not known. The $\star$ marks at R$_{\rm off}$ indicate that the galaxy is not in the centre of the cluster but closer to a local maximum X-ray density, different from the X-ray coordinates given in the literature. The references are: a = Chen et al.\ (2007); b = Bohringer et al.\ (2004); c = Bohringer et al.\ (2000); d = White (2000); f = Peterson et al.\ (2003); g = Sivanandam et al.\ (2009); h = David et al.\ (2003); i = Vikhlinin et al.\ (2009); j = Osmond $\&$ Ponman (2004); k = Sakelliou, Hardcastle $\&$ Jetha (2008); l = Finoguenov, Arnaud $\&$ David (2001); n = Ikebe et al.\ (2002); o = Sun et al.\ (2009); q = Cavagnolo et al.\ (2008); r = Fukazawa, Makishima $\&$ Ohashi (2004);  w = White, Jones $\&$ Forman (1997); e = Edwards et al.\ (2007); cc = Calculated from Bohringer et al.\ (2000); cl = Calculated from Ledlow et al.\ (2003); cb = Calculated from Bohringer et al.\ (2004); m = Mahdavi $\&$ Geller (2001); st = Struble $\&$ Rood (1999); s = Struble $\&$ Rood (1991); p = Patel et al.\ (2006); pe = Peres et al.\ (1998). All X-ray luminosity (L$_{\rm X}$) values were measured in the 0.1--2.4 keV band. \end{footnotesize}}
\label{Xray}
\end{table*} 

\section{Discussion and summary}
\label{conclusions}

The luminosity and photometric uniformity of BCGs imply that they are not just the bright extension of the luminosity function of cluster galaxies (Bernstein $\&$ Bhavsar 2001). The position of these galaxies at the centres of clusters and their unique properties link their formation and evolution to that of their environment. However, the mechanisms behind their growth are still poorly understood. If environment has an effect on the star formation, the number of interactions a galaxy undergoes, or the dissipation of gas within the
galaxy's potential well, we would expect to see an environmental dependence on the inferred stellar population gradients. We have derived the gradients of the SSP-equivalent parameters (age, [Z/H] and [E/Fe]) and the velocity dispersion for 24 BCGs in the nearby Universe. The host clusters of these galaxies comprise a large range of richness, ranging from groups to massive clusters. This sample is the largest BCG sample for which spatially resolved spectroscopy has been obtained.


We find very shallow gradients in all three considered parameters: age, metallicity and $\alpha$-enhancement. For normal elliptical galaxies, shallow gradients in age and $\alpha$-enhancement have also been found, and the mean metallicity gradient is normally around $-0.3$ dex per decade of variation in the radius, similar to what we found here. We note that Gorgas et al.\ (1990) found that the Mg$_{2}$ gradients of three BCGs are shallower than the mean gradient of normal ellipticals. However, Davidge $\&$ Grinder (1995) found the D4000 gradients of six BCGs to be steeper than that for non-BCGs. The contradictory results could be the consequence of the limited sample sizes. We do not know if the large fraction of null metallicity gradients that we found is a consequence of the privileged position of the BCGs in the centre of the clusters or if it is due to their large masses.

Up to now, there have been contradicting results among different studies about the existence of a relation between the metallicity gradients and the mass (usually parameterised by the velocity dispersion) of elliptical galaxies (Peletier et al.\ 1990; Gorgas et al.\ 1990; Gonz\'{a}lez 1993; Davies, Sadler $\&$ Peletier 1993; Carollo $\&$ Danziger 1994; Mehlert et al.\ 2003; Forbes et al.\ 2005), although a consensus started to appear that the the behaviour is different in different mass ranges (see Carollo, Danziger $\&$ Buson 1993; S\'{a}nchez-Bl\'{a}zquez et al.\ 2007; Spolaor et al.\ 2009; and Kuntschner et al.\ 2010). For elliptical galaxies, gradients get steeper with mass for galaxies with masses below $\sim$10$^{11}$ M$_{\odot}$, while the opposite behaviour is observed for more massive galaxies (although the slope is much flatter and the scatter also increases). We find a correlation with a slope of --1.616 $\pm$ 0.539 ($P$=0.007) for the metallicity gradients against central velocity dispersion of the 24 galaxies in this sample. This correlation becomes slightly stronger when we also include the two BCGs from Mehlert et al.\ (2003) and five from Brough et al.\ (2007) in the same mass range. Contrary to the previous findings for elliptical galaxies not in the centres of clusters, the gradients seem to become steeper as the velocity dispersion increases. 

We notice that this steepening of BCG gradients was also shown in Spolaor et al. (2009) using a literature compilation of eight BCGs. We also find a correlation between the metallicity gradients and $K$-band luminosity supporting that there is a trend between the strength of the gradient and the mass of the BCGs that it is not present, or at least not so evident in elliptical galaxies.

To study the possible influence of the cluster on the properties of the central galaxy, we investigated the metallicity gradients against the host cluster properties (X-ray luminosity $L_{\rm X}$; X-ray temperature $T_{\rm X}$; cluster velocity dispersion $\sigma_{cluster}$; the BCG offset from the X-ray peak $R_{\rm off}$; and whether or not the host cluster is a 
cooling flow cluster or not). We find a strong correlation between the metallicity gradients and $R_{\rm off}$ with a slope of --7.546 $\pm$ 2.752 ($P$=0.009). 
This agrees with our findings for the Mg$_{2}$ gradients in Paper 4. We find no other correlations with host cluster properties. Notice that Cardiel et al.\ (1998a) found differences in the Mg$_2$ gradients between galaxies in clusters with and without cooling flows.

In summary, we have found hints of differences in the spatially resolved stellar population properties
of BCGs and normal elliptical galaxies pointing to an influence of the cluster in the way these systems
assemble. The trends showed here are, admittedly, weak, but they are interesting as this is the first
time they have been found. These correlations are worth exploring in more detail, with stellar population gradients 
reaching larger radii.

\section*{Acknowledgments}
We thank the anonymous referee for constructive comments which contributed to the improvement of this paper. PSB is supported by the Ministerio de Ciencia e Innovaci\'on (MICINN) of Spain through the  Ramon y Cajal programme. This work has been supported by the Programa Nacional de Astronom\'{\i}a y Astrof\'{\i}sica of the Spanish Ministry of Science and Innovation under the grant AYA2007-67752-C03-01.

Based on observations obtained on the Gemini North and South telescopes. The Gemini Observatory is operated by the Association of Universities for Research in Astronomy, Inc., under cooperative agreement with the NSF on behalf of the Gemini Partnership: the National Science Foundation (USA), the Science and Technology Facilities Council (UK), the National Research Council (Canada), CONICYT (Chile), the Australian Research Council (Australia), CNPq (Brazil) and CONICET (Argentina). This research has made use of the NASA/IPAC Extragalactic Database (NED) which is operated by the Jet Propulsion Laboratory, California Institute of Technology.

\appendix

\section{Comparison with previous measurements}
\label{previous}

\begin{figure*}
   \centering
   \includegraphics[scale=0.5]{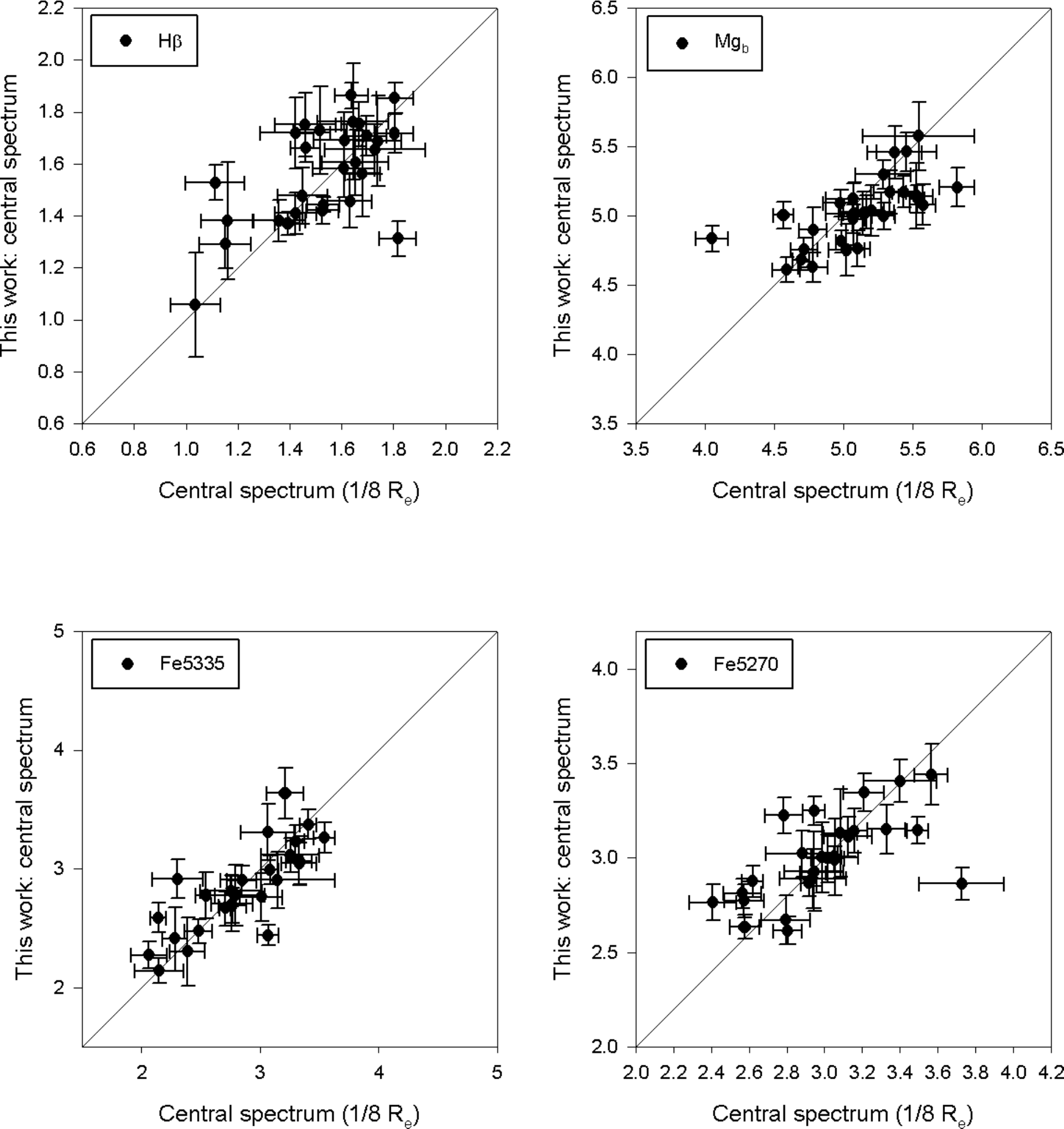}
   \caption{Comparison between the central index measurements from Paper 2 (measured in $a_{\rm e}/8$ apertures) and the central bins measured in this work (minimum S/N 40 per \AA{} at H$\beta$) for four key indices.}
   \label{fig:Central_comp}
\end{figure*}

To test self-consistency, we compared the indices measured in central bins with the central measurements (from Paper 2, measured in $a_{\rm e}/8$ apertures) in Figure \ref{fig:Central_comp}.

\begin{figure*}
   \centering
   \includegraphics[scale=0.6]{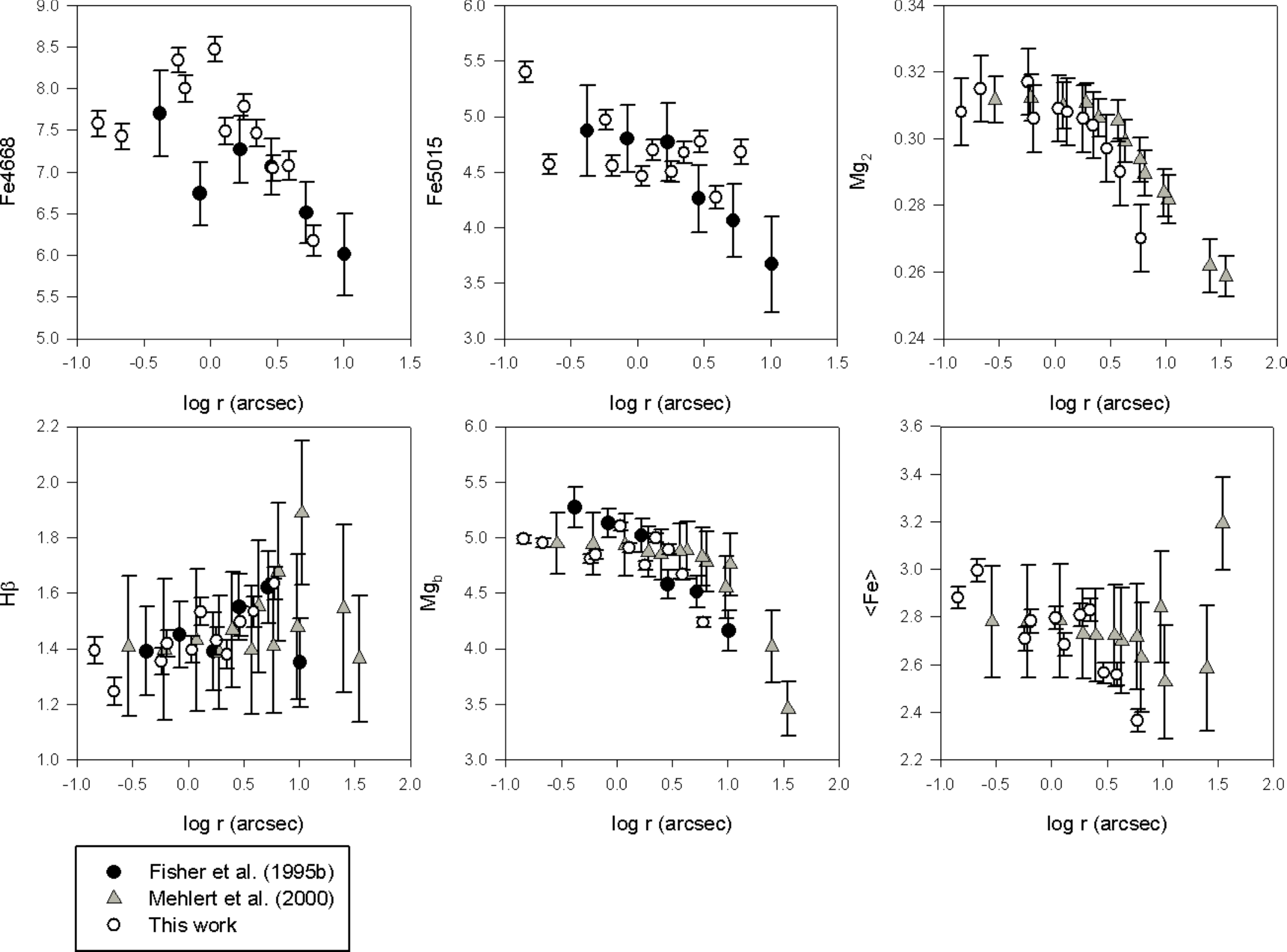}
   \caption{Comparison to previous data (NGC4839). The data are folded with respect to the centre of the galaxy. Mg$_{2}$ is measured in magnitude, and all the other indices shown are measured in \AA{}.}
   \label{fig:Previous_comp}
\end{figure*}

The index gradients of one of the galaxies could also be compared to previous measurements in the literature from Fisher et al.\ (1995b) and Mehlert et al.\ (2000), as shown in Figure \ref{fig:Previous_comp}. The profiles compare well within the errors and small differences can be as a result of spectral resolution and slit width.

\section{Index and stellar population gradients}
\label{GradsFigures}

\begin{figure*}
   \centering
   \mbox{\subfigure[ESO146-028]{\includegraphics[width=3.9cm,height=3.9cm]{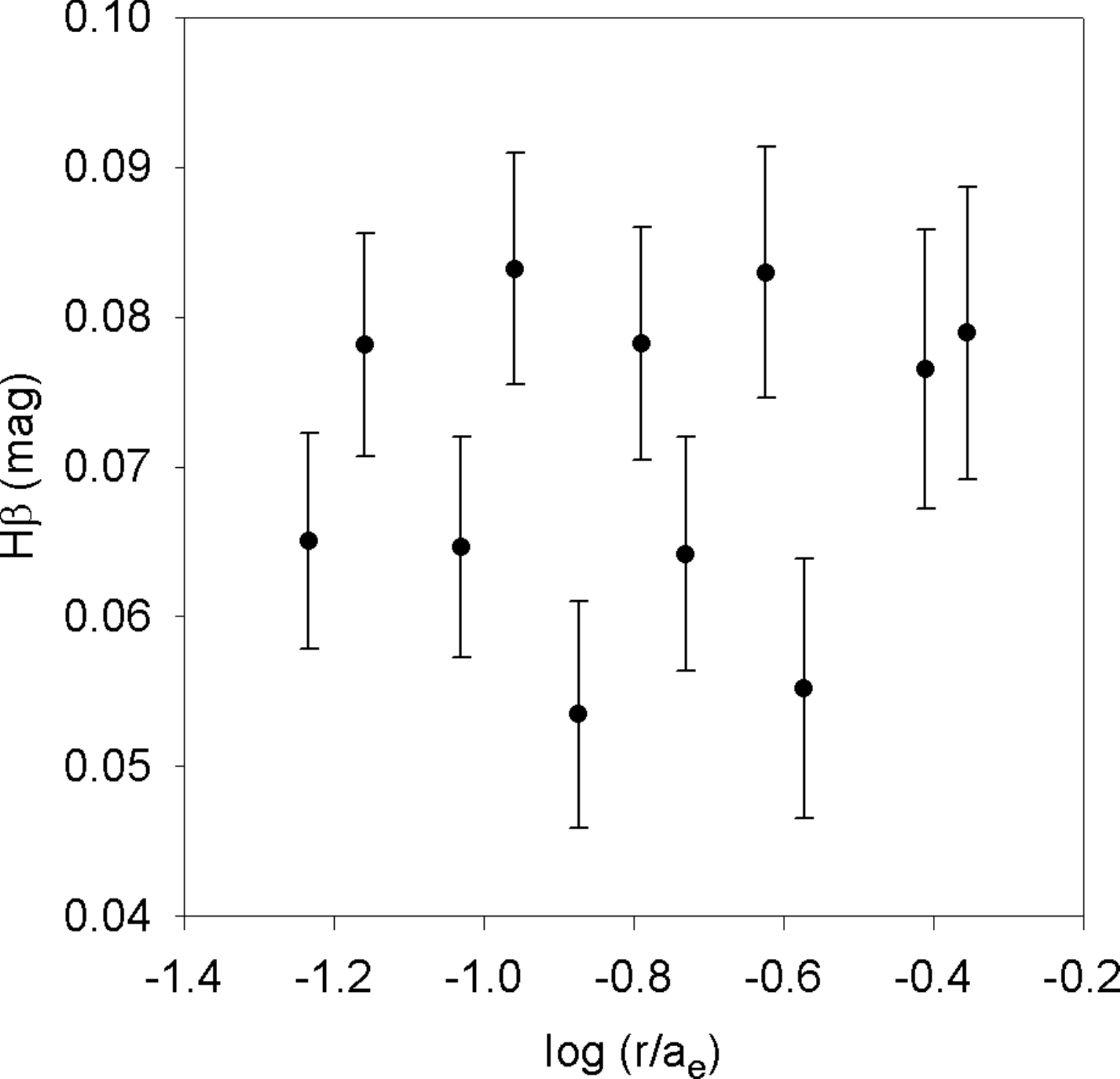}}\quad
         \subfigure[ESO303-005]{\includegraphics[width=3.9cm,height=3.9cm]{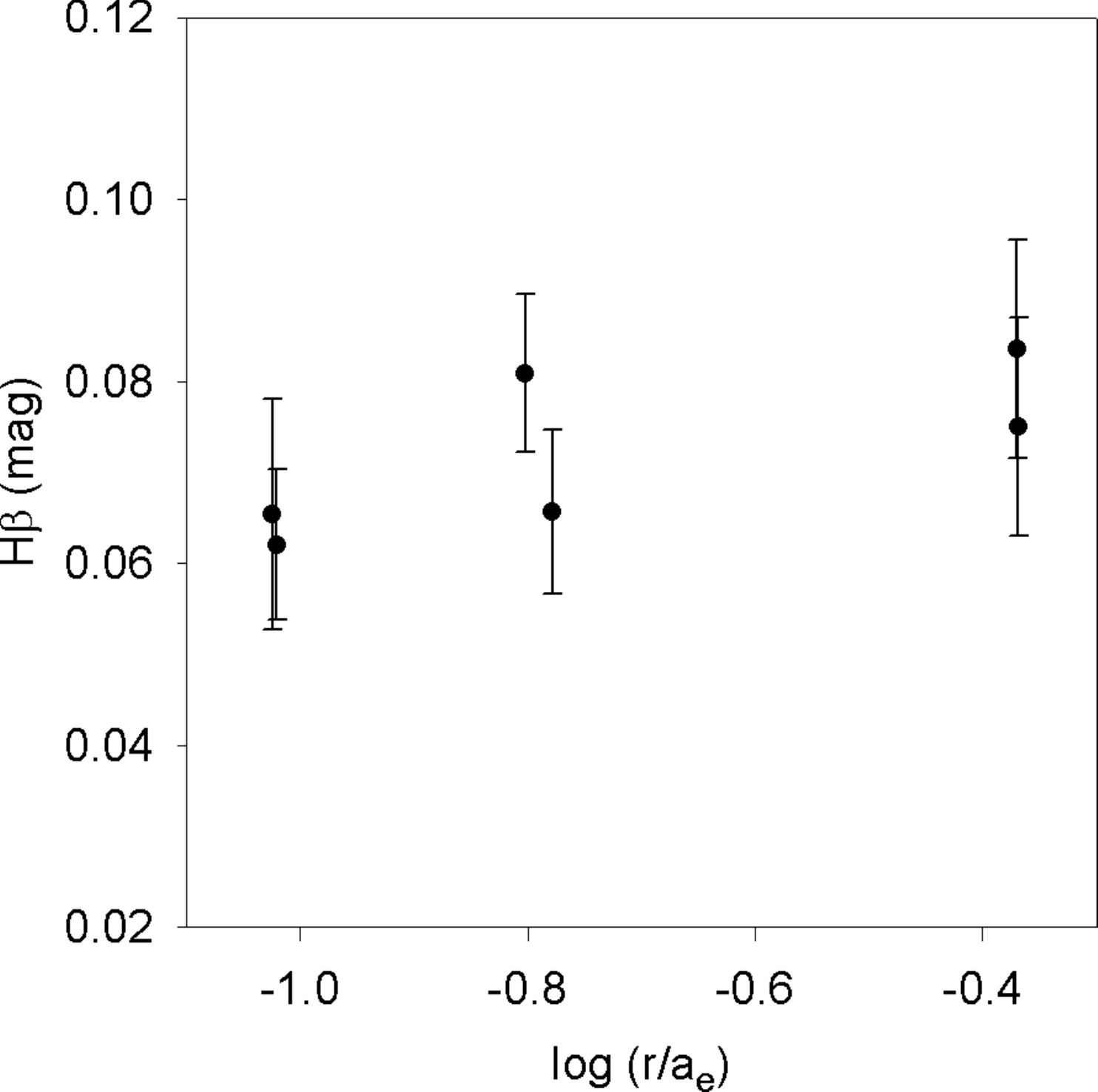}}\quad
         \subfigure[ESO488-027]{\includegraphics[width=3.9cm,height=3.9cm]{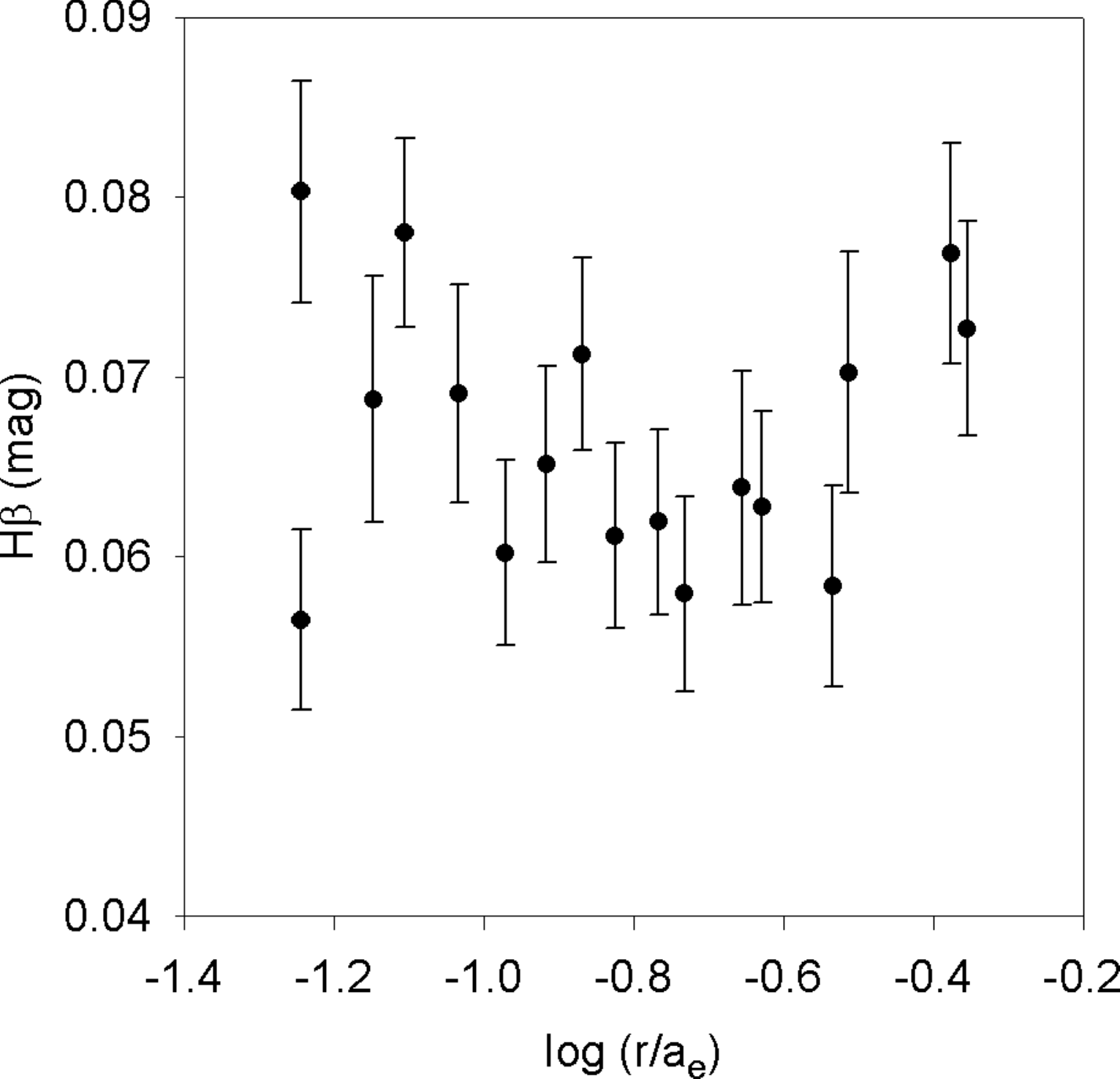}}\quad
         \subfigure[ESO552-020]{\includegraphics[width=3.9cm,height=3.9cm]{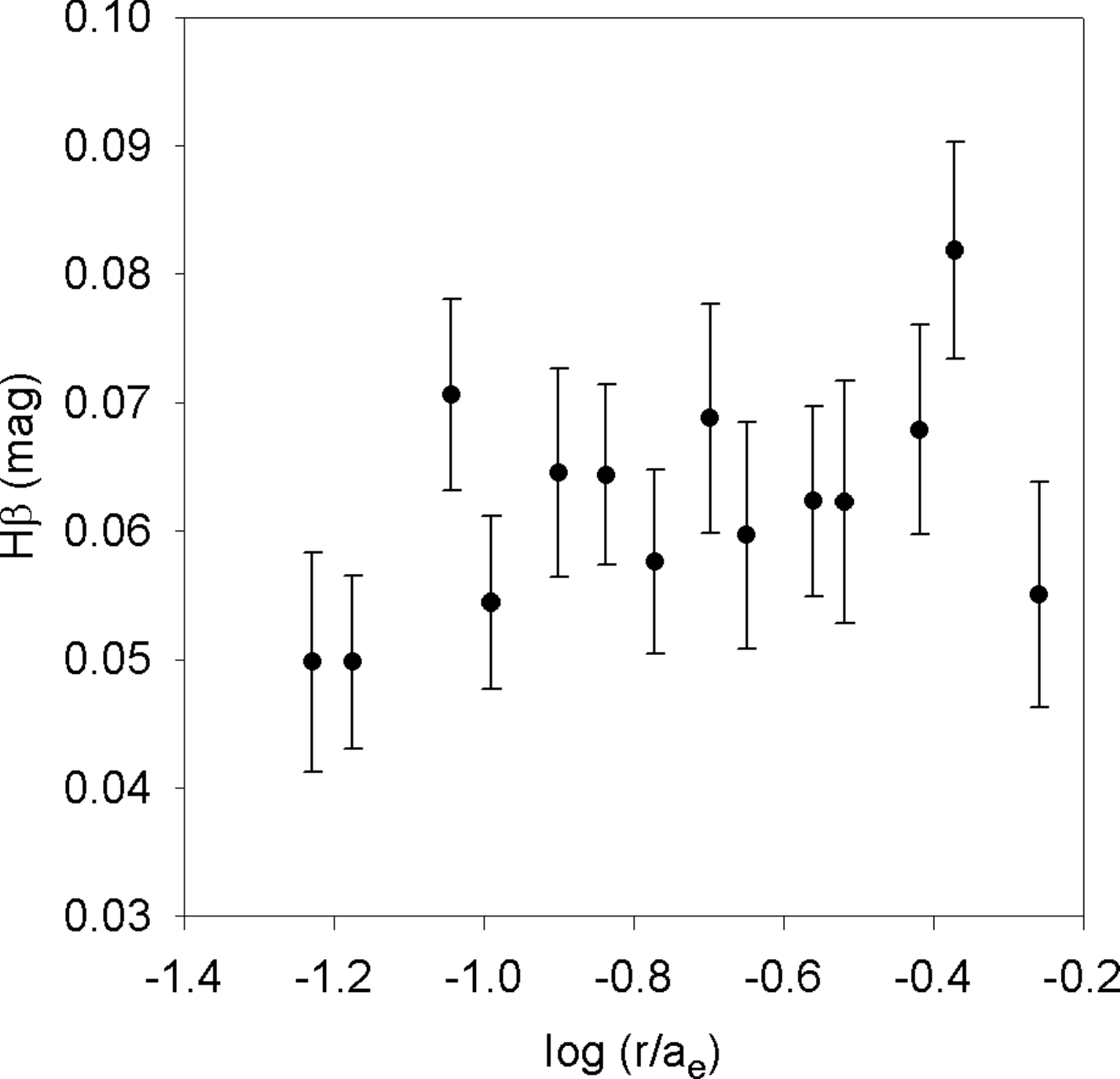}}}
  \mbox{\subfigure[GSC555700266]{\includegraphics[width=3.9cm,height=3.9cm]{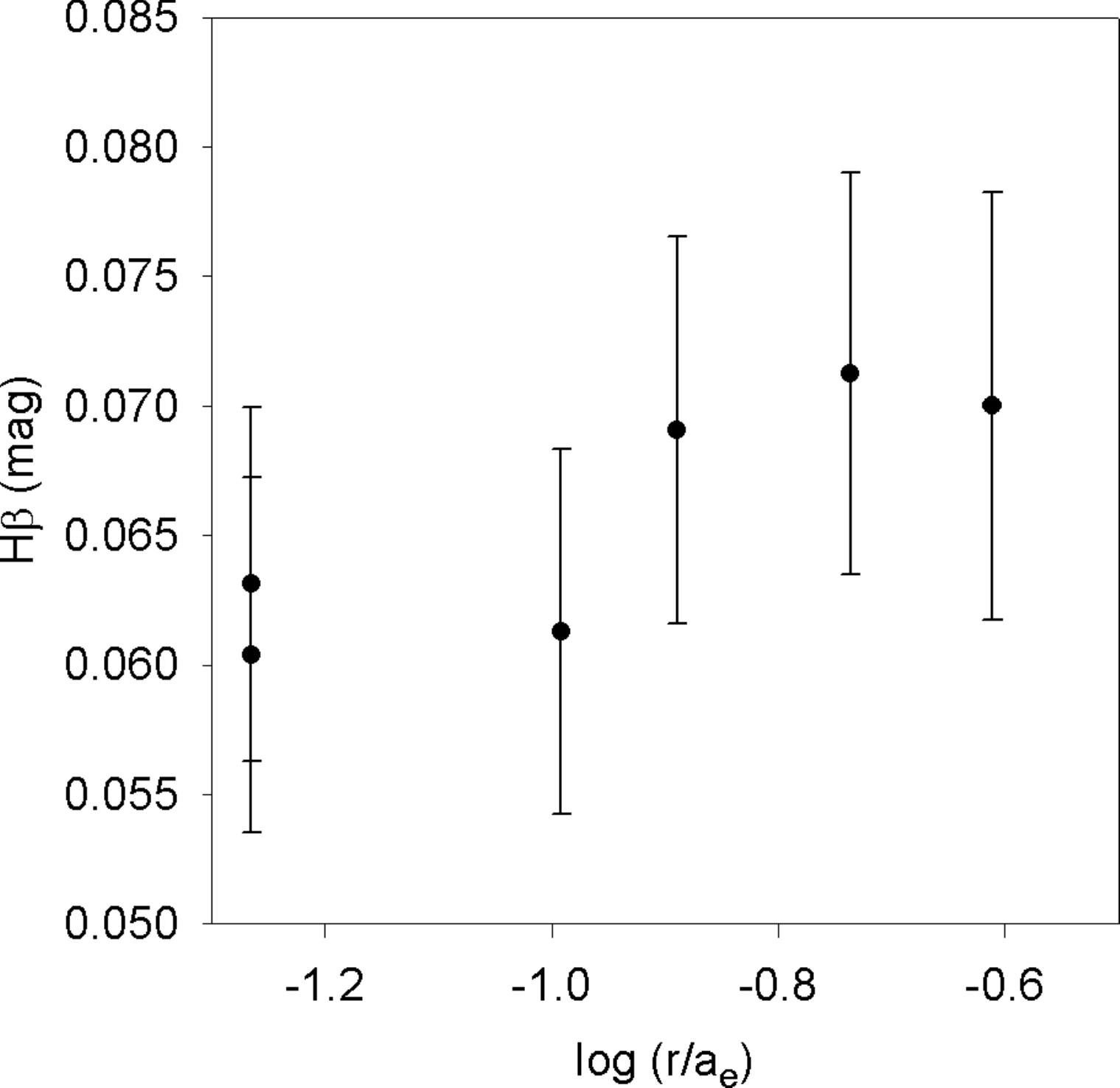}}\quad
 \subfigure[IC1633]{\includegraphics[width=3.9cm,height=3.9cm]{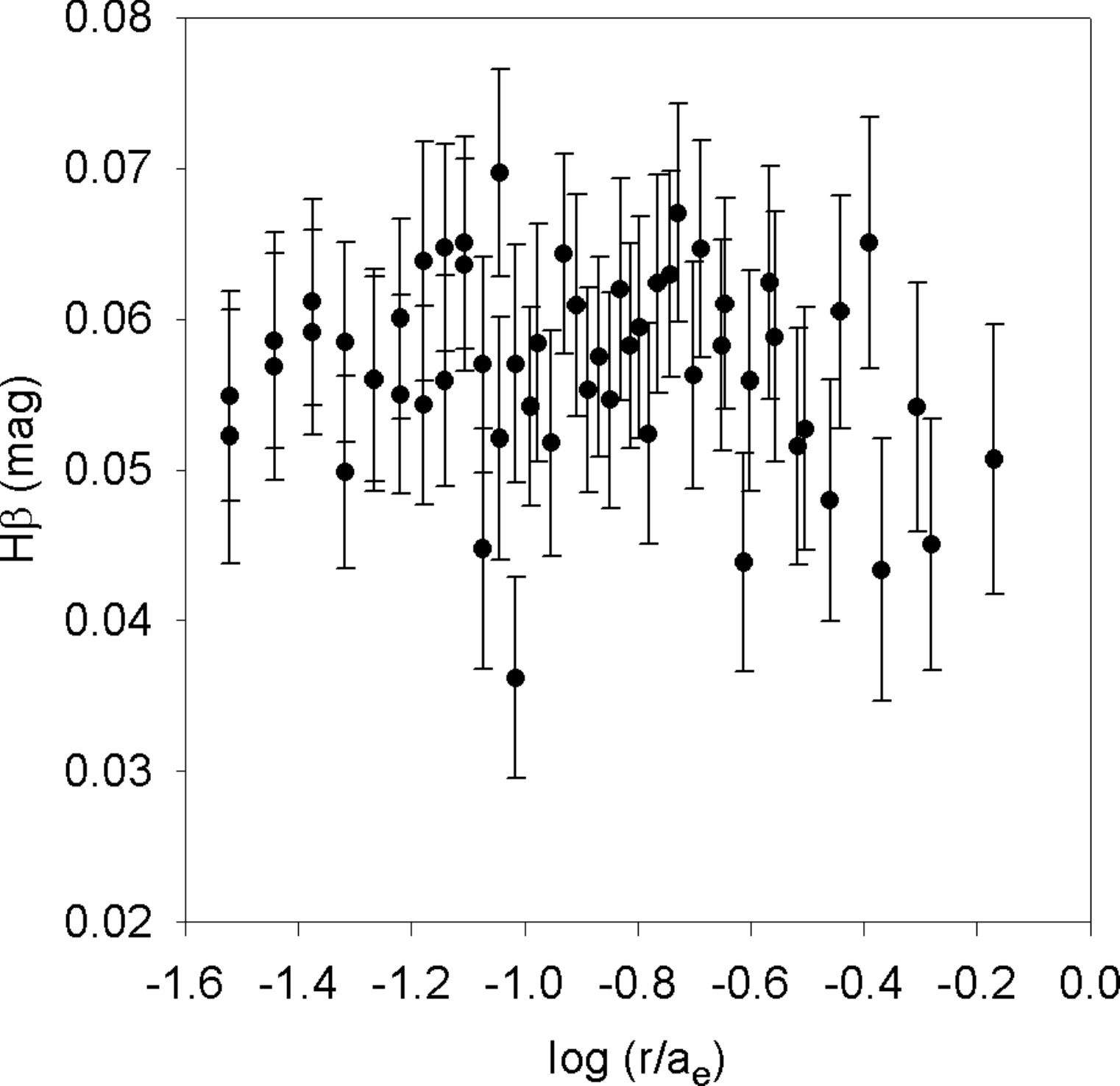}}\quad
 \subfigure[IC4765]{\includegraphics[width=3.9cm,height=3.9cm]{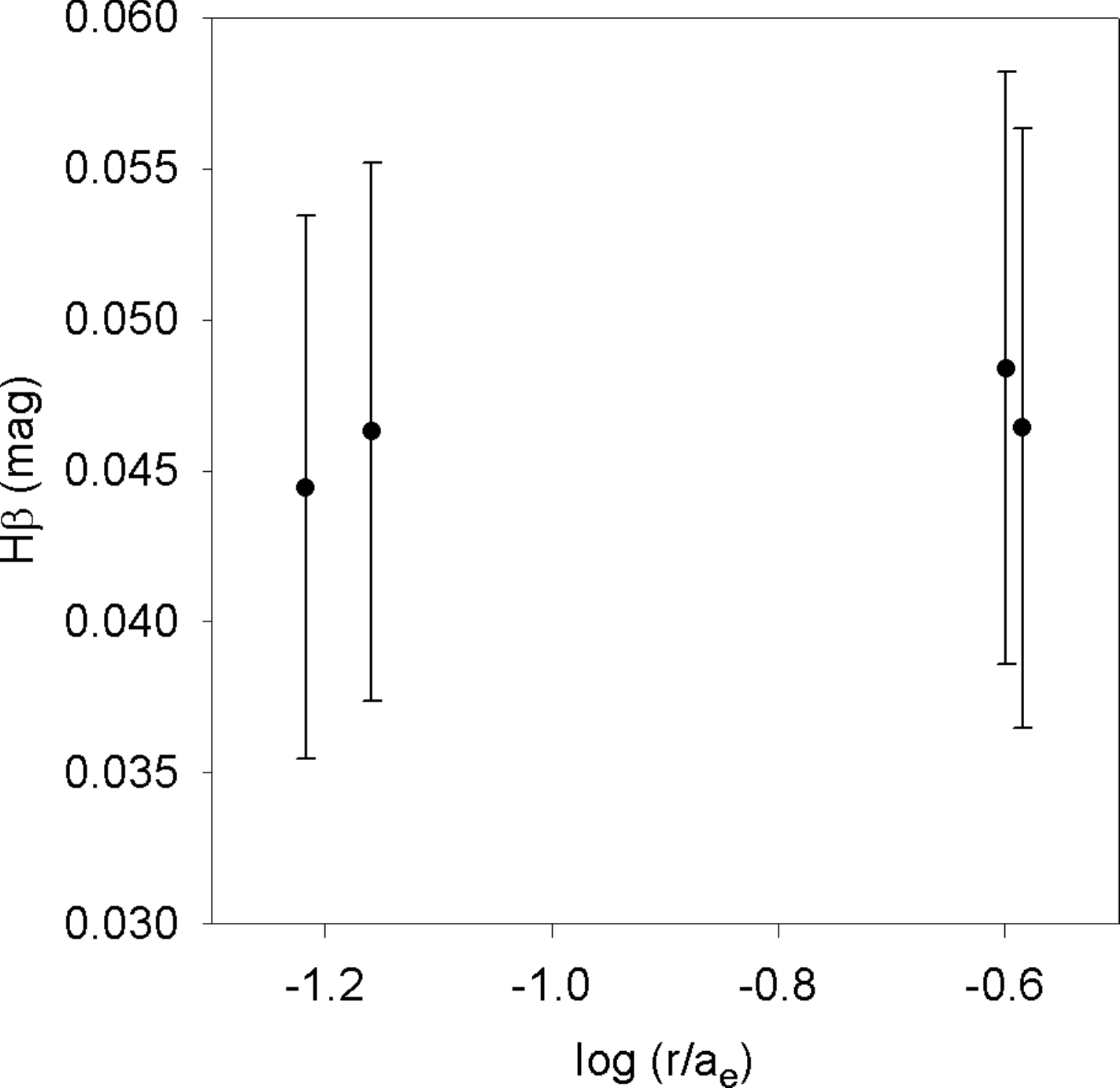}}\quad
\subfigure[IC5358]{\includegraphics[width=3.9cm,height=3.9cm]{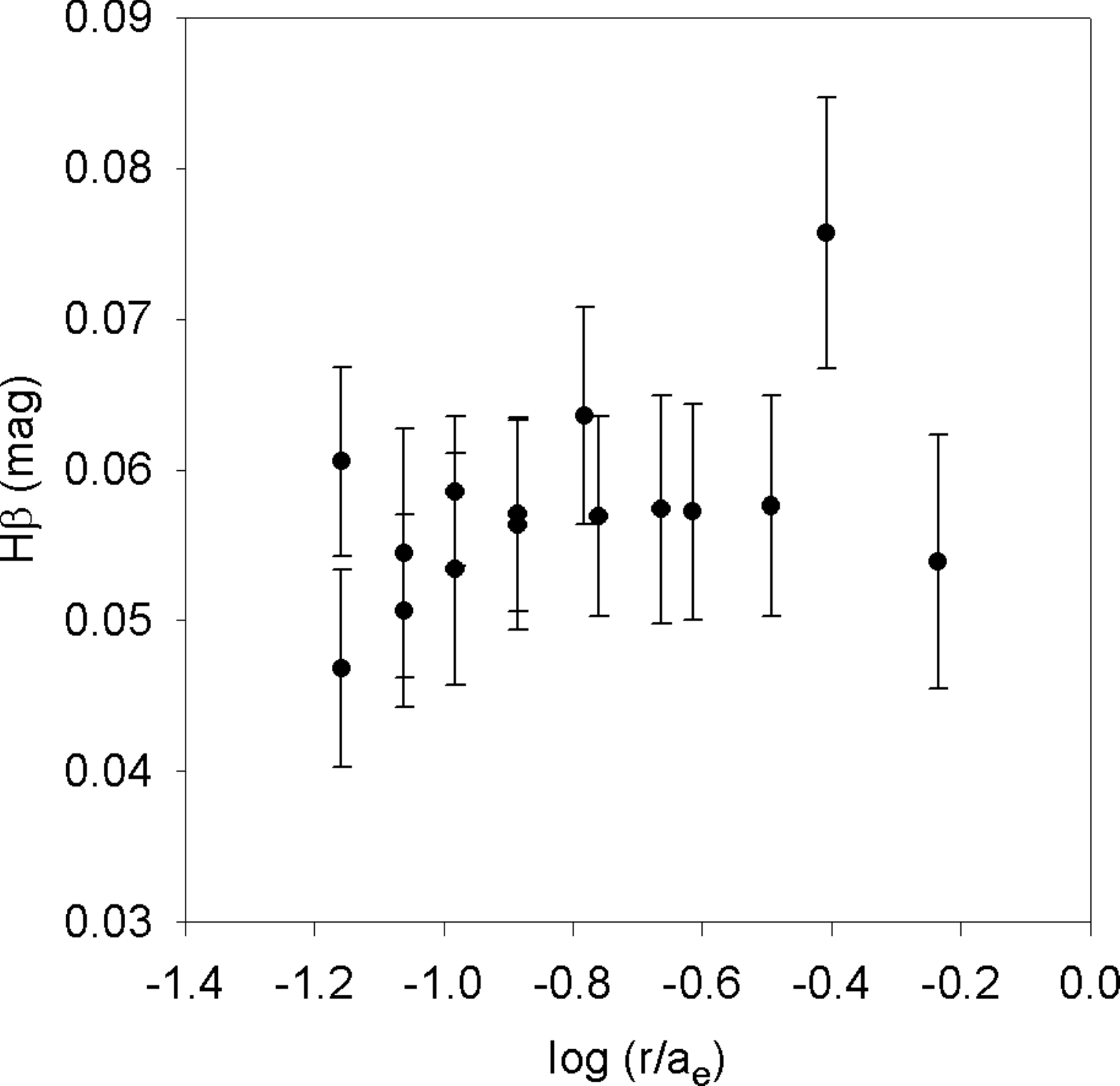}}}
 \mbox{\subfigure[Leda094683]{\includegraphics[width=3.9cm,height=3.9cm]{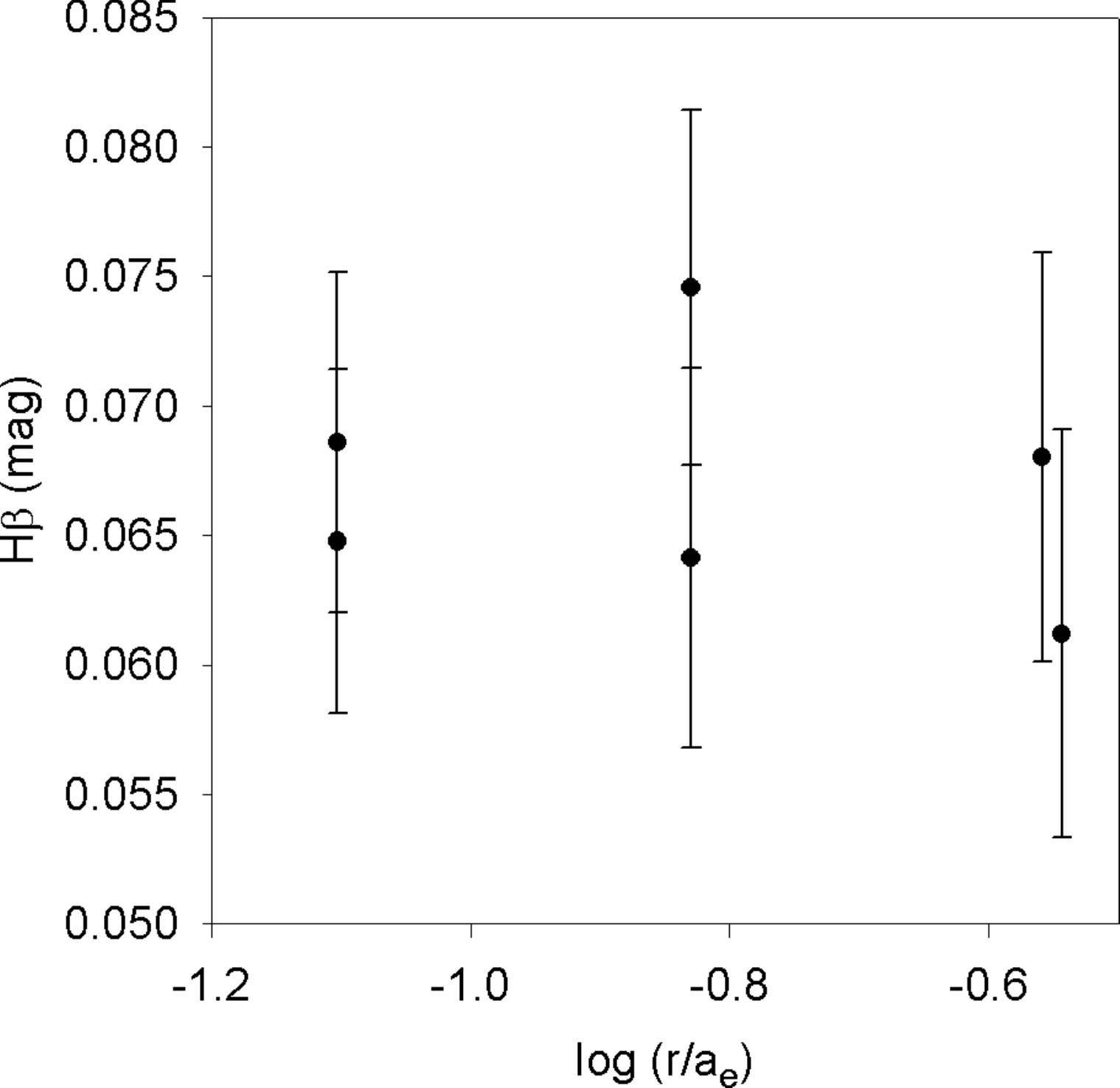}}\quad
         \subfigure[NGC0533]{\includegraphics[width=3.9cm,height=3.9cm]{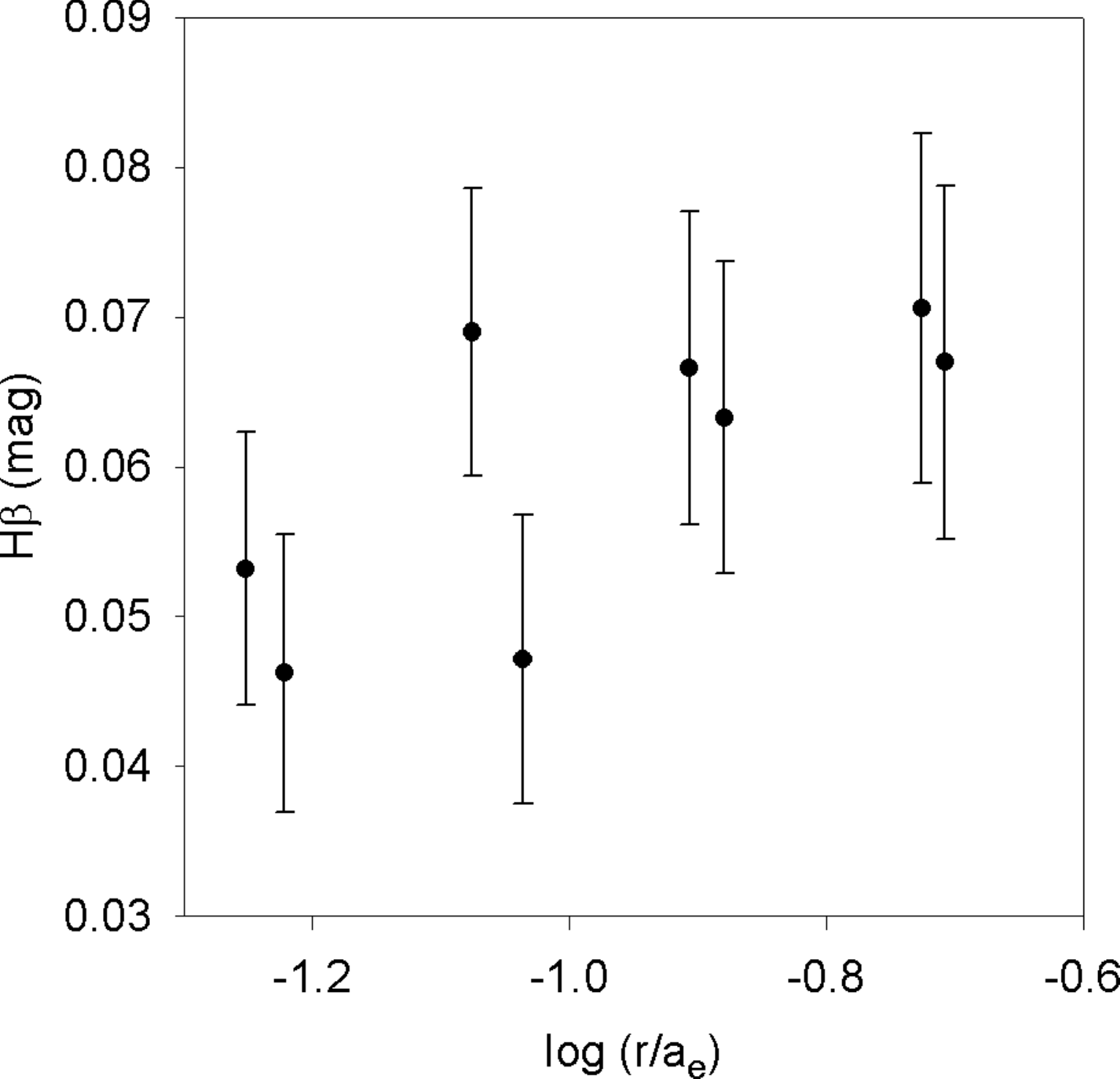}}\quad
         \subfigure[NGC0541]{\includegraphics[width=3.9cm,height=3.9cm]{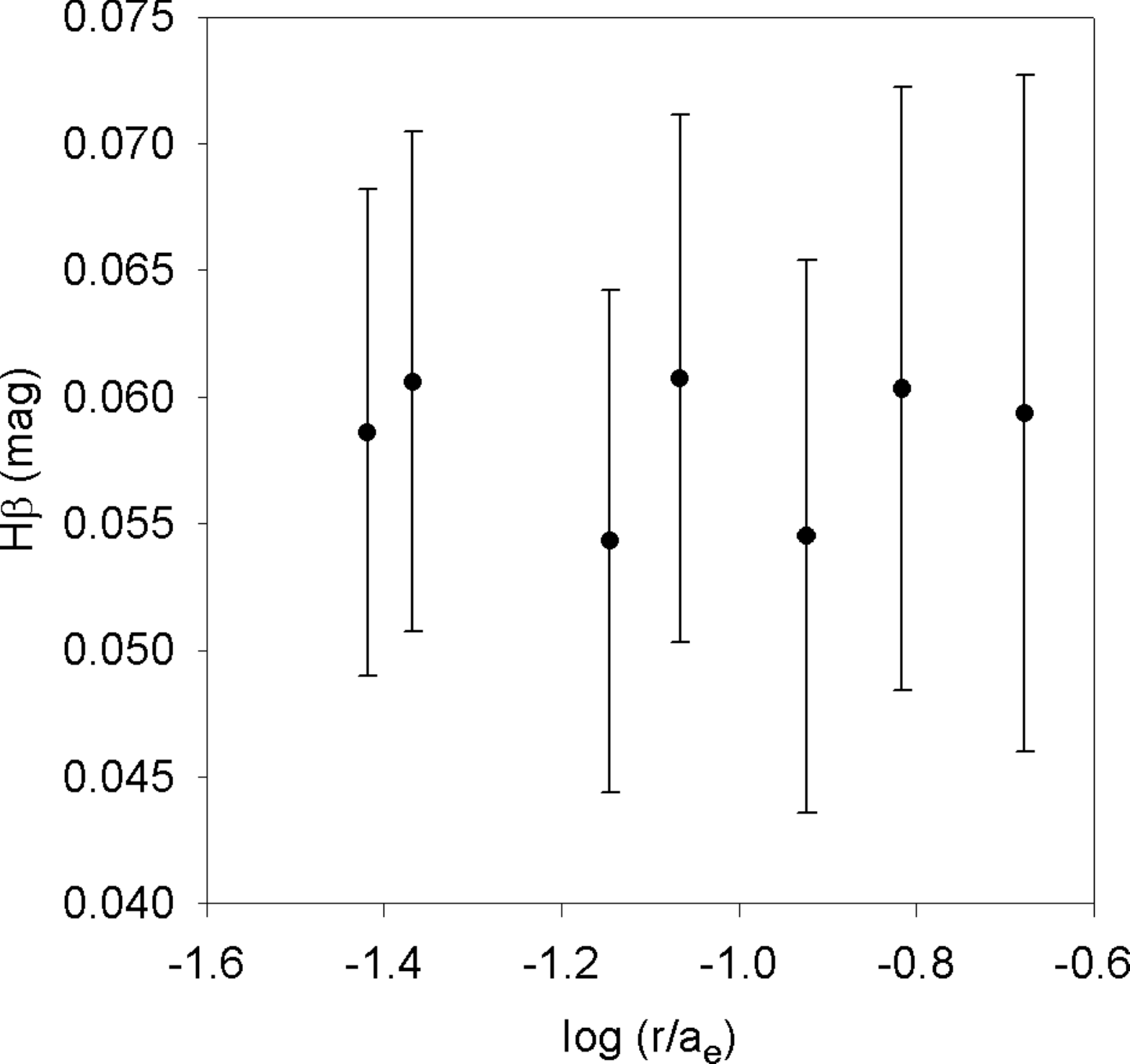}}\quad
\subfigure[NGC1399]{\includegraphics[width=3.9cm,height=3.9cm]{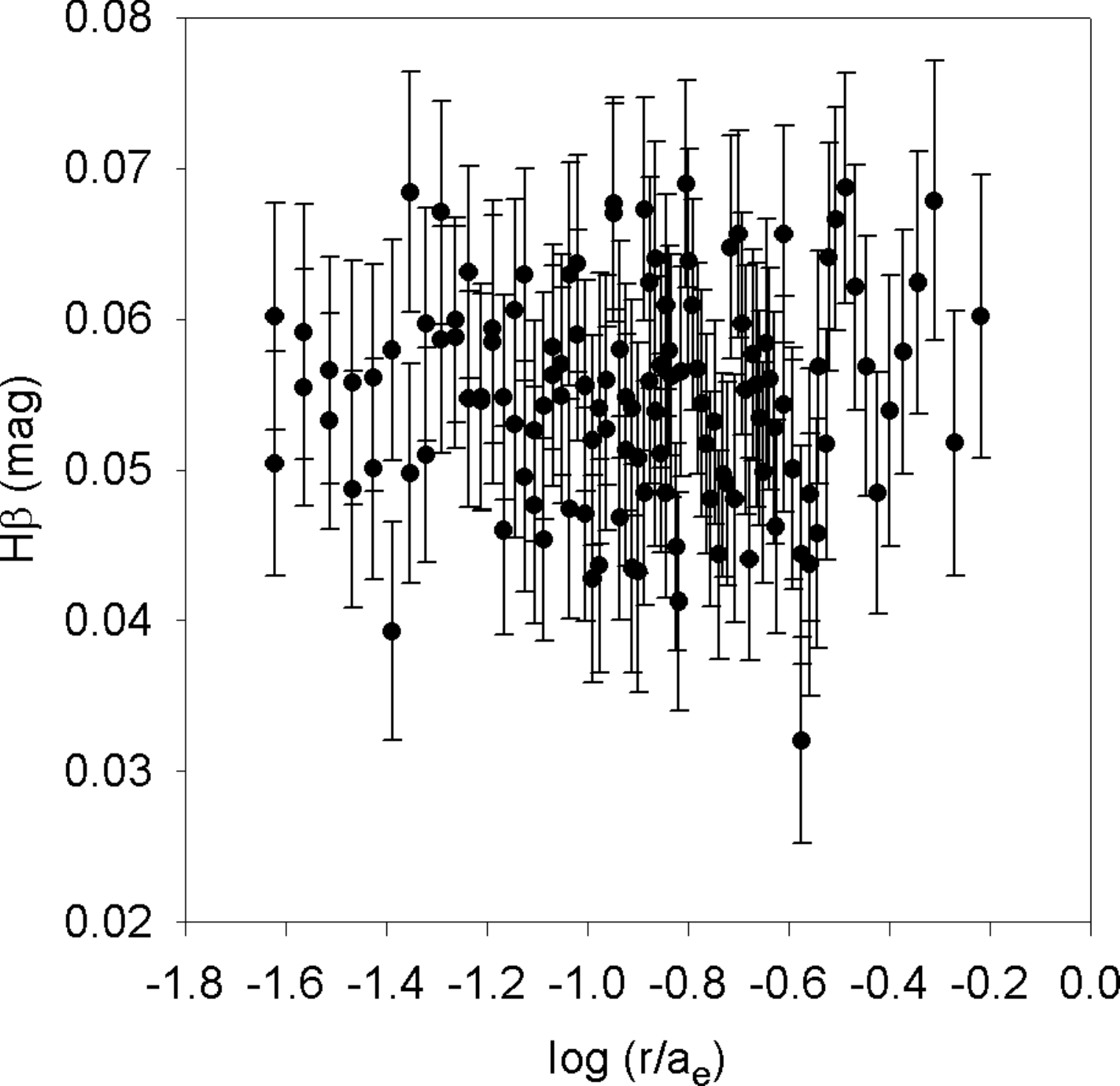}}}
\mbox{\subfigure[NGC1713]{\includegraphics[width=3.9cm,height=3.9cm]{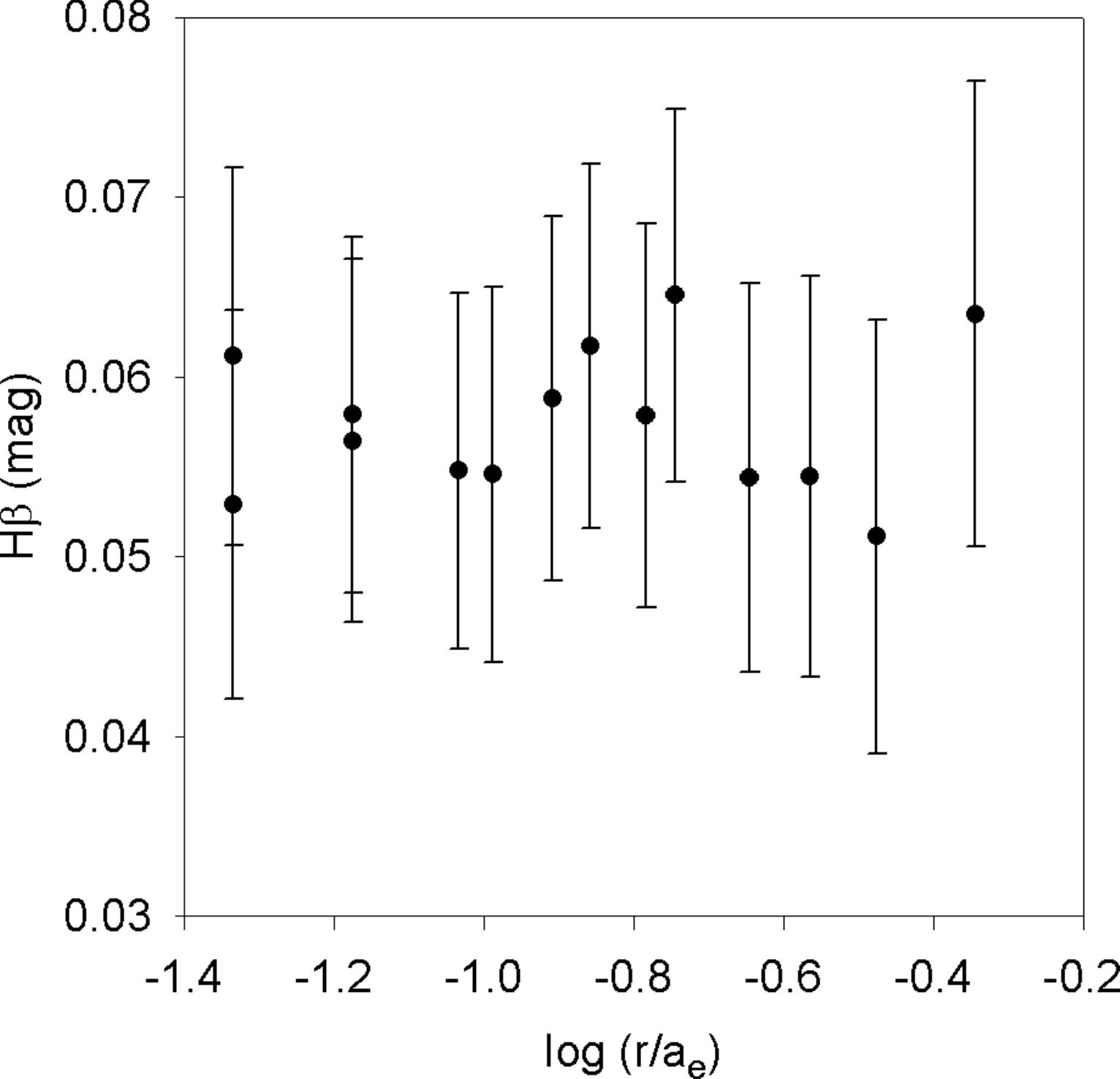}}\quad
         \subfigure[NGC2832]{\includegraphics[width=3.9cm,height=3.9cm]{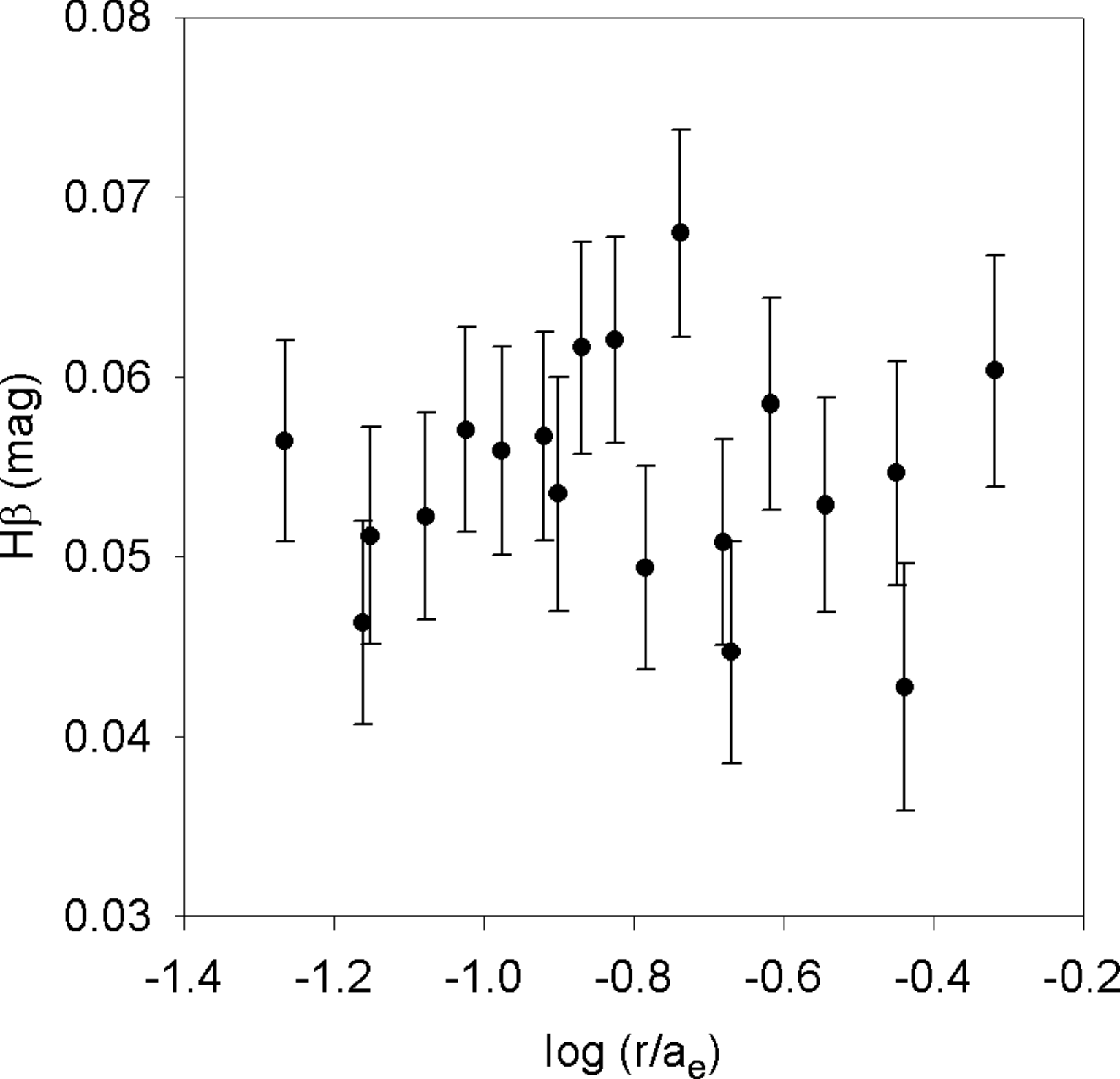}}\quad
         \subfigure[NGC3311]{\includegraphics[width=3.9cm,height=3.9cm]{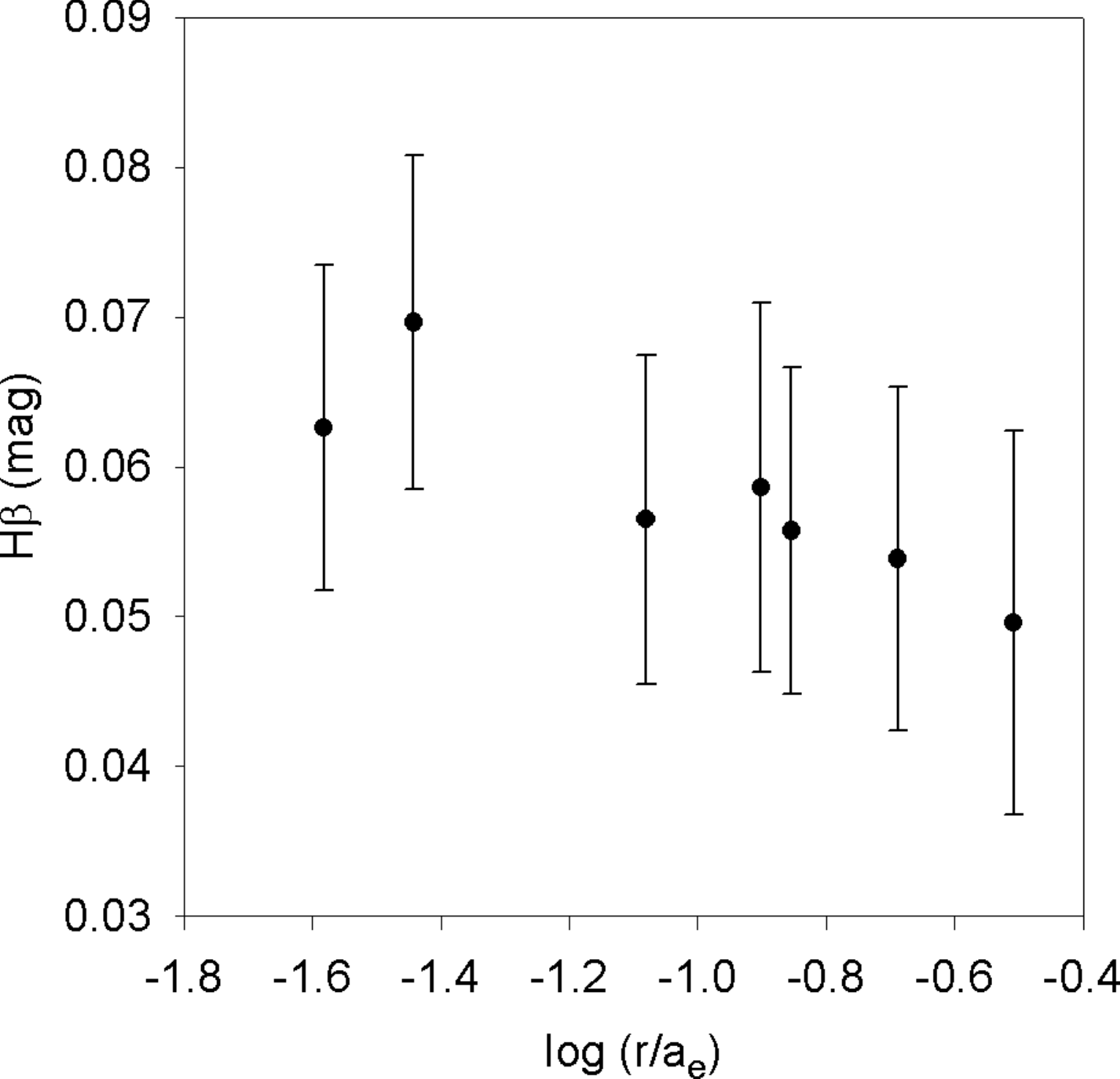}}\quad
\subfigure[NGC4839]{\includegraphics[width=3.9cm,height=3.9cm]{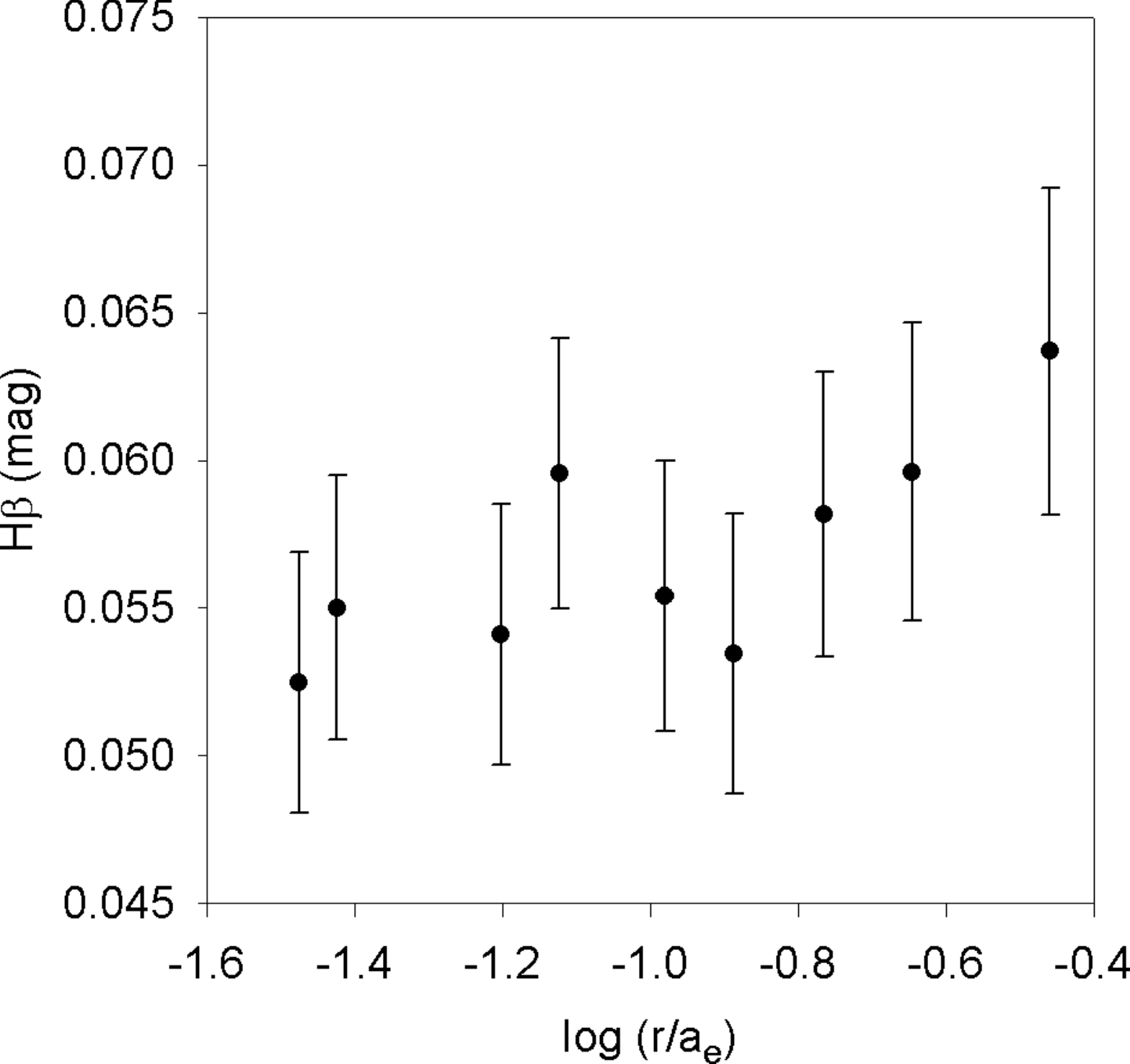}}}
\caption[]{BCG H$\beta$ gradients. All data is folded with respect to the galaxy centres. The central 0.5 arcsec, to each side (thus in total the central 1 arcsec, comparable to the  seeing), were excluded in the figure as well as in the fitted correlations.}
   \label{Fig:Hbeta1}
\end{figure*}

\begin{figure*}
   \centering
\mbox{\subfigure[NGC6173]{\includegraphics[width=3.9cm,height=3.9cm]{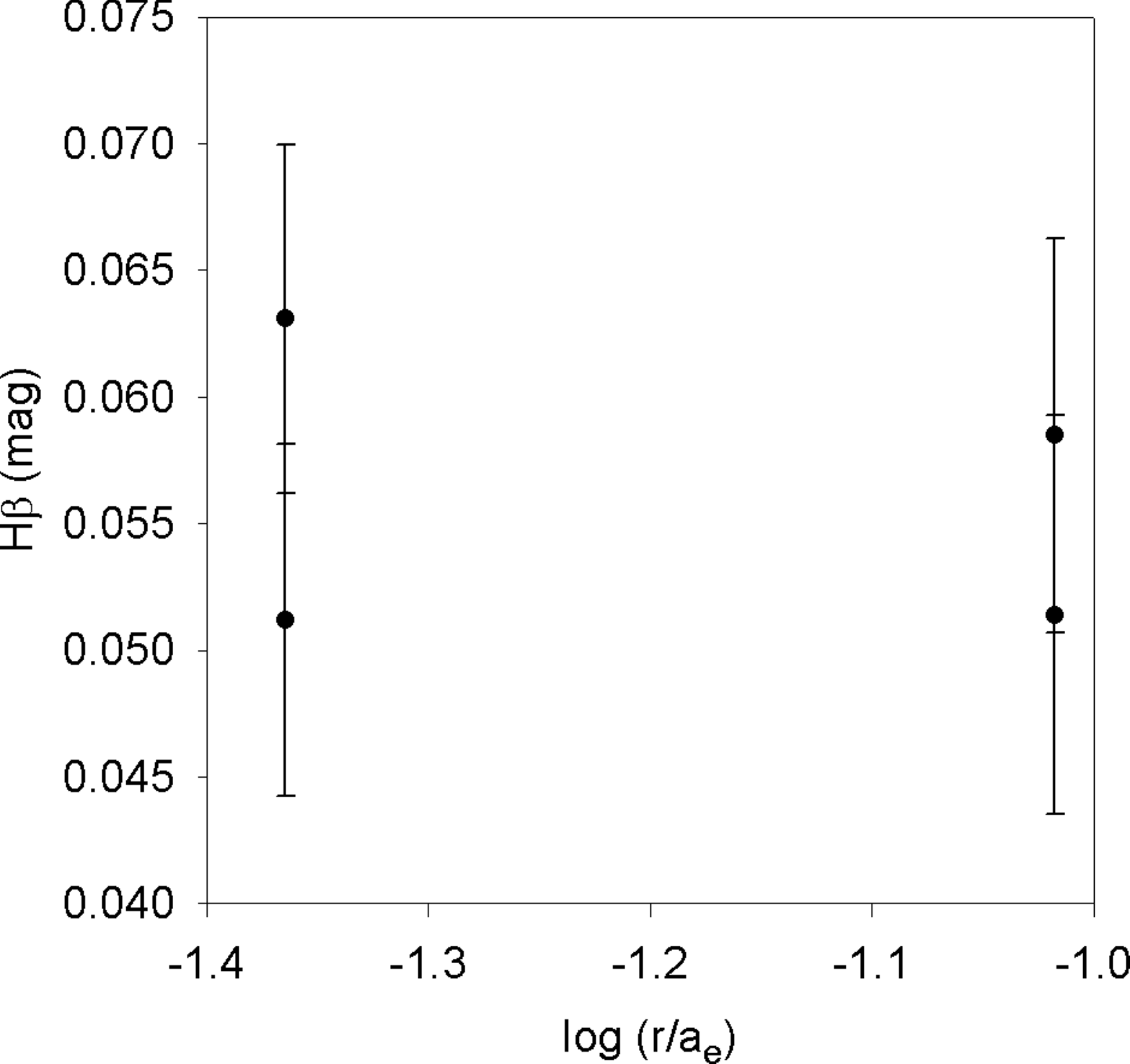}}\quad
         \subfigure[NGC6269]{\includegraphics[width=3.9cm,height=3.9cm]{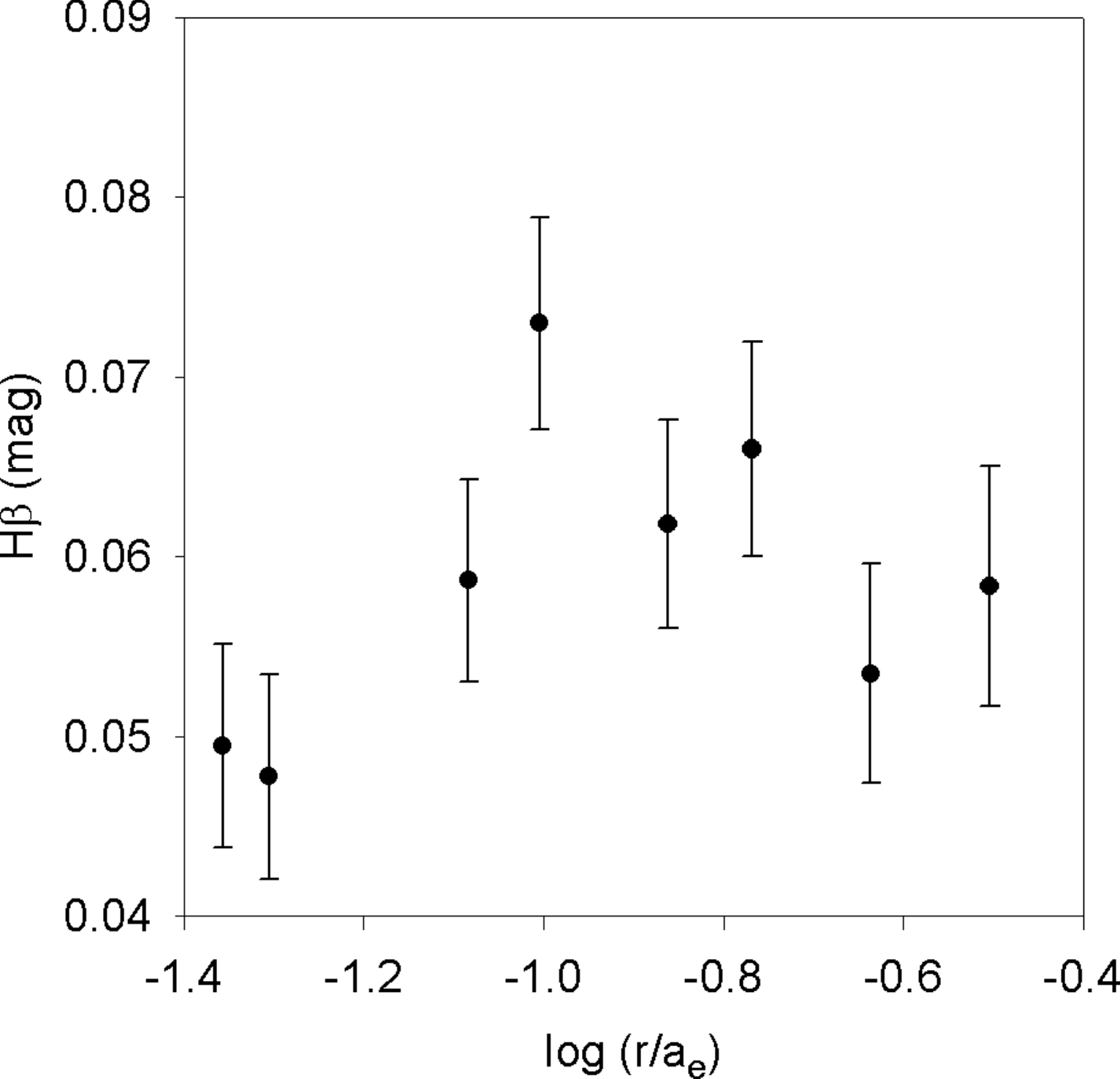}}\quad
         \subfigure[NGC7012]{\includegraphics[width=3.9cm,height=3.9cm]{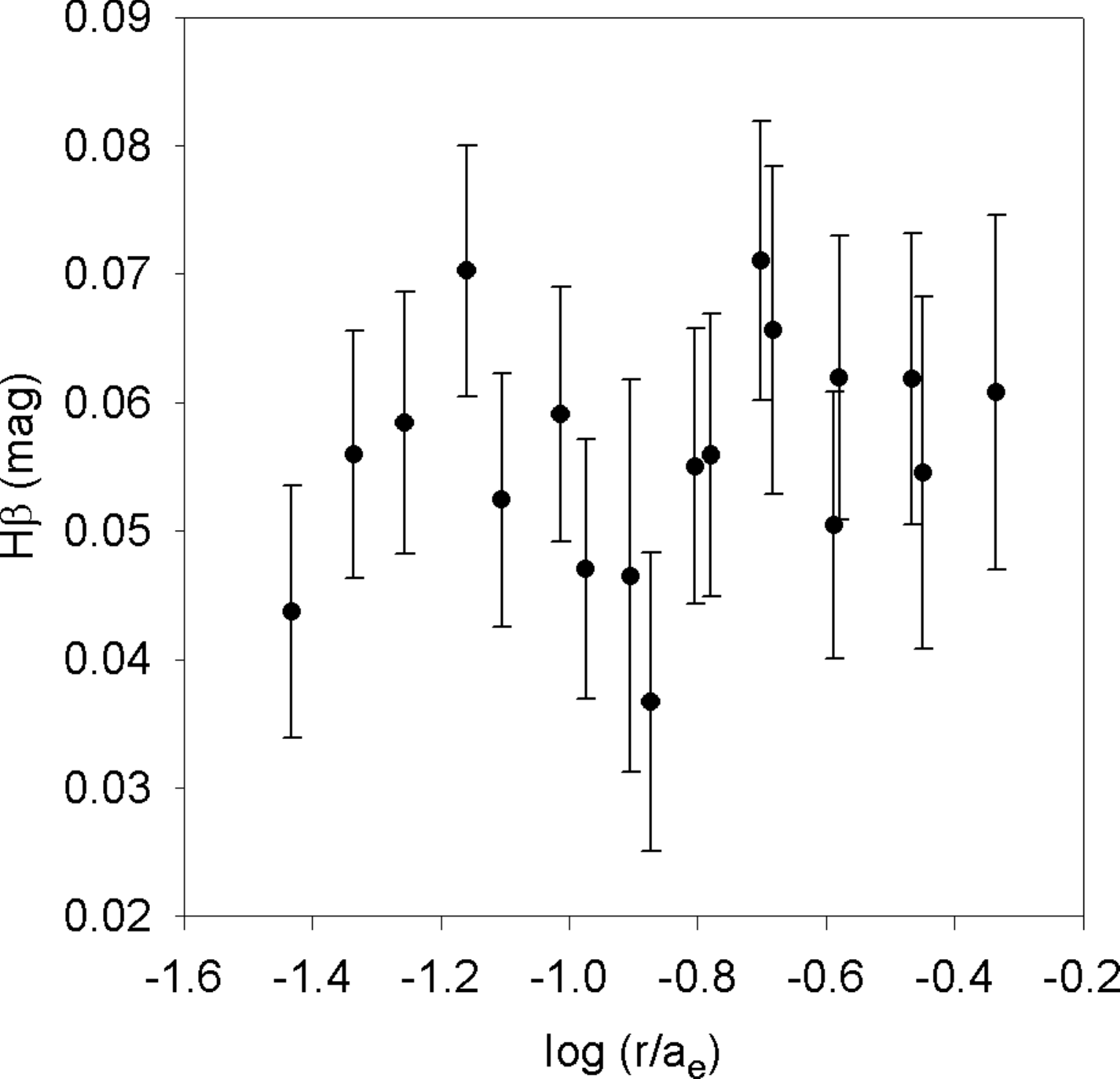}}\quad
\subfigure[PGC004072]{\includegraphics[width=3.9cm,height=3.9cm]{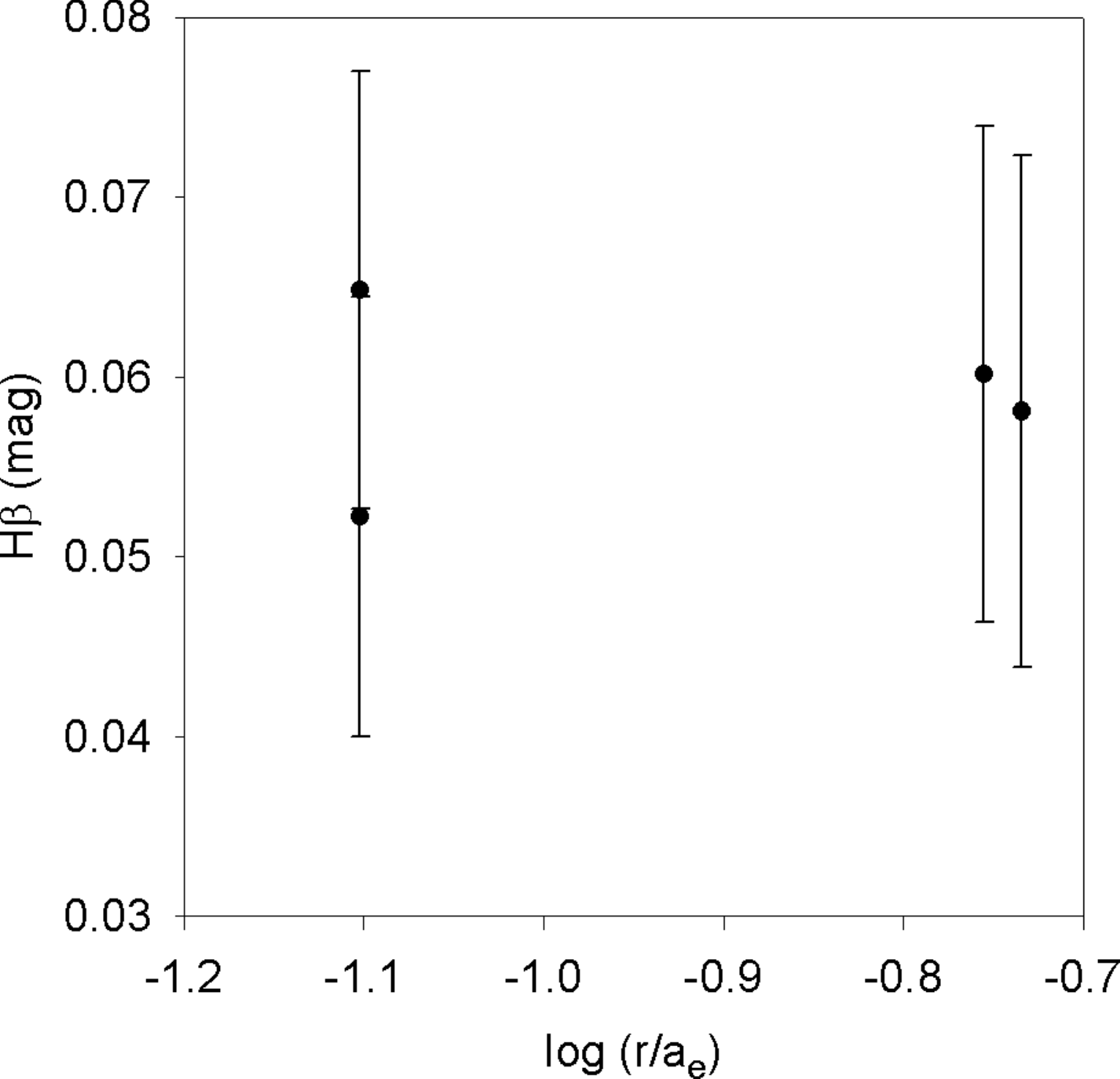}}}
\mbox{\subfigure[PGC030223]{\includegraphics[width=4.0cm,height=4.0cm]{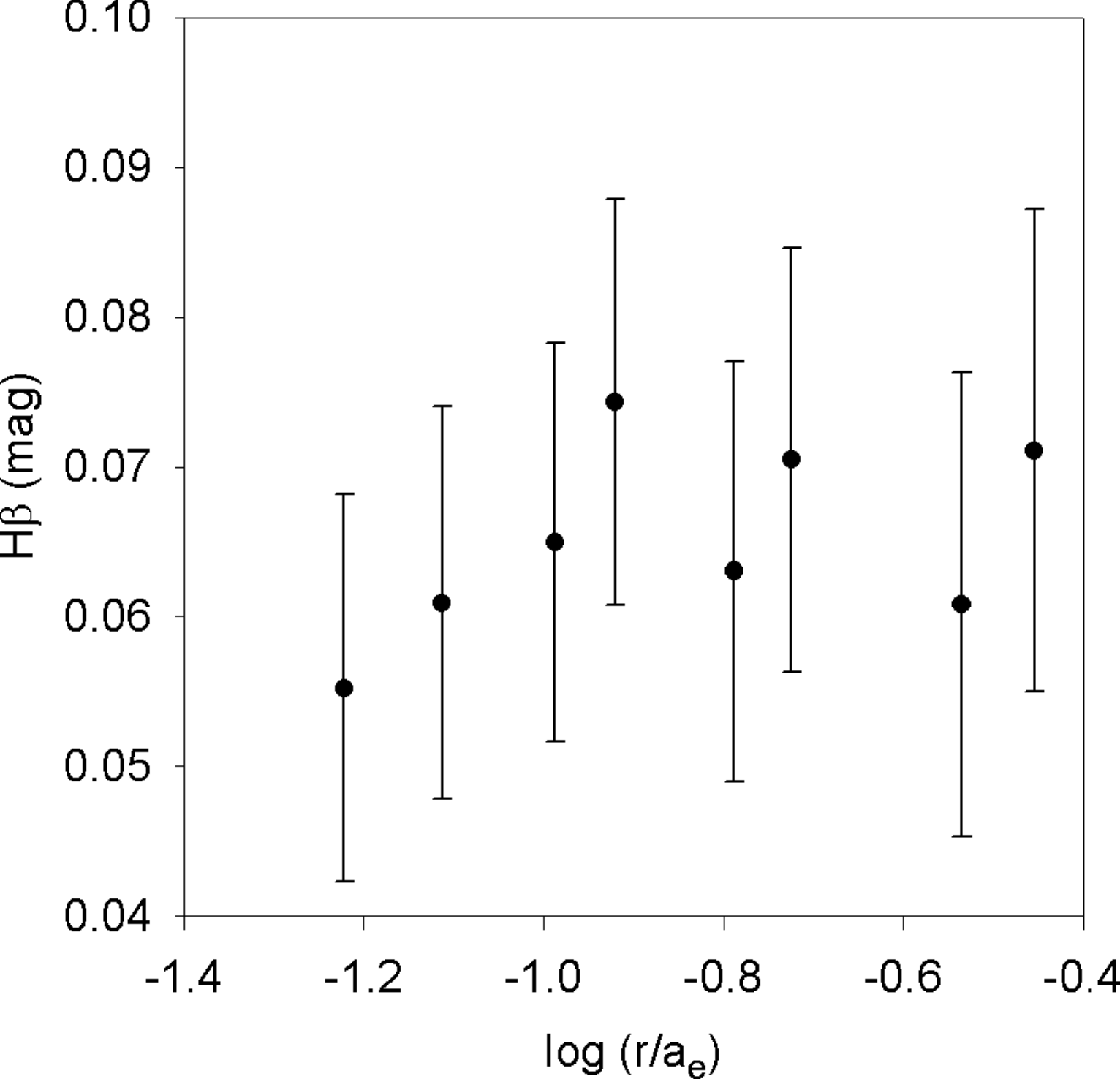}}\quad
         \subfigure[PGC072804]{\includegraphics[width=4.0cm,height=4.0cm]{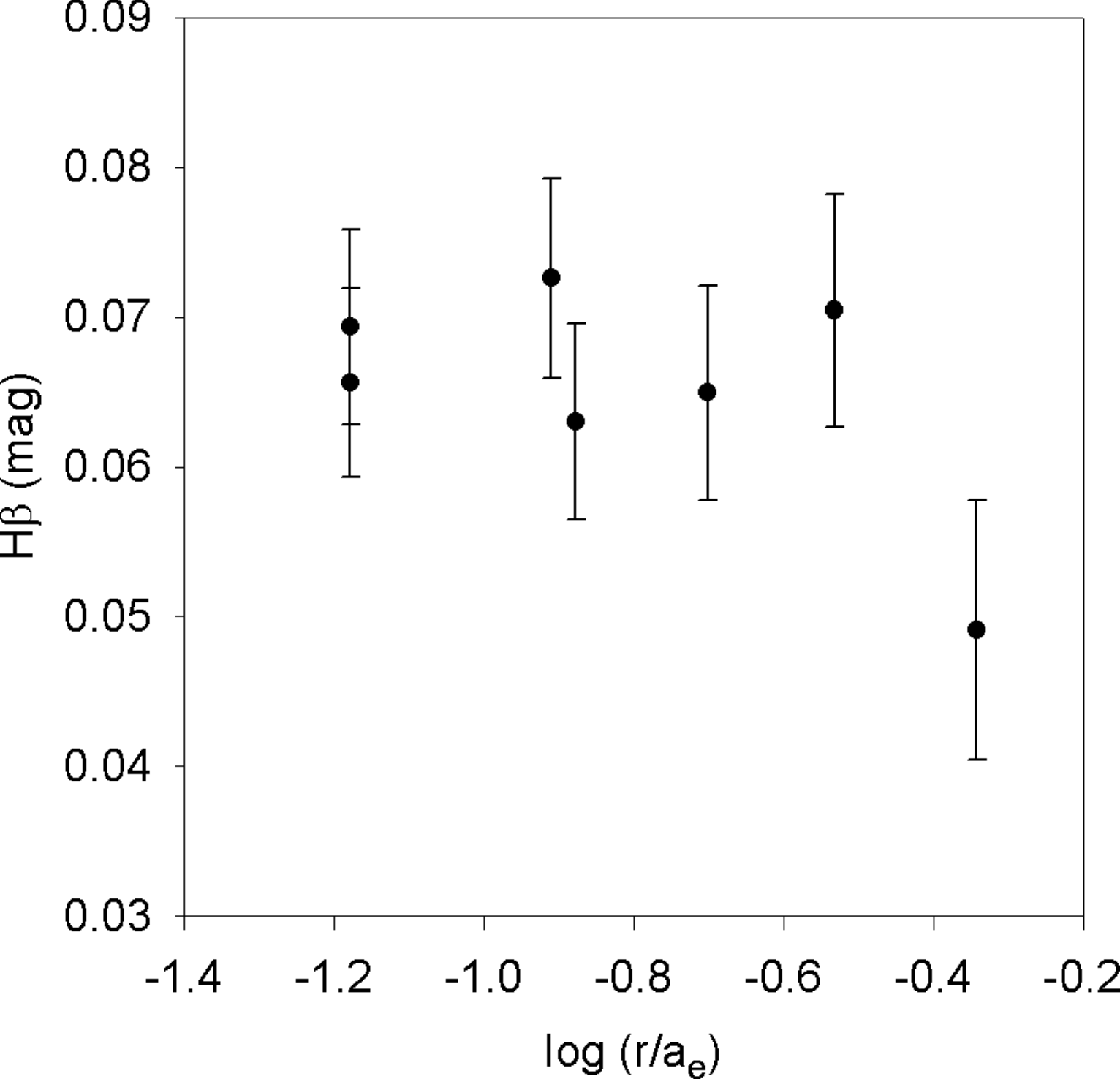}}\quad
         \subfigure[UGC02232]{\includegraphics[width=4.0cm,height=4.0cm]{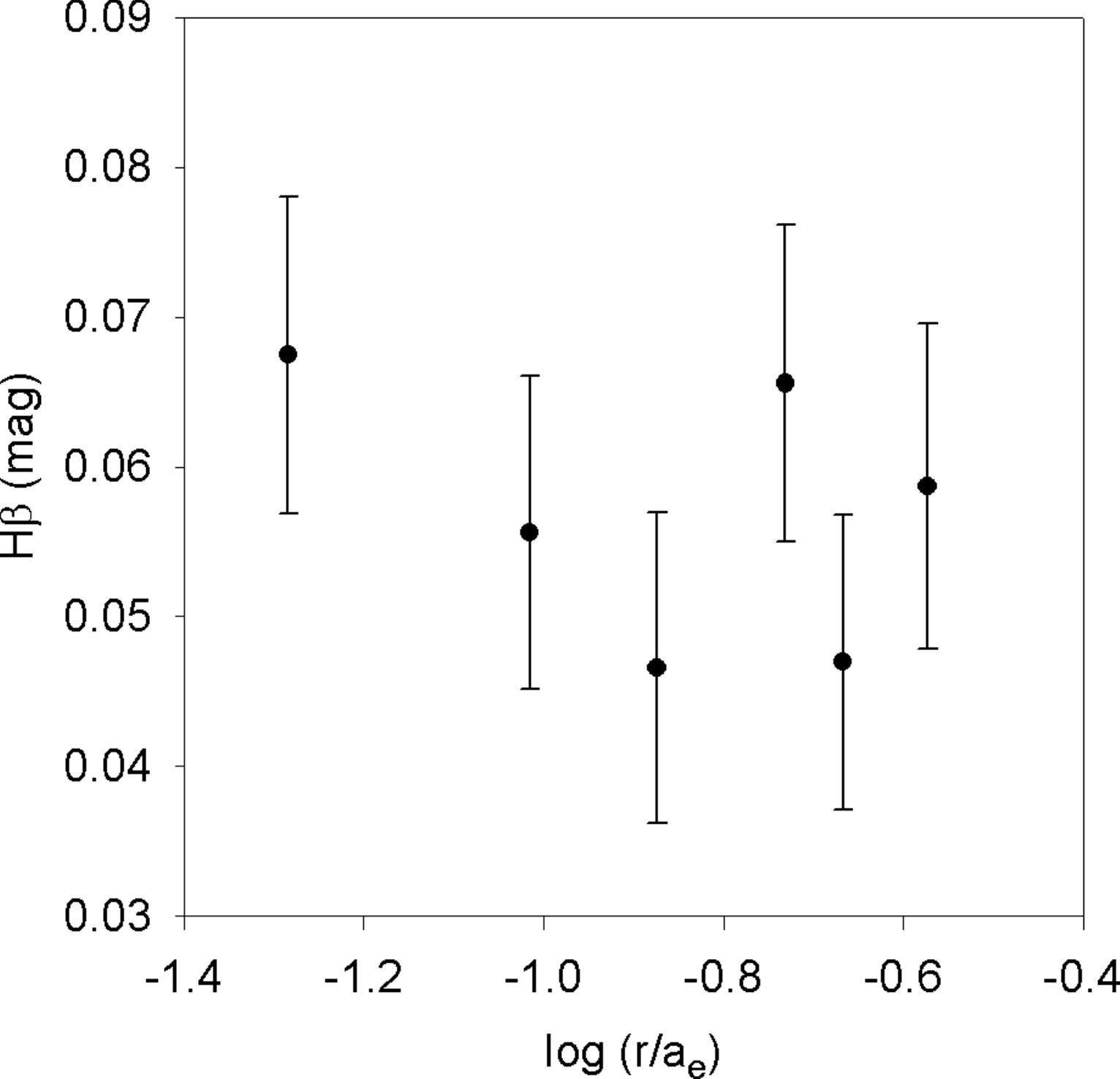}}\quad
\subfigure[UGC05515]{\includegraphics[width=4.0cm,height=4.0cm]{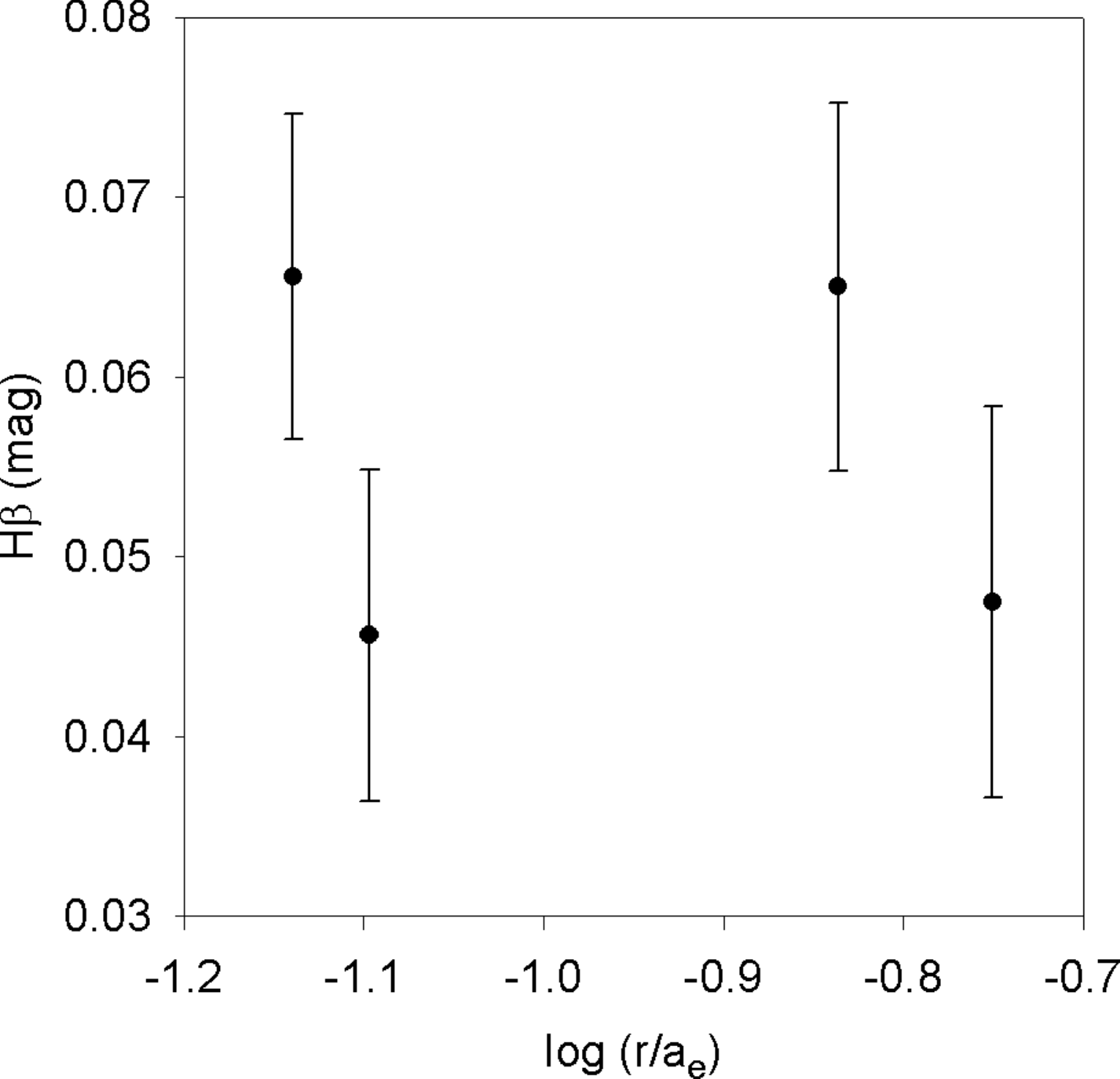}}}
\caption[]{BCG H$\beta$ gradients continue.}
   \label{Fig:Hbeta2}
\end{figure*}

\begin{figure*}
   \centering
   \mbox{\subfigure[ESO146-028]{\includegraphics[scale=0.5]{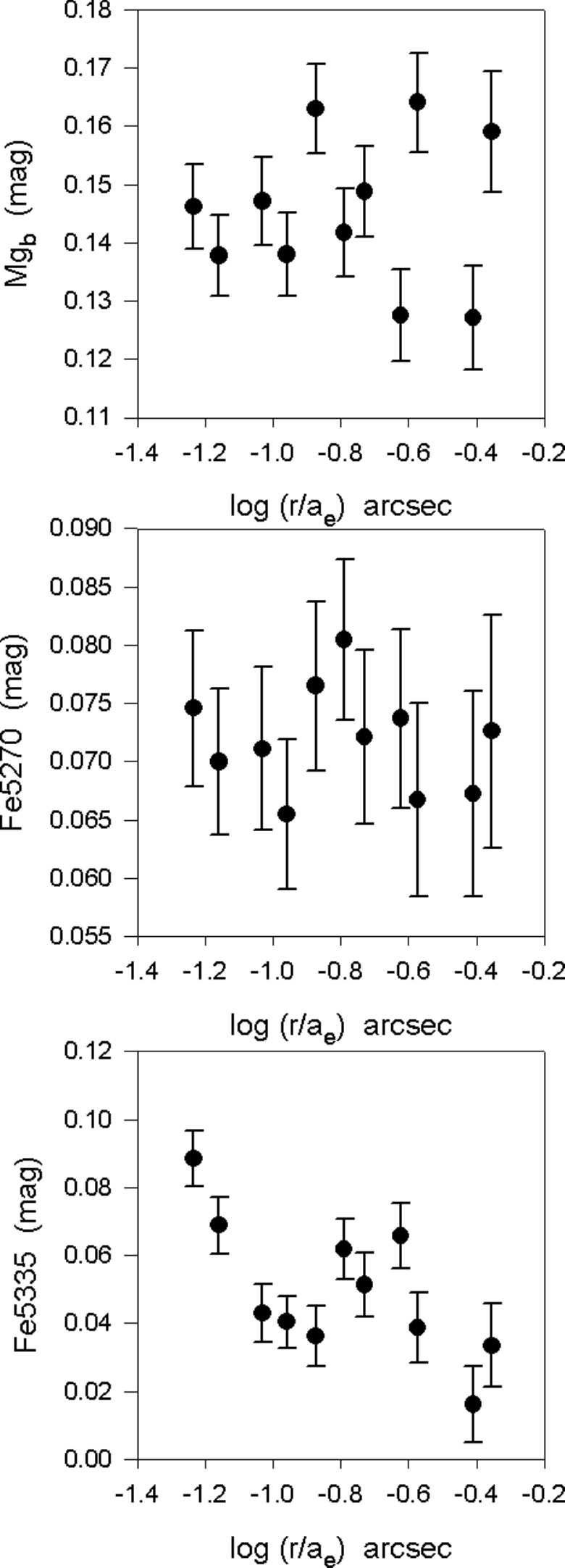}}\quad
         \subfigure[ESO303-005]{\includegraphics[scale=0.5]{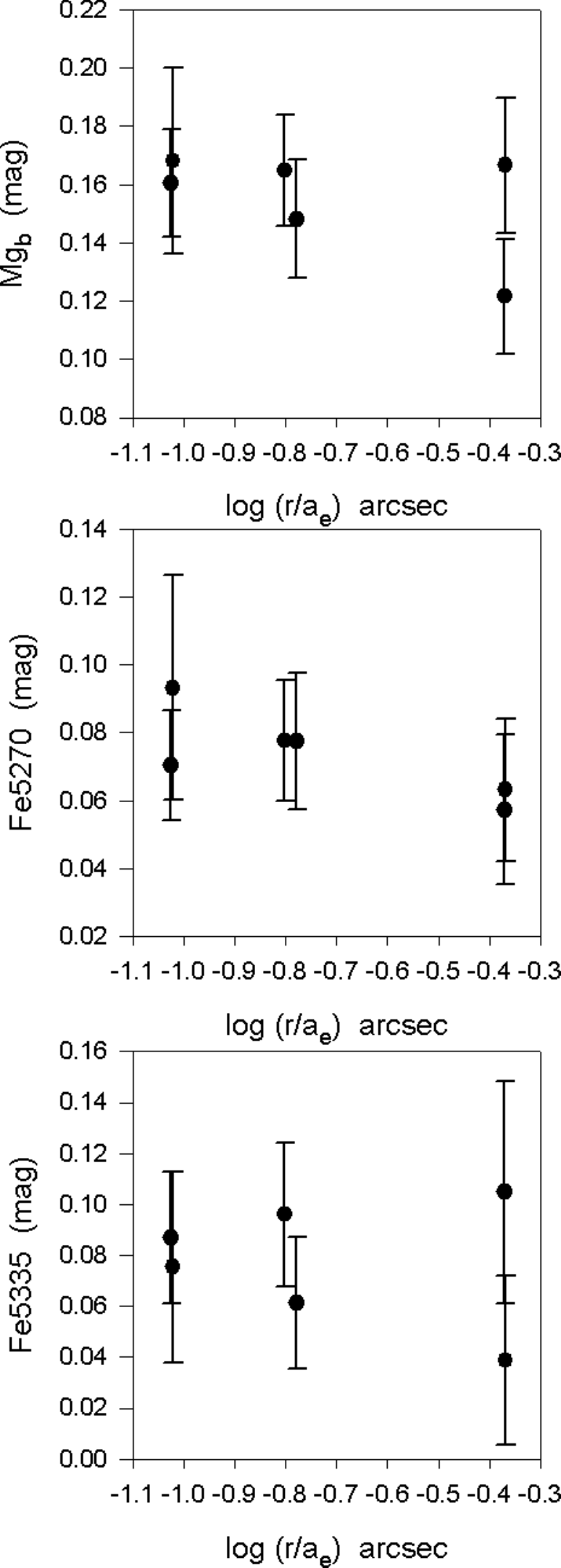}}\quad
         \subfigure[ESO488-027]{\includegraphics[scale=0.5]{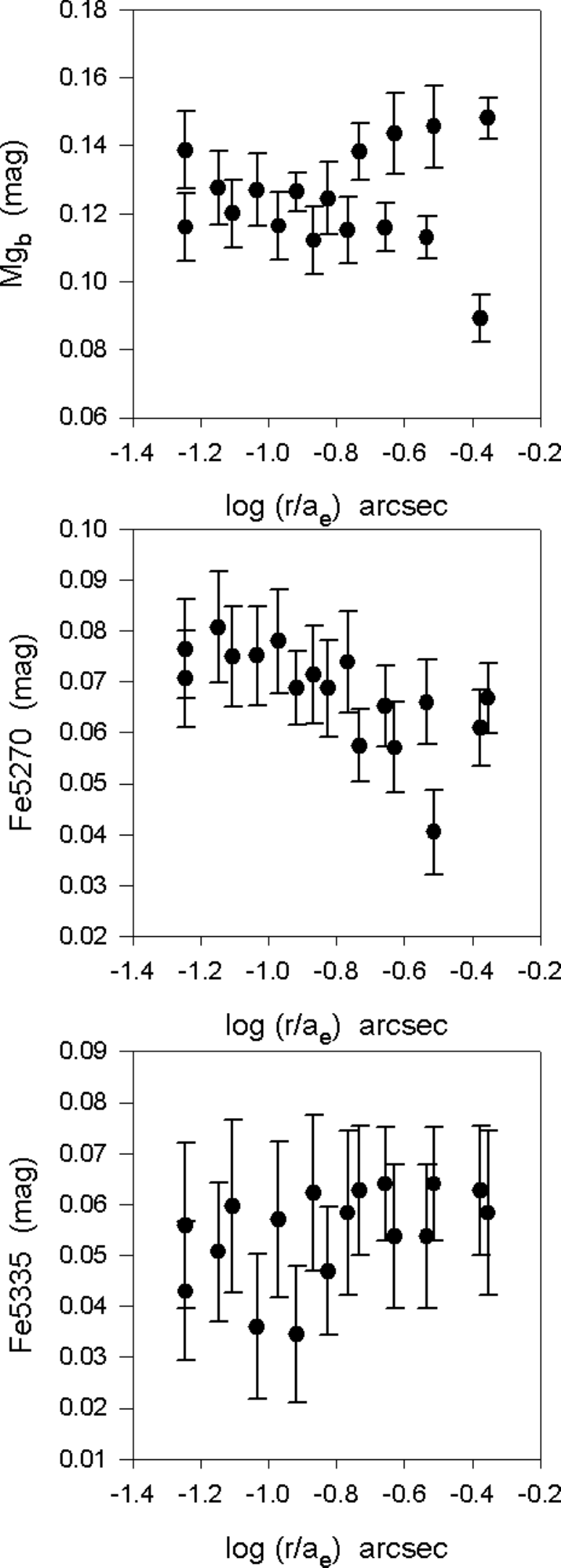}}\quad
         \subfigure[ESO552-020]{\includegraphics[scale=0.5]{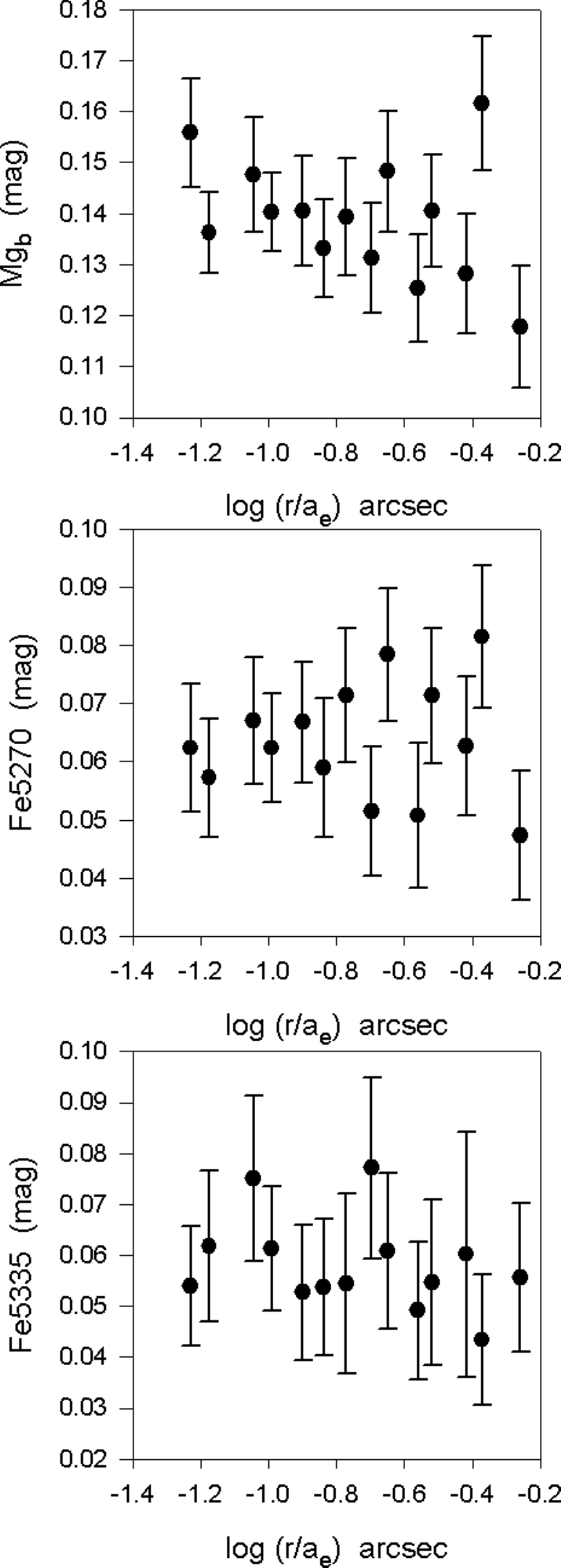}}}
\caption[]{BCG Mg$_{b}$, Fe5270 and Fe5335 gradients. All data is folded with respect to the galaxy centres. The central 0.5 arcsec, to each side (thus in total the central 1 arcsec, comparable to the  seeing), were excluded in the figure as well as in the fitted correlations.}
   \label{Fig:Indices}
\end{figure*}

\begin{figure*}
   \centering
   \mbox{\subfigure[GSC555700266]{\includegraphics[scale=0.47]{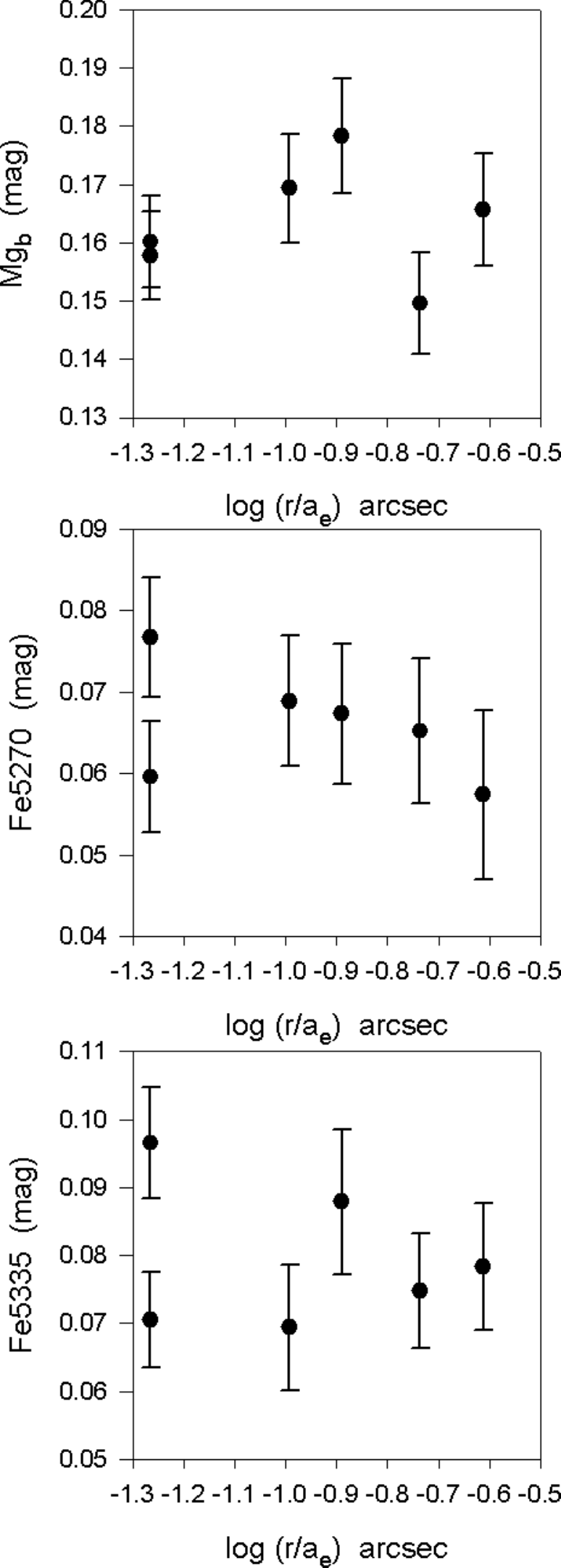}}\quad
         \subfigure[IC1633]{\includegraphics[scale=0.47]{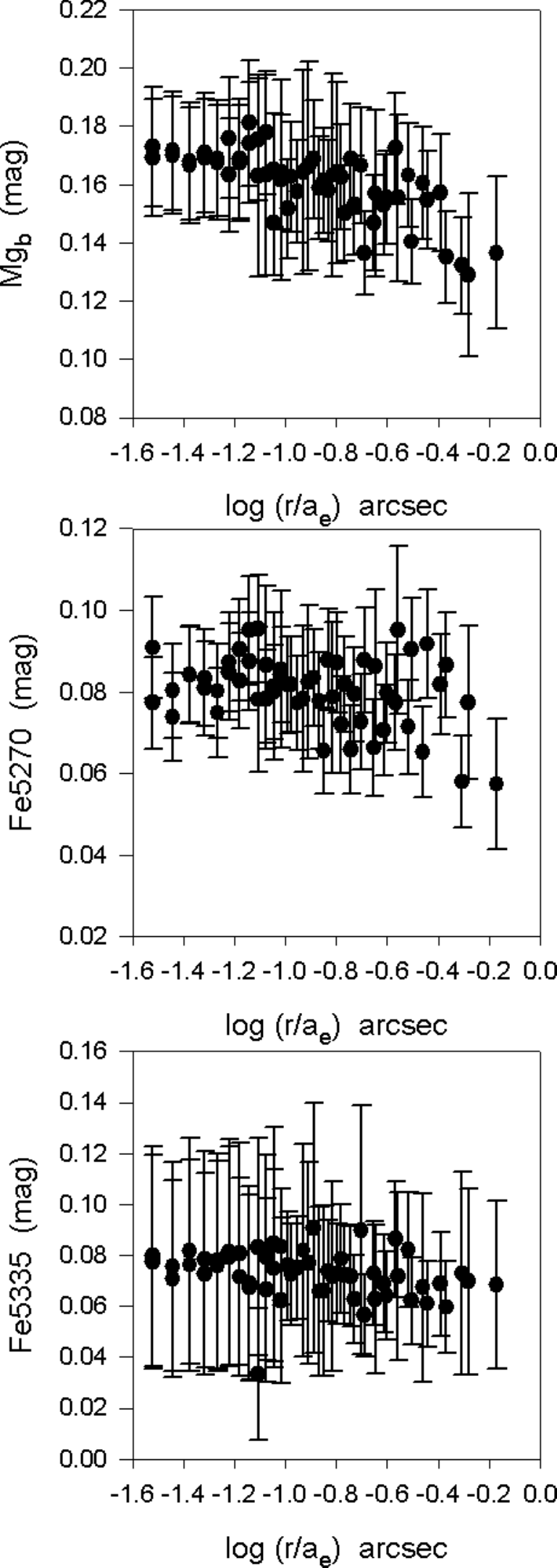}}\quad
         \subfigure[IC4765]{\includegraphics[scale=0.47]{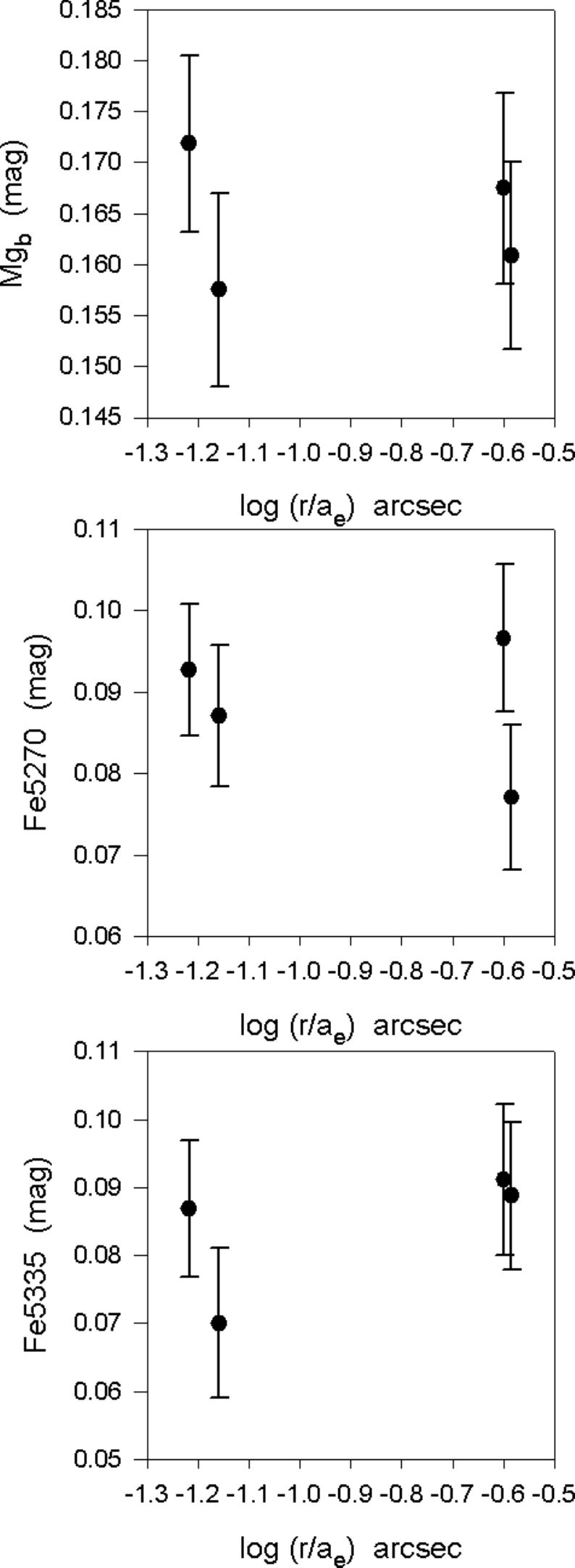}}\quad
         \subfigure[IC5358]{\includegraphics[scale=0.47]{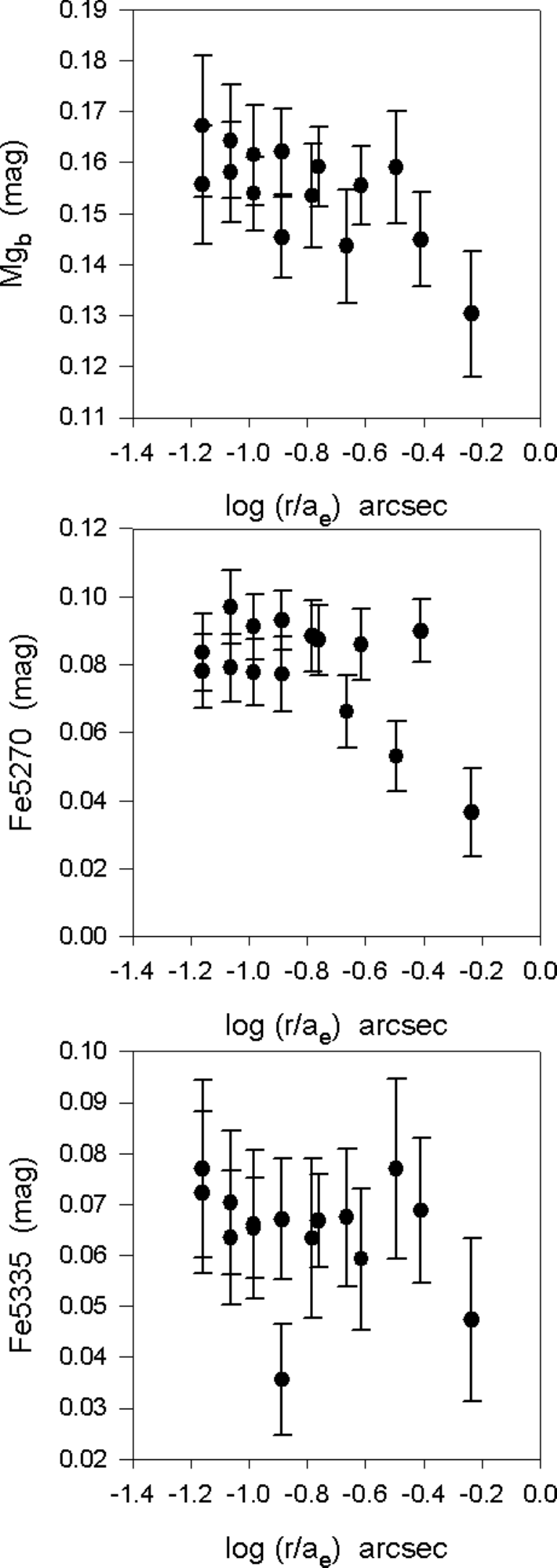}}}
\mbox{\subfigure[Leda094683]{\includegraphics[scale=0.47]{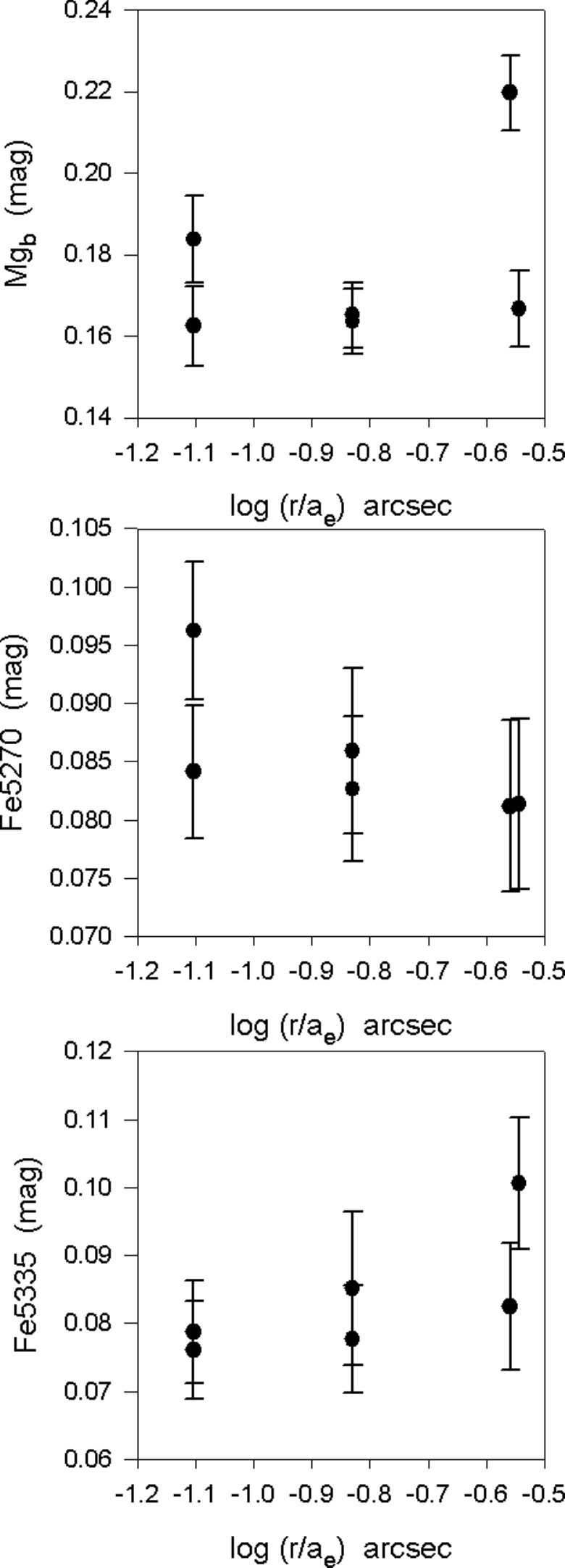}}\quad
         \subfigure[NGC0533]{\includegraphics[scale=0.47]{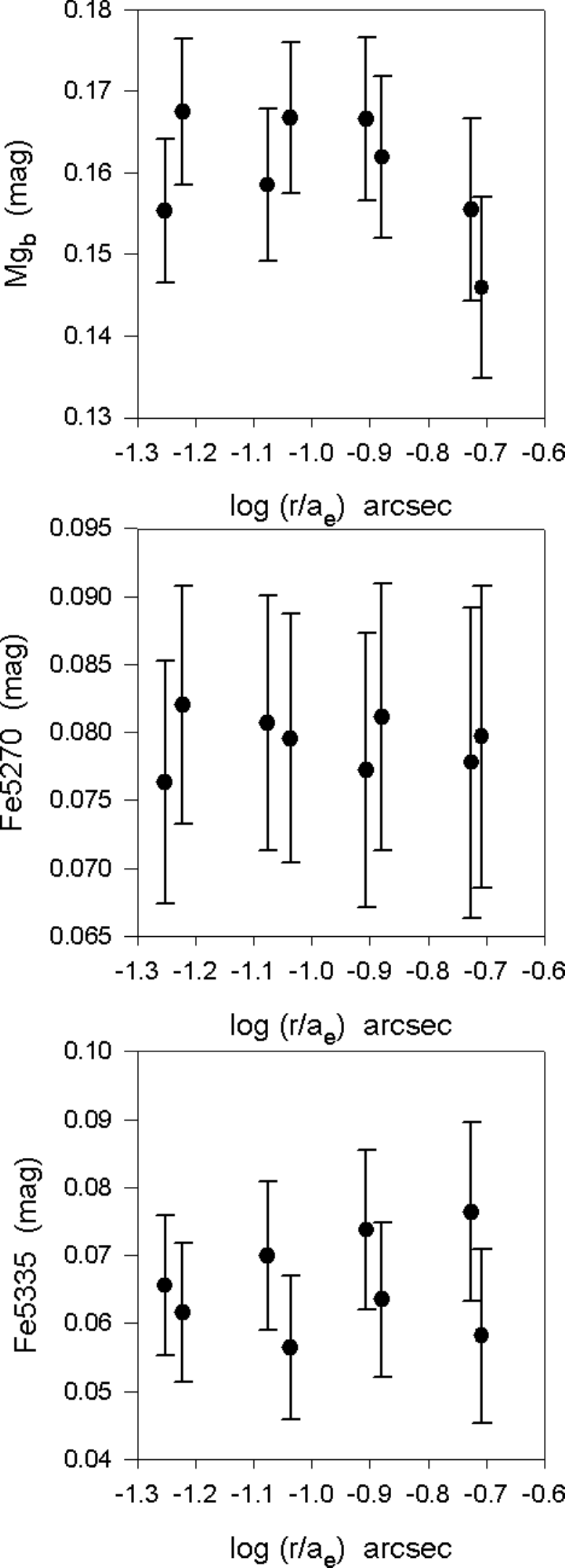}}\quad
         \subfigure[NGC0541]{\includegraphics[scale=0.47]{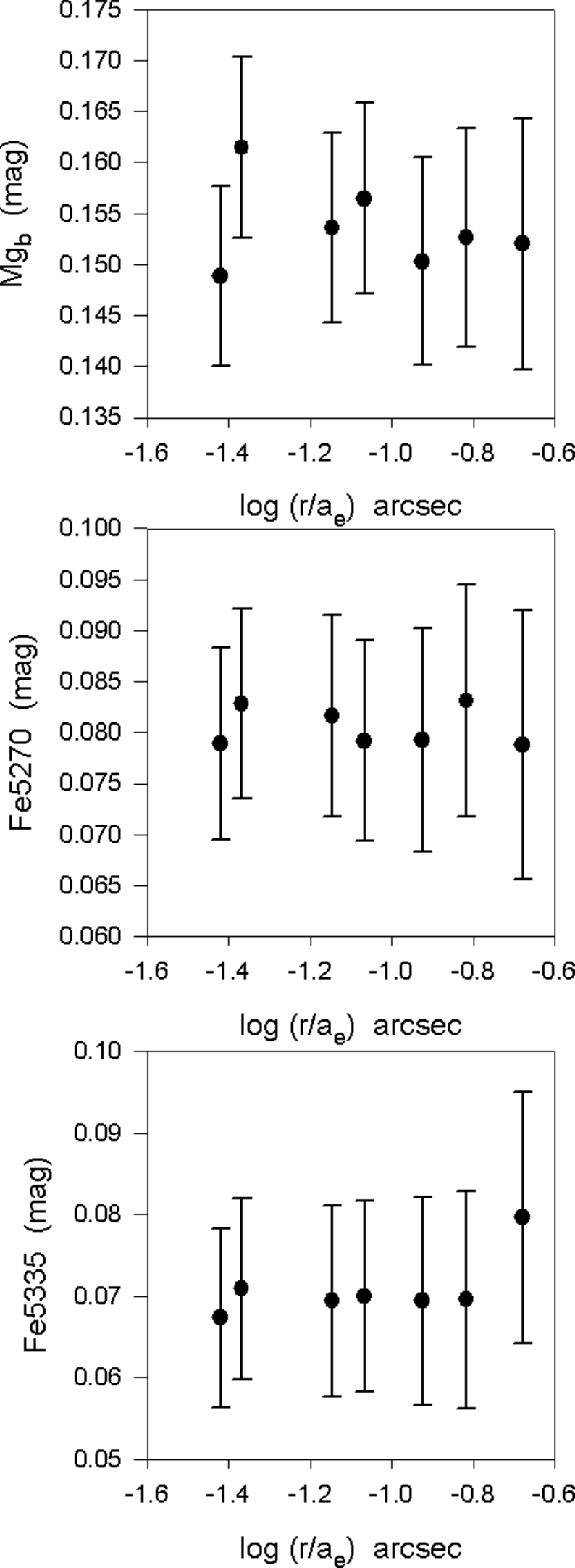}}\quad
         \subfigure[NGC1399]{\includegraphics[scale=0.47]{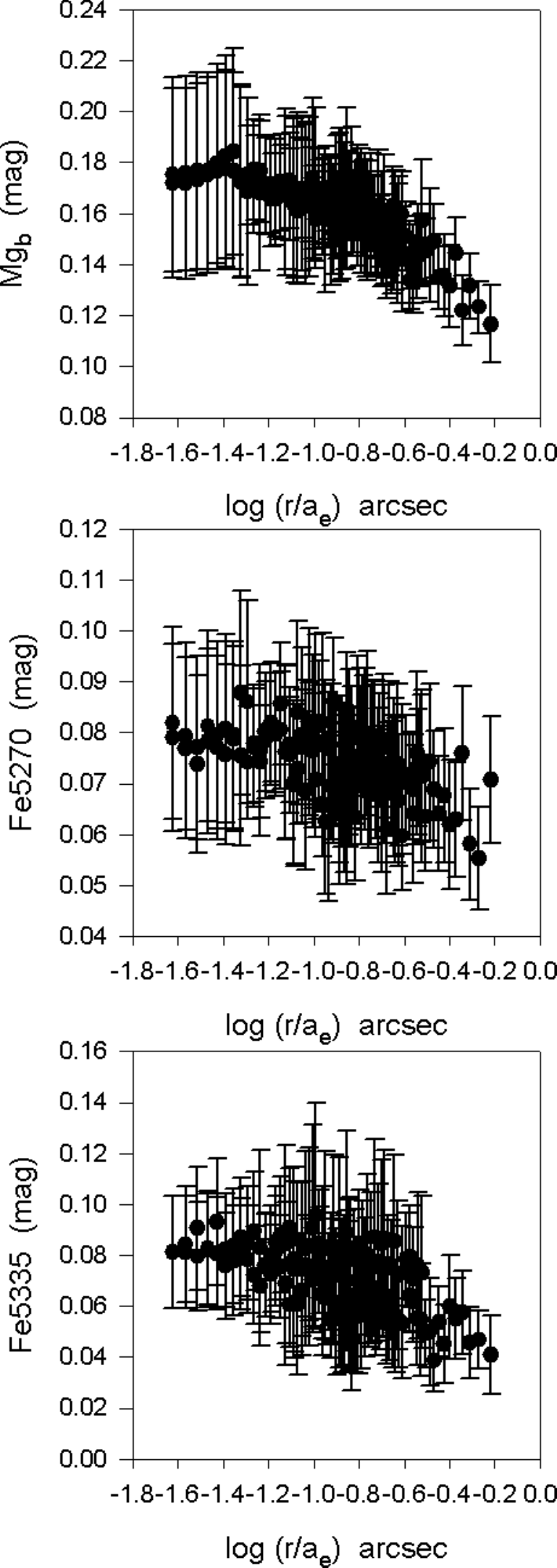}}}
\caption[]{BCG Mg$_{b}$, Fe5270 and Fe5335 gradients continue.}
   \label{Fig:Indices2}
\end{figure*}

\begin{figure*}
   \centering
   \mbox{\subfigure[NGC1713]{\includegraphics[scale=0.47]{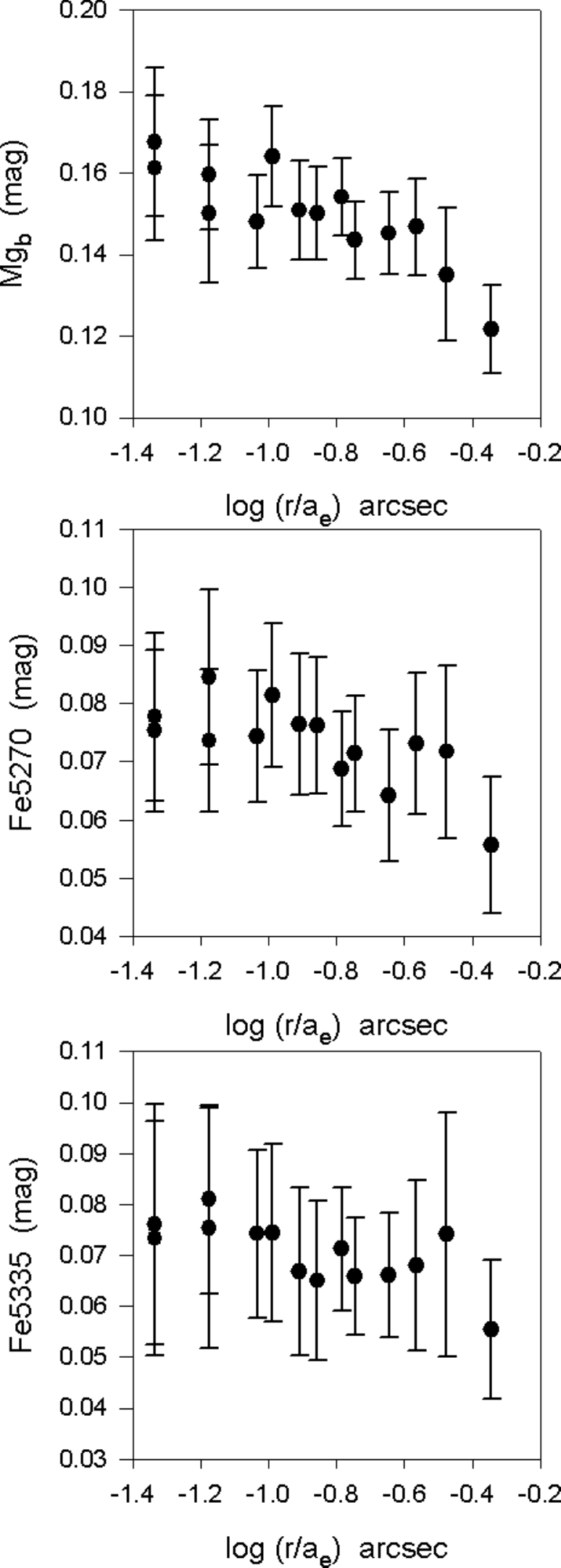}}\quad
         \subfigure[NGC2832]{\includegraphics[scale=0.47]{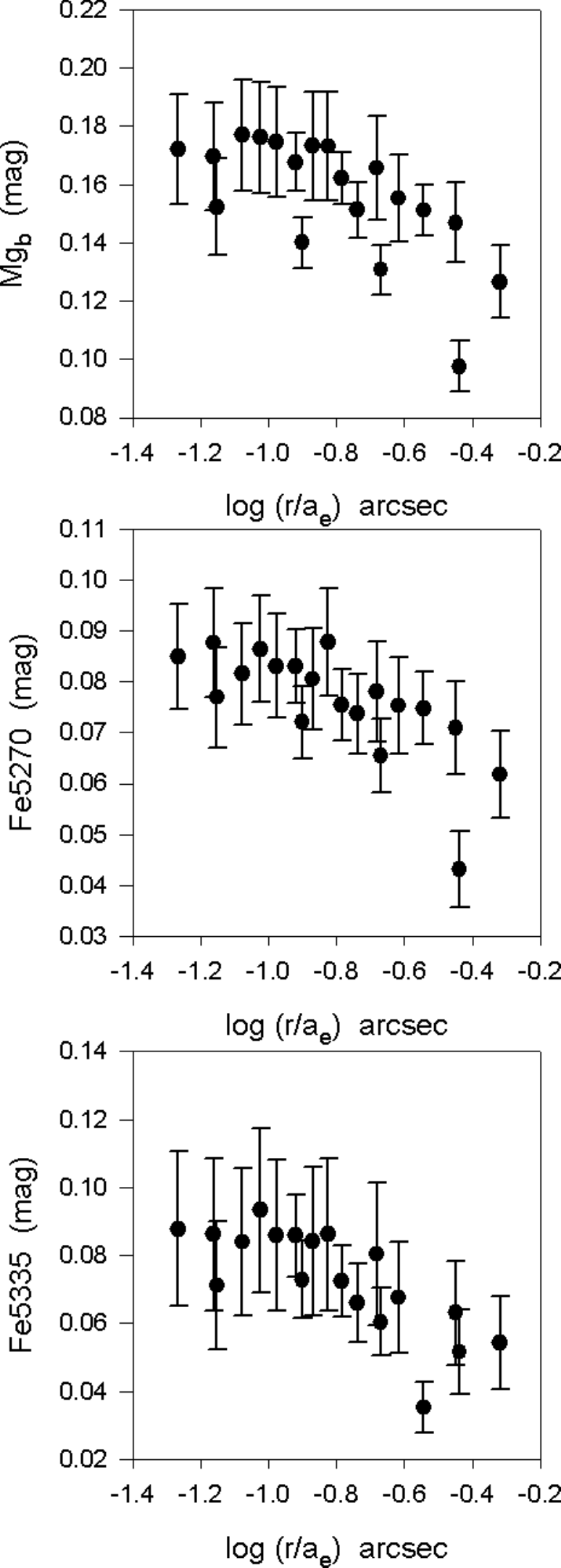}}\quad
         \subfigure[NGC3311]{\includegraphics[scale=0.47]{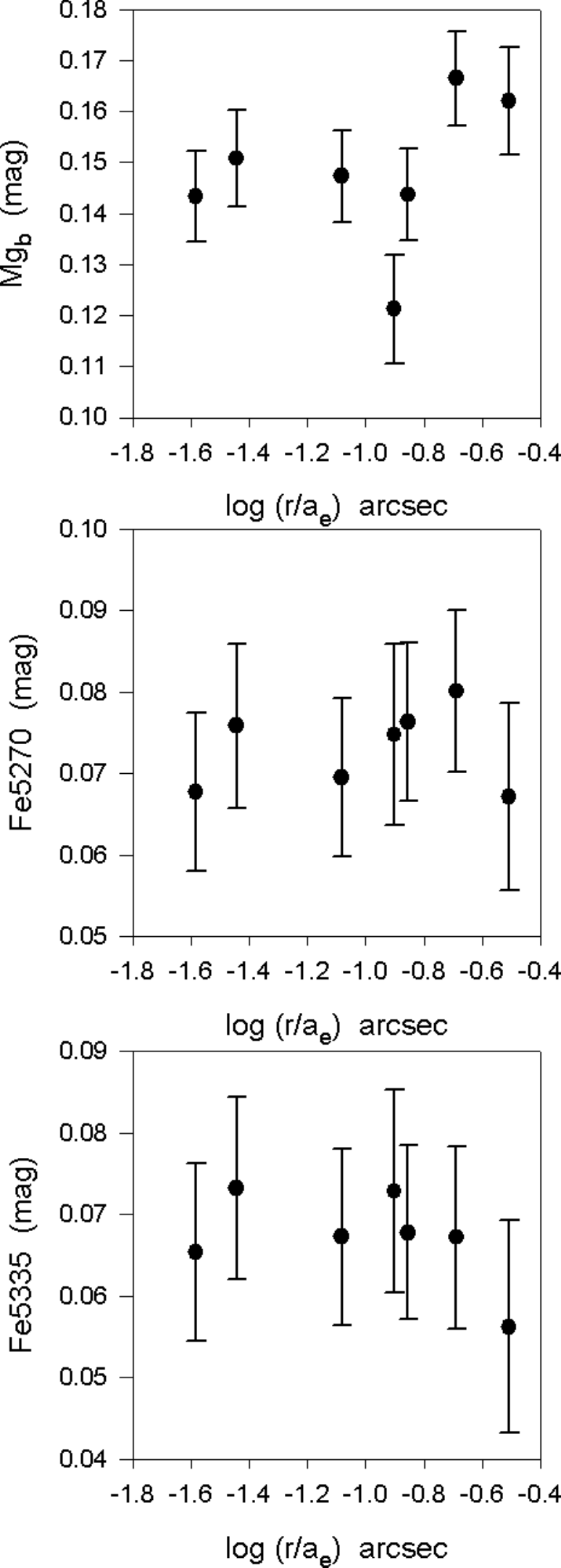}}\quad
         \subfigure[NGC4839]{\includegraphics[scale=0.47]{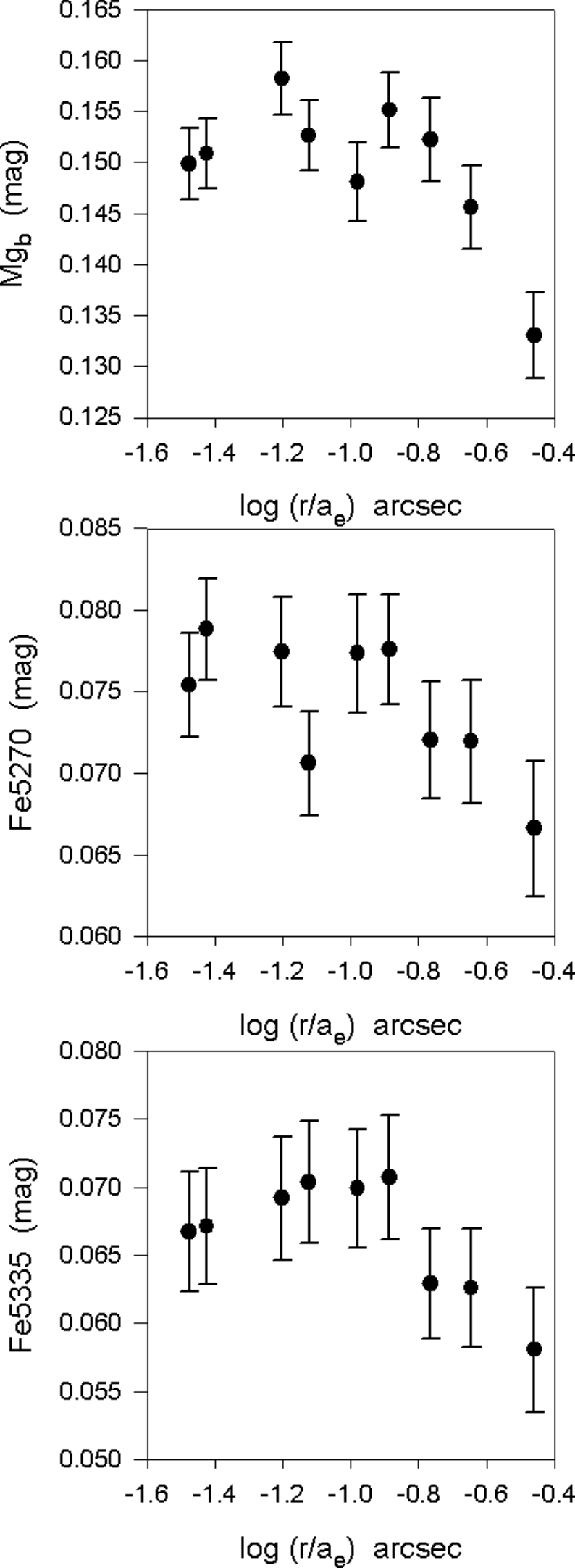}}}
\mbox{\subfigure[NGC6173]{\includegraphics[scale=0.47]{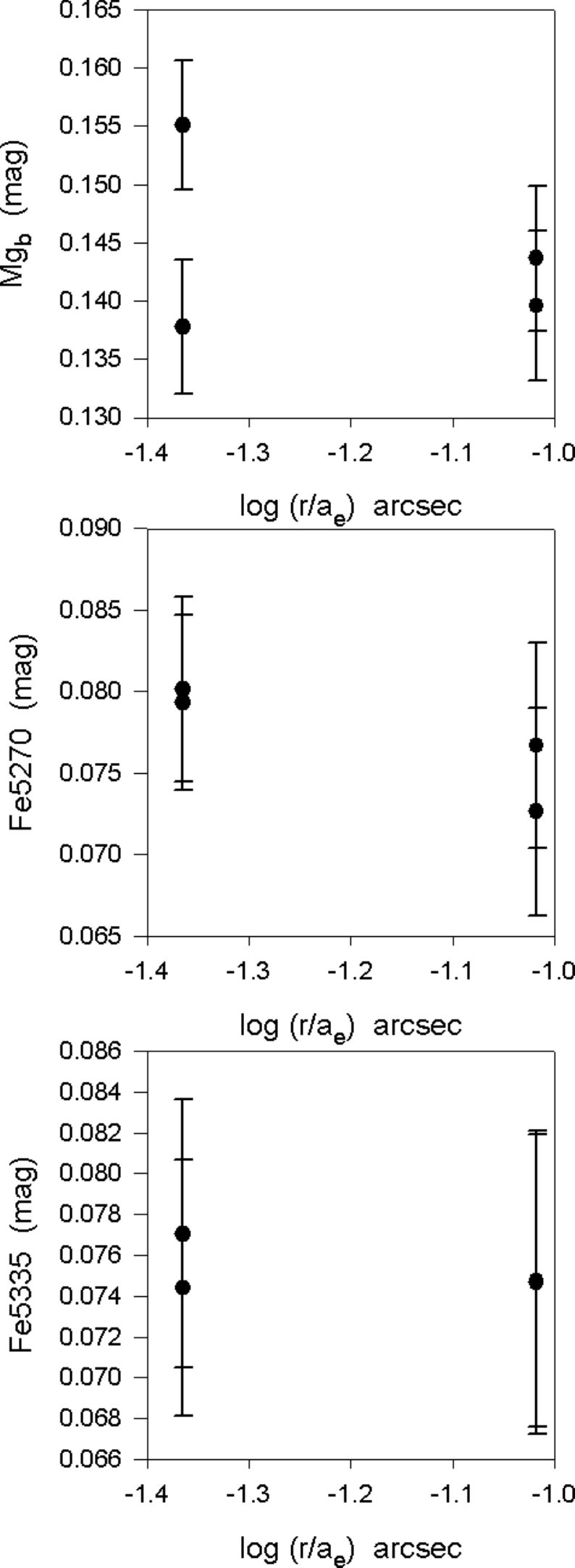}}\quad
         \subfigure[NGC6269]{\includegraphics[scale=0.47]{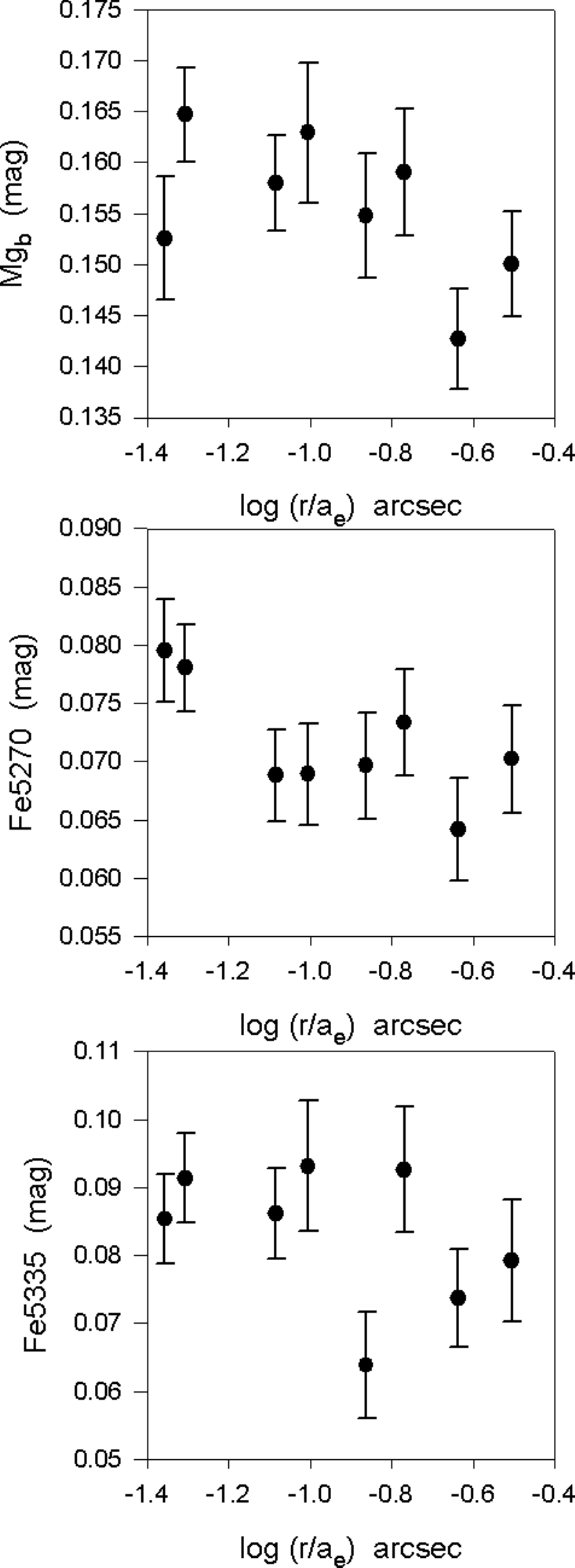}}\quad
         \subfigure[NGC7012]{\includegraphics[scale=0.47]{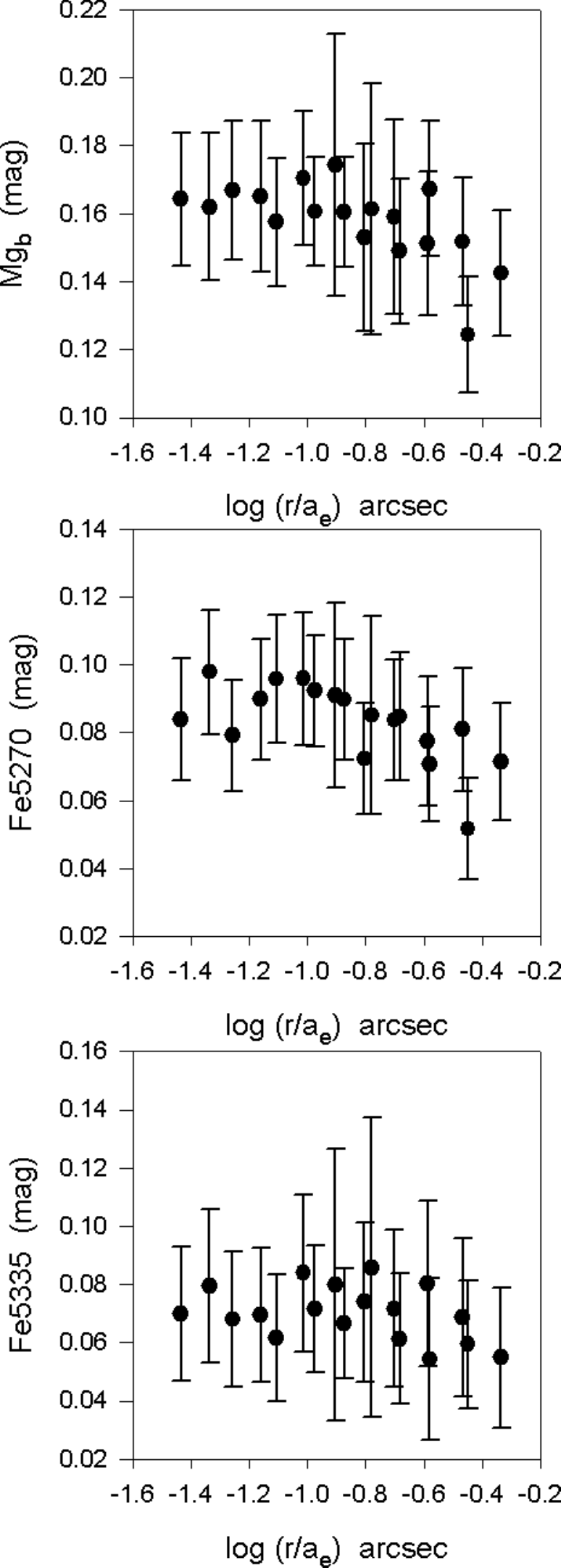}}\quad
         \subfigure[PGC004072]{\includegraphics[scale=0.47]{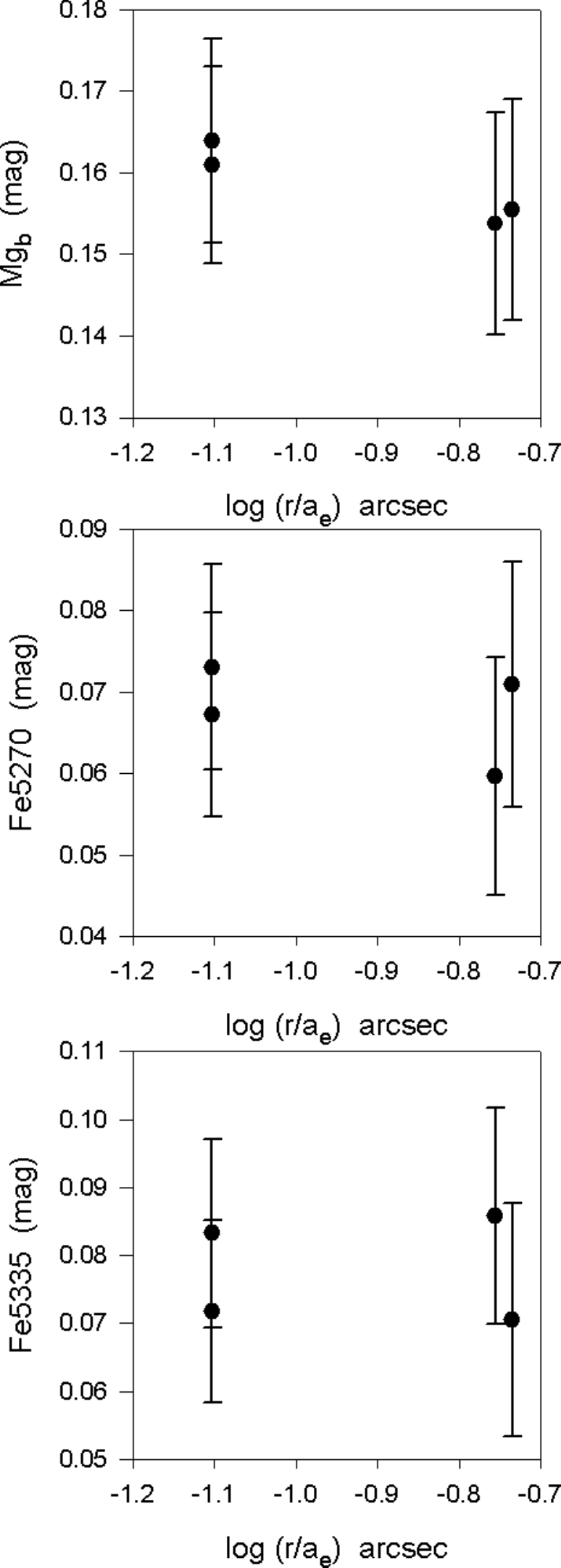}}}
\caption[]{BCG Mg$_{b}$, Fe5270 and Fe5335 gradients continue.}
   \label{Fig:Indices3}
\end{figure*}

\begin{figure*}
   \centering
   \mbox{\subfigure[PGC030223]{\includegraphics[scale=0.5]{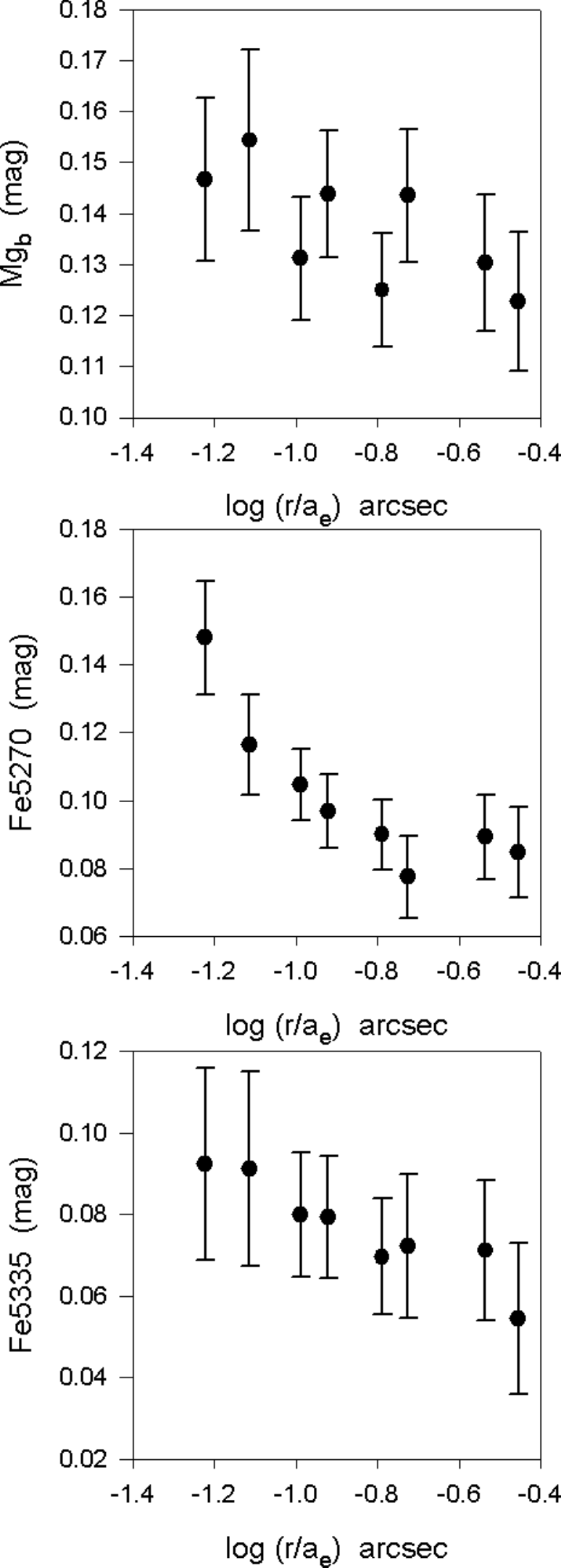}}\quad
         \subfigure[PGC072804]{\includegraphics[scale=0.5]{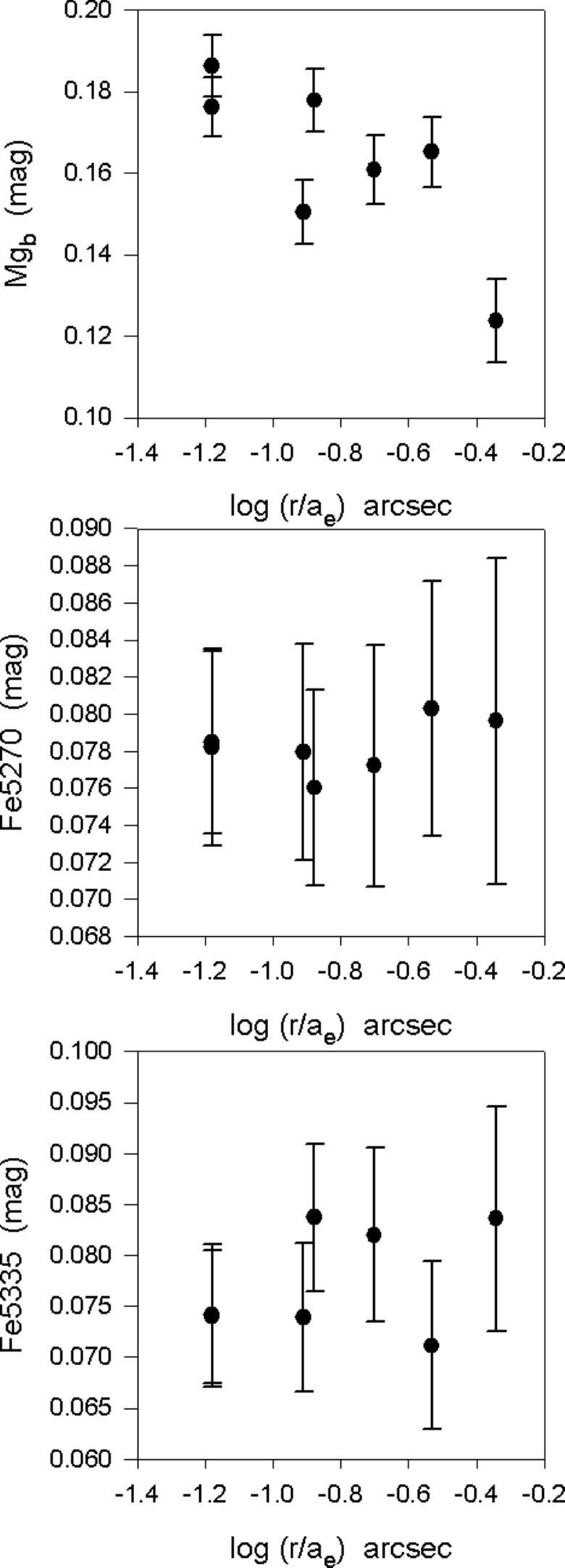}}\quad
         \subfigure[UGC02232]{\includegraphics[scale=0.5]{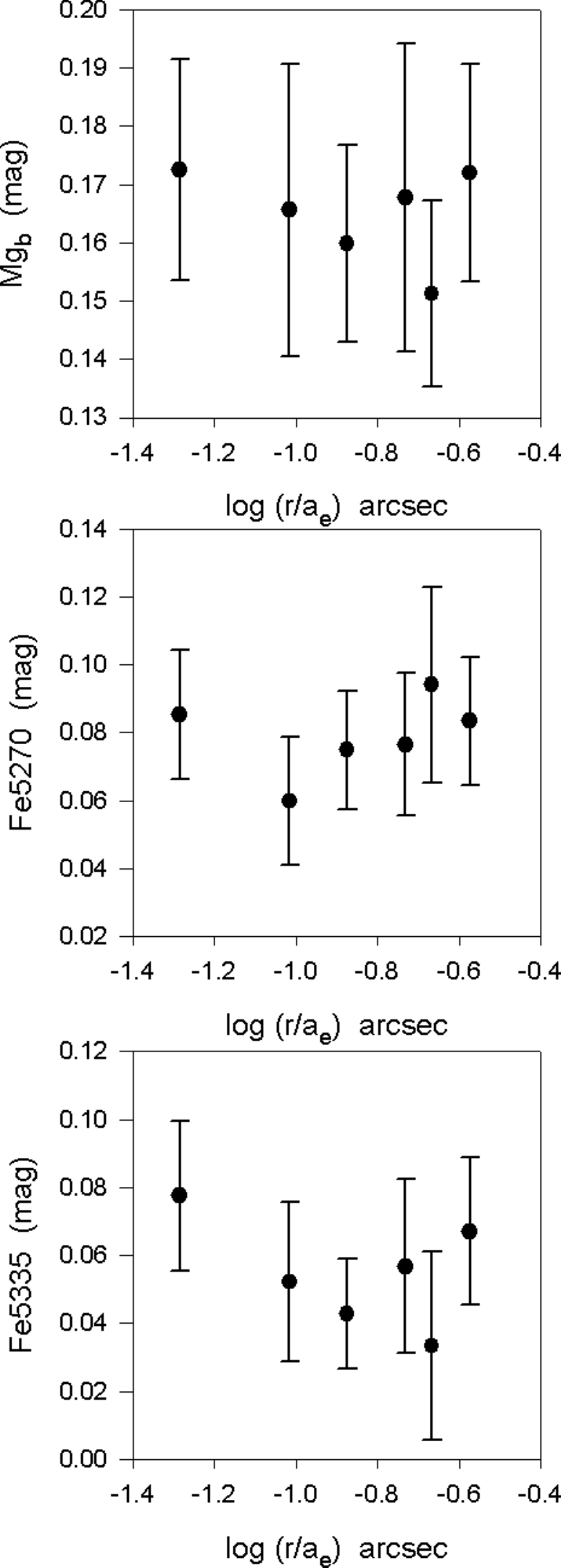}}\quad
         \subfigure[UGC05515]{\includegraphics[scale=0.5]{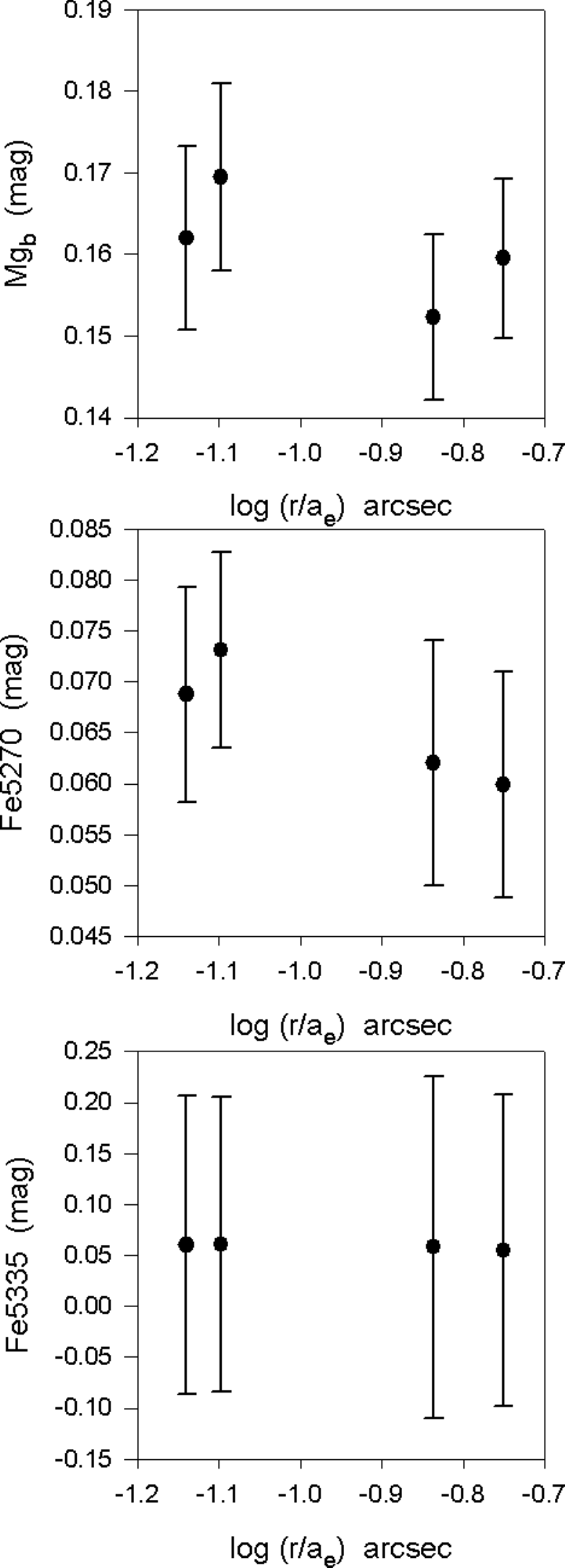}}}
\caption[]{BCG Mg$_{b}$, Fe5270 and Fe5335 gradients continue.}
   \label{Fig:Indices4}
\end{figure*}

\begin{figure*}
\centering
   \mbox{\subfigure{\includegraphics[height=7.5cm,width=5.6cm]{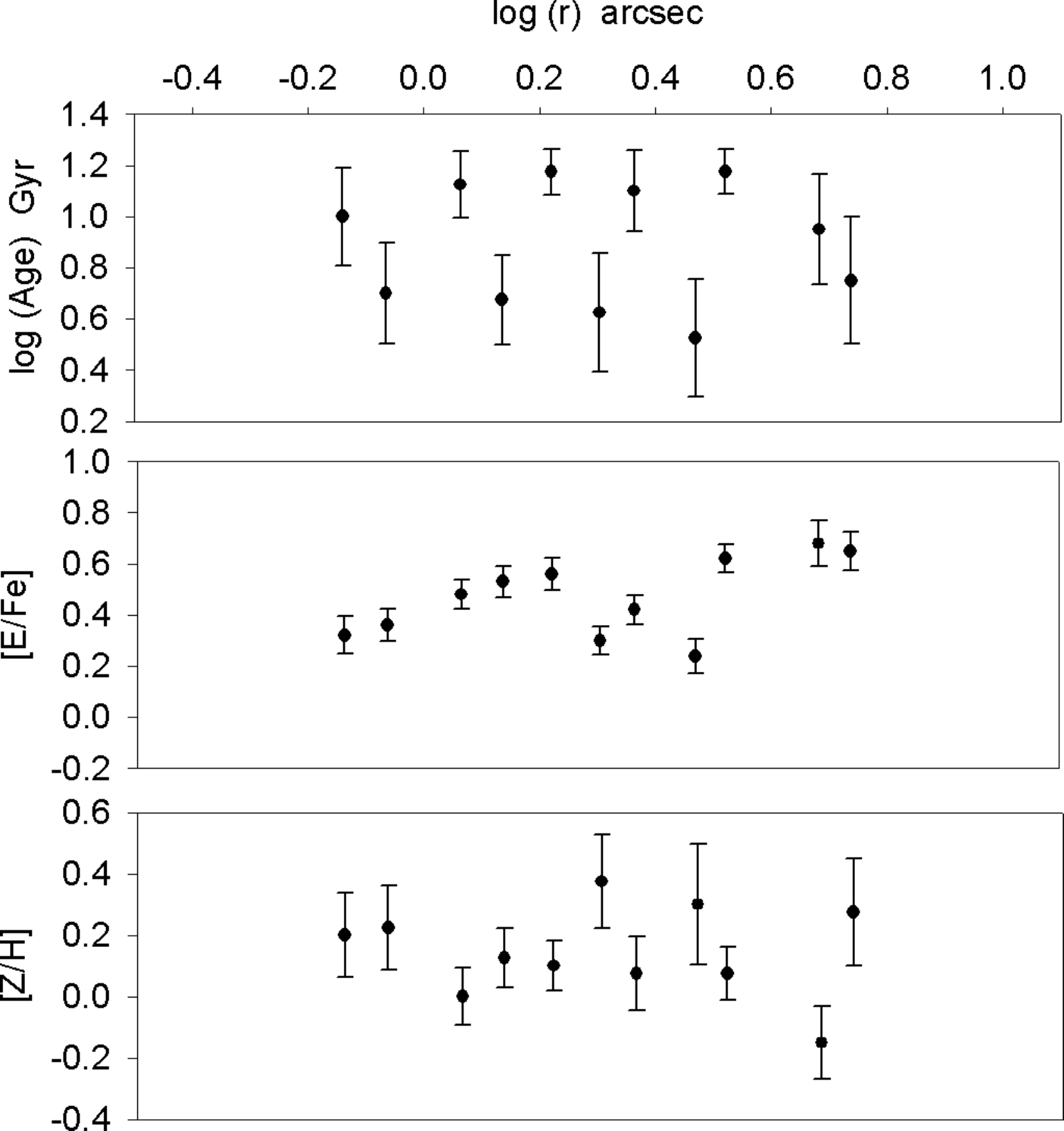}}\quad
         \subfigure{\includegraphics[height=7.5cm,width=5.6cm]{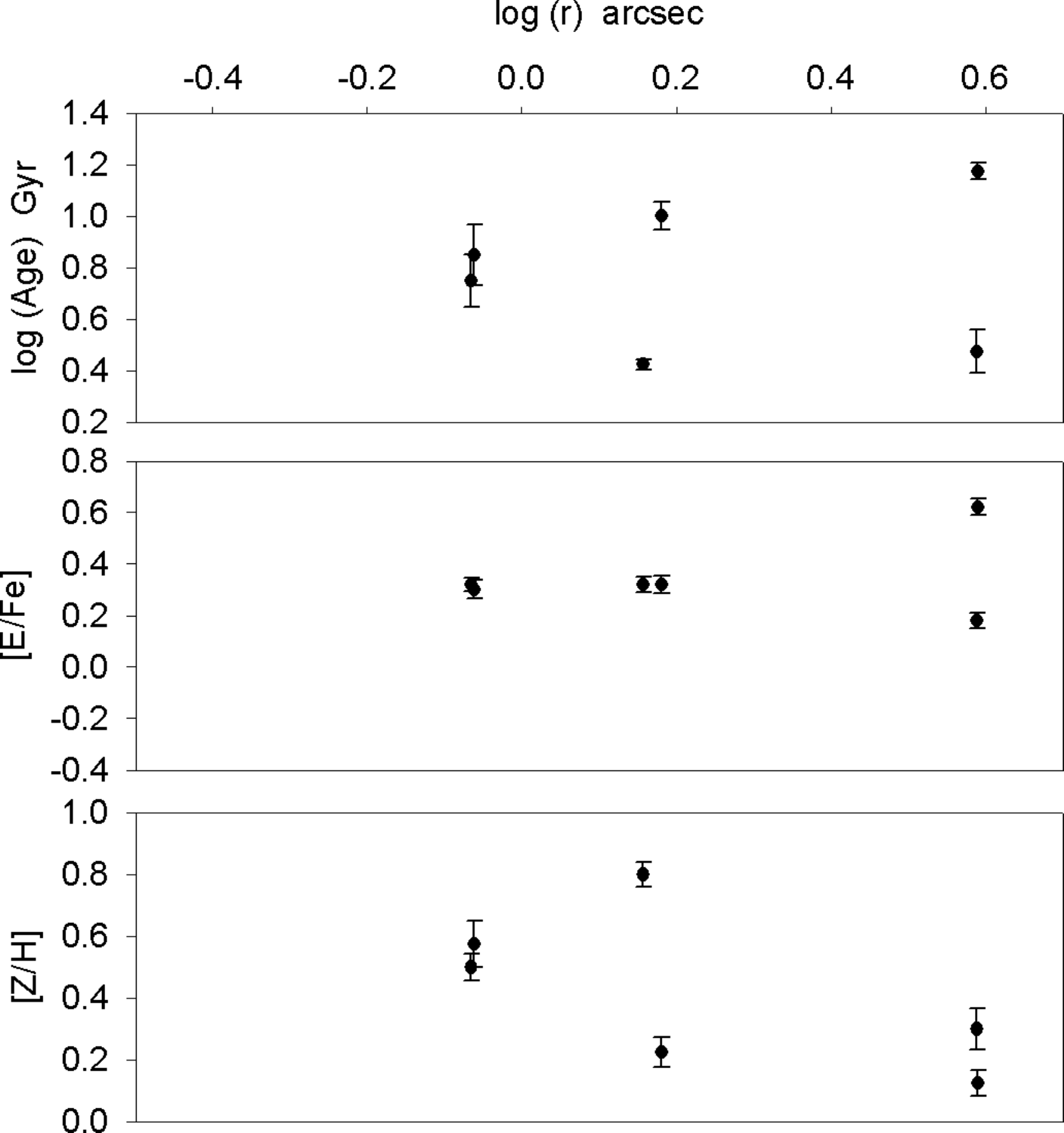}}\quad
         \subfigure{\includegraphics[height=7.5cm,width=5.6cm]{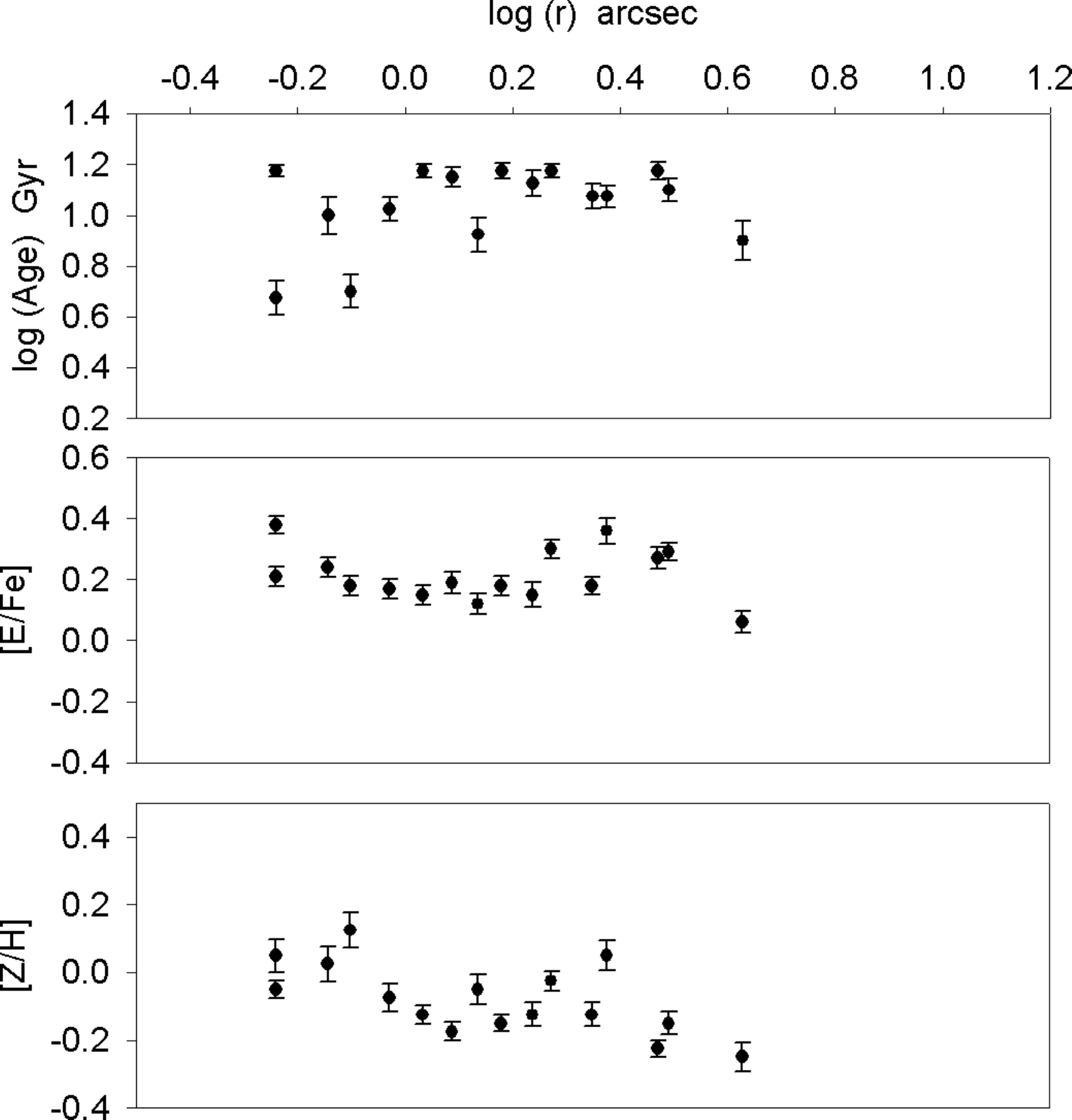}}}
   \mbox{\subfigure[ESO146-028]{\includegraphics[height=2.6cm,width=5.6cm]{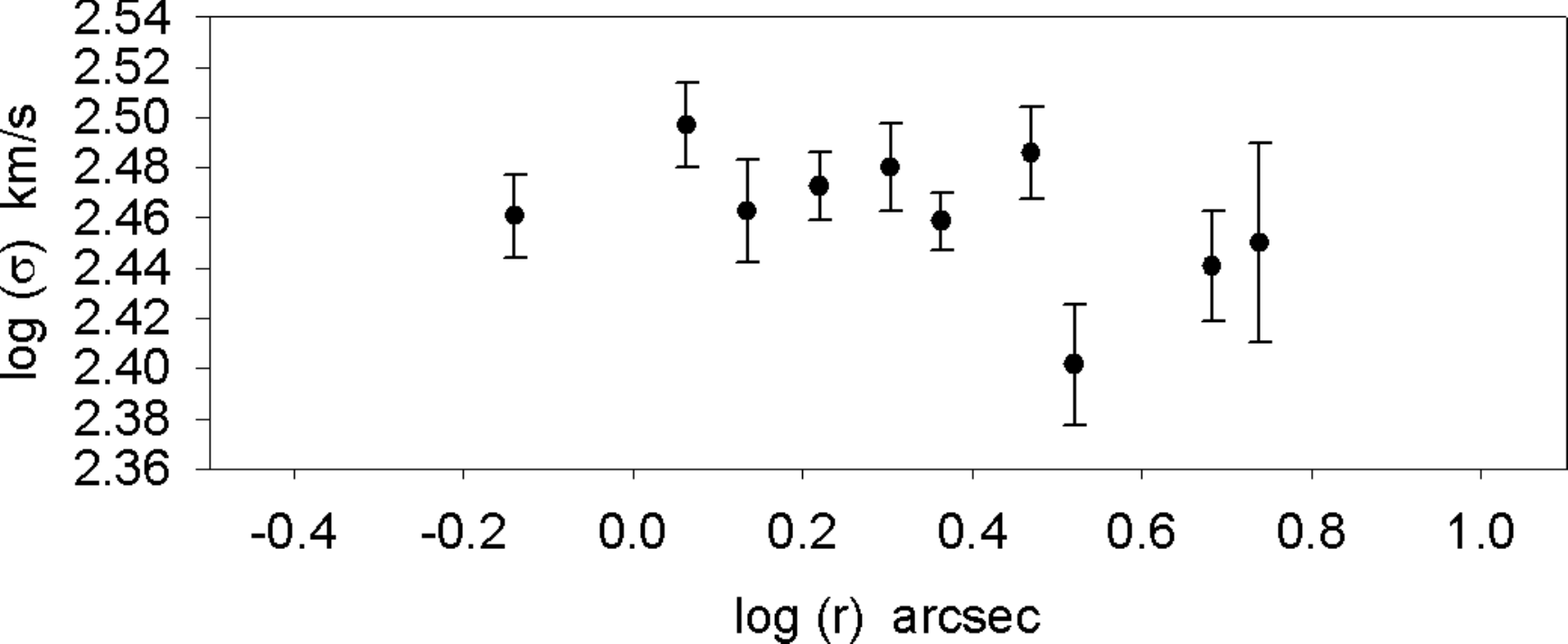}}\quad
         \subfigure[ESO303-005]{\includegraphics[height=2.6cm,width=5.6cm]{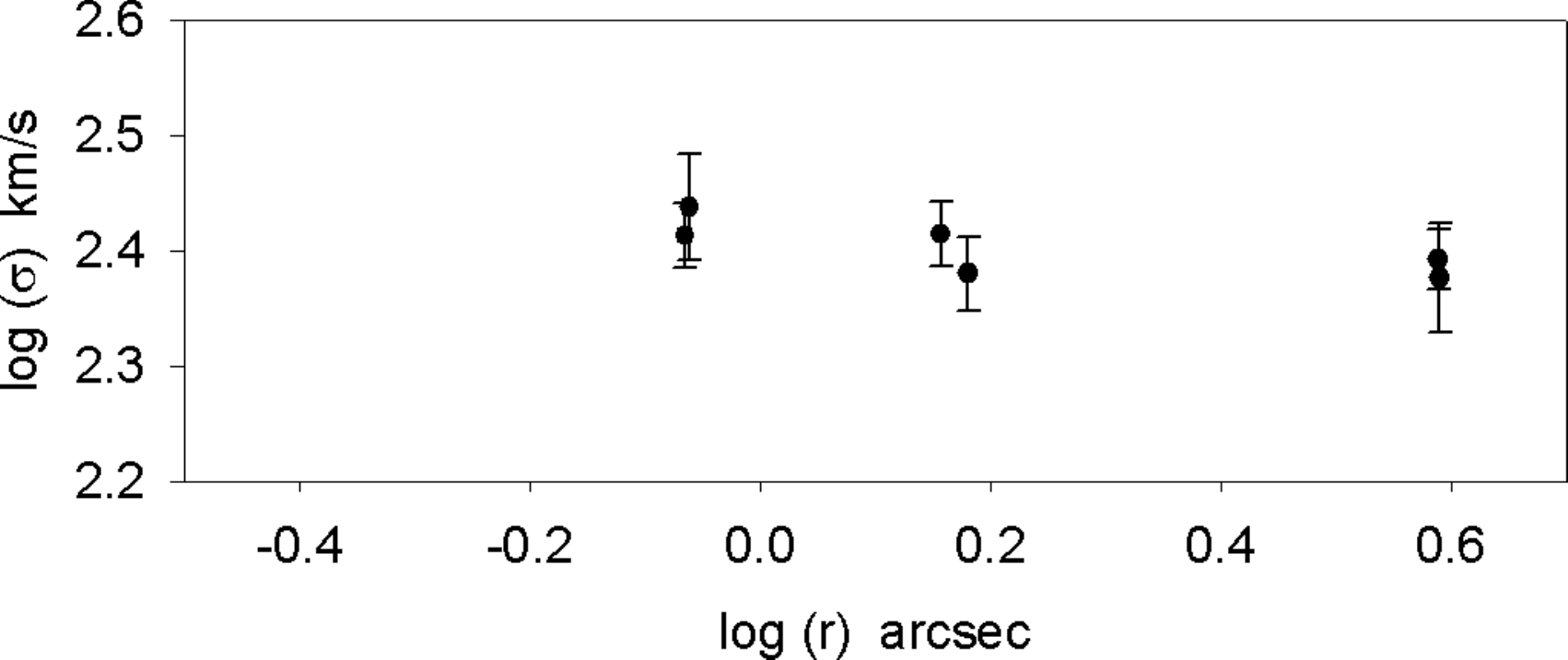}}\quad
         \subfigure[ESO488-027]{\includegraphics[height=2.6cm,width=5.6cm]{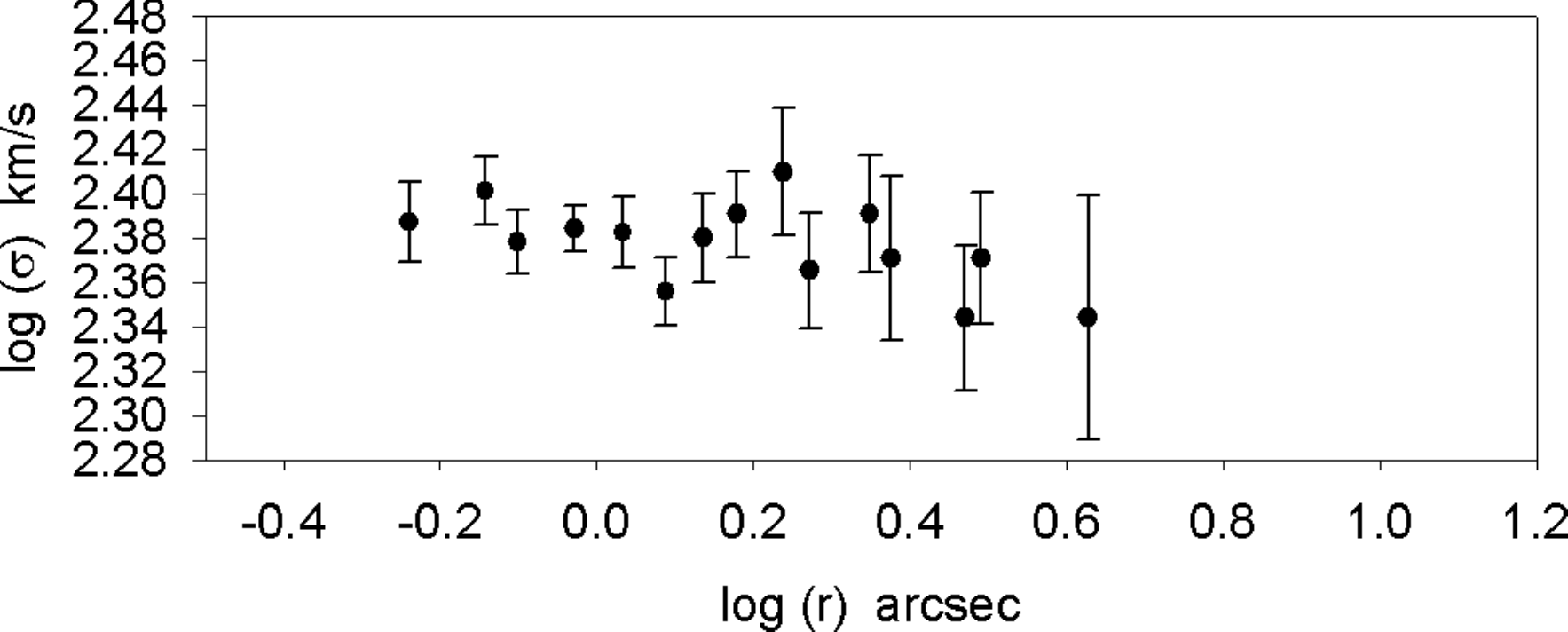}}}
 \mbox{\subfigure{\includegraphics[height=7.5cm,width=5.6 cm]{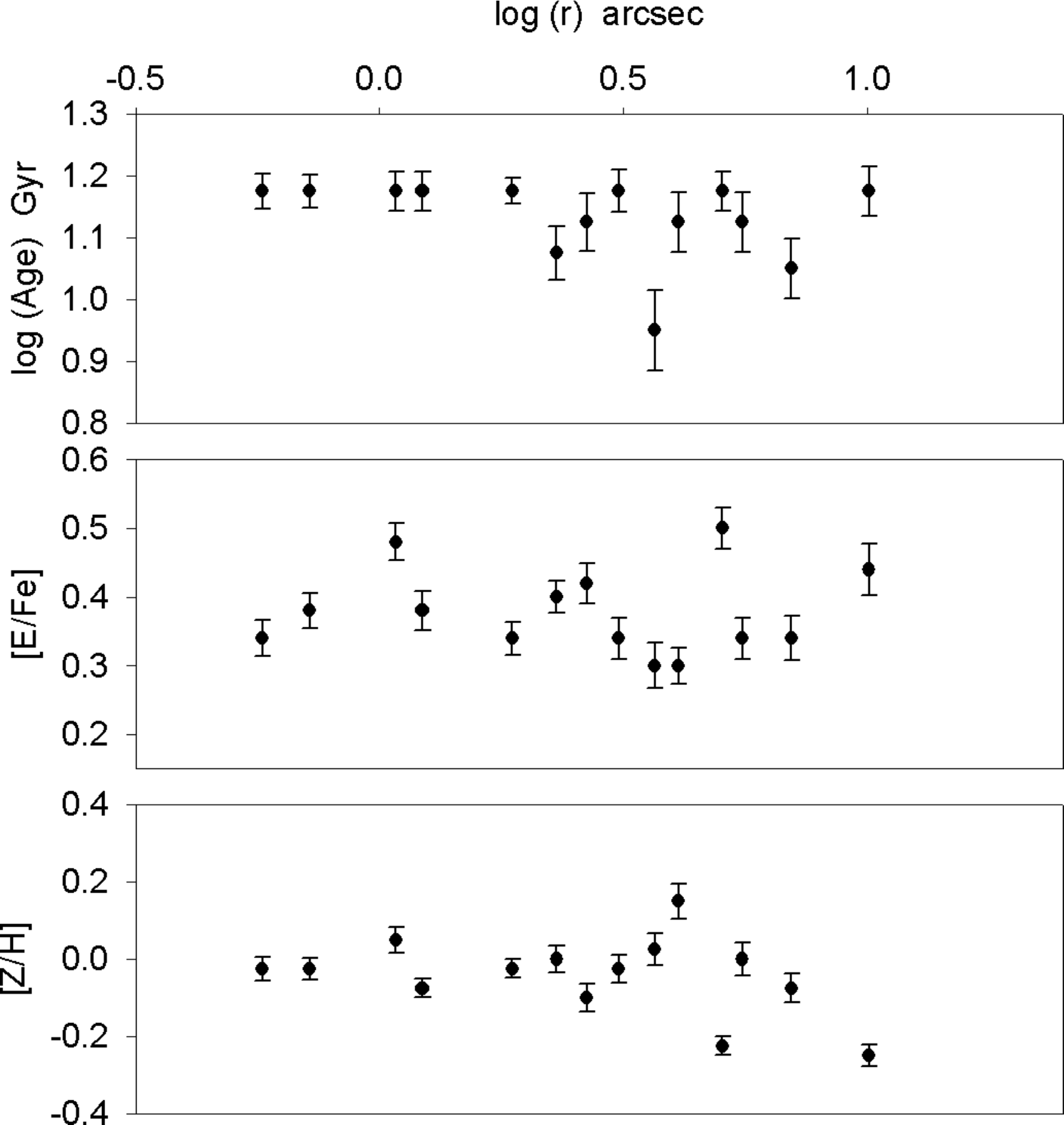}}\quad
         \subfigure{\includegraphics[height=7.5cm,width=5.6cm]{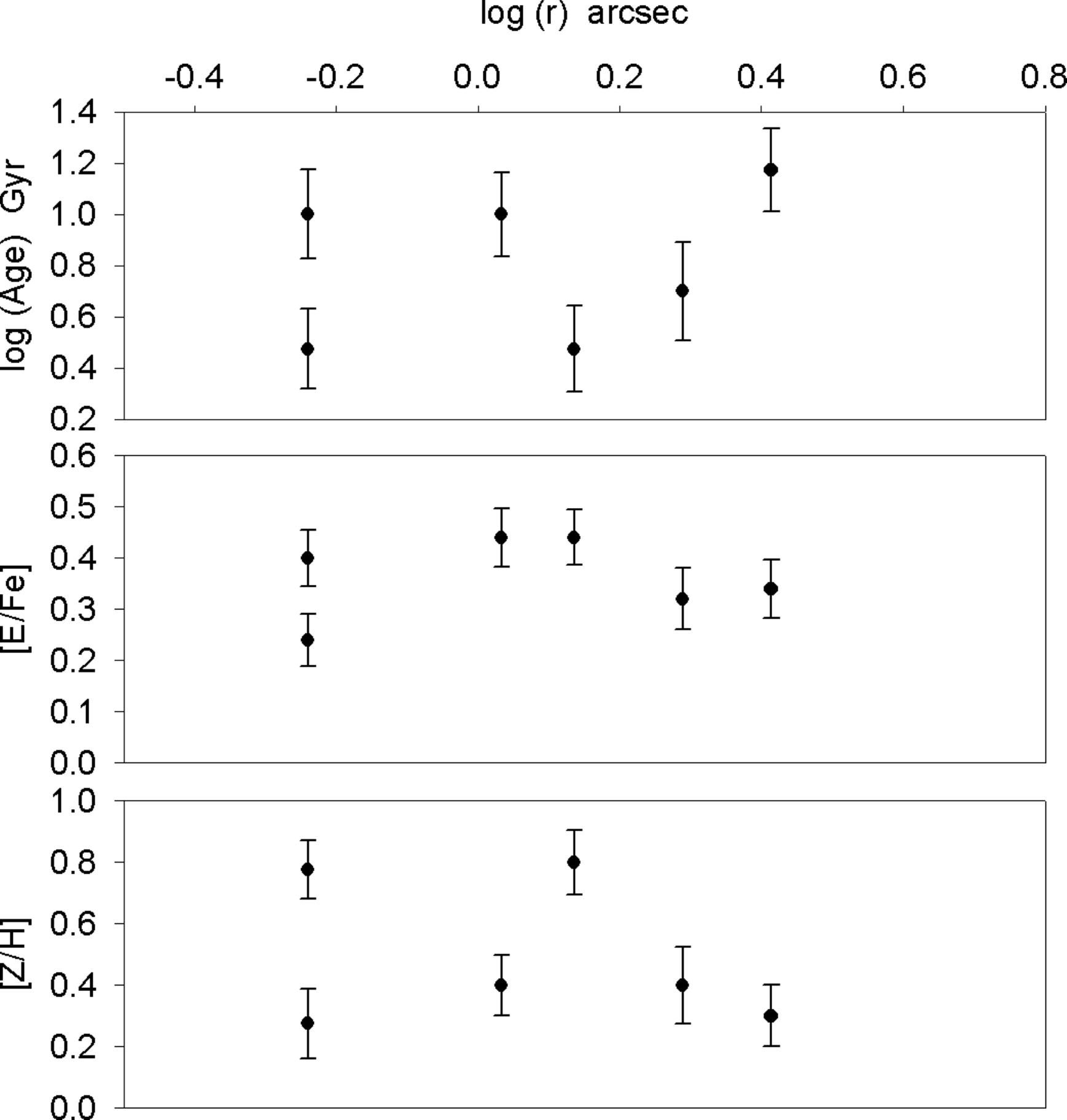}}\quad
         \subfigure{\includegraphics[height=7.5cm,width=5.6cm]{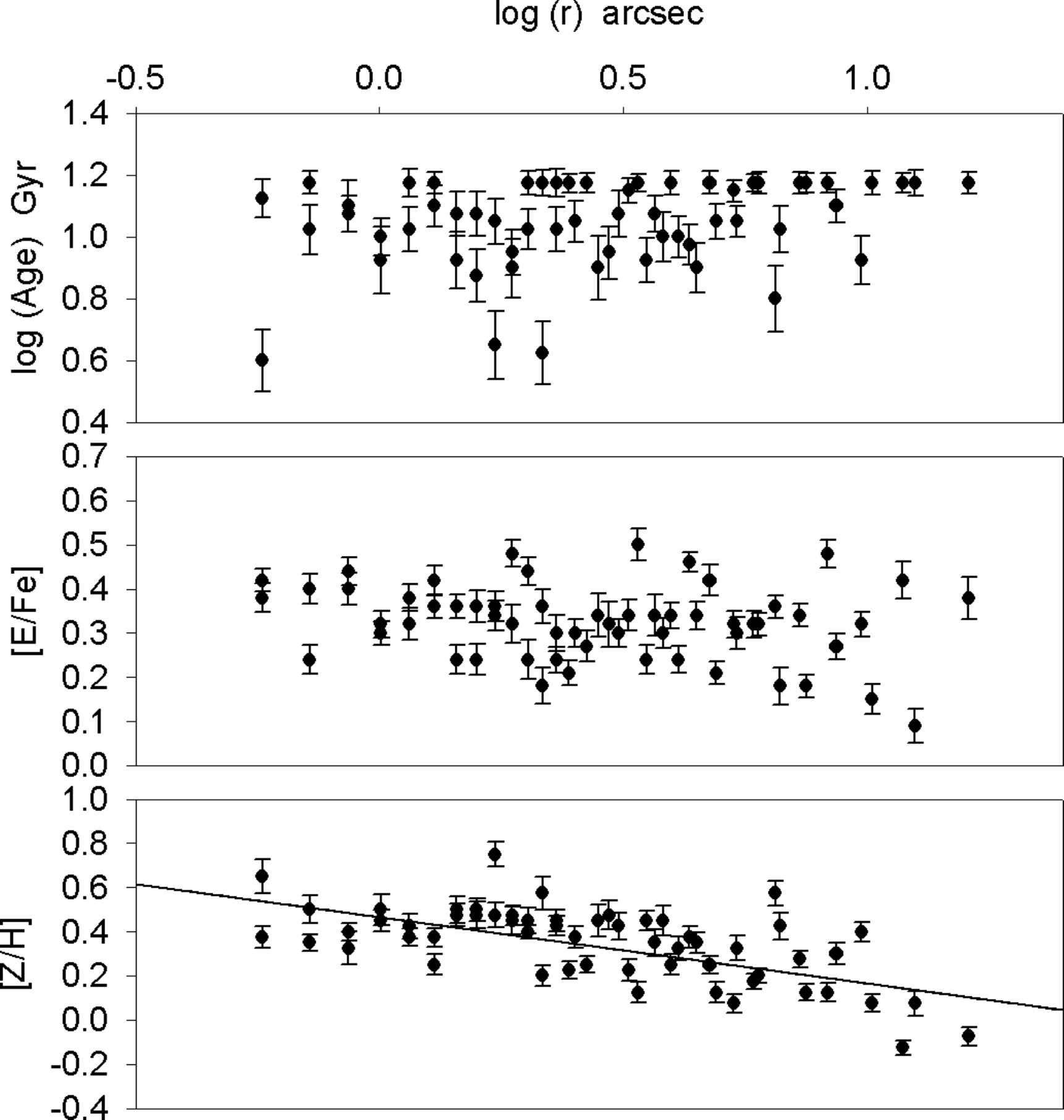}}}
   \mbox{\subfigure[ESO552-020]{\includegraphics[height=2.6cm,width=5.6cm]{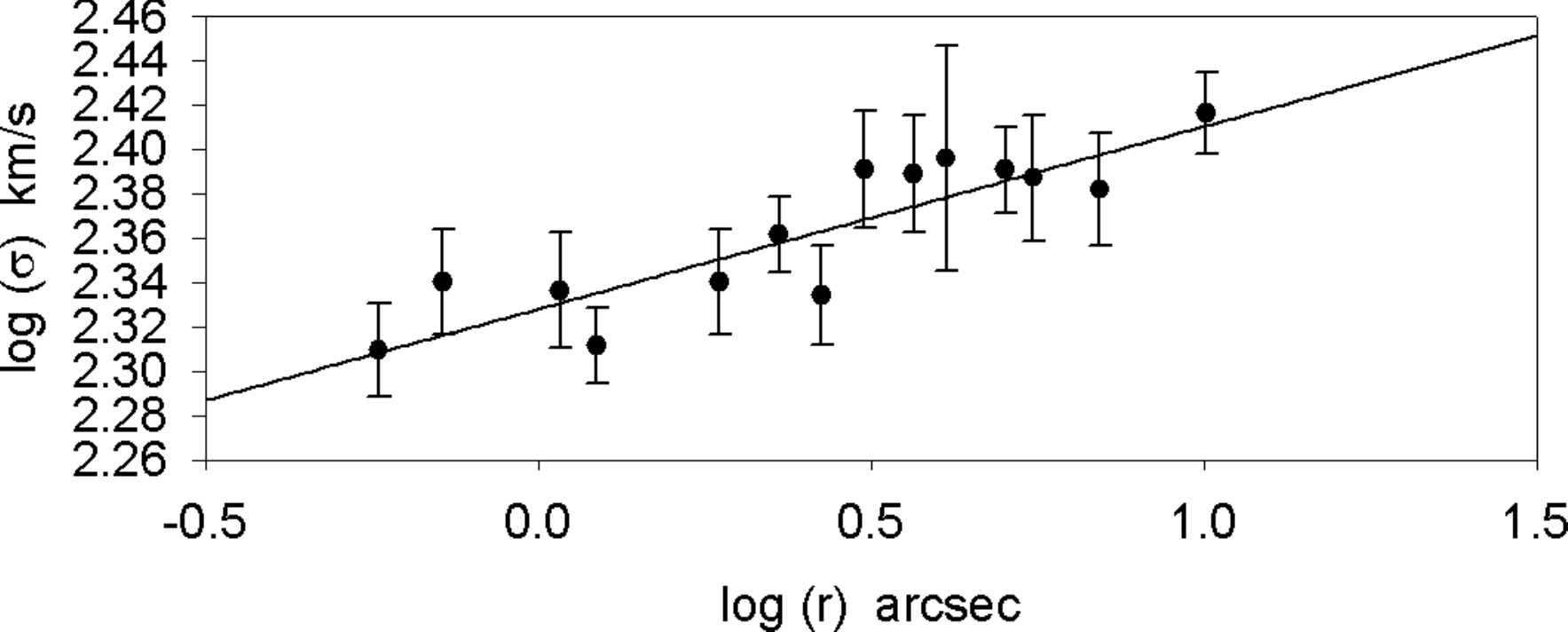}}\quad
         \subfigure[GSC555700266]{\includegraphics[height=2.6cm,width=5.6cm]{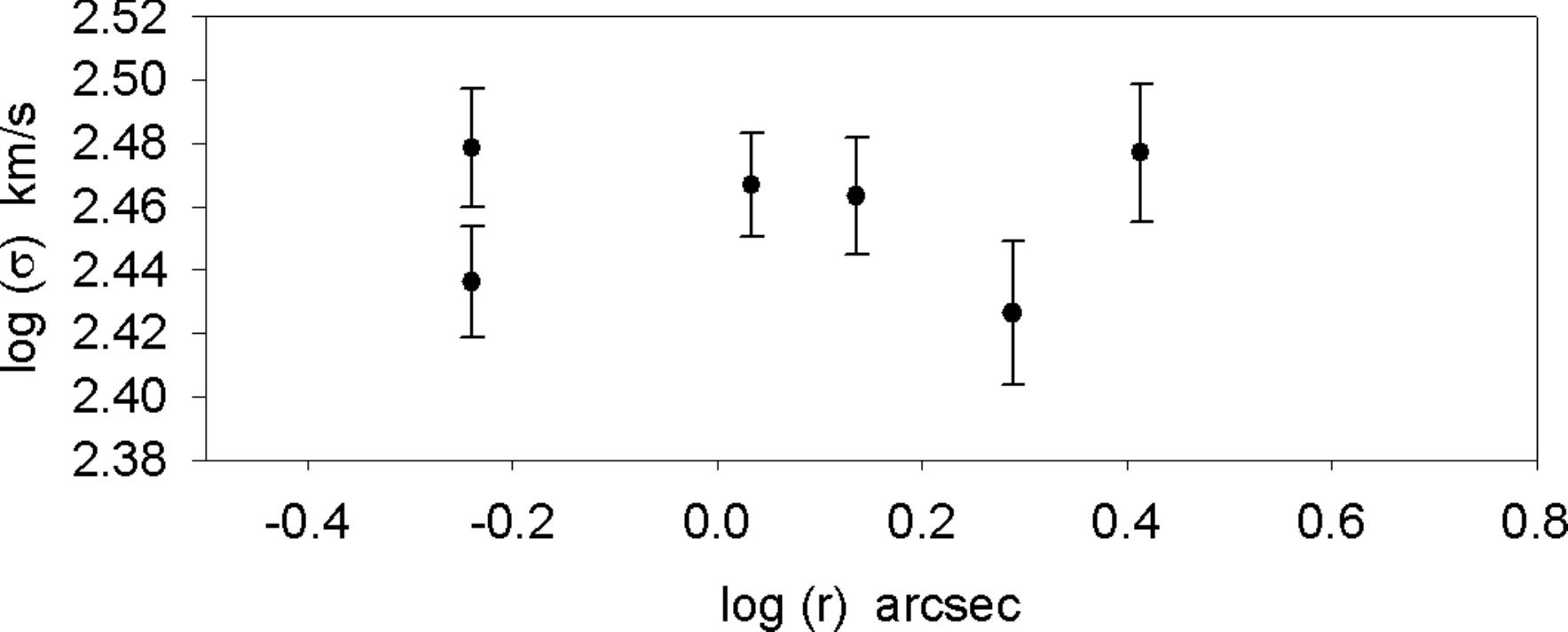}}\quad
         \subfigure[IC1633]{\includegraphics[height=2.6cm,width=5.6cm]{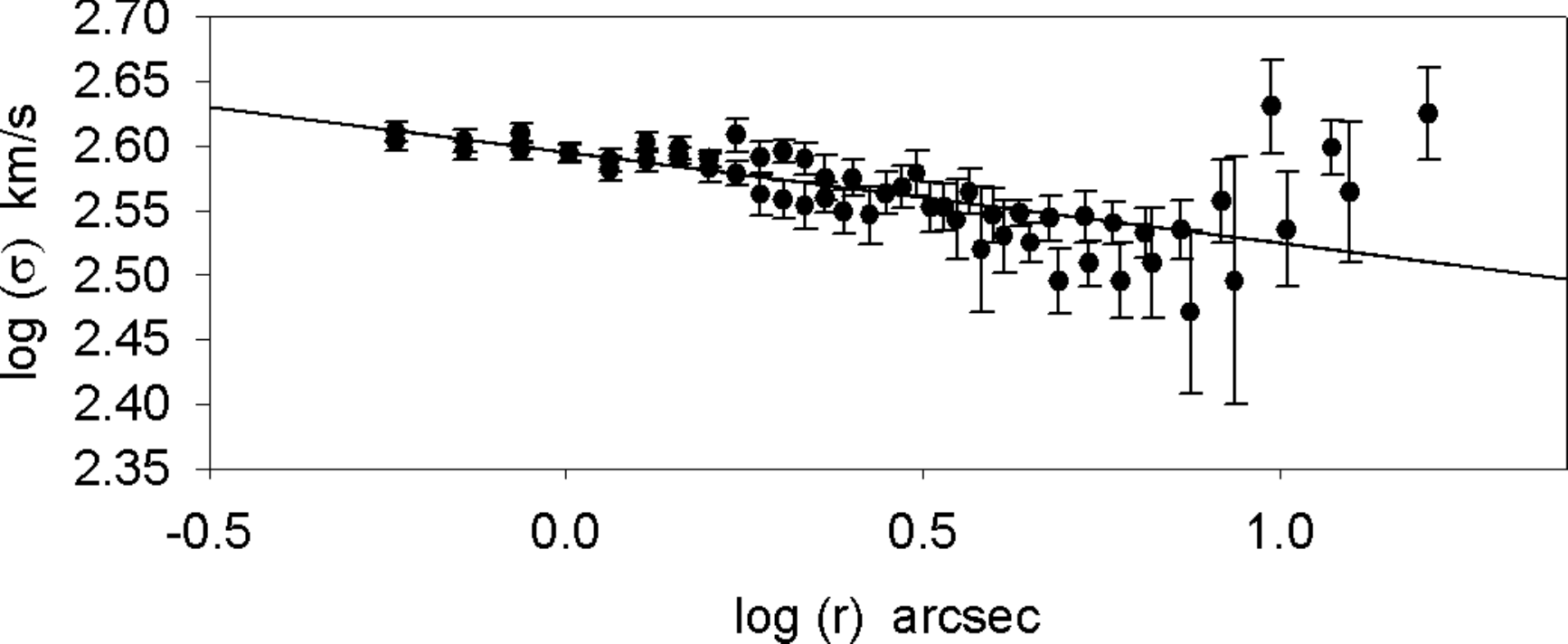}}}
\caption[]{Age, $\alpha$-enhancement, metallicity and velocity dispersion profiles of the BCGs. The galaxies were binned in the spatial direction to a minimum S/N of 40 per \AA{} in the H$\beta$ region of the spectrum. The central 0.5 arcsec, to each side, are excluded. In all the gradient plots, the data were folded with respect to the galaxy centres.}
\label{fig:Profiles}
\end{figure*}

\begin{figure*}
\centering
\mbox{\subfigure{\includegraphics[height=7.6cm,width=5.6 cm]{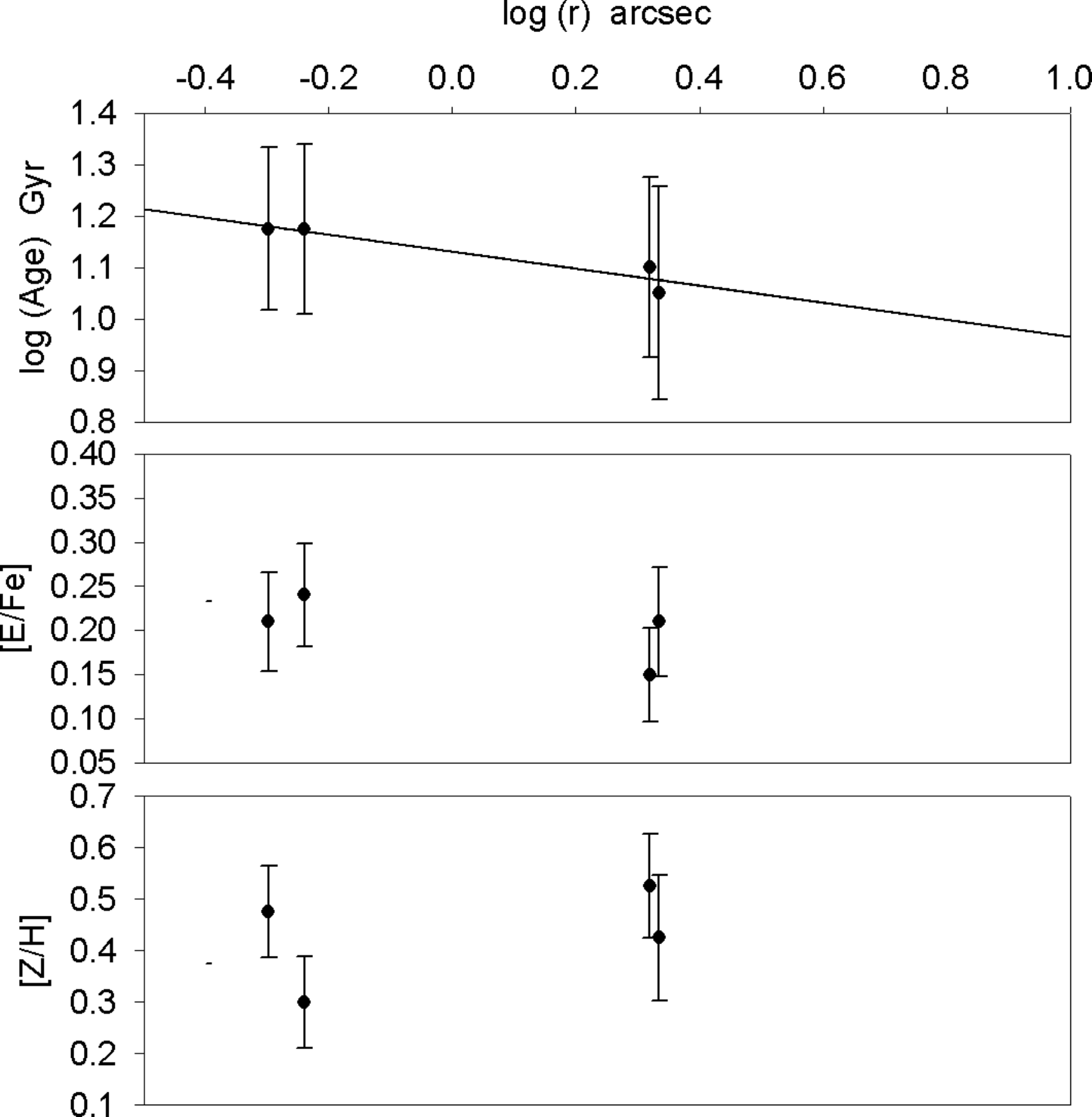}}\quad
         \subfigure{\includegraphics[height=7.6cm,width=5.6cm]{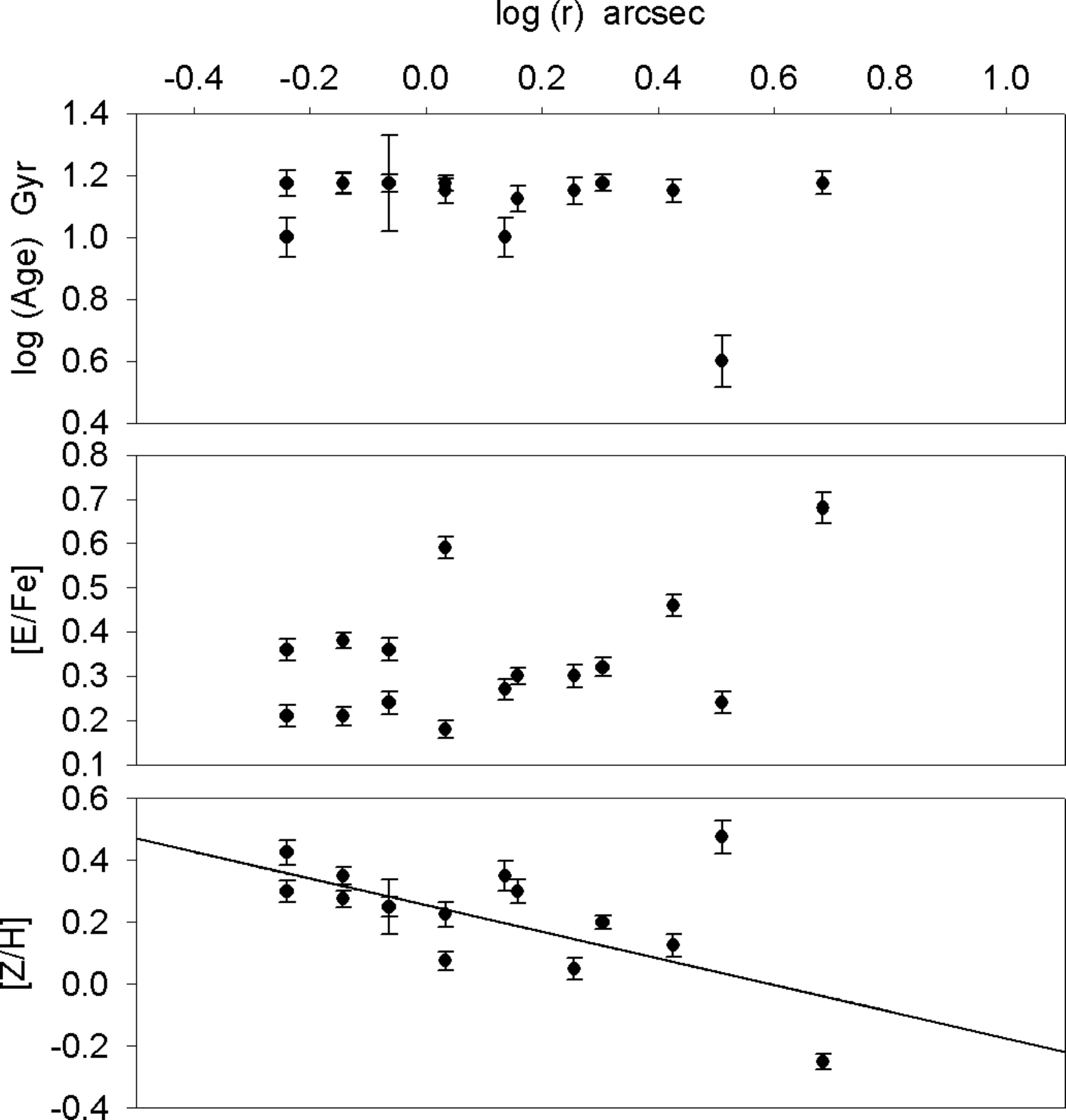}}\quad
         \subfigure{\includegraphics[height=7.6cm,width=5.6cm]{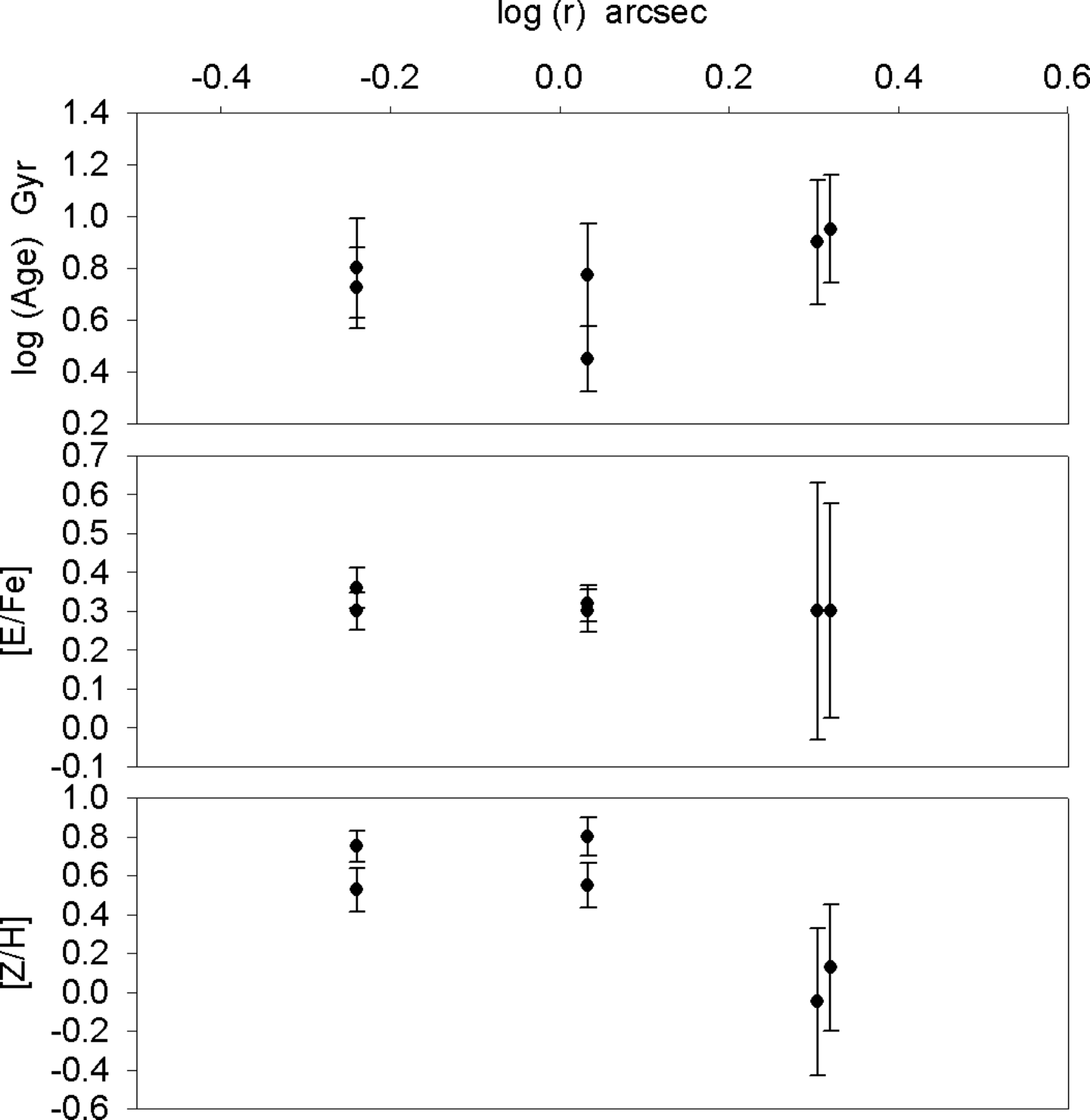}}}
   \mbox{\subfigure[IC4765]{\includegraphics[height=2.7cm,width=5.6cm]{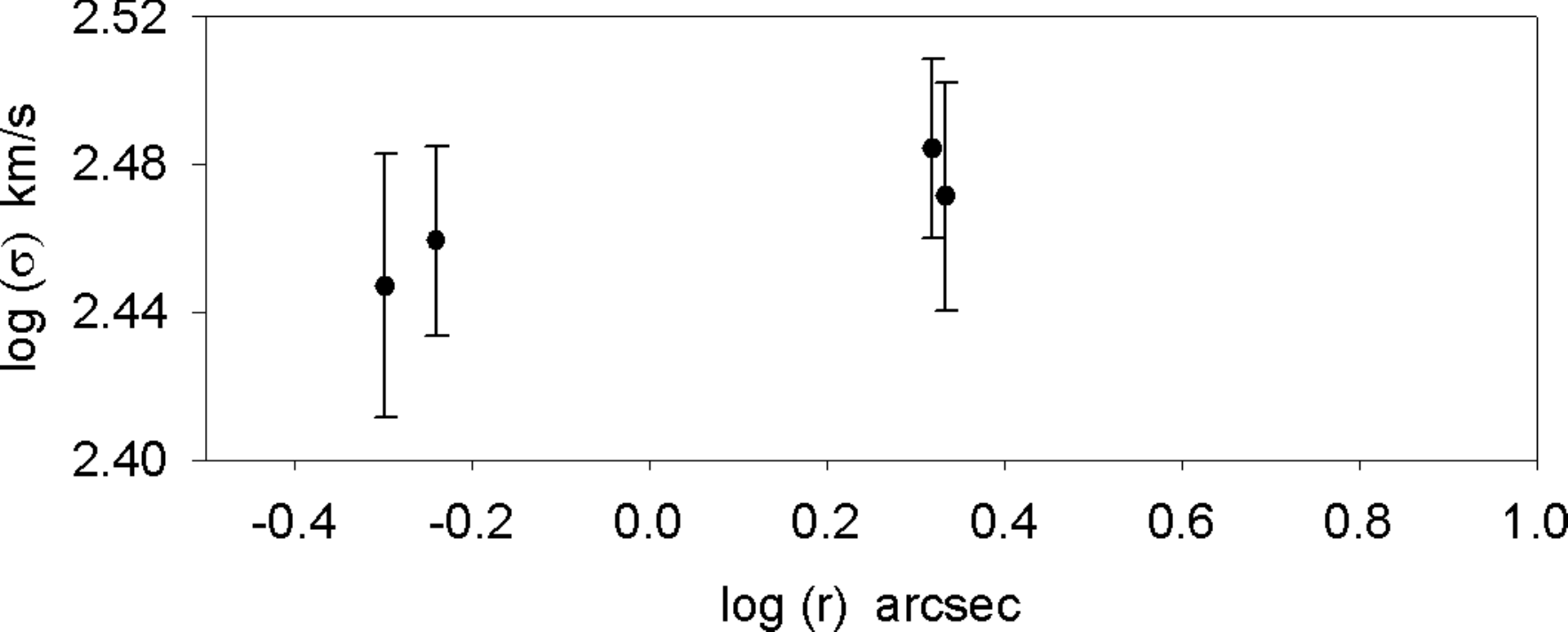}}\quad
         \subfigure[IC5358]{\includegraphics[height=2.7cm,width=5.6cm]{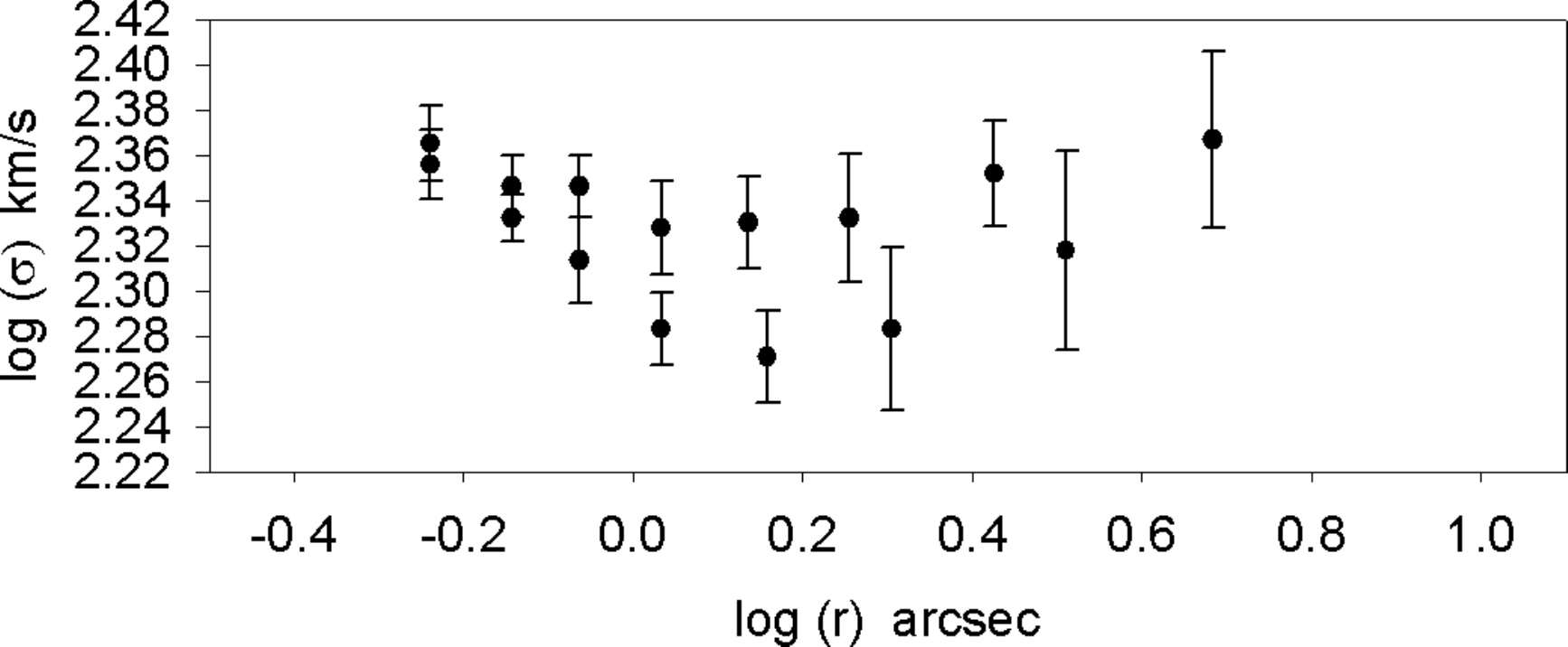}}\quad
         \subfigure[Leda094683]{\includegraphics[height=2.7cm,width=5.6cm]{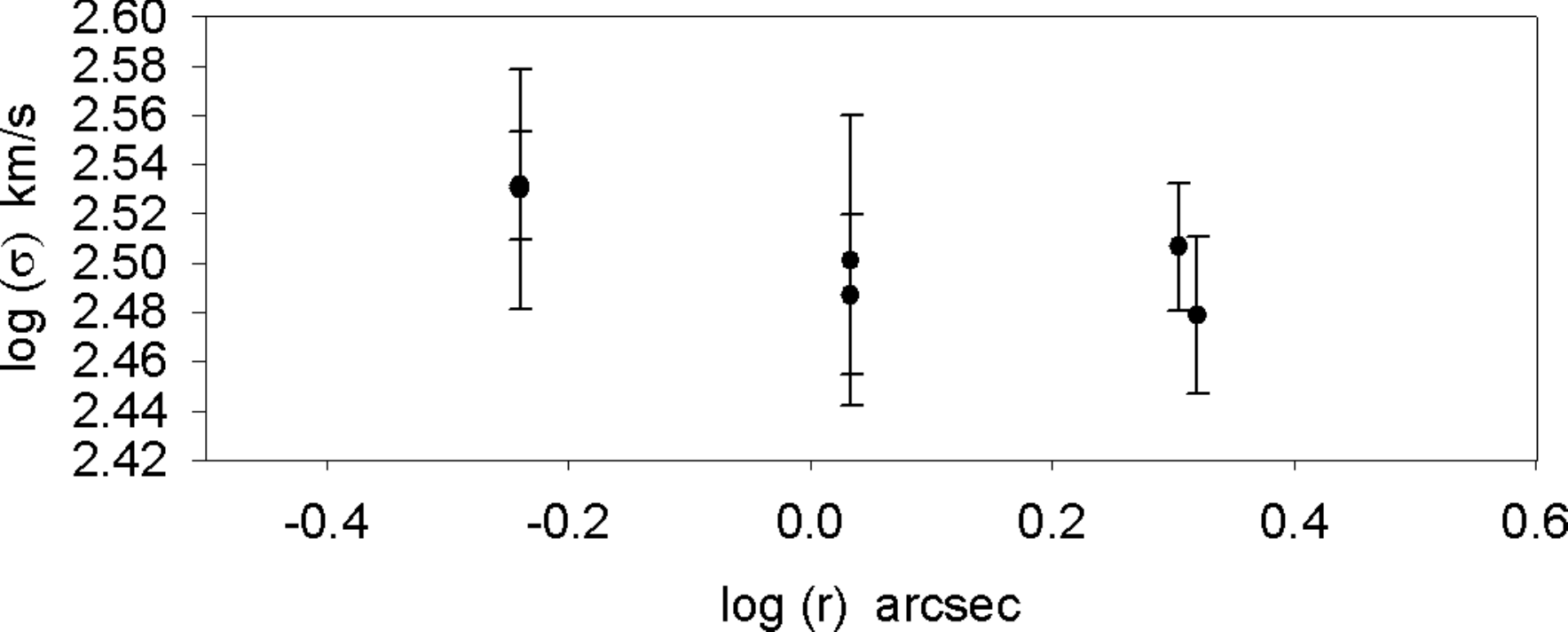}}}
   \mbox{\subfigure{\includegraphics[height=7.6cm,width=5.6 cm]{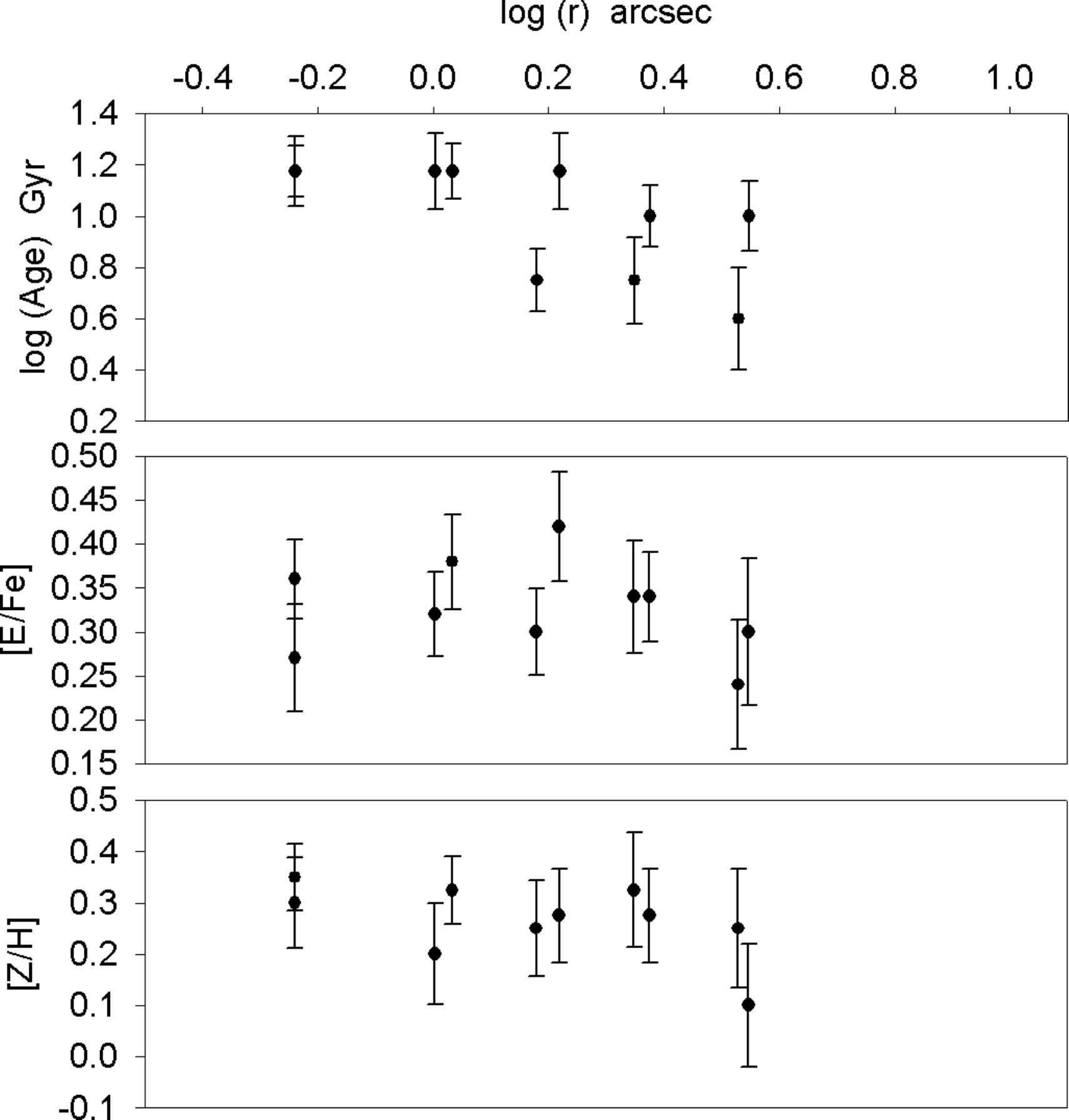}}\quad
         \subfigure{\includegraphics[height=7.6cm,width=5.6cm]{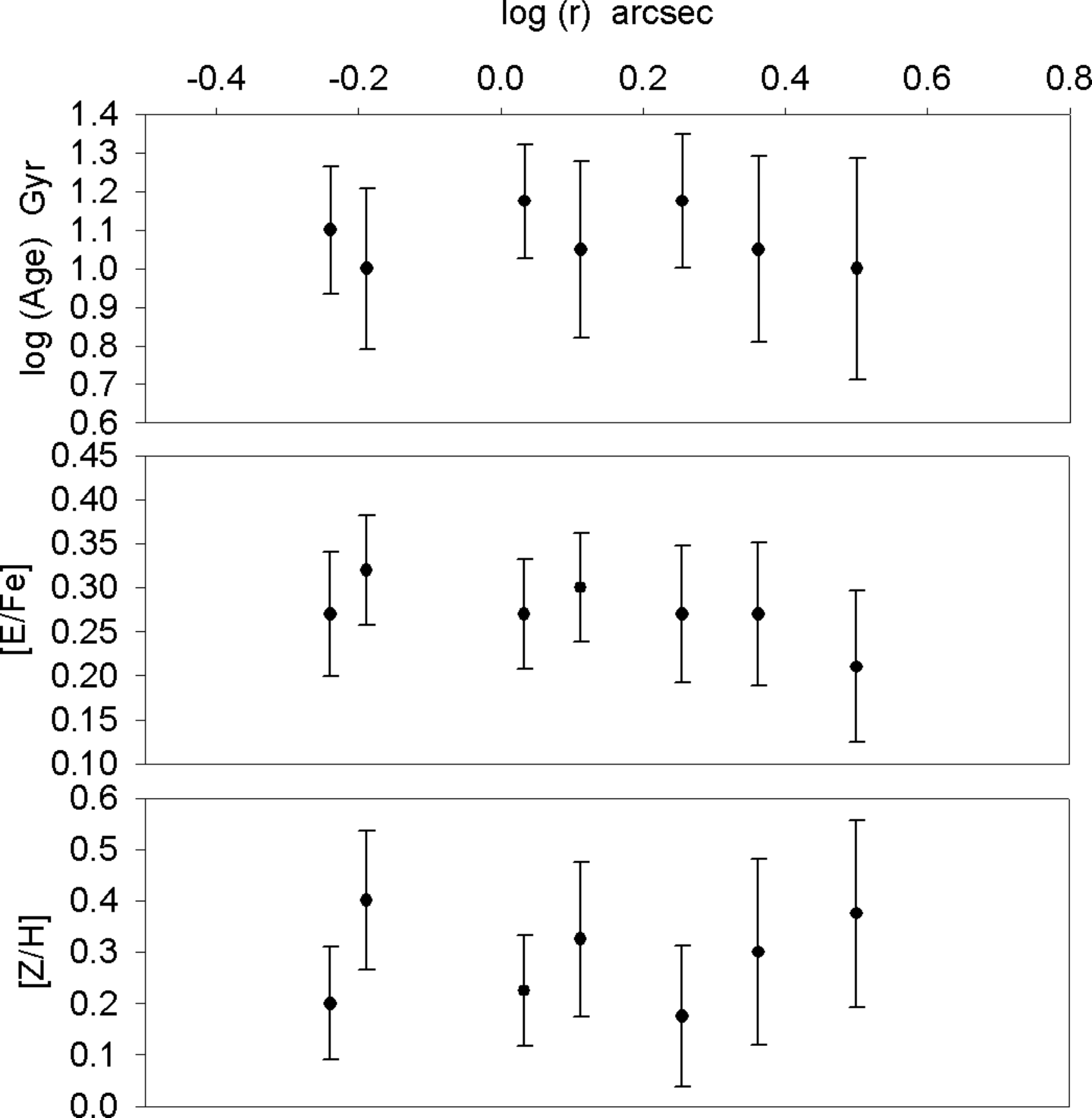}}\quad
         \subfigure{\includegraphics[height=7.6cm,width=5.6cm]{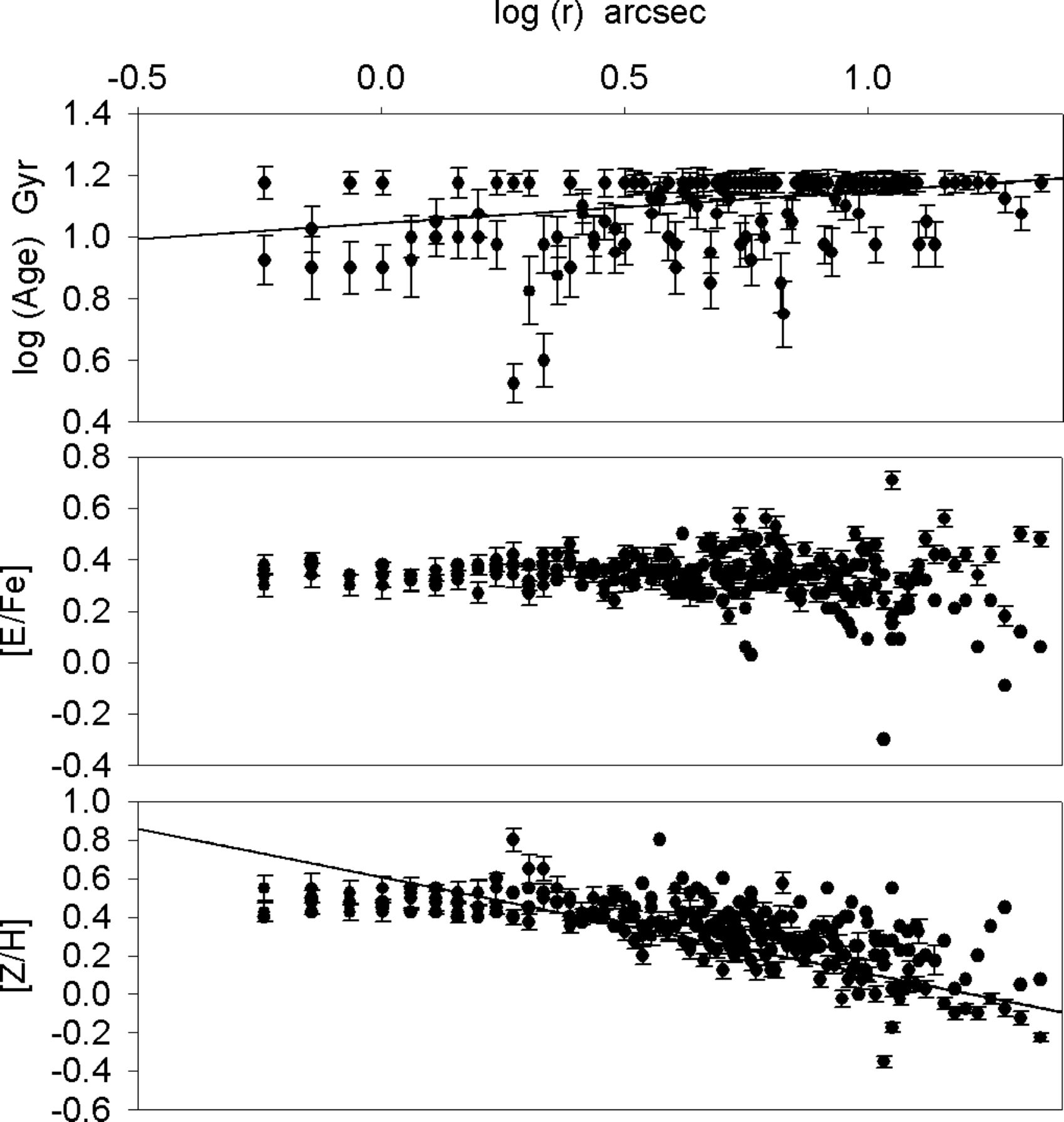}}}
   \mbox{\subfigure[NGC0533]{\includegraphics[height=2.7cm,width=5.6cm]{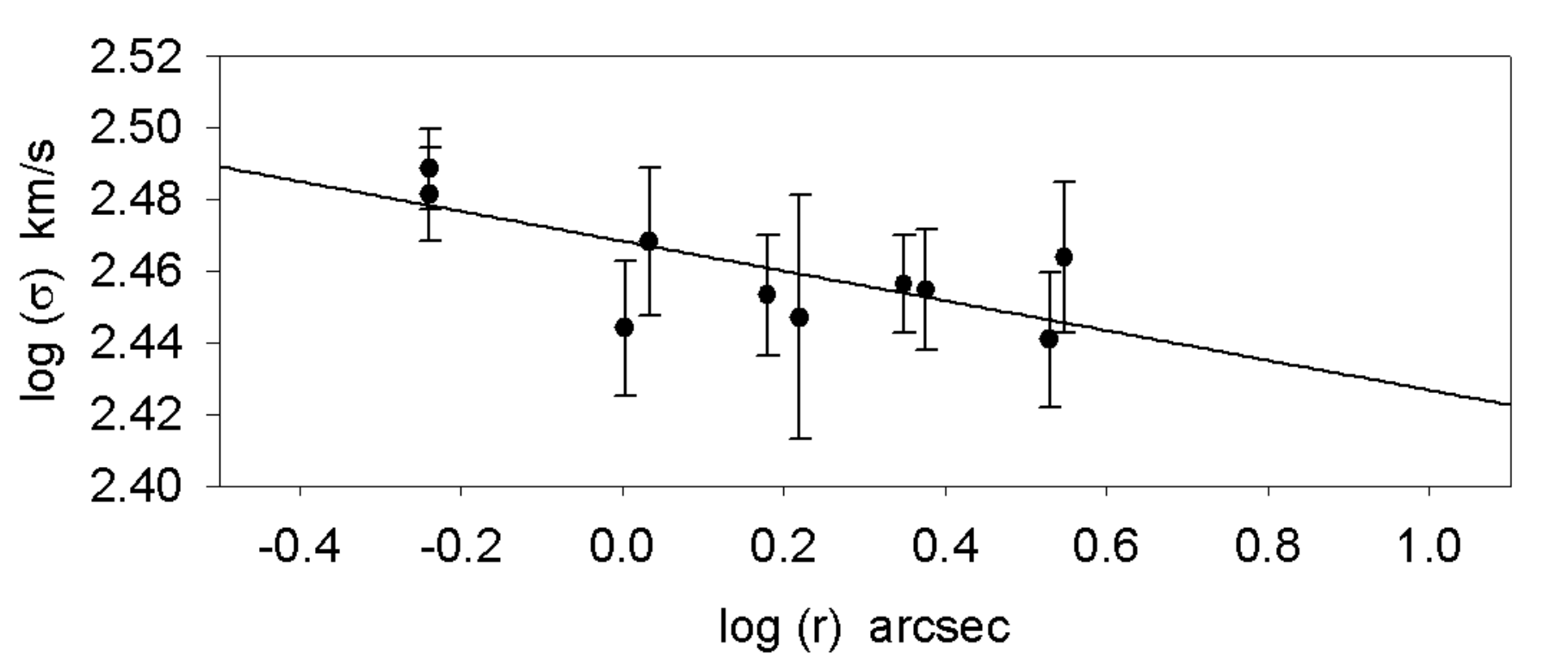}}\quad
         \subfigure[NGC0541]{\includegraphics[height=2.7cm,width=5.6cm]{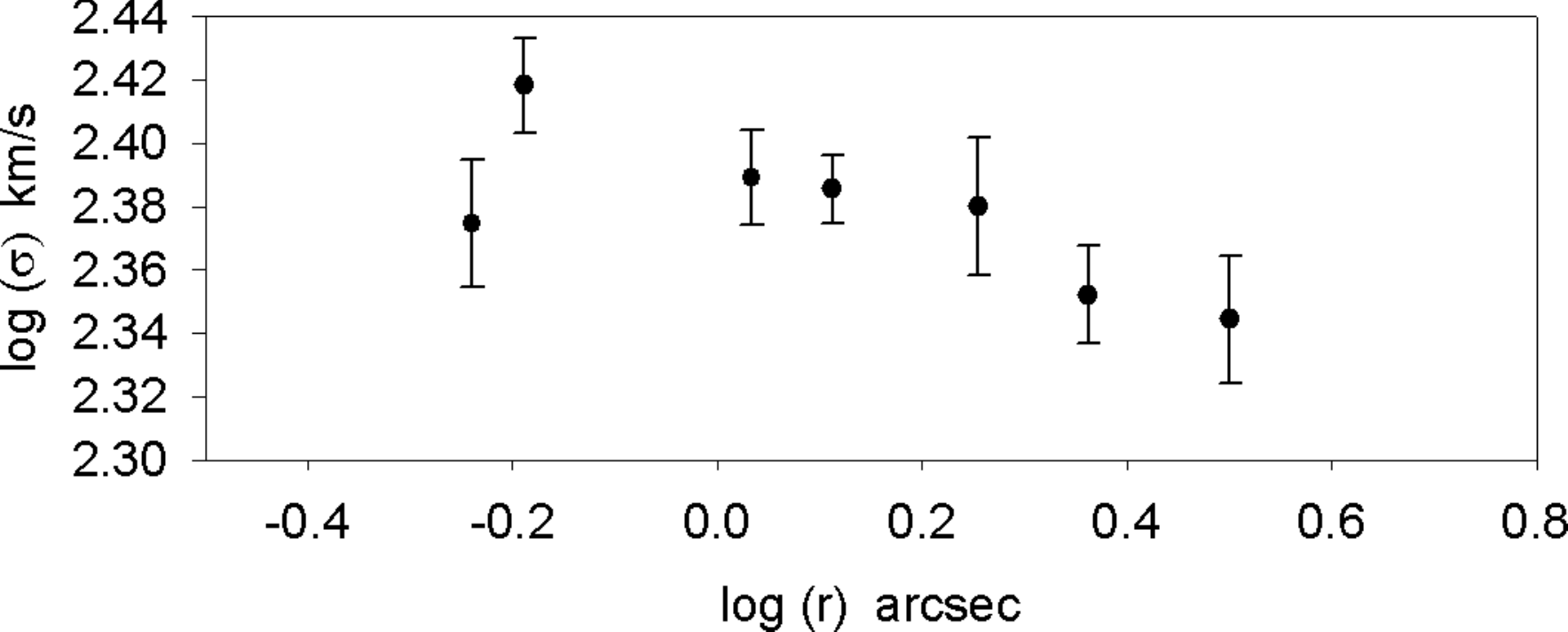}}\quad
         \subfigure[NGC1399]{\includegraphics[height=2.7cm,width=5.6cm]{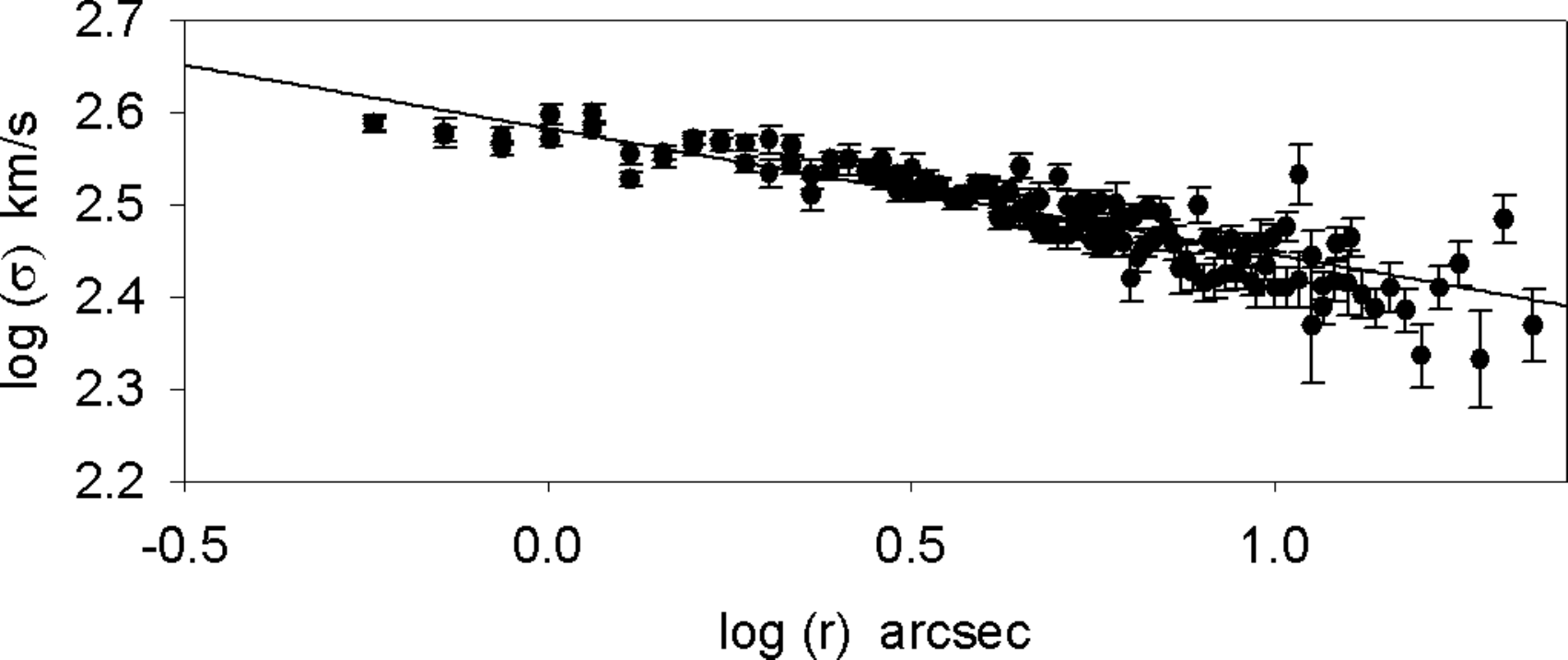}}}
\caption[]{Age, $\alpha$-enhancement, metallicity and velocity dispersion profiles of the BCGs continue.}
\label{fig:Profiles2}
\end{figure*}

\begin{figure*}
\centering
\mbox{\subfigure{\includegraphics[height=7.6cm,width=5.6cm]{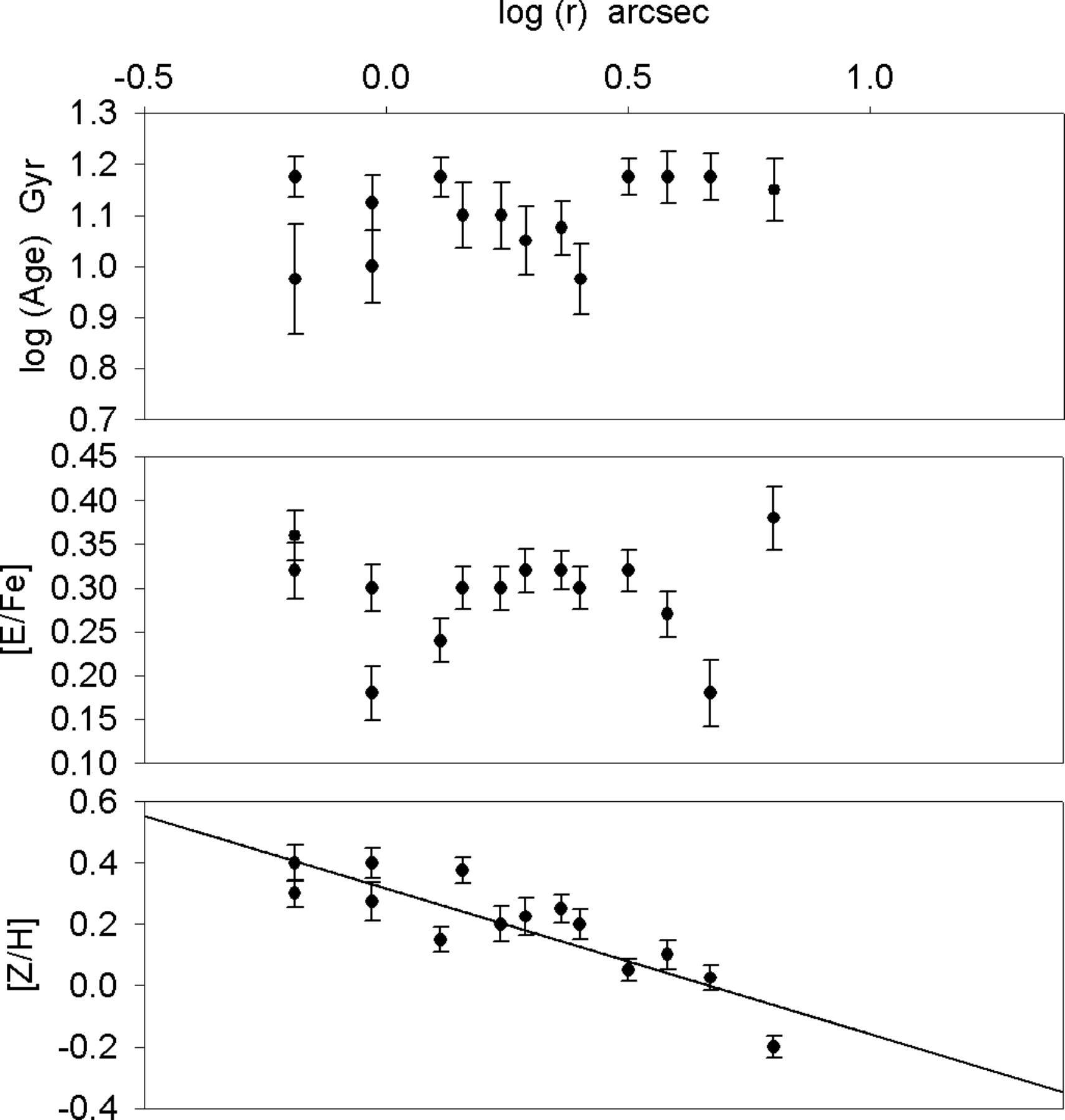}}\quad
         \subfigure{\includegraphics[height=7.6cm,width=5.6cm]{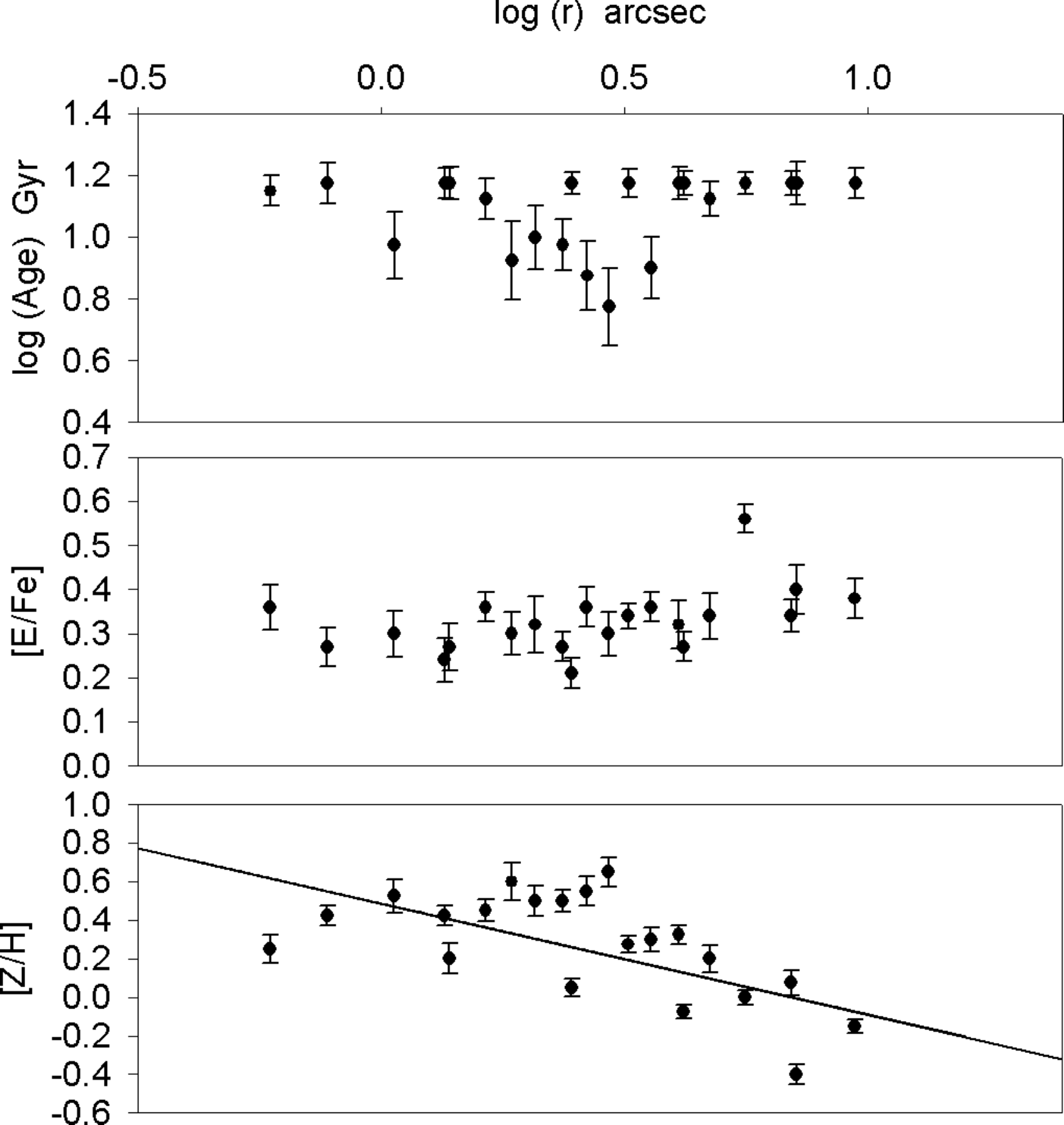}}\quad
         \subfigure{\includegraphics[height=7.6cm,width=5.6cm]{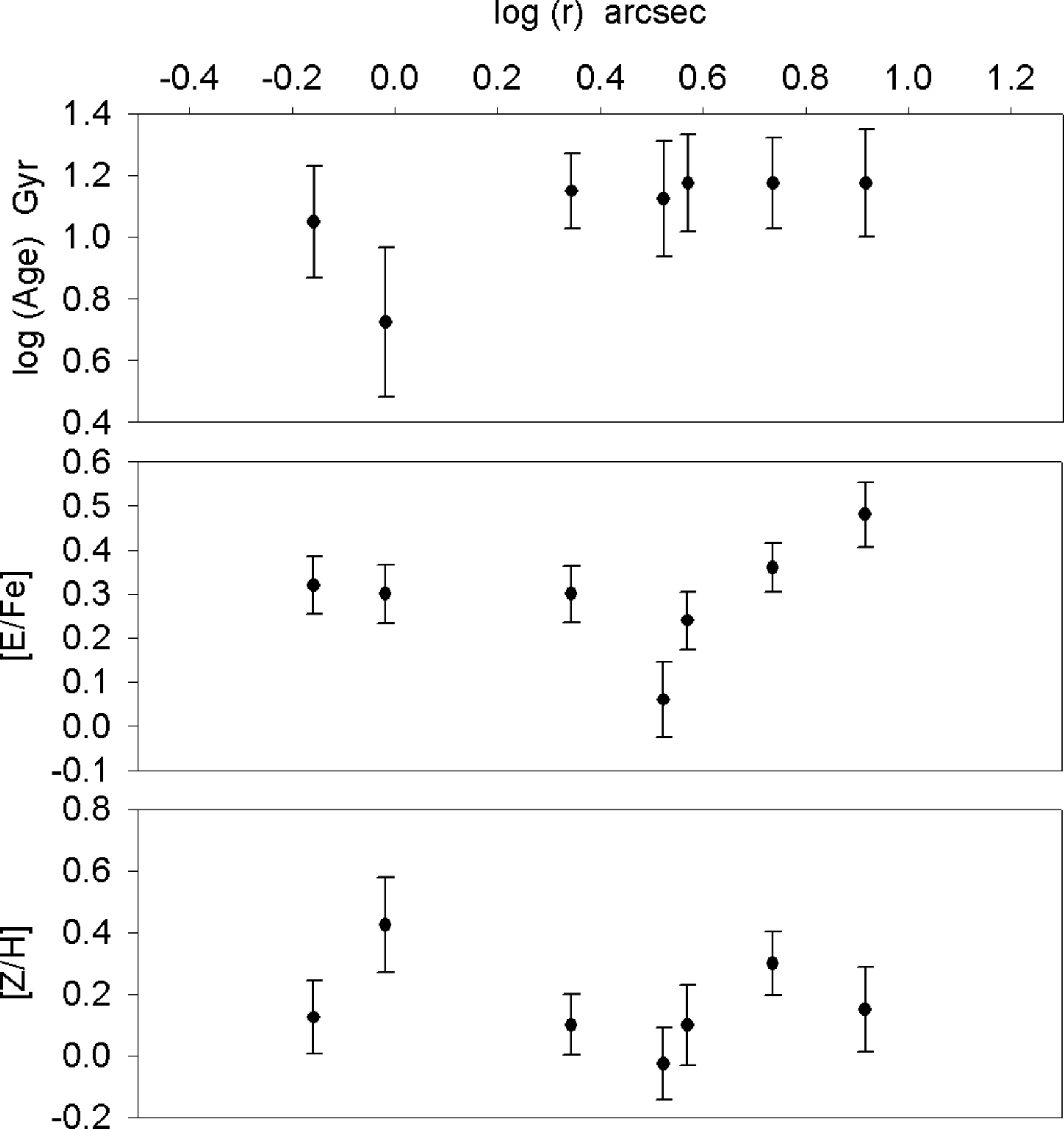}}}
   \mbox{\subfigure[NGC1713]{\includegraphics[height=2.7cm,width=5.6cm]{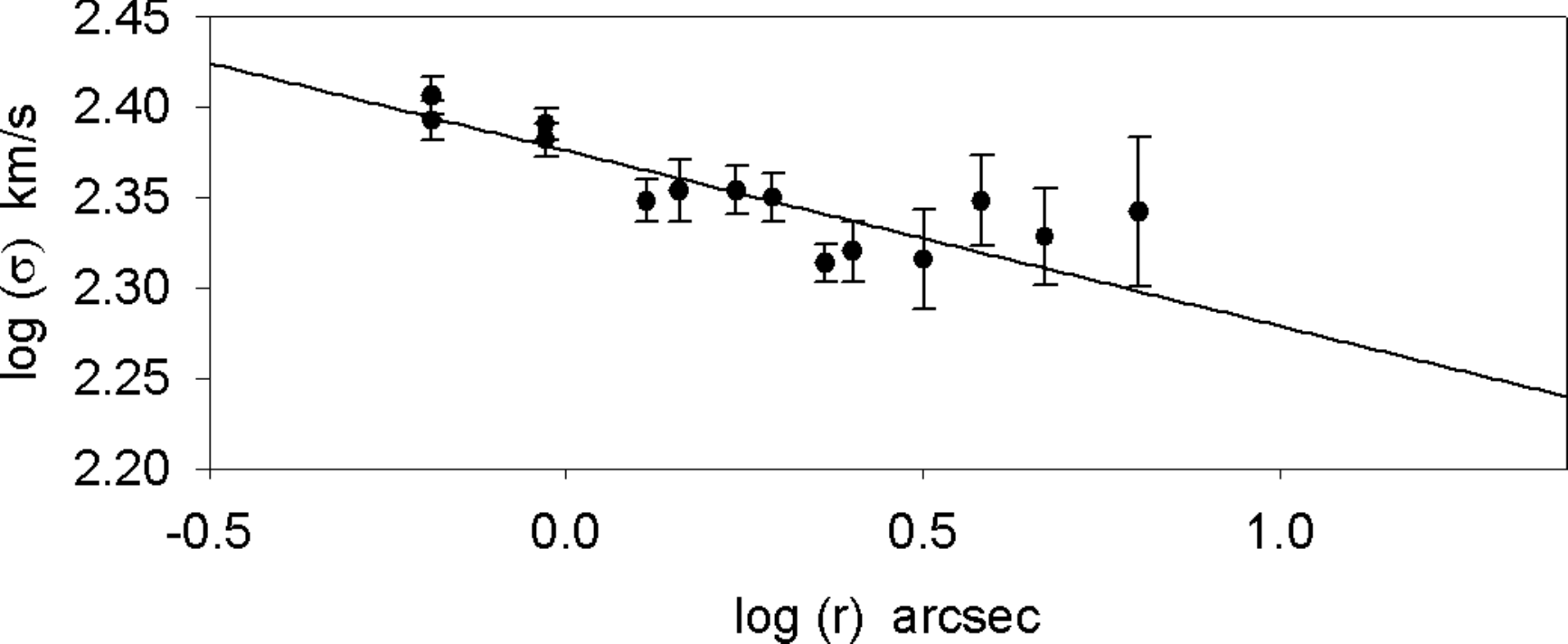}}\quad
         \subfigure[NGC2832]{\includegraphics[height=2.7cm,width=5.6cm]{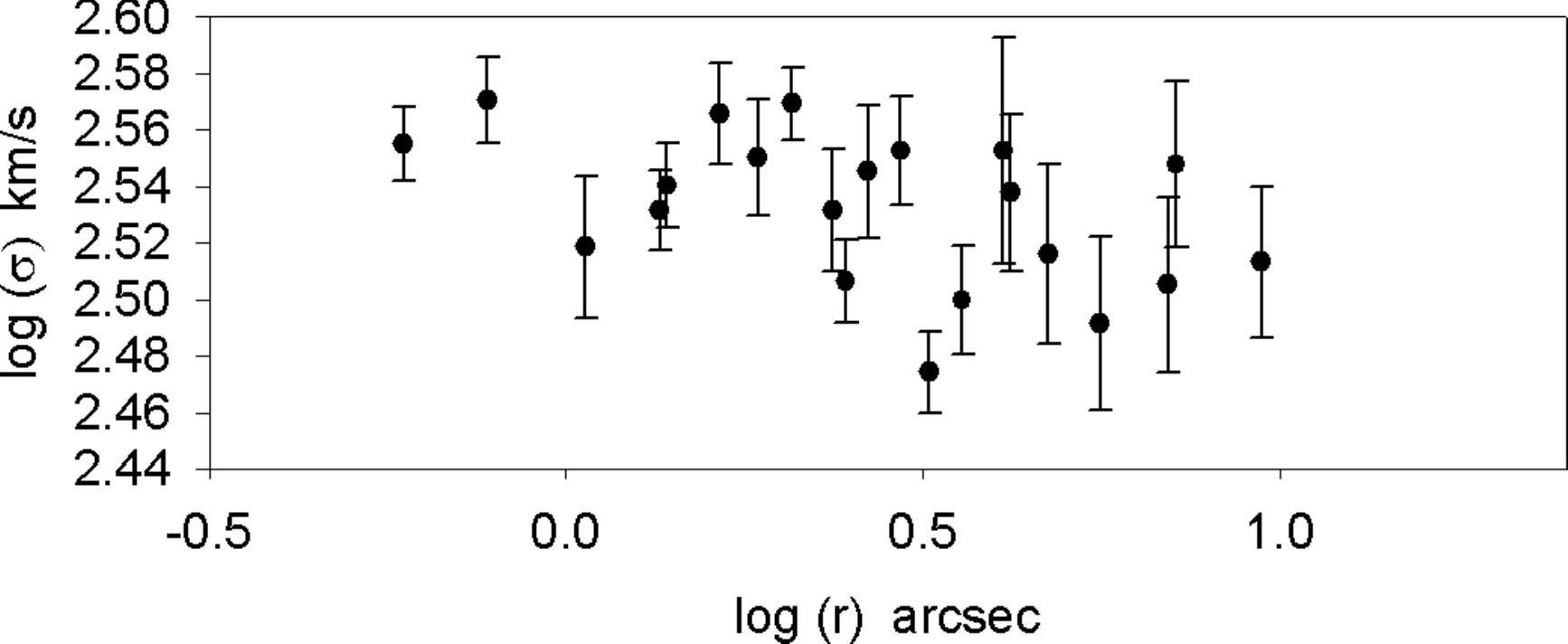}}\quad
         \subfigure[NGC3311]{\includegraphics[height=2.7cm,width=5.6cm]{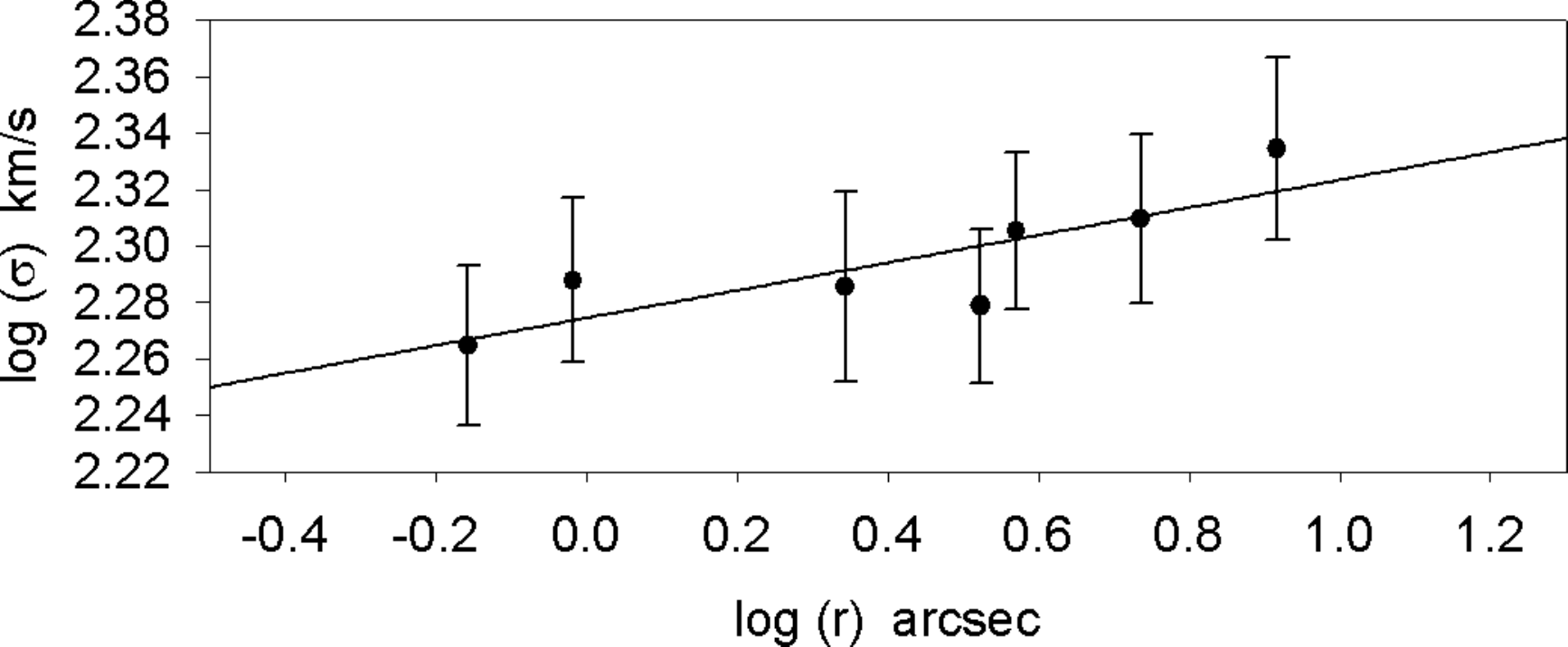}}}
   \mbox{\subfigure{\includegraphics[height=7.6cm,width=5.6cm]{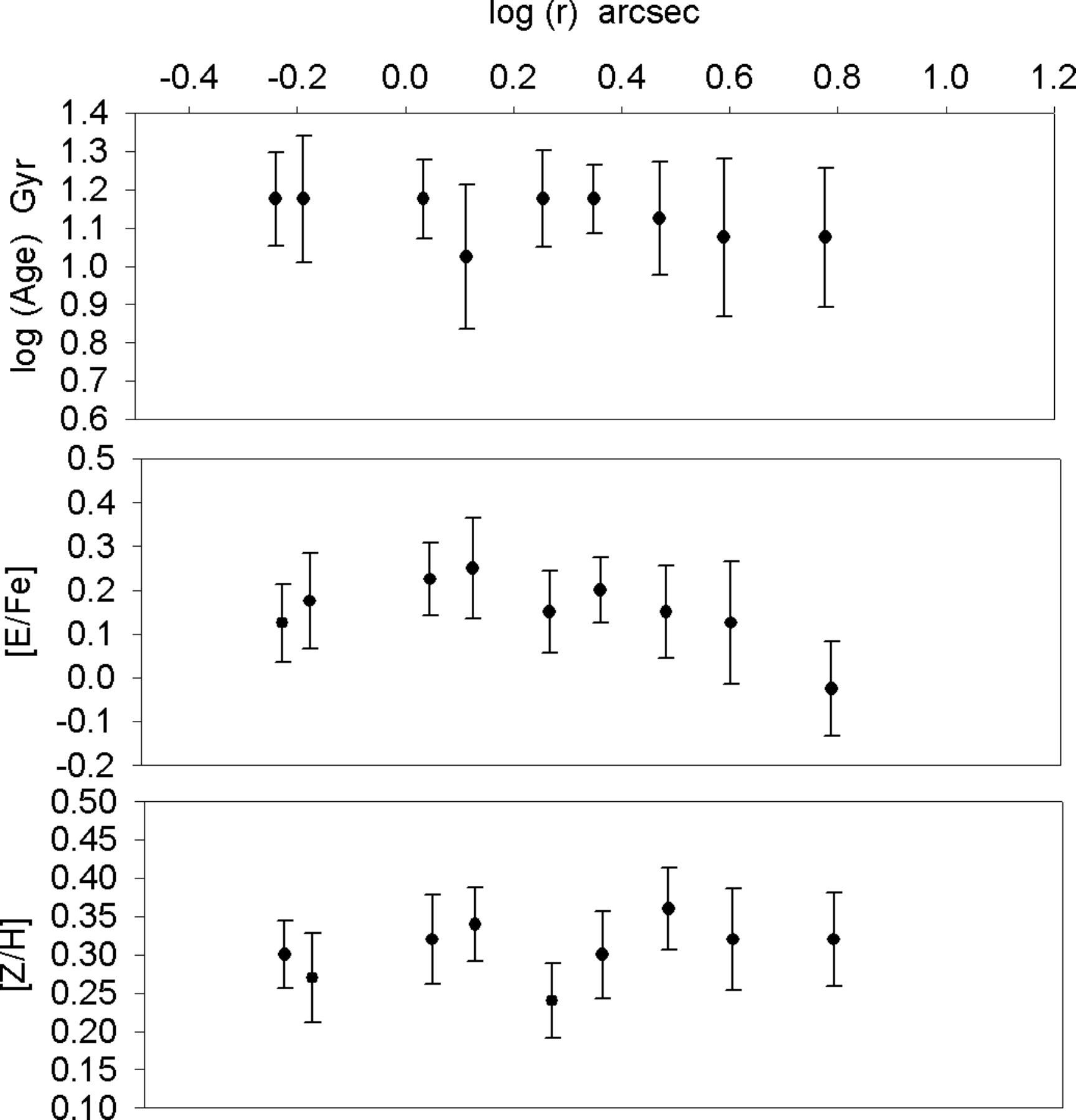}}\quad
         \subfigure{\includegraphics[height=7.6cm,width=5.6cm]{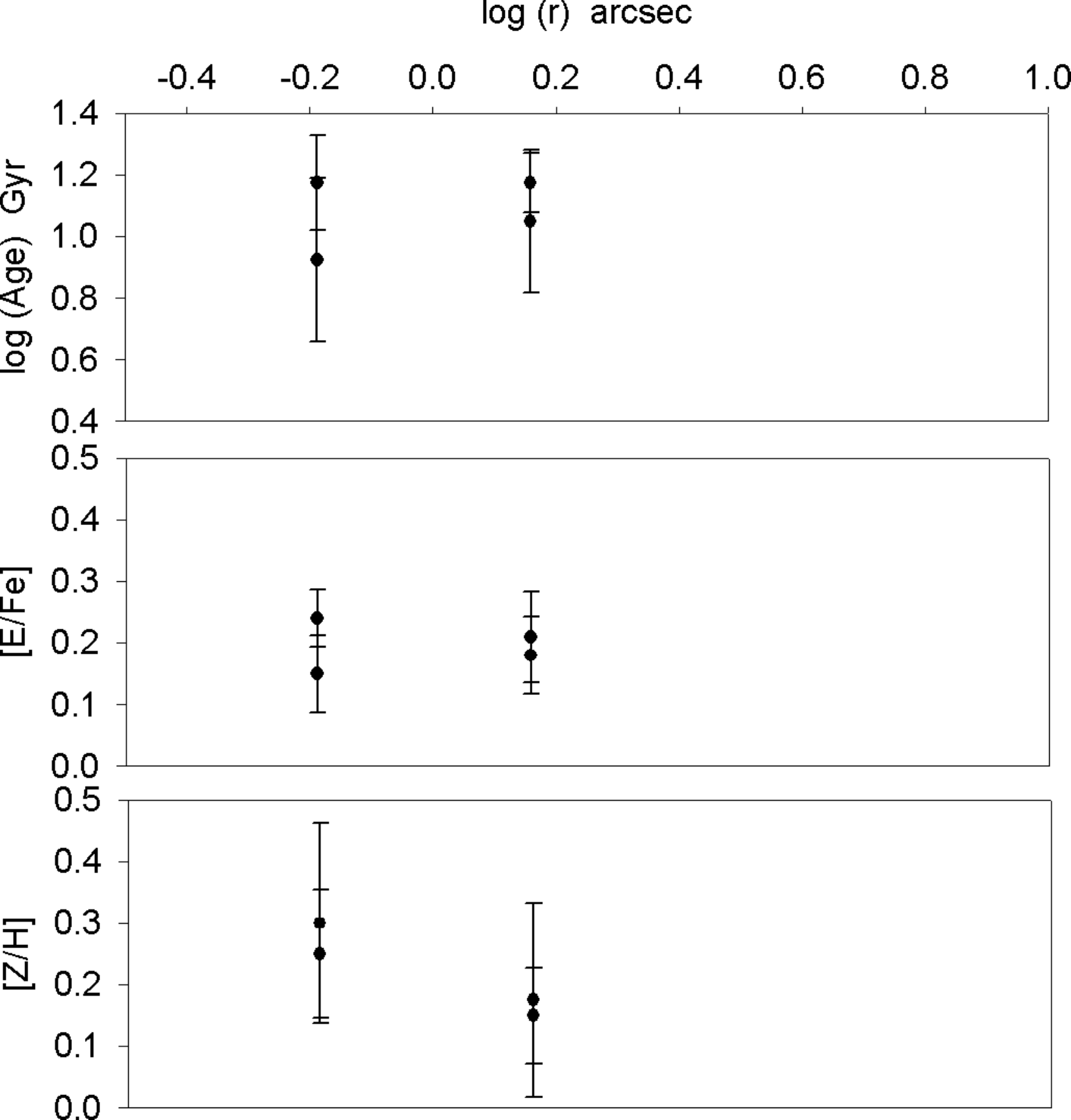}}\quad
         \subfigure{\includegraphics[height=7.6cm,width=5.6cm]{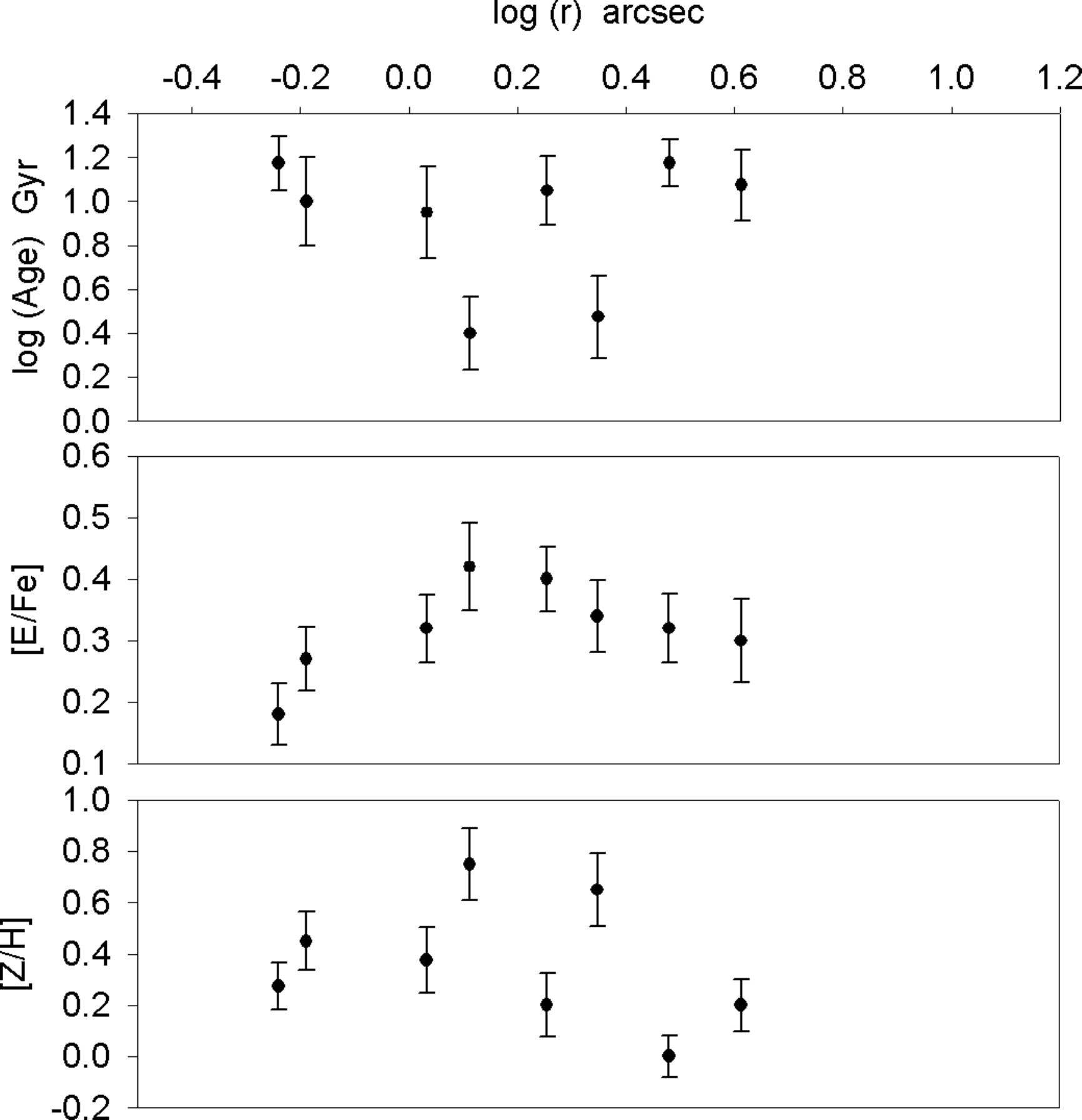}}}
   \mbox{\subfigure[NGC4839]{\includegraphics[height=2.7cm,width=5.6cm]{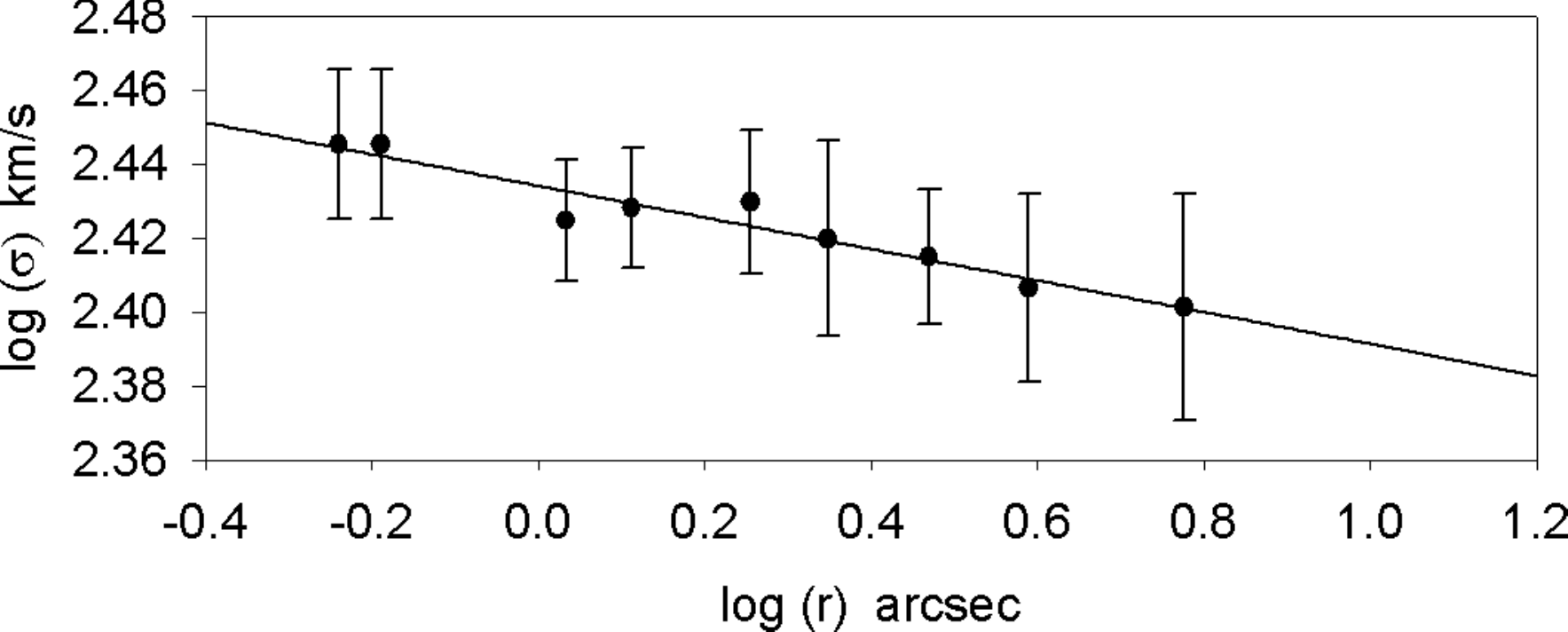}}\quad
         \subfigure[NGC6173]{\includegraphics[height=2.7cm,width=5.6cm]{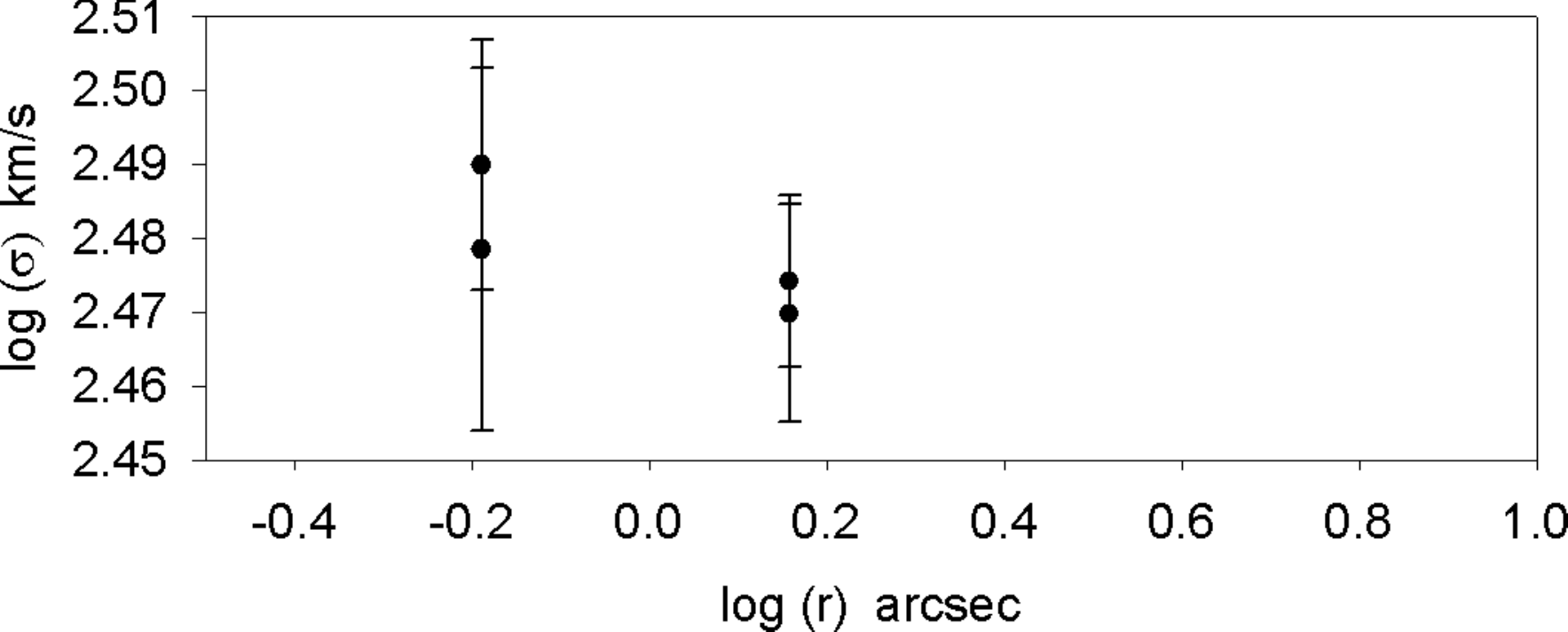}}\quad
         \subfigure[NGC6269]{\includegraphics[height=2.7cm,width=5.6cm]{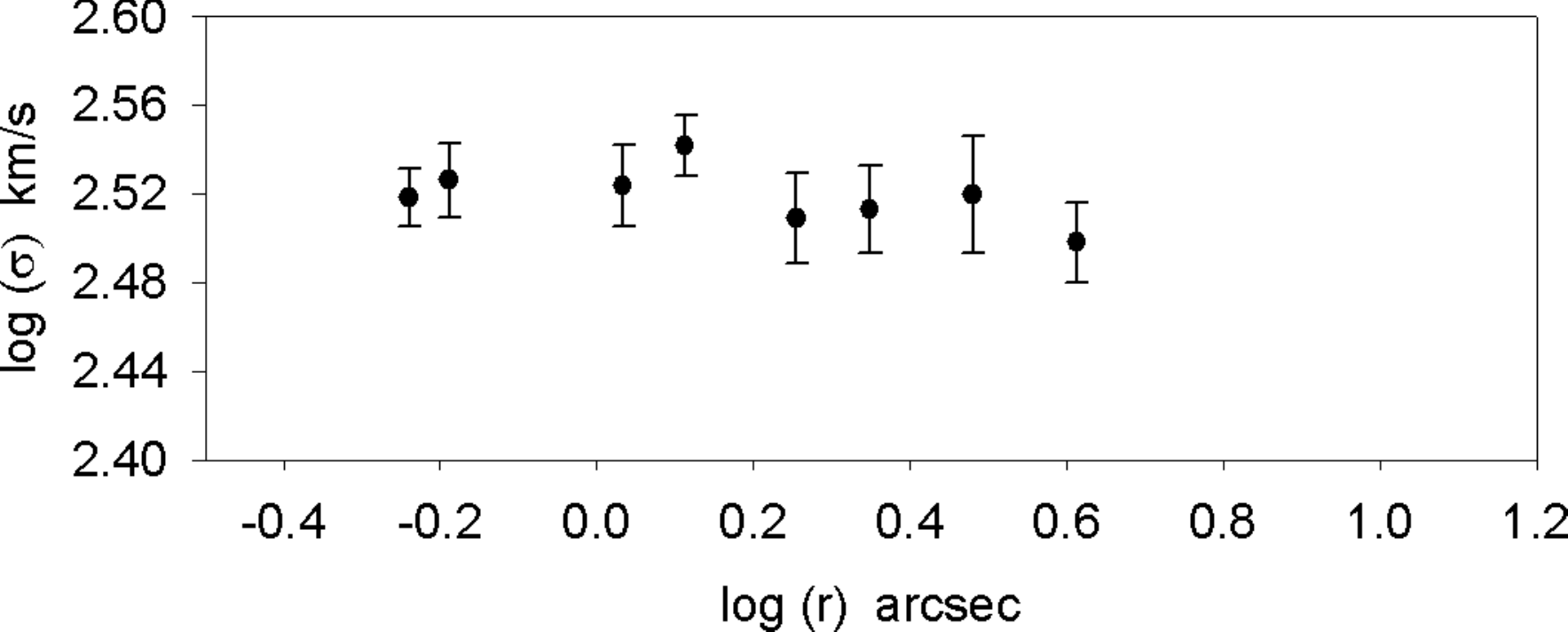}}}
\caption[]{Age, $\alpha$-enhancement, metallicity and velocity dispersion profiles of the BCGs continue.}
\label{fig:Profiles4}
\end{figure*}

\begin{figure*}
\centering
 \mbox{\subfigure{\includegraphics[height=7.6cm,width=5.6cm]{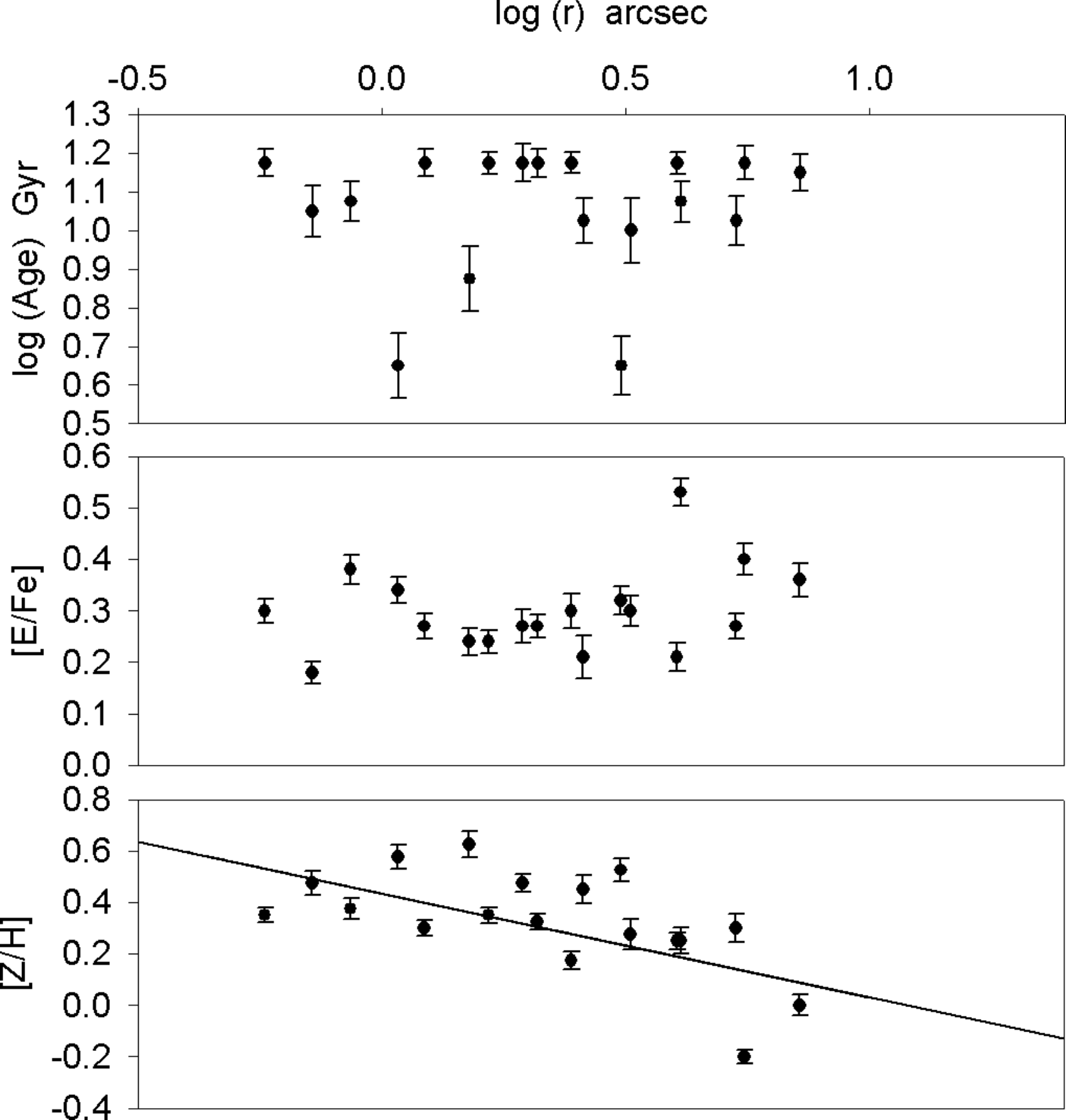}}\quad
         \subfigure{\includegraphics[height=7.6cm,width=5.6cm]{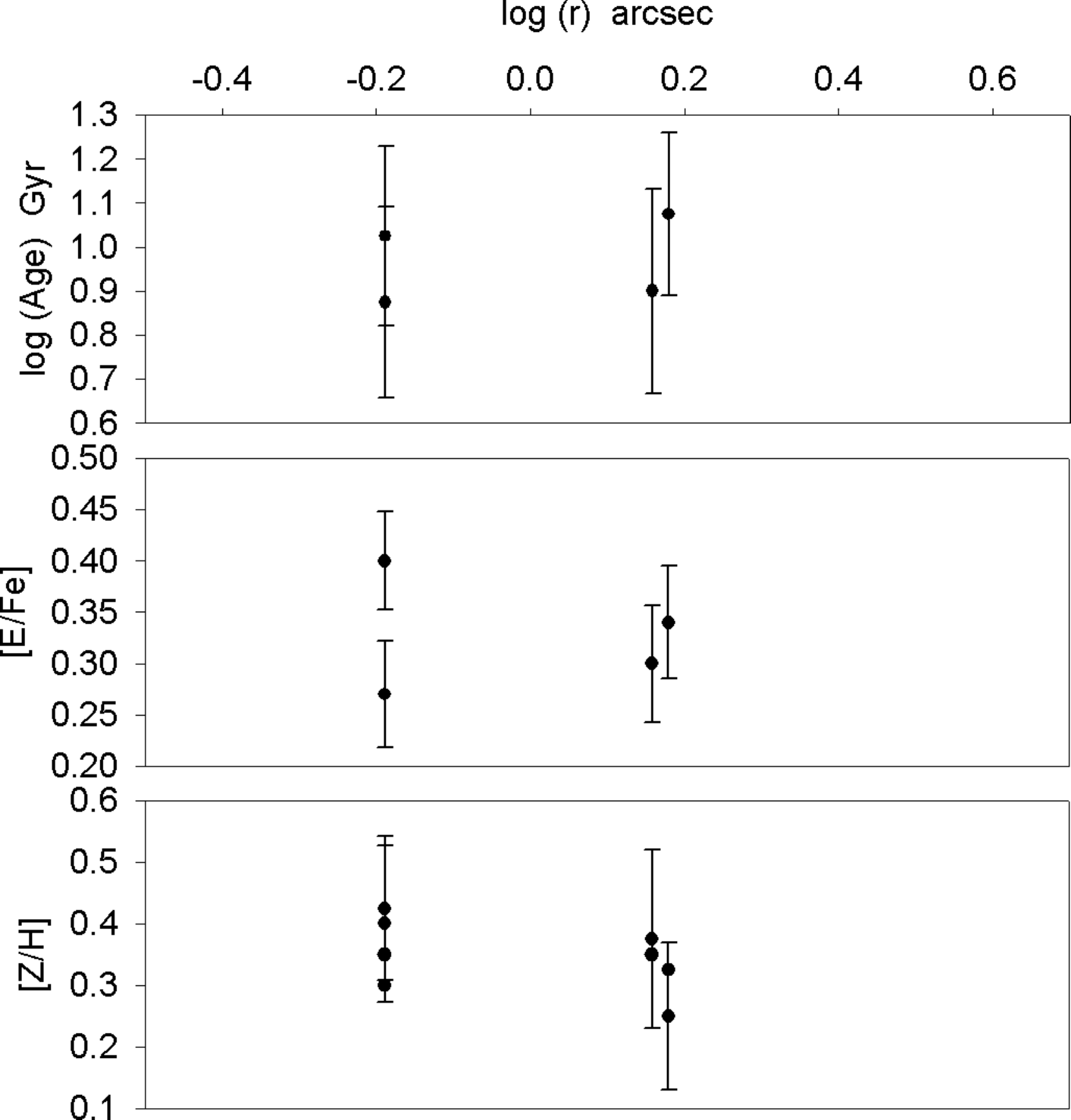}}\quad
         \subfigure{\includegraphics[height=7.6cm,width=5.6cm]{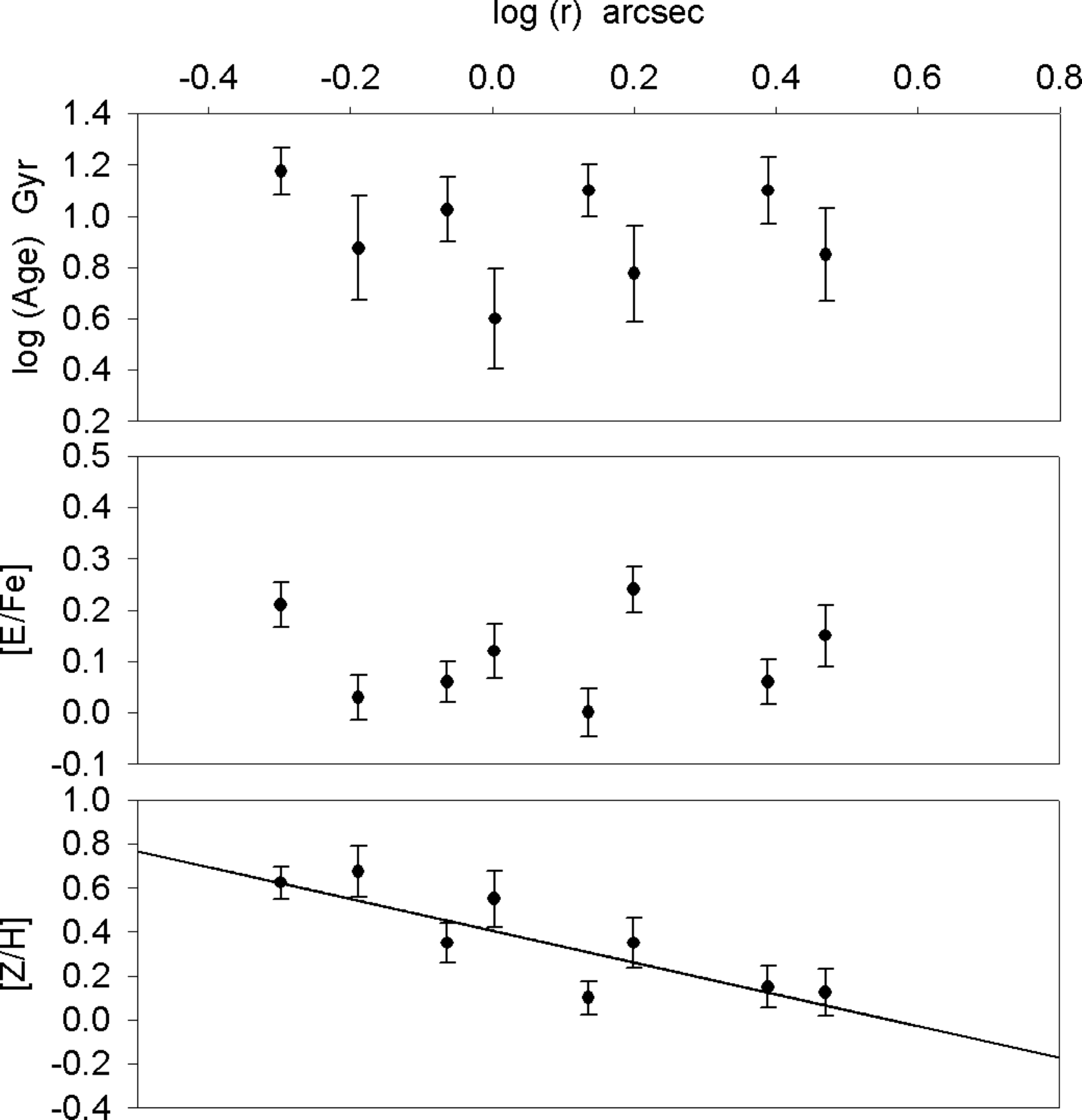}}}
   \mbox{\subfigure[NGC7012]{\includegraphics[height=2.7cm,width=5.6cm]{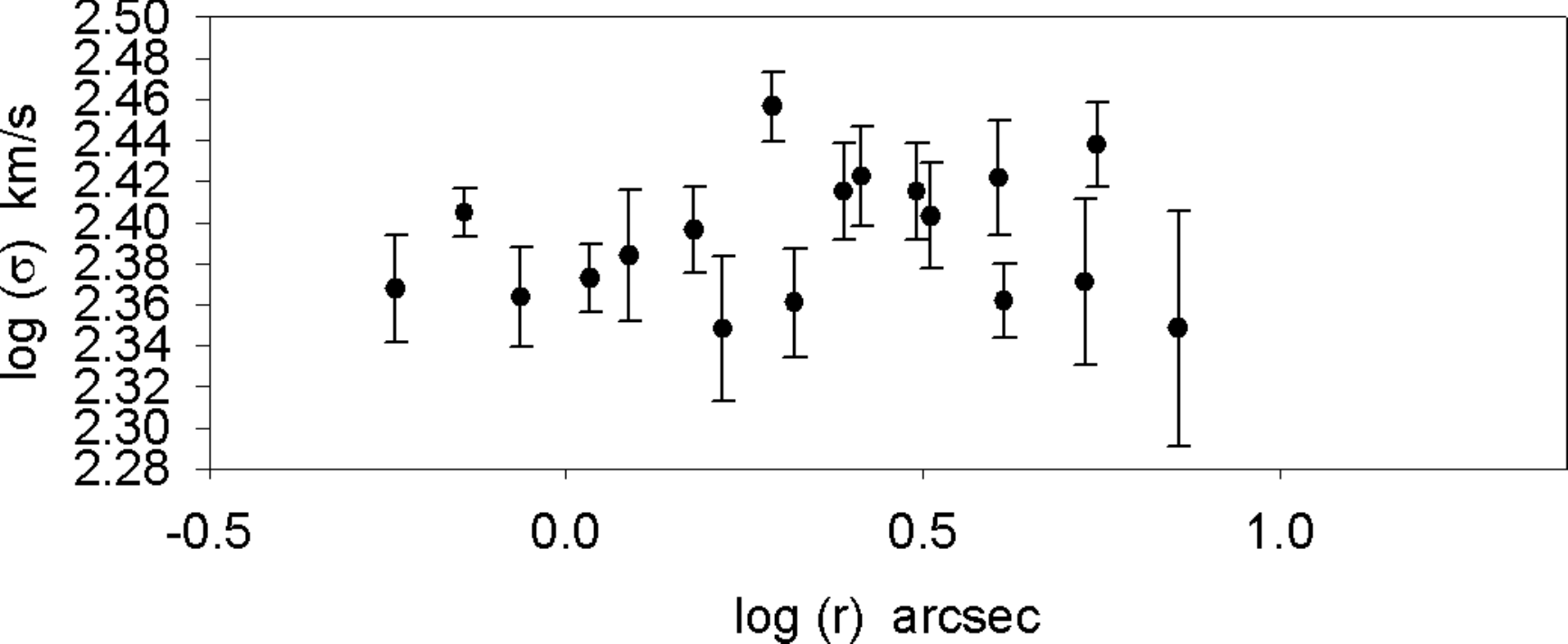}}\quad
         \subfigure[PGC004072]{\includegraphics[height=2.7cm,width=5.6cm]{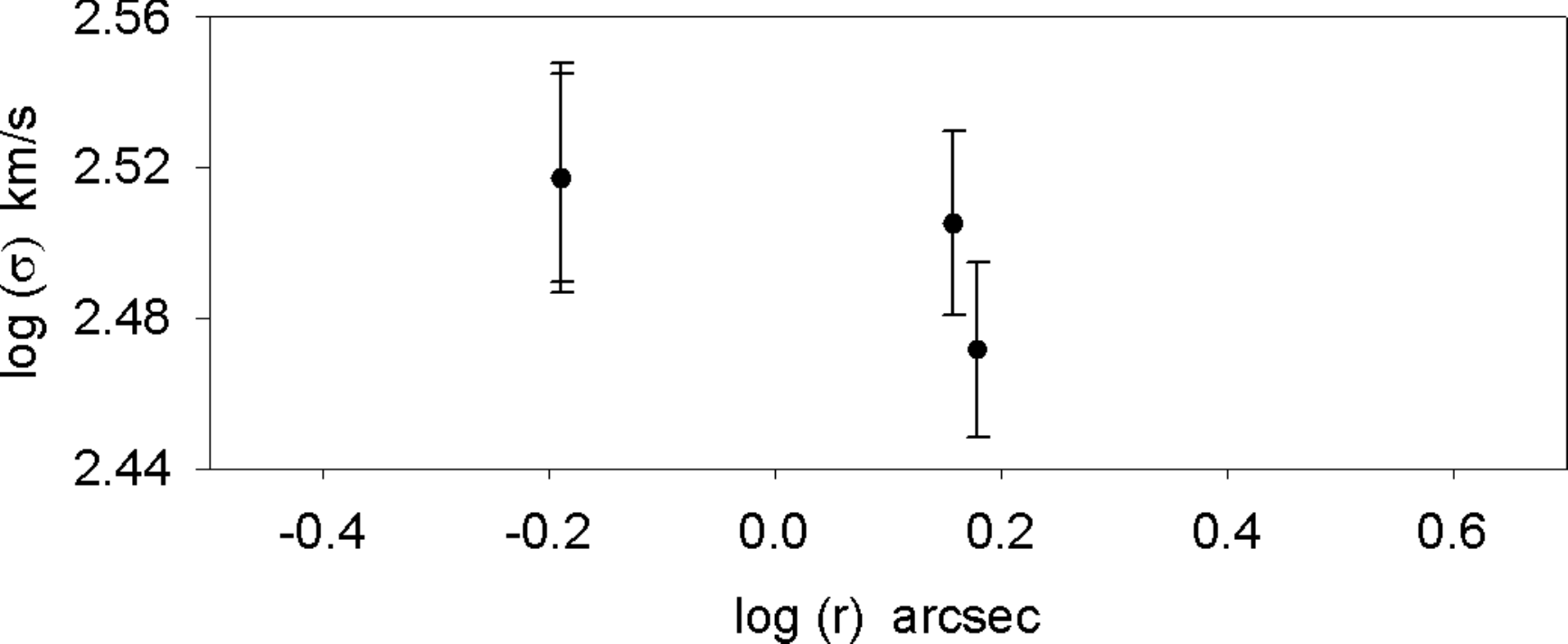}}\quad
         \subfigure[PGC030223]{\includegraphics[height=2.7cm,width=5.6cm]{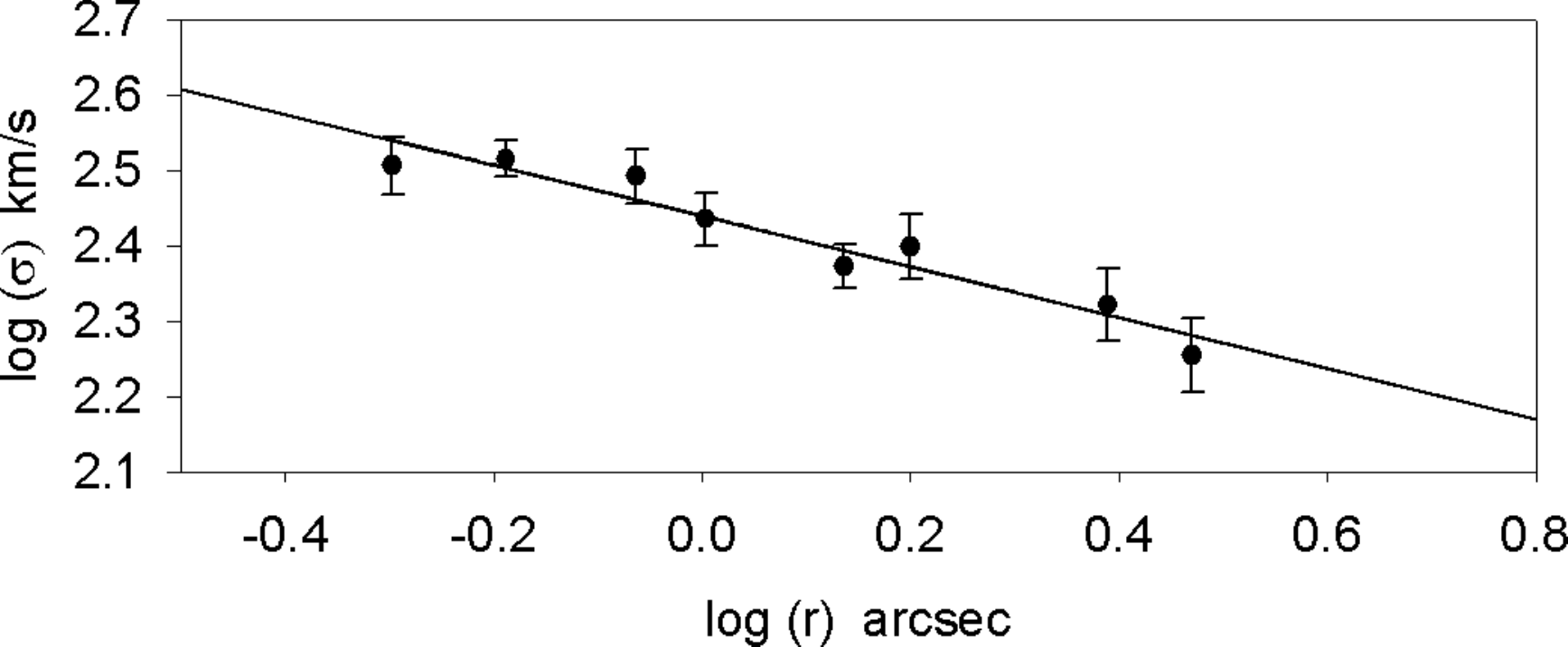}}}
   \mbox{\subfigure{\includegraphics[height=7.6cm,width=5.6 cm]{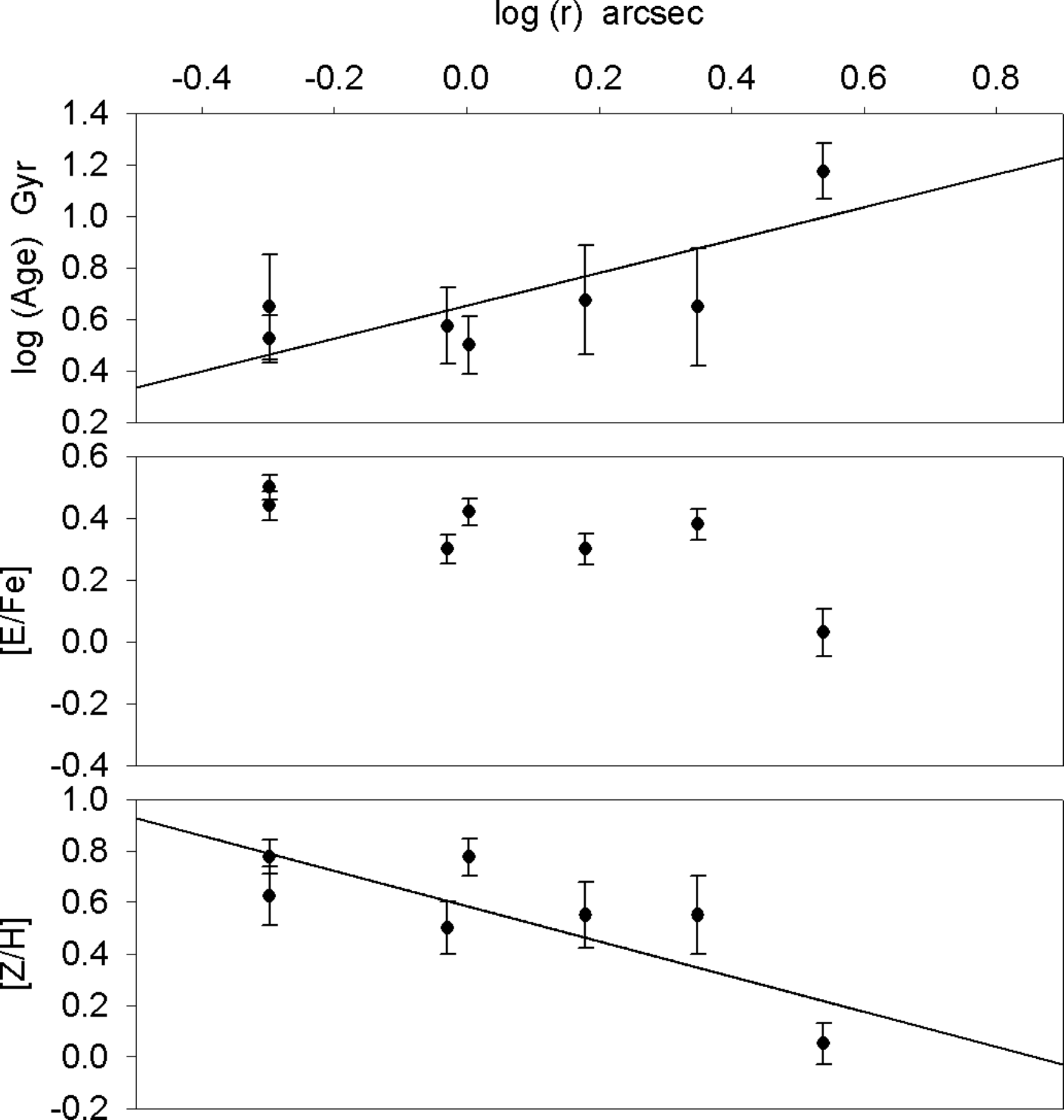}}\quad
         \subfigure{\includegraphics[height=7.6cm,width=5.6cm]{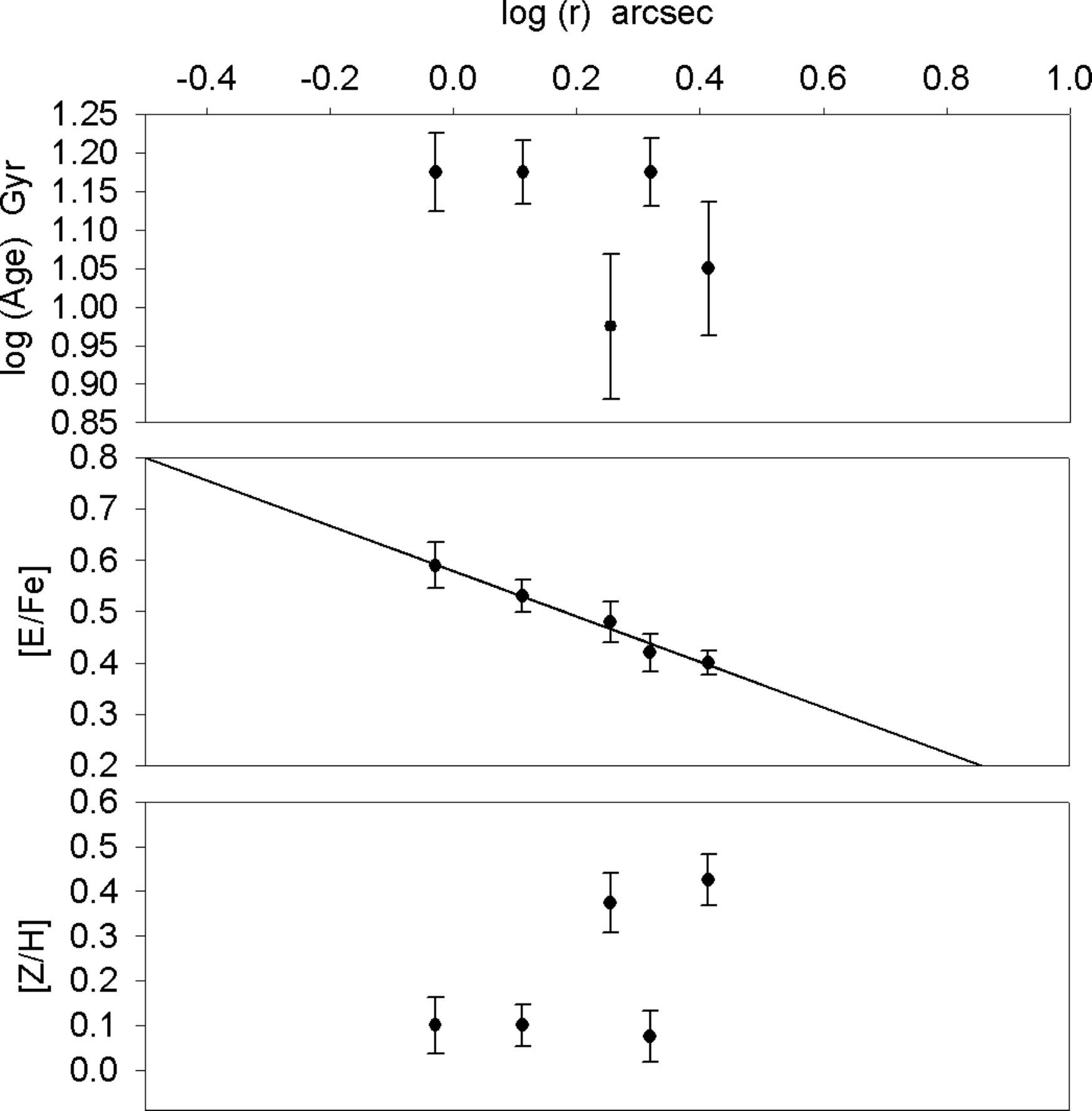}}\quad
         \subfigure{\includegraphics[height=7.6cm,width=5.6cm]{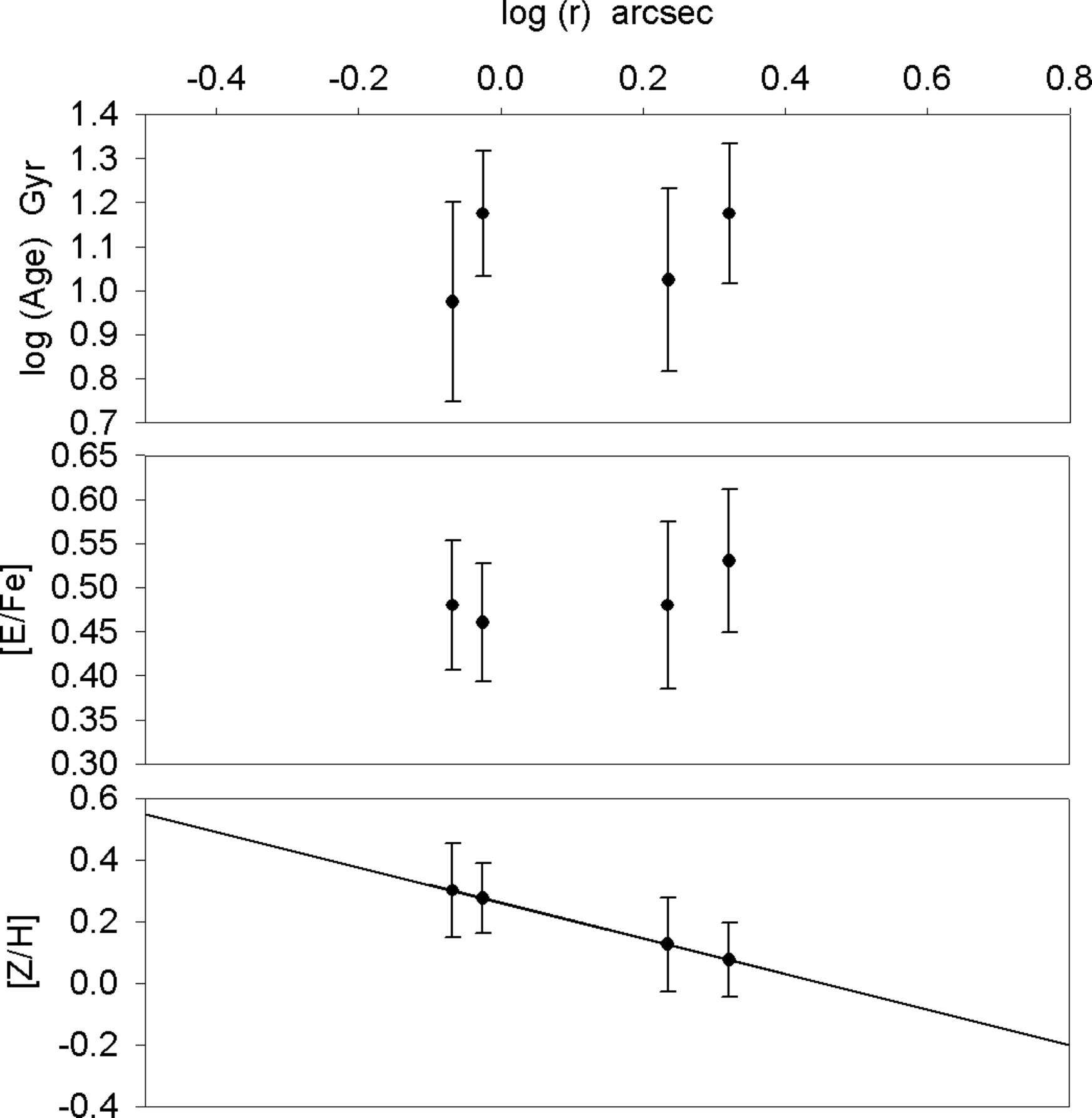}}}
   \mbox{\subfigure[PGC072804]{\includegraphics[height=2.7cm,width=5.6cm]{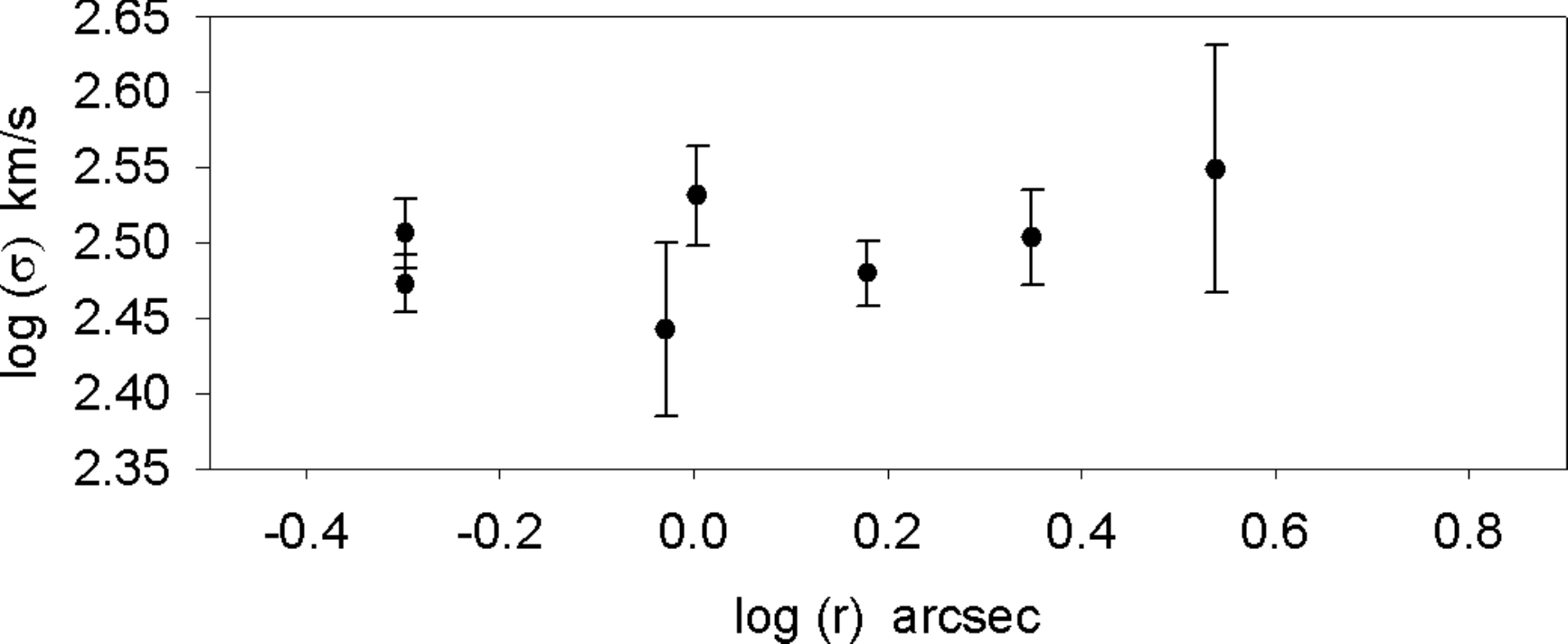}}\quad
         \subfigure[UGC02232]{\includegraphics[height=2.7cm,width=5.6cm]{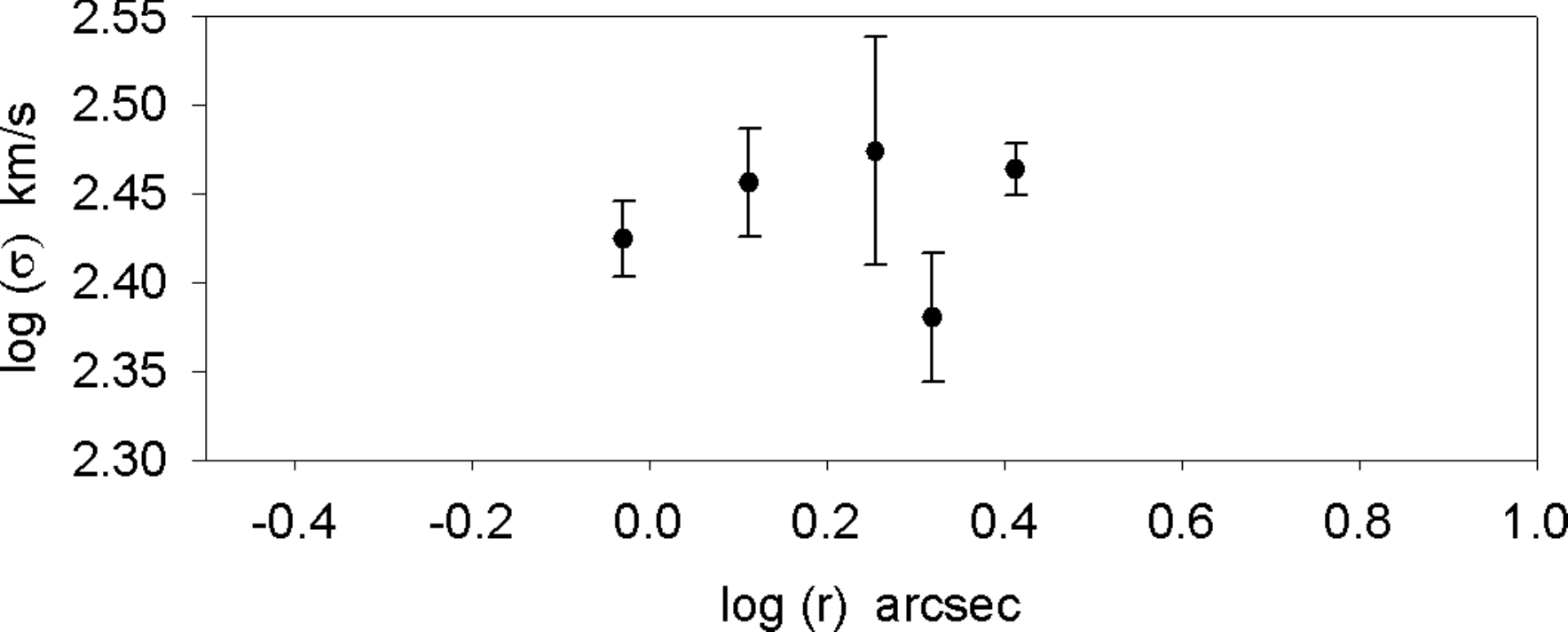}}\quad
         \subfigure[UGC05515]{\includegraphics[height=2.7cm,width=5.6cm]{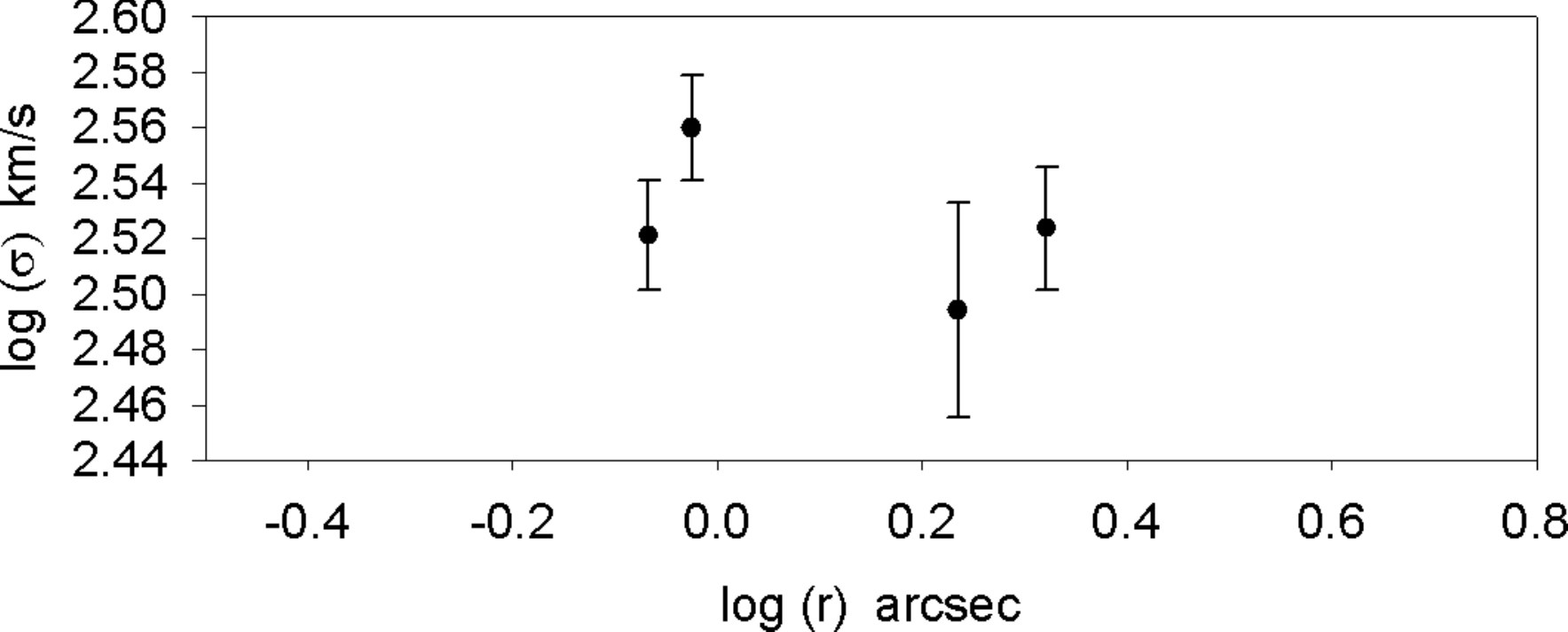}}}
\caption[]{Age, $\alpha$-enhancement, metallicity and velocity dispersion profiles of the BCGs continue.}
\label{fig:Profiles6}
\end{figure*}

The H$\beta$ index gradients are plotted in Figures \ref{Fig:Hbeta1} and \ref{Fig:Hbeta2}, and the Mg$_{b}$, Fe5270 and Fe5335 index gradients are plotted in Figures \ref{Fig:Indices} to \ref{Fig:Indices4}. Figures \ref{fig:Profiles} to \ref{fig:Profiles6} show the age, $\alpha$-enhancement, metallicity and velocity dispersion profiles of the BCGs.

\renewcommand*{\thesubfigure}{}

\renewcommand*{\thesubfigure}{}

\bsp

\label{lastpage}

\end{document}